\documentstyle[fleqn,psfig,epsf]{elsart}

\begin{document}

\newcommand \ga{\raisebox{-.5ex}{$\stackrel{>}{\sim}$}}
\newcommand \la{\raisebox{-.5ex}{$\stackrel{<}{\sim}$}}

\begin{frontmatter}

\title{PHASES OF DENSE MATTER IN NEUTRON STARS}

\author[cph]{Henning Heiselberg}
\and
\author[oslo]{Morten Hjorth-Jensen}

\address[cph]{NORDITA, Blegdamsvej 17, DK-2100 K\o benhavn \O, Denmark}

\address[oslo]{Department of Physics, University of Oslo, N-0316 Oslo, Norway}

\maketitle

\begin{abstract}

Recent equations of state for dense nuclear matter are discussed
with possible phase transitions arising in neutron stars 
such as pion, kaon and hyperon kondensation, superfluidity
and quark matter. Specifically, we treat the nuclear to quark
matter phase transition, the possible mixed phase and its structure.
A number of numerical calculations of
rotating neutron stars with and without phase transitions are given
and compared to observed masses, radii, temperatures and glitches.

\end{abstract}

%\pacs{PACS numbers: 12.38.Mh, 21.30.-x, 21.65.+f, 26.60.+c, 97.60.Gb, 97.60.Jd}

\end{frontmatter}

\newpage

\tableofcontents

\section{INTRODUCTION}\label{sec:intro}

\subsection{The past, present and future of neutron stars}

The discovery of the neutron by Chadwick in 1932 prompted Landau to
predict the existence of neutron stars. The birth of such stars
in supernovae explosions was suggested by Baade and Zwicky 1934. First
theoretical neutron star calculations were performed by Tolman,
Oppenheimer and Volkoff in 1939 and Wheeler in 1960. Bell and Hewish
were the first to discover a neutron star in 1967 as a {\it radio
pulsar}.  The discovery of the rapidly rotating Crab pulsar in the
remnant of the Crab supernova observed by the chinese in 1054 A.D. confirmed
the link to supernovae. Radio pulsars are rapidly rotating with
periods in the range $0.033$ s $ \le P\le 4.0$ s. They are believed to be
powered by rotational energy loss and are rapidly spinning down with
period derivatives of order $\dot{P}\sim 10^{-12}-10^{-16}$ s.  Their high
magnetic field $B$ leads to dipole magnetic braking radiation
proportional to the magnetic field squared. One estimates magnetic
fields of the order of $B\sim 10^{11}-10^{13}$ G.  
The total number of pulsars discovered sofar has just exceeded
1000 ahead of the turn of the millenium and the number is increasing
rapidly.

A distinct subclass of radio pulsars are {\it millisecond pulsars}
with periods between $1.56$ ms$ \le P\le 100$ ms. The period derivatives
are very small corresponding to very small magnetic fields $B\sim
10^8-10^{10}$G.  They are believed to be recycled pulsars, i.e.\  old
pulsars with low magnetic fields that have been spun up by accretion
preserving their low magnetic field and therefore only slowly spinning
down.  About 20 - almost half of the millisecond pulsars - 
are found in binaries
where the companion is either a white dwarf or a neutron star. Six
double neutron stars are known sofar including the Hulse-Taylor PSR
1913+16.  The first binary pulsar was found by Hulse and Taylor in
1973 and by measuring the inward spiralling one could determine all
parameters in the binary system as both masses, orbital periods and
period derivatives, orbital distances and inclination. Hereby general
relativity could be tested to an unprecedented accuracy. The binary
neutron stars all have masses in the narrow interval $1.3-1.5
M_{\odot}$, which may either be due to the creation process or that
heavier neutron stars are unstable.

With X-ray detectors on board satellites since 1971
almost two hundred X-ray pulsars and bursters have been found of which
the orbital period has been determined for about sixty. The X-ray
pulsars and bursters
are believed to be accreting neutrons stars from high ($M\ga 10M_\odot$)
and low mass ($M\la 1.2M_\odot$)
companions respectively. The X-ray pulses are most probably due to
strong accretion on the magnetic poles emitting X-ray (as northern
lights) with orbital frequency. The X-ray burst are due to slow
accretion spreading all over the neutron star surface before
igniting in a thermonuclear flash. 
The resulting (irregular) bursts have periods depending on
accretion rates rather than orbital periods. 
Recently, the ``missing link'' between bursters and pulsars has been
discovered. It is the low mass X-ray burster XTE J1808-369 where also
401Hz X-ray pulsations have been detected \cite{Klis}.
The radiation from X-ray bursters is not
blackbody and therefore only upper limits on temperatures can be extracted from
observed luminosities in most cases. Masses are less accurately
measured than for binary pulsars.
We mention recent mass determinations for the X-ray pulsar
Vela X-1: $M=(1.9\pm 0.1)M_\odot$, and the burster Cygnus X-2: 
$M=1.8\pm0.2)M_\odot$, which will be discussed below.
A subclass of six anomalous X-ray pulsars are slowly rotating but rapidly
spinning down indicating that they are young with enormous magnetic
fields, $B\sim 10^{13}$G, and thus named ``Magnetars'' \cite{Duncan}.
Recently, quasi-periodic
oscillations (QPO) have been found in 11 low mass X-ray binaries.
The QPO's set strict limits on masses and radii of neutrons stars, but 
if the periodic oscillations arrive from the
innermost stable orbit \cite{zss97}, it implies definite neutron star
masses up to $M\simeq 2.3M_\odot$.

Non-rotating and non-accreting neutron stars are virtually undetectable.
With the Hubble space telescope one single thermally radiating neutron
star has been found \cite{Walter}. Its distance is only 160 pc from
Earth and its surface temperature is $T\simeq 60$ eV. From its
luminosity one deduces a radius of the neutron star $R\le 14$ km.  In
our galaxy astrophysicists expect a large abundance $\sim 10^8$ of
neutron stars.  At least as many supernova explosions have occurred
since Big Bang which are responsible for all heavier elements present
in the Universe today. The scarcity of neutron stars in the solar
neighborhood may be due to a high initial velocity (asymmetric
``kick'') during their birth in supernovae. Recently, many neutron
stars have been found far away from their supernova remnants.  Such
``invisible'' neutron stars have probably been detected by
gravitational microlensing experiments. Future microlensing
observation will determine the population of such dark matter objects
in the galactic halo.

 From the view of physicists (and mass extinctionists) supernova
explosions are unfortunately rare in our and neighboring galaxies.
The predicted rate is 1-3 per century in our galaxy but the most recent
one was 1987A in LMC.
With luck we may observe one in the near future which produces a
rapidly rotating pulsar. Light curves and neutrino counts will test
supernova and neutron star models. The rapid spin down may be
exploited to test the structure and possible phase transitions in
the cores of neutron stars \cite{glendenning92,gpw97,rot}.

The recent discovery of afterglow in Gamma Ray Bursters (GRB) allows
determination of the very high redshifts ($z\ge 1$) and thus the
enormous distance and energy output $E\sim 10^{53}$ ergs in GRB. Most
previous GRB models cannot produce these energies. Only neutron
star mergers or black holes may be able to produce such violent
events. In comparison supernovae produce $\sim10^{51}$ erg's
mainly in neutrinos whereas a long distance missile 
has (only) $\sim10^{31}$ ergs of explosive energy.

The marvelous discoveries made in the past few decades will continue
as numerous satellite experiments are running at present and will be
launched. History tells us that the future will bring great surprises
and discoveries in this field.

\subsection{Physics of neutron stars}

\begin{figure}
   \setlength{\unitlength}{1mm}
   \begin{picture}(100,60)
   \put(25,0){\epsfxsize=10cm \epsfbox{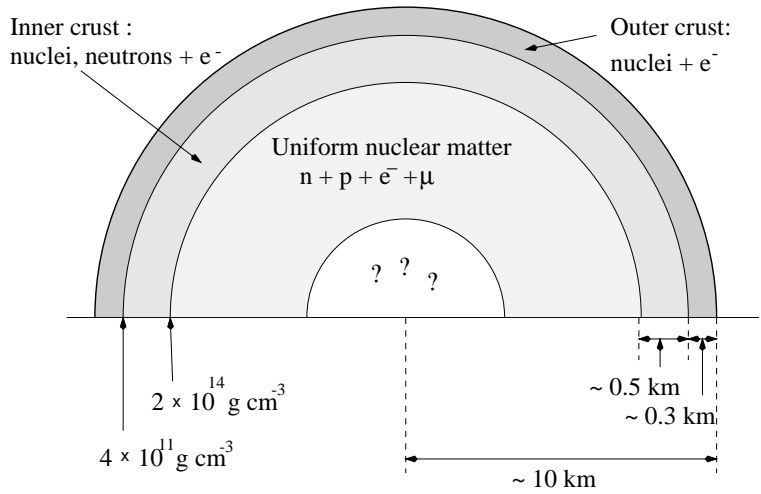}}
   \end{picture}
   \caption{Possible structure of a neutron star.}
   \label{fig:sec1fig_phases}
\end{figure}

The physics of compact objects like neutron stars offers
an intriguing interplay between nuclear processes  and
astrophysical observables.
Neutron stars exhibit conditions far from those 
encountered on earth; typically, expected densities $\rho$ 
of a neutron star interior are of the
order of $10^3$ or more times the density  
$\rho_d\approx 4\cdot 10^{11}$ g/cm$^{3}$ at 'neutron drip',
the density at which nuclei begin to 
dissolve and merge together.
Thus, the determination of an equation of state (EoS) 
for dense matter is essential to calculations of neutron 
star properties. The EoS determines properties  such as 
the mass range, the mass-radius relationship, the crust 
thickness and the cooling rate.
The same EoS is also crucial
in calculating the energy released in a supernova explosion.
Clearly,
the relevant degrees of freedom will not be the same in 
the crust region of a neutron star, 
where the density is much smaller than the 
saturation density of nuclear matter, and in the center
of the star, where density is so high that models based 
solely on interacting nucleons are questionable.
These features are pictorially displayed 
in Fig.\ \ref{fig:sec1fig_phases}. 
Neutron star models including various so-called 
realistic equations
of state result in the following general picture of the 
interior of a neutron star.  
The surface region, with typical densities 
$\rho< 10^6$ g/cm$^3$,  is a region in which
temperatures and magnetic fields may affect the equation of 
state. The outer crust for  
$10^6$ g/cm$^3$ $< \rho < 4\cdot 10^{11}$g/cm$^3$
is a solid region where a Coulomb lattice of heavy nuclei 
coexist in $\beta$-equilibrium  with a relativistic 
degenerate electron gas. The    inner crust for   
$4\cdot10^{11}$ g/cm$^3$ $< \rho < 2\cdot10^{14}$g/cm$^3$
consists of a lattice of neutron-rich nuclei together 
with a superfluid
neutron gas and an electron gas. The neutron liquid for  
$2\cdot  10^{14}$ g/cm$^3$ $< \rho < \cdot10^{15}$g/cm$^3$ 
contains mainly
superfluid neutrons with a smaller concentration of 
superconducting protons and normal electrons \cite{eeho96}. At higher
densities, typically $2-3$ times nuclear matter saturation
density, interesting phase transitions from a phase
with just nucleonic degrees of freedom to quark matter may
take place \cite{qmref}. Furthermore, one may have 
a mixed phase of quark and nuclear matter \cite{glendenning92,hps93},
kaon \cite{kn87} or pion condensates \cite{ap97,apr98}, 
hyperonic matter
\cite{glendenning92,schulze97,rueber93,sm96,kpe96,stoks99a,stoks99b}, 
strong magnetic fields in young stars 
\cite{kutschera,pal97} etc. 
 
The first aim of this work is therefore to attempt at
a review of various approaches to the equation
of state for dense neutron star matter relevant for stars
which have achieved thermal equilibrium.
Various approaches to the EoS and phases 
which may occur in a neutron star are discussed in section
\ref{sec:eos} while an overview of the 
thermodynamical
properties of the mixed phase and possible phases in neutron stars 
are presented  
in sections \ref{sec:phases} and \ref{sec:structure}. 

Our second aim is to discuss the relation between the 
EoS and various neutron star observables when a phase transition in the 
interior of the star occurs.
Astronomical observations leading to global neutron 
star parameters
such as the total mass, radius, or moment of inertia,  
are important since they
are sensitive to microscopic model calculations.
The mass, together with the 
moment of inertia, are also the gross structural 
parameters of a neutron star which are most accesible to 
observation. It is the mass which controls the gravitational
interaction of the star with other systems such as a binary
companion. The moment of inertia controls the energy stored
in rotation and thereby the energy available to the pulsar
emission mechanism.          
Determining the possible ranges
of neutron star is not only important in constraining the
EoS, but has important theoretical consequences for the
observational prediction of black holes
in the  universe.
Examples are the galactic black 
hole candidates Cyg X-1 \cite{Gies_Bolton_1986} and 
LMC X-3 \cite{Cowley_etal_1983}, which are massive X-ray 
binaries. Their mass functions (0.25 $M_{\odot}$ 
and 2.3 $M_{\odot}$) are, however, smaller 
than for some low-mass X-ray binaries like A0620-00 
\cite{McClintock_Remillard_1986} 
and V404 Cyg \cite{Casares_etal_1992}, 
which make better black hole candidates with mass 
functions in excess 
of three solar masses. There is a maximum mass a non-rotating
neutron star can have. There is however no upper limit on the mass
a black hole can have. If, therefore, one can find a dense,
highly compact object and can argue that its
rotation is slow, and can deduce that its mass is greater 
than the allowed maximum mass allowed to non-rotating
(or slowly rotating) neutron stars, then one has a
candidate for a black hole. 

Since neutron stars are objects of highly compressed matter,
this means that the geometry of space-time is changed
considerably form that of a flat space. Stellar models 
must therefore be based on Einstein's theory of general
relativity.  
Based on several of the theoretical equations of state
and possible phases of matter
discussed in sections \ref{sec:eos}, \ref{sec:phases} and
\ref{sec:structure},
various properties
of non-rotating and rotating neutron stars are presented in
section \ref{sec:starproperties}.
The relevant equations needed for the study
of the structure of a neutron star are summarized
in this section as well,
for both non-rotating
and rotating stellar structures.
There we also discuss the 
obervational implications when phase transitions occur
in the interior of the star. In addition to studies of the mass-radius
relationship and the moment of inertia, we extract also analytical
properties of quantities like the braking index and the rate of slow down 
near the critical angular velocity where the pressure inside the
star just exceeds that needed to make a phase transition.
The observational properties for first- and second-order 
phase transitions are also discussed. 
Other properties like glicthes and cooling of stars are also discussed
in section \ref{sec:starproperties}.
Summary and perspectives are given in section 
\ref{sec:conclusions}.
        
Finally, we mention several excellent and recent review articles
covering various aspects of neutron stars properties in the literature
addressing the interesting physics of the neutron star crust
\cite{pr95}, the nuclear equation of state \cite{apr98}, hot neutron
star matter \cite{pbpelk97} in connection with protoneutron stars, and
cooling calculations \cite{tsuruta98}.  However, as previously
mentioned, our aim will be to focus on the connection between the
various possible phases of dense neutron star matter in chemical
equilibrium and the implications of first- and second-order phase
transitions for various observables.

\section{PHASES OF DENSE MATTER} \label{sec:eos}

Several theoretical approaches to the EoS for the interior
of a neutron star have been considered.
Over the past two decades many authors \cite{qmref}
have considered the existence
of quark matter in neutron stars. Assuming a first 
order phase transition one
has, depending on the equation of states, found  either
complete strange quark matter stars or neutron stars 
with a core of quark matter
surrounded by a mantle of nuclear matter and a crust on top. 
Recently, the
possibility of a mixed phase of quark and nuclear matter 
was considered \cite{glendenning92}
and found to be energetically favorable. 
Including surface and Coulomb energies
this mixed phase was still found to be favored for 
reasonable bulk and interface
properties \cite{hps93}. The structure of the mixed phase of  
quark matter embedded
in nuclear matter with a uniform background of electrons 
was studied and
resembles that in the neutron drip region in the crust. 
Starting
from the outside, the crust consists of the outer layer, 
which is  a dense solid of neutron rich nuclei, and the 
inner layer in which
neutrons have dripped and form a neutron gas coexisting 
with the nuclei.
The structure of the latter mixed phase has recently 
been calculated in detail
\cite{pr95} and is found to exhibit rod-, plate- 
and bubble-like structures. At
nuclear saturation density $\rho_0\simeq 
2.8\cdot  10^{14}$ g/cm$^3$
there is only one phase of
uniform nuclear matter consisting of mainly neutrons, 
a small fraction of protons and
the same amount of electrons to achieve charge neutrality. 
A mixed phase of quark matter (QM) and nuclear matter (NM) 
appears already around a few times nuclear
saturation density - lower than the phase transition in 
hybrid stars. In the
beginning only few droplets of quark matter 
appear but at higher densities their
number increase and they merge into:  QM rods, QM plates, 
NM rods, NM bubbles, and finally pure QM at very high 
densities if the neutron stars 
have not become unstable towards gravitational collapse.

In this section we review various attempts at describing 
the above possible phases of dense neutron star matter. 
For the part of the neutron star that can be described
in terms of  nucleonic degrees of freedom only, i.e.\  $\beta$-stable matter
with protons, neutrons and electrons (and also muons),
we will try to shed light on recent advances within the 
frawework of various many-body approaches.  
This review is presented in subsection \ref{subsec:nucdeg}.
For the more exotic states of matter such as hyperonic
degrees of freedom 
we will point to recent studies of hyperonic
matter in terms of more microscopic models in subsection \ref{subsec:hyper}.
The problem however with e.g.\  hyperonic degrees of freedom is that
knowledge of the hyperon-nucleon or hyperon-hyperon interactions
has not yet reached the level
of sophistication encountered in the nucleon-nucleon sector.
Mean fields methods have however been much favoured in studies
of hyperonic matter. A discussion of pion and kaon condensation will also
be presented in the two subsequent subsections. 
Superfluidity is addressed in subsection \ref{subsec:superf}.

In general, we will avoid a discussion of non-relativistic and
relativistic mean fields methods
of relevance for neutron matter studies, mainly since such aspects
have been covered in depth in the literature see e.g.\  Refs.\ 
\cite{sm96,kpe96,glendenning91,sw86}.
Moreover, as pointed out by Akmal et al.\  \cite{apr98}, albeit
exhibiting valuable tutorial features, the main problem with 
relativistic mean field methods is that they rely on the approximation 
$\mu r \ll 1$, with $\mu$ the inverse Compton wave length of the meson
and $r$ the interparticle spacing. For nuclear and neutron matter
densities ranging from saturation density to five times saturation density,
$\mu r$ is in the range 1.4 to 0.8 for the pion and 7.8 to 4.7 for vector
mesons.
Clearly, these values are far from being small. The relativistic mean field
approximation can however be based on effective values for the coupling 
constants,
taking thereby into account correlation effects. These coupling constants have 
however a density dependence and a more microscopic theory is needed
to calculate them.  

Our knowledge of quark matter is however limited, and we will resort
to phenomenological models in subsection \ref{subsec:qmeos} in our
description of this phase of matter. Typical models are the so-called
Bag model \cite{bagmodel} or the Color-Dielectric model \cite{cdm}.

However, before proceeding with the above
more specific aspects of neutron star matter,
we need to introduce some general properties and features which will enter 
our description of dense matter.
These are introduced in the first subsection.
The reader should also note that we will omit a discussion of the 
properties of matter in the crust of the star since this is covered
in depth by the review of Ravenhall and Pethick \cite{pr95}.
Morever, for
neutron stars with masses  $\approx 1.4M_{\odot}$ or greater, 
the mass fraction contained in the crust of the star
is less than about 2\%.  We will therefore in our final EoS employ
results from earlier works 
\cite{lrp93,prl95} for matter at densities $\le 0.05$ fm$^{-3}$.

\subsection{Prerequisites and definitions} \label{subsec:conditions}

At densities of  0.1 fm$^{-3}$ and greater, we will in this 
work require properties of 
charge neutral uniform matter
to be made of mainly neutrons, protons, electrons and 
muons in beta equilibrium, although the presence of 
other baryons will be discussed as well\footnote{
In this work we will also set $G=c=\hbar=1$, where
$G$ is the gravitational constant.}.

In this section we will merely 
focus on distinct phases of matter, such as pure baryonic
matter or quark matter.
The composition of matter is then 
determined by the requirements of chemical and electrical equilibrium.
Furthermore, we will also consider matter at temperatures much lower
than the typical Fermi energies.
The equilibrium conditions are governed by the weak processes 
(normally referred to as the processes
for $\beta$-equilibrium)
\begin{equation} 
      b_1 \rightarrow b_2 + l +\bar{\nu}_l \hspace{1cm} b_2 +l \rightarrow b_1 
+\nu_l,
      \label{eq:betadecay}
\end{equation}
where $b_1$ and $b_2$ refer to e.g.\  the baryons being a neutron and a proton, 
respectively, 
$l$ is either an electron or a muon and  $\bar{\nu}_l $
and $\nu_l$ their respective anti-neutrinos and neutrinos. Muons typically 
appear at
a density close to nuclear matter saturation density, the latter being
\[
     n_0 \approx 0.16 \pm 0.02 \hspace{1cm} \mathrm{fm}^{-3},
\]
with a corresponding binding energy ${\cal E}_0$ 
for symmetric nuclear matter (SNM) at saturation density of
\[
     {\cal E}_0 = B/A=-15.6\pm 0.2 \hspace{1cm} \mathrm{MeV}.
\]
In this work the energy per baryon ${\cal E}$ will always be in units of MeV, 
while
the energy density $\varepsilon$ will 
be in units of MeVfm$^{-3}$ and the number density\footnote{We will often 
loosely just use density in our discussions.}
$n$ in units of fm$^{-3}$. The pressure $P$ is 
defined through the relation
\begin{equation}
    P=n^2\frac{\partial {\cal E}}{\partial n}=
      n\frac{\partial \varepsilon}{\partial n}-\varepsilon,
\end{equation}
with 
dimension MeVfm$^{-3}$. 
Similarly, the chemical potential for particle species $i$
is given by
\begin{equation}
     \mu_i = \left(\frac{\partial \varepsilon}{\partial n_i}\right),
     \label{eq:chemicalpotdef}
\end{equation}
with dimension MeV.
In our calculations of properties of neutron star matter in $\beta$-equilibrium,
we will need to calculate the energy per baryon ${\cal E}$ for e.g.\  several 
proton fractions $x_p$, which corresponds to
the ratio of protons as
compared to the total nucleon number ($Z/A$), 
 defined as
\begin{equation}
    x_p = \frac{n_p}{n},
\end{equation}
where $n=n_p+n_n$, the total baryonic density if neutrons and
protons are the only baryons present. In that case,
the total Fermi momentum $k_F$ and the Fermi momenta $k_{Fp}$,
$k_{Fn}$ for protons and neutrons are related to the total nucleon density
$n$ by
\begin{eqnarray}
     n & = & \frac{2}{3\pi^2} k_F^3 \nonumber \\
       & = & x_p n + (1-x_p) n \nonumber \\
       & = & \frac{1}{3\pi^2} k_{Fp}^3 + \frac{1}{3\pi^2} k_{Fn}^3.
    \label{eq:densi}
\end{eqnarray}
The energy per baryon will thus be
labelled as ${\cal E}(n,x_p)$.
${\cal E}(n,0)$ will then refer to the energy per baryon for pure neutron
matter (PNM) while ${\cal E}(n,\frac{1}{2})$ is the corresponding value for 
SNM. Furthermore, in this work, subscripts $n,p,e,\mu$
will always refer to neutrons, protons, electrons and muons, respectively.

Since the mean free path of a neutrino in a neutron star is bigger
than the typical radius of such a star ($\sim 10$ km), 
we will throughout assume that neutrinos escape freely from the neutron star,
see e.g.\  the work of Prakash et al.\  in Ref.\ \cite{pbpelk97} 
for a discussion
on trapped neutrinos. Eq.\ (\ref{eq:betadecay}) yields then the following
conditions for matter in $\beta$ equilibrium with e.g.\  nucleonic degrees 
freedom only
\begin{equation}
    \mu_n=\mu_p+\mu_e,
     \label{eq:npebetaequilibrium}
\end{equation}
and 
\begin{equation}
     n_p = n_e,
     \label{eq:chargeconserv}
\end{equation}
where $\mu_i$ and $n_i$ refer to the chemical potential and number density
in fm$^{-3}$ of particle species $i$. 
If muons are present as well,  we need to modify the equation for 
charge conservation, Eq.\ (\ref{eq:chargeconserv}), to read 
\[
     n_p = n_e+n_{\mu},
\]
and require that $\mu_e = \mu_{\mu}$.
With more particles present, the equations read
\begin{equation}
    \sum_i\left(n_{b_i}^+ +n_{l_i}^+\right) =   
    \sum_i\left(n_{b_i}^- +n_{l_i}^-\right),
    \label{eq:generalcharge}
\end{equation}
and 
\begin{equation} 
     \mu_n=b_i\mu_i+q_i\mu_l,
     \label{eq:generalbeta}
\end{equation}
where $b_i$ is the baryon number, $q_i$ the lepton charge and the superscripts 
$(\pm)$ on 
number densities $n$ represent particles with positive or negative charge.
To give an example, it is possible to have baryonic matter with hyperons like
$\Lambda$ 
and $\Sigma^{-,0,+}$ and isobars $\Delta^{-,0,+,++}$ as well in addition
to the nucleonic degrees of freedom.
In this case the chemical equilibrium condition of Eq.\ (\ref{eq:generalbeta}  ) 
becomes,
excluding muons,
\begin{eqnarray}
    \mu_{\Sigma^-} = \mu_{\Delta^-} = \mu_n + \mu_e , \nonumber \\ 
    \mu_{\Lambda} = \mu_{\Sigma^0} = \mu_{\Delta^0} = \mu_n , \nonumber \\
    \mu_{\Sigma^+} = \mu_{\Delta^+} = \mu_p = \mu_n - \mu_e ,\nonumber \\
    \mu_{\Delta^{++}} = \mu_n - 2 \mu_e .
    \label{eq:beta_baryonicmatter}
\end{eqnarray}
      
A transition from hadronic to quark matter is expected at high densities. 
The high-density quark matter phase
in the interior of neutron stars is also described by
requiring the system to be locally neutral
\begin{equation} 
    \label{eq:quarkneut}
    (2/3)n_u -(1/3)n_d - (1/3)n_s - n_e = 0,
\end{equation}
where $n_{u,d,s,e}$ 
are the densities of the $u$, $d$ and $s$ quarks and of the
electrons (eventually muons as well), respectively. 
Morover, the system must be in $\beta$-equilibrium, i.e.\ 
the chemical potentials have to satisfy the following equations:
\begin{equation}
      \label{eq:ud}
      \mu_d=\mu_u+\mu_e,
\end{equation}
and
\begin{equation}
      \label{eq:us}
      \mu_s=\mu_u+\mu_e .
\end{equation}
Eqs.\ (\ref{eq:quarkneut})-(\ref{eq:us}) have to be solved 
self-consistently together with
e.g.\  the field equations for quarks 
at a fixed density $n=n_u+n_d+n_s$.
In this section we will mainly deal with distinct phases of matter,
the additional constraints coming from the existence of a mixed
phase of hadrons and quarks and the related thermodynamics will be discussed
in Section \ref{sec:phases}.

An important ingredient in the discussion of the EoS and the criteria for
matter in $\beta$-equilibrium is the so-called symmetry energy ${\cal S} (n)$, 
defined as
the difference in energy for symmetric nuclear matter
and pure neutron matter 
\begin{equation}
      {\cal S} (n) = {\cal E} (n,x_p=0) - {\cal E} (n,x_p=1/2 ).
      \label{eq:symenergy}
\end{equation}
If we expand the energy per baryon in the case of nucleonic degrees of freedom 
only
in the proton concentration $x_p$ about the value of the energy 
for SNM ($x_p=\frac{1}{2}$), we obtain,
\begin{equation}
     {\cal E} (n,x_p)={\cal E} (n,x_p=\frac{1}{2})+
     \frac{1}{2}\frac{d^2 {\cal E}}{dx_p^2} (n)\left(1-2x_p\right)^2+\dots ,
     \label{eq:energyexpansion}
\end{equation}
where the term $d^2 {\cal E}/dx_p^2$ 
is to be associated with the symmetry energy ${\cal S} (n)$ in the empirical
mass formula. If
we assume that higher order derivatives in the above expansion are small
(we will see examples of this in the next subsection), then through the 
conditions
for $\beta$-equilbrium of Eqs.\ (\ref{eq:npebetaequilibrium}) and 
(\ref{eq:chargeconserv})
and Eq.\ (\ref{eq:chemicalpotdef}) we can define the proton
fraction by the symmetry energy as
\begin{equation}  
    \hbar c\left(3\pi^2nx_p\right)^{1/3} = 4{\cal S} (n)\left(1-2x_p\right),
    \label{eq:crudeprotonfraction}
\end{equation}
where the electron chemical potential is given
by $\mu_e = \hbar c k_F$, i.e.\  ultrarelativistic electrons are assumed.
Thus, the symmetry energy is of paramount importance for studies 
of neutron star matter in $\beta$-equilibrium.
One can extract information about the value of the symmetry energy at saturation 
density
$n_0$ from systematic studies of the masses of atomic nuclei. However, these 
results
are limited to densities around $n_0$ and for proton fractions close to 
$\frac{1}{2}$.
Typical values for ${\cal S} (n)$ at $n_0$ are in the range $27-38$ MeV.
For densities greater than $n_0$ it is more difficult to get a reliable 
information on the symmetry energy, and thereby the related proton fraction.
We will shed more light on this topic in the next subsection.

Finally, another property of interest in the discussion of the various 
equations of state  
is the incompressibility modulus $K$ at non-zero pressure
\begin{equation}
    K=9\frac{\partial P}{\partial n}.
    \label{eq:incompressibility}
\end{equation}
The sound speed $v_s$ depends as well on the density
of the nuclear medium through the relation
\begin{equation}
    \left(\frac{v_s}{c}\right)^2=\frac{dP}{d\varepsilon}=
    \frac{dP}{dn}\frac{dn}{d\varepsilon}=
    \left(\frac{K}{9(m_nc^2+{\cal E}+P/n)}\right).
    \label{eq:speedofsound}
\end{equation}
It is important to keep track of the dependence on density of $v_s$
since a superluminal behavior can occur at higher densities for most
non-relativistic EoS.
Superluminal behavior would
not occur with a fully relativistic theory, and it is necessary to
gauge the magnitude of the effect it introduces at the higher densities.
This will be discussed at the end of this section.
The adiabatic constant $\Gamma$ can also be extracted from the EoS
by 
\begin{equation}
    \Gamma = \frac{n}{P}\frac{\partial P}{\partial n}.
    \label{eq:adiabaticconstant}
\end{equation}

\subsection{Nucleonic degrees of freedom} \label{subsec:nucdeg}

A major part of the densities inside neutron stars can be well
represented by nucleonic degrees of freedom only, namely the inner
part of the crust to the outer part of the core, i.e.\  densities
ranging from $0.5$ to $2-3$ times nuclear matter saturation density,
There is a wealth of experimental and theoretical data, see e.g.\ 
Ref.\ \cite{migdal90} for an overview, which lend support to the
assumption that nucleons do not loose their individuality in dense
matter, i.e.\  that properties of the nucleon at such densities are
rather close to those of free nucleons. The above density range would
correspond to internucleon distances of the order of $\sim 1$ fm.  At
such interparticle distances there is little overlap between the
various nucleons and we may therefore assume that they still behave as
individual nucleons and that one can absorb the effects of overlap
into the two nucleon interaction.  The latter, when embedded in a
nuclear medium, is also different from the free nucleon-nucleon
interaction. In the medium there are interaction mechanisms which
obviously are absent in vacuum. As an example, the one-pion exchange
potential is modified in nuclear matter due to ``softening'' of pion
degrees of freedom in matter.

In order to illustrate how the nucleon-nucleon interaction is
renormalized in a nuclear medium, we will start with the simplest
possible many-body approach, namely the so-called
Brueckner-Hartree-Fock (BHF) approach.  This is done since the
Lippmann-Schwinger equation used to construct the scattering matrix
$T$, which in turn relates to the phase shifts, is rather similar to
the $G$-matrix which enters the BHF approach. The
difference resides in the introduction of a Pauli-blocking operator in
order to prevent scattering to intermediate particle states prohibited
by the Pauli principle. In addition, the single-particle energies of
the interacting particle are no longer given by kinetic energies only.
However, several of the features seen at the level of the scattering
matrix, pertain to the $G$-matrix as well. Therefore, if one employs
different nucleon-nucleon interactions in the calculation of the energy
per baryon in pure neutron matter with the BHF $G$-matrix, eventual
differences can be retraced at the level of the $T$-matrix.  We will
illustrate these aspects in the next subsection. More complicated
many-body terms and relativistic effects will be discussed in
subsections
\ref{subsubsec:manybody} and \ref{subsubsec:relativistic}

\subsubsection{From the NN interaction to the nuclear $G$-matrix}

The NN interactions we will employ here are the recent models
of the Nijmegen group \cite{nim}, the Argonne $V_{18}$ potential \cite{v18}
and the charge-dependent Bonn interaction (CD-Bonn \cite{cdbonn}).
In 1993, the Nijmegen group
presented a phase-shift analysis of all proton-proton and neutron-proton
data below $350$ MeV with a $\chi^2$ per datum of $0.99$
for 4301 data entries. The above potentials have all been constructed
based on these data. The CD-Bonn interaction has 
a $\chi^2$ per datum of $1.03$ and the same is true for the Nijm-I, Nijm-II
and Reid93 potential
versions of the Nijmegen group \cite{nim}. The new Argonne
potential $V_{18}$ \cite{v18} has a $\chi^2$ per datum of $1.09$. 

Although all these potentials predict almost identical phase shifts,
their mathematical structure is quite different. The Argonne 
potential, the Nijm-II and the Reid93  potentials 
are non-relativistic potential models defined in terms of local potential
functions, which are attached to various (non-relativistic)
operators of the spin, isospin
and/or angular momentum operators of the interacting pair of nucleons.
Such approaches to the NN interaction have traditionally been quite
popular since they
are numerically easy to use in configuration space calculations.
The Nijm-I model is similar to the Nijm-II model, but it  
includes also a $\bf p^2$ term,
see Eq.\ (13) of Ref.\ \cite{nim},  
which may be interpreted as a non-local contribution to the central
force.
The CD-Bonn
potential is based on the relativistic meson-exchange model
of Ref.\ \cite{mac89} which is non-local and cannot be
described correctly in terms of local potential functions.

For a given NN interaction $V$, the $R$-matrix (or $K$-matrix) for free-space
two-nucleon scattering is obtained from the Lippmann-Schwinger
equation, which reads in the center-of-mass (c.m.) system 
and in a partial-wave decomposition
\begin{equation}
   R_{ll'}^{\alpha T_z}(kk'\omega)=V_{ll'}^{\alpha T_z}(kk')
   +\sum_{l''}\int \frac{d^3 q}{(2\pi )^3}V_{ll''}^{\alpha T_z}(kq)
   \frac{1}{\omega-H_0}
   R_{l''l'}^{\alpha T_z}(qk'\omega),
\label{eq:tmatrix}
\end{equation}
with $ll'$ and $kk'$ the orbital angular momentum and the linear 
momentum of the
relative motion, respectively. $T_z$ is the total isospin projection. 
The angular momentum $J$ and total spin $S$  are represented
by the variable $\alpha$. 
The term $H_0$ represents the kinetic energy of the intermediate states.
The phase-shifts for a given partial wave can be calculated from the on-shell
matrix element of $R$, which is obtained by setting
$q = q' = q_0$ with $\omega=\frac{q_0^2}{m_n}$, $m_n$ being the mass
of the nucleon.
Since all of the above interactions reproduce the same
phase-shifts, the corresponding on-shell matrix elements of $R$
calculated from these various potentials 
are identical as well. 
However, due to the way the potentials are constructed, their off-shell
properties may be different. This was discussed in detail in
Refs.\ \cite{ehmmp97,pmmh98}. In those works the authors showed that especially
for the $^1S_0$ and $^3S_1$-$^3D_1$ channels, the CD-Bonn and Nijm-I 
interactions
which include the effects of  non-localities, yield a more attractive 
interaction for the free scattering case.
For $D$-waves and higher partial waves
the various potentials were almost equal while there were still differences
for $P$-waves.

We now turn the attention to the application of such 
NN interactions in a nuclear medium. First we will therefore 
employ a as simple as possible many-body scheme, in order to preserve
a link between the preceeding discussion on the NN interaction and the solution
of the Lippmann-Schwinger equation. As stated above, 
it will thus suffice to employ 
the BHF method.

Following the conventional many-body approach, we divide the full
Hamiltonian $H=T+V$, with $T$ being the kinetic energy
and $V$ the bare NN interaction,
into an unperturbed part $H_0 =T+U$ and an interacting part $H_I = V-U$,
such that
\[
   H=T+V=H_0 + H_I,
\]
where we have introduced an auxiliary single-particle (sp)  potential $U$. If
$U$ is chosen such that $H_I$ becomes small, then perturbative
many-body techniques can presumably be applied.
A serious obstacle to any perturbative treatment is the fact that the
bare NN interaction $V$ is very large at short inter-nucleonic distances,
which renders a perturbative approach highly prohibitive. To overcome
this problem, we introduce the reaction matrix $G$ given
by the solution of
the Bethe-Goldstone equation (in operator form)
\begin{equation}
    G(\omega)=V+VQ\frac{1}{\omega - QH_0Q}QG,
    \label{eq:bg}
\end{equation}
where $\omega$ is the unperturbed energy of the interacting  nucleons and
 $Q$ is the the Pauli operator which prevents scattering into occupied states. 
The Pauli operator is given by 
\begin{equation}
    Q(k_m\tau_m , k_n \tau_n) =
    \left\{\begin{array}{cc}1,&k_m >k_F^{\tau_m} ,k_n>k_F^{\tau_n}  ,\\
    0,&otherwise,\end{array}\right.
    \label{eq:pauli}
\end{equation}
in the laboratory system, where $k_F^{\tau_i}$ defines the Fermi 
momenta of the 
proton ($\tau_i=1/2$) and neutron ($\tau_i=-1/2$).
For notational economy, we set $|{\bf k}_m|=k_m$.

The above expression for the Pauli operator is in the
laboratory frame. In the calculations of the $G$-matrix,
we will employ a Pauli operator in the center-of-mass and relative
coordinate system. Further, this Pauli operator
will be given by the so-called angle-average approximation,
for details see Ref.\ \cite{hko95}.  Eq. (\ref{eq:bg}) reads 
then (in a partial wave representation)
\begin{eqnarray}
     G_{ll'}^{\alpha T_z}(kk'K\omega)&=&V_{ll'}^{\alpha T_z}(kk') 
\label{eq:gnonrel}\\
     &&+\sum_{l''}\int \frac{d^3 q}{(2\pi )^3}V_{ll''}^{\alpha T_z}(kq)
     \frac{Q^{T_z}(q,K)}{\omega-H_0}
     G_{l''l'}^{\alpha T_z}(qk'K\omega).\nonumber
\end{eqnarray}
The variable $K$ is the momentum of the center-of-mass 
motion. Since we are going to use an angular
average for the Pauli operator, the $G$-matrix is diagonal in total
angular momentum $J$.
Further, the $G$-matrix is diagonal in the center-of-mass orbital momentum
$L$ and the total spin $S$, all three variables represented by the index 
$\alpha$.
The variable $\alpha$ differs therefore from the definiton of the $R$-matrix,
where $K=0$.
Three different $G$-matrices  have to be evaluated, 
depending on the individual isospins ($\tau_1\tau_2$)  
of the interacting nucleons
($\frac{1}{2}\frac{1}{2}$, $-\frac{1}{2}-\frac{1}{2}$ 
and $-\frac{1}{2}\frac{1}{2})$. 
These quantities are represented by the total isospin projection $T_z$ 
in Eq. (\ref{eq:gnonrel}).
The different $G$-matrices  originate from the discrimination 
between protons and neutrons in Eq. (\ref{eq:pauli}).
The term $H_0$ in the denominator of Eq. (\ref{eq:gnonrel}) is 
the unperturbed energy of the intermediate states and depends
on $k$, $K$ and the individual isospin of the interacting particles.
Only ladder diagrams with 
intermediate two-particle states are included in Eq.\ 
(\ref{eq:bg}). The structure of the $G$-matrix equation in Eq.\ 
(\ref{eq:gnonrel})
can  then be directly
compared to the $R$-matrix for free NN scattering, Eq.~(\ref{eq:tmatrix}).
Therefore, as discussed below, eventual differences
between various potentials in a finite medium should be easily retraced
to the structure of the $R$-matrix.
It is also obvious that one expects the matrix elements of $G$ to be rather 
close to
those of $R$ with only small deviations. These deviations originate from two
effects which reduce the contributions of second and higher order in $V$ to the
$G$-matrix as compared to their contributions to $R$. One is the 
above-mentioned
Pauli quenching effect: 
the Pauli operator $Q$ in (\ref{eq:gnonrel}) restricts the
intermediate particle states to states above the Fermi energy. 
The second one is
the dispersive effect: the energy denominators in (\ref{eq:gnonrel}) are defined 
in
terms of the single-particle energies of nucleons in the medium while the
corresponding denominators of (\ref{eq:tmatrix}) are differences between the
energies of free nucleons. Since the absolute values for the energy differences
between nucleons, which feel the mean field of the nuclear system, are larger
than the energy differences between the kinetic energies, also this dispersive
correction reduces the attractive contributions of the non-Born terms. 
As a result, the matrix elements of $G$ tend to be less attractive
than the corresponding matrix elements of $R$, see e.g.\  Refs.\ 
\cite{ehmmp97,pmmh98}
for further details.

We use a continuous single particle (sp) spectrum advocated by 
Mahaux et al.\ \cite{mahaux}. It is  defined by the 
self-consistent solution of the following equations:
\begin{equation}
   \varepsilon_i    = t_i + u_i=\frac{k_i^2}{2m}
   +u_i,
   \label{eq:spnrel}
\end{equation}
where $m$ is the bare nucleon mass,
and     
\begin{equation}
    u_i = \sum_{h\leq k_F}
   \left\langle ih \right|    
    G(E = \varepsilon_{i} + \varepsilon_h )
   \left|  ih \right\rangle _{\mathrm{AS}}. 
   \label{eq:selfcon}
\end{equation}
In Eqs.\ (\ref{eq:spnrel})-(\ref{eq:selfcon}), the subscripts $i$ and $h$
represent the quantum numbers of the single-particle states, such as
isospin projections $\tau_i$ and $\tau_h$, momenta $k_i$ and $k_h$, etc.
The sp kinetic energy is given by $t_{i}$ and similarly the sp potential by
$u_{i}$. 

Finally, the non-relativistic
energy per nucleon ${\cal E}$ is formally given as
\begin{eqnarray} 
   {\cal E} &= &
    \frac{1}{A} \sum_{h\leq k_F}
    \frac{k_h^2}{2m}+ 
   \label{eq:enrel} \\
 & &   \frac{1}{2A}
     \sum_{h \leq k_F,h'\leq k_F}
   \left\langle hh'\right | G(E=\varepsilon_h +\varepsilon_{h'})\left |
   hh'\right \rangle_{\mathrm{AS}} .\nonumber  
\end{eqnarray}
In this equation  we have suppressed the isospin indices for the Fermi momenta.
Eq.\ (\ref{eq:enrel}) is actually calculated for various proton fractions
$x_p$, and is thereby a function of both density $n$ and $x_p$.
We will therefore in the following discussion always label the energy
per particle as 
${\cal E}(n,x_p)$.

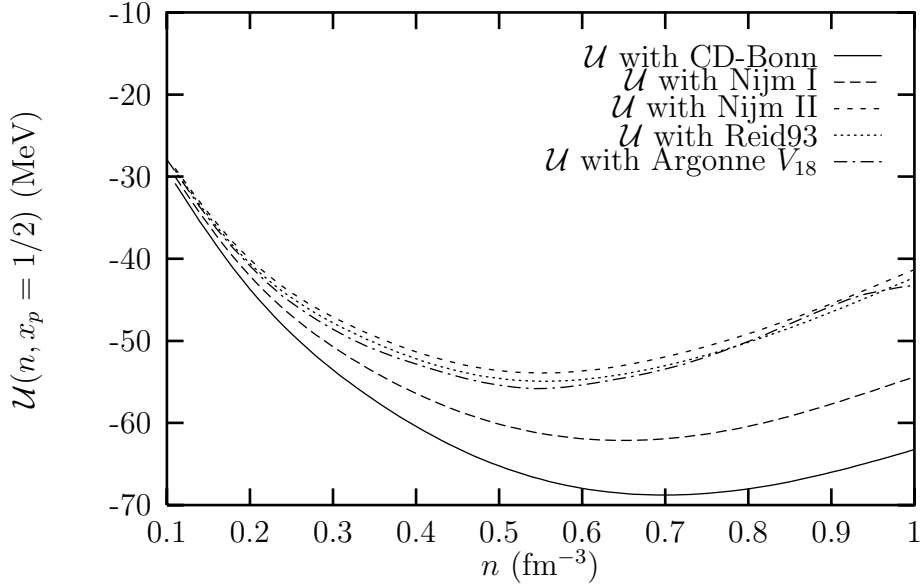
\begin{figure}\begin{center} 
     % GNUPLOT: LaTeX picture with Postscript
\setlength{\unitlength}{0.1bp}
\special{!
%!PS-Adobe-2.0
%%Creator: gnuplot
%%DocumentFonts: Helvetica
%%BoundingBox: 50 50 770 554
%%Pages: (atend)
%%EndComments
/gnudict 40 dict def
gnudict begin
/Color false def
/Solid false def
/gnulinewidth 5.000 def
/vshift -33 def
/dl {10 mul} def
/hpt 31.5 def
/vpt 31.5 def
/M {moveto} bind def
/L {lineto} bind def
/R {rmoveto} bind def
/V {rlineto} bind def
/vpt2 vpt 2 mul def
/hpt2 hpt 2 mul def
/Lshow { currentpoint stroke M
  0 vshift R show } def
/Rshow { currentpoint stroke M
  dup stringwidth pop neg vshift R show } def
/Cshow { currentpoint stroke M
  dup stringwidth pop -2 div vshift R show } def
/DL { Color {setrgbcolor Solid {pop []} if 0 setdash }
 {pop pop pop Solid {pop []} if 0 setdash} ifelse } def
/BL { stroke gnulinewidth 2 mul setlinewidth } def
/AL { stroke gnulinewidth 2 div setlinewidth } def
/PL { stroke gnulinewidth setlinewidth } def
/LTb { BL [] 0 0 0 DL } def
/LTa { AL [1 dl 2 dl] 0 setdash 0 0 0 setrgbcolor } def
/LT0 { PL [] 0 1 0 DL } def
/LT1 { PL [4 dl 2 dl] 0 0 1 DL } def
/LT2 { PL [2 dl 3 dl] 1 0 0 DL } def
/LT3 { PL [1 dl 1.5 dl] 1 0 1 DL } def
/LT4 { PL [5 dl 2 dl 1 dl 2 dl] 0 1 1 DL } def
/LT5 { PL [4 dl 3 dl 1 dl 3 dl] 1 1 0 DL } def
/LT6 { PL [2 dl 2 dl 2 dl 4 dl] 0 0 0 DL } def
/LT7 { PL [2 dl 2 dl 2 dl 2 dl 2 dl 4 dl] 1 0.3 0 DL } def
/LT8 { PL [2 dl 2 dl 2 dl 2 dl 2 dl 2 dl 2 dl 4 dl] 0.5 0.5 0.5 DL } def
/P { stroke [] 0 setdash
  currentlinewidth 2 div sub M
  0 currentlinewidth V stroke } def
/D { stroke [] 0 setdash 2 copy vpt add M
  hpt neg vpt neg V hpt vpt neg V
  hpt vpt V hpt neg vpt V closepath stroke
  P } def
/A { stroke [] 0 setdash vpt sub M 0 vpt2 V
  currentpoint stroke M
  hpt neg vpt neg R hpt2 0 V stroke
  } def
/B { stroke [] 0 setdash 2 copy exch hpt sub exch vpt add M
  0 vpt2 neg V hpt2 0 V 0 vpt2 V
  hpt2 neg 0 V closepath stroke
  P } def
/C { stroke [] 0 setdash exch hpt sub exch vpt add M
  hpt2 vpt2 neg V currentpoint stroke M
  hpt2 neg 0 R hpt2 vpt2 V stroke } def
/T { stroke [] 0 setdash 2 copy vpt 1.12 mul add M
  hpt neg vpt -1.62 mul V
  hpt 2 mul 0 V
  hpt neg vpt 1.62 mul V closepath stroke
  P  } def
/S { 2 copy A C} def
end
}
\begin{picture}(3600,2160)(0,0)
\special{"
gnudict begin
gsave
50 50 translate
0.100 0.100 scale
0 setgray
/Helvetica findfont 100 scalefont setfont
newpath
-500.000000 -500.000000 translate
LTa
LTb
600 251 M
63 0 V
2754 0 R
-63 0 V
600 561 M
63 0 V
2754 0 R
-63 0 V
600 870 M
63 0 V
2754 0 R
-63 0 V
600 1180 M
63 0 V
2754 0 R
-63 0 V
600 1490 M
63 0 V
2754 0 R
-63 0 V
600 1799 M
63 0 V
2754 0 R
-63 0 V
600 2109 M
63 0 V
2754 0 R
-63 0 V
600 251 M
0 63 V
0 1795 R
0 -63 V
913 251 M
0 63 V
0 1795 R
0 -63 V
1226 251 M
0 63 V
0 1795 R
0 -63 V
1539 251 M
0 63 V
0 1795 R
0 -63 V
1852 251 M
0 63 V
0 1795 R
0 -63 V
2165 251 M
0 63 V
0 1795 R
0 -63 V
2478 251 M
0 63 V
0 1795 R
0 -63 V
2791 251 M
0 63 V
0 1795 R
0 -63 V
3104 251 M
0 63 V
0 1795 R
0 -63 V
3417 251 M
0 63 V
0 1795 R
0 -63 V
600 251 M
2817 0 V
0 1858 V
-2817 0 V
600 251 L
LT0
3114 1946 M
180 0 V
631 1464 M
32 -49 V
31 -48 V
31 -48 V
32 -45 V
31 -45 V
31 -43 V
31 -42 V
32 -40 V
31 -39 V
31 -36 V
32 -35 V
31 -34 V
31 -32 V
32 -31 V
31 -29 V
31 -28 V
31 -28 V
32 -26 V
31 -25 V
31 -25 V
32 -23 V
31 -23 V
31 -22 V
32 -22 V
31 -21 V
31 -20 V
31 -20 V
32 -19 V
31 -18 V
31 -18 V
32 -17 V
31 -17 V
31 -16 V
32 -15 V
31 -15 V
31 -14 V
31 -13 V
32 -13 V
31 -11 V
31 -12 V
32 -10 V
31 -10 V
31 -10 V
32 -8 V
31 -9 V
31 -7 V
31 -7 V
32 -6 V
31 -6 V
31 -5 V
32 -4 V
31 -4 V
31 -3 V
32 -3 V
31 -3 V
31 -1 V
31 -1 V
32 -1 V
31 0 V
31 0 V
32 1 V
31 1 V
31 2 V
32 2 V
31 3 V
31 3 V
31 3 V
32 4 V
31 4 V
31 5 V
32 5 V
31 5 V
31 6 V
32 6 V
31 7 V
31 7 V
31 7 V
32 7 V
31 8 V
31 7 V
32 8 V
31 9 V
31 8 V
32 9 V
31 8 V
31 9 V
31 9 V
32 9 V
31 10 V
LT1
3114 1846 M
180 0 V
631 1489 M
32 -47 V
31 -46 V
31 -45 V
32 -43 V
31 -43 V
31 -40 V
31 -38 V
32 -37 V
31 -35 V
31 -33 V
32 -32 V
31 -29 V
31 -29 V
32 -27 V
31 -25 V
31 -25 V
31 -23 V
32 -22 V
31 -22 V
31 -20 V
32 -21 V
31 -18 V
31 -19 V
32 -18 V
31 -17 V
31 -17 V
31 -16 V
32 -15 V
31 -15 V
31 -14 V
32 -14 V
31 -13 V
31 -13 V
32 -12 V
31 -11 V
31 -10 V
31 -10 V
32 -10 V
31 -9 V
31 -8 V
32 -7 V
31 -7 V
31 -7 V
32 -6 V
31 -5 V
31 -4 V
31 -4 V
32 -4 V
31 -3 V
31 -2 V
32 -2 V
31 -1 V
31 -1 V
32 0 V
31 0 V
31 1 V
31 1 V
32 2 V
31 2 V
31 3 V
32 3 V
31 4 V
31 4 V
32 4 V
31 5 V
31 6 V
31 5 V
32 6 V
31 7 V
31 7 V
32 7 V
31 8 V
31 8 V
32 8 V
31 9 V
31 9 V
31 9 V
32 9 V
31 10 V
31 9 V
32 11 V
31 10 V
31 10 V
32 10 V
31 11 V
31 10 V
31 11 V
32 10 V
31 10 V
LT2
3114 1746 M
180 0 V
631 1521 M
32 -44 V
31 -43 V
31 -41 V
32 -41 V
31 -38 V
31 -37 V
31 -35 V
32 -33 V
31 -31 V
31 -29 V
32 -26 V
31 -25 V
31 -24 V
32 -22 V
31 -20 V
31 -20 V
31 -18 V
32 -17 V
31 -17 V
31 -15 V
32 -15 V
31 -15 V
31 -14 V
32 -13 V
31 -13 V
31 -12 V
31 -12 V
32 -11 V
31 -11 V
31 -10 V
32 -9 V
31 -9 V
31 -8 V
32 -8 V
31 -7 V
31 -6 V
31 -5 V
32 -5 V
31 -4 V
31 -3 V
32 -3 V
31 -1 V
31 -1 V
32 -1 V
31 1 V
31 0 V
31 2 V
32 2 V
31 3 V
31 3 V
32 3 V
31 5 V
31 4 V
32 6 V
31 5 V
31 7 V
31 6 V
32 7 V
31 7 V
31 7 V
32 8 V
31 8 V
31 8 V
32 9 V
31 9 V
31 9 V
31 9 V
32 10 V
31 10 V
31 10 V
32 11 V
31 11 V
31 11 V
32 11 V
31 11 V
31 12 V
31 12 V
32 12 V
31 12 V
31 13 V
32 12 V
31 13 V
31 13 V
32 13 V
31 13 V
31 13 V
31 13 V
32 13 V
31 13 V
LT3
3114 1646 M
180 0 V
631 1513 M
32 -45 V
31 -43 V
31 -43 V
32 -41 V
31 -40 V
31 -37 V
31 -36 V
32 -34 V
31 -32 V
31 -29 V
32 -28 V
31 -26 V
31 -24 V
32 -22 V
31 -21 V
31 -20 V
31 -19 V
32 -18 V
31 -17 V
31 -16 V
32 -15 V
31 -15 V
31 -15 V
32 -13 V
31 -14 V
31 -12 V
31 -13 V
32 -11 V
31 -11 V
31 -11 V
32 -10 V
31 -9 V
31 -8 V
32 -8 V
31 -7 V
31 -7 V
31 -6 V
32 -4 V
31 -4 V
31 -4 V
32 -3 V
31 -2 V
31 -1 V
32 -1 V
31 0 V
31 1 V
31 1 V
32 2 V
31 2 V
31 4 V
32 3 V
31 5 V
31 4 V
32 5 V
31 6 V
31 6 V
31 6 V
32 7 V
31 7 V
31 7 V
32 8 V
31 8 V
31 9 V
32 8 V
31 9 V
31 10 V
31 9 V
32 10 V
31 10 V
31 10 V
32 11 V
31 11 V
31 11 V
32 11 V
31 12 V
31 11 V
31 12 V
32 13 V
31 12 V
31 12 V
32 13 V
31 13 V
31 13 V
32 13 V
31 13 V
31 14 V
31 13 V
32 13 V
31 14 V
LT4
3114 1546 M
180 0 V
-2694 8 R
63 -90 V
62 -90 V
63 -82 V
62 -72 V
63 -67 V
63 -62 V
62 -54 V
63 -45 V
62 -41 V
63 -37 V
63 -34 V
62 -29 V
63 -25 V
62 -23 V
63 -20 V
63 -20 V
62 -18 V
63 -17 V
62 -14 V
63 -11 V
63 -8 V
62 -5 V
63 1 V
62 5 V
63 8 V
63 10 V
62 11 V
63 11 V
62 13 V
63 15 V
63 16 V
62 20 V
63 21 V
62 23 V
63 24 V
63 25 V
62 28 V
63 27 V
62 27 V
63 26 V
63 25 V
62 20 V
63 13 V
62 12 V
63 9 V
stroke
grestore
end
showpage
}
\put(3054,1546){\makebox(0,0)[r]{${\cal U}$ with Argonne $V_{18}$}}
\put(3054,1646){\makebox(0,0)[r]{${\cal U}$ with Reid93}}
\put(3054,1746){\makebox(0,0)[r]{${\cal U}$ with Nijm II}}
\put(3054,1846){\makebox(0,0)[r]{${\cal U}$ with Nijm I}}
\put(3054,1946){\makebox(0,0)[r]{${\cal U}$ with CD-Bonn}}
\put(2008,21){\makebox(0,0){$n$ (fm$^{-3}$)}}
\put(100,1180){%
\special{ps: gsave currentpoint currentpoint translate
270 rotate neg exch neg exch translate}%
\makebox(0,0)[b]{\shortstack{${\cal U}(n,x_p=1/2)$ (MeV)}}%
\special{ps: currentpoint grestore moveto}%
}
\put(3417,151){\makebox(0,0){1}}
\put(3104,151){\makebox(0,0){0.9}}
\put(2791,151){\makebox(0,0){0.8}}
\put(2478,151){\makebox(0,0){0.7}}
\put(2165,151){\makebox(0,0){0.6}}
\put(1852,151){\makebox(0,0){0.5}}
\put(1539,151){\makebox(0,0){0.4}}
\put(1226,151){\makebox(0,0){0.3}}
\put(913,151){\makebox(0,0){0.2}}
\put(600,151){\makebox(0,0){0.1}}
\put(540,2109){\makebox(0,0)[r]{-10}}
\put(540,1799){\makebox(0,0)[r]{-20}}
\put(540,1490){\makebox(0,0)[r]{-30}}
\put(540,1180){\makebox(0,0)[r]{-40}}
\put(540,870){\makebox(0,0)[r]{-50}}
\put(540,561){\makebox(0,0)[r]{-60}}
\put(540,251){\makebox(0,0)[r]{-70}}
\end{picture}
     \caption{Potential energy per particle ${\cal U}$ 
              for symmetric nuclear matter as function 
              of total baryonic density $n$.}
     \label{fig:sec2fig3}
\end{center}\end{figure}

In the limit of pure neutron matter
only those partial waves contribute where the pair of interacting nucleons
is coupled to isospin $T=1$. Due to the antisymmetry of the matrix elements
this implies that only partial waves with even values for the sum $l+S$, like
$^1S_0$, $^3P_0$ etc.\ need to be considered in this case. For proton fractions
different from zero, in particular the case of symmetric nuclear matter, also
the other partial waves, like $^3S_1 - ^3D_1$ and $^1P_1$ contribute.
In a BHF calculation
the kinetic energy is
independent of the NN interaction chosen. We will then restrict the following
discussion to the potential energy per nucleon ${\cal U}$, the second
term in the RHS of Eq.\ (\ref{eq:enrel}).
Putting the contributions from 
various channels together\footnote{In our 
calculations we include all partial waves with $l < 10$.}, one obtains 
the total potential energy per nucleon ${\cal  U}$ for symmetric
nuclear matter and neutron matter. These results are displayed
in Figs.\ \ref{fig:sec2fig3} and 
\ref{fig:sec2fig4} as functions of density $n$ for the 
the CD-Bonn interaction 
\cite{cdbonn}(solid line), the three Nijmegen potentials, Nijm-I (long 
dashes), Nijm-II (short dashes) and Reid93 (dotted line)\cite{nim} and 
the Argonne $V_{18}$ \cite{v18} (dot-dashed line).   

The differences between the various potentials are larger for the 
energy in nuclear matter. This is mainly due to the importance of the 
$^3S_1$-$^3D_1$ 
contribution which is absent in pure neutron matter. This is in line with
previous investigations, which showed that the predicted binding energy of
nuclear matter is correlated with the strength of the tensor force, expressed in
terms of the $D$-state probability obtained for the deuteron (see e.g.\
\cite{hko95}). This importance of the strength of the tensor force is also 
seen in the calculation of the binding energy of the triton in Ref.\ 
\cite{cdbonn,nhkg97}. 
The CD-Bonn interaction yields a binding energy of $8.00$ MeV, 
the Nijm-I potential gives $7.72$ MeV while the Nijm-II 
yields $7.62$ MeV, the same as does the new Argonne potential \cite{v18}. 
Typically, potentials with a smaller $D$-state probability have a weaker 
tensor force and exhibit therefore a smaller quenching of the non-born
terms in Eq.\ (\ref{eq:gnonrel}). Moreover, the fact that the CD-Bonn 
interaction
and the Nijm I potential include effects of non-localities, yields also
a further attraction from the central force both in the $^3S_1$-$^3D_1$ and
the singlet $^1S_0$ channels. The latter explains the additional difference 
in nuclear matter between the Nijm I and the Nijm II, Reid93 and Argonne 
$V_{18}$
potentials as well as part of the difference seens in Fig.\ \ref{fig:sec2fig4}
for pure neutron matter (PNM). For both SNM and PNM there are also additional
differences arising from $P$ waves, notably 
for the Argonne potential in PNM, 
where the difference between the Argonne $V_{18}$ interaction
the Reid93 and the Nijm II potentials
is mainly due to more repulsive 
contributions from $P$-waves.
For higher partial waves, the differences are rather small, typically of the 
order
of few per cent.
\begin{figure}\begin{center} 
     \input{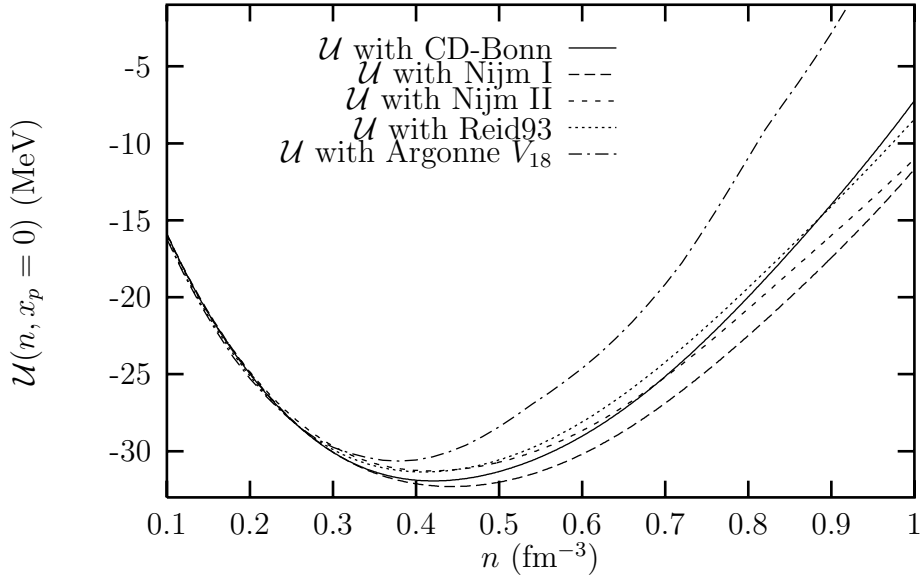}
     \caption{Potential energy per particle ${\cal U}$ 
              for pure neutron matter.}
     \label{fig:sec2fig4}
\end{center}\end{figure}
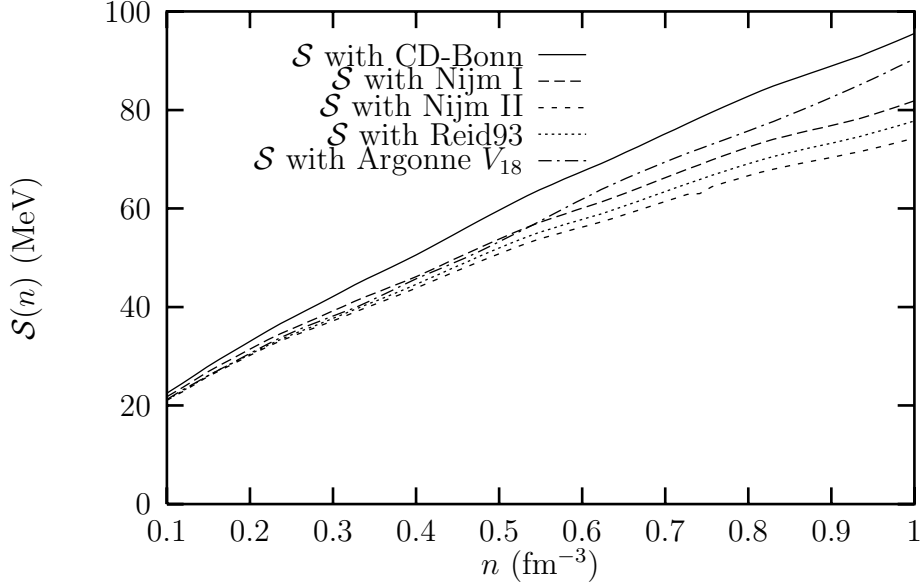
\begin{figure}\begin{center} 
     % GNUPLOT: LaTeX picture with Postscript
\setlength{\unitlength}{0.1bp}
\special{!
%!PS-Adobe-2.0
%%Creator: gnuplot
%%DocumentFonts: Helvetica
%%BoundingBox: 50 50 770 554
%%Pages: (atend)
%%EndComments
/gnudict 40 dict def
gnudict begin
/Color false def
/Solid false def
/gnulinewidth 5.000 def
/vshift -33 def
/dl {10 mul} def
/hpt 31.5 def
/vpt 31.5 def
/M {moveto} bind def
/L {lineto} bind def
/R {rmoveto} bind def
/V {rlineto} bind def
/vpt2 vpt 2 mul def
/hpt2 hpt 2 mul def
/Lshow { currentpoint stroke M
  0 vshift R show } def
/Rshow { currentpoint stroke M
  dup stringwidth pop neg vshift R show } def
/Cshow { currentpoint stroke M
  dup stringwidth pop -2 div vshift R show } def
/DL { Color {setrgbcolor Solid {pop []} if 0 setdash }
 {pop pop pop Solid {pop []} if 0 setdash} ifelse } def
/BL { stroke gnulinewidth 2 mul setlinewidth } def
/AL { stroke gnulinewidth 2 div setlinewidth } def
/PL { stroke gnulinewidth setlinewidth } def
/LTb { BL [] 0 0 0 DL } def
/LTa { AL [1 dl 2 dl] 0 setdash 0 0 0 setrgbcolor } def
/LT0 { PL [] 0 1 0 DL } def
/LT1 { PL [4 dl 2 dl] 0 0 1 DL } def
/LT2 { PL [2 dl 3 dl] 1 0 0 DL } def
/LT3 { PL [1 dl 1.5 dl] 1 0 1 DL } def
/LT4 { PL [5 dl 2 dl 1 dl 2 dl] 0 1 1 DL } def
/LT5 { PL [4 dl 3 dl 1 dl 3 dl] 1 1 0 DL } def
/LT6 { PL [2 dl 2 dl 2 dl 4 dl] 0 0 0 DL } def
/LT7 { PL [2 dl 2 dl 2 dl 2 dl 2 dl 4 dl] 1 0.3 0 DL } def
/LT8 { PL [2 dl 2 dl 2 dl 2 dl 2 dl 2 dl 2 dl 4 dl] 0.5 0.5 0.5 DL } def
/P { stroke [] 0 setdash
  currentlinewidth 2 div sub M
  0 currentlinewidth V stroke } def
/D { stroke [] 0 setdash 2 copy vpt add M
  hpt neg vpt neg V hpt vpt neg V
  hpt vpt V hpt neg vpt V closepath stroke
  P } def
/A { stroke [] 0 setdash vpt sub M 0 vpt2 V
  currentpoint stroke M
  hpt neg vpt neg R hpt2 0 V stroke
  } def
/B { stroke [] 0 setdash 2 copy exch hpt sub exch vpt add M
  0 vpt2 neg V hpt2 0 V 0 vpt2 V
  hpt2 neg 0 V closepath stroke
  P } def
/C { stroke [] 0 setdash exch hpt sub exch vpt add M
  hpt2 vpt2 neg V currentpoint stroke M
  hpt2 neg 0 R hpt2 vpt2 V stroke } def
/T { stroke [] 0 setdash 2 copy vpt 1.12 mul add M
  hpt neg vpt -1.62 mul V
  hpt 2 mul 0 V
  hpt neg vpt 1.62 mul V closepath stroke
  P  } def
/S { 2 copy A C} def
end
}
\begin{picture}(3600,2160)(0,0)
\special{"
gnudict begin
gsave
50 50 translate
0.100 0.100 scale
0 setgray
/Helvetica findfont 100 scalefont setfont
newpath
-500.000000 -500.000000 translate
LTa
600 251 M
2817 0 V
LTb
600 251 M
63 0 V
2754 0 R
-63 0 V
600 623 M
63 0 V
2754 0 R
-63 0 V
600 994 M
63 0 V
2754 0 R
-63 0 V
600 1366 M
63 0 V
2754 0 R
-63 0 V
600 1737 M
63 0 V
2754 0 R
-63 0 V
600 2109 M
63 0 V
2754 0 R
-63 0 V
600 251 M
0 63 V
0 1795 R
0 -63 V
913 251 M
0 63 V
0 1795 R
0 -63 V
1226 251 M
0 63 V
0 1795 R
0 -63 V
1539 251 M
0 63 V
0 1795 R
0 -63 V
1852 251 M
0 63 V
0 1795 R
0 -63 V
2165 251 M
0 63 V
0 1795 R
0 -63 V
2478 251 M
0 63 V
0 1795 R
0 -63 V
2791 251 M
0 63 V
0 1795 R
0 -63 V
3104 251 M
0 63 V
0 1795 R
0 -63 V
3417 251 M
0 63 V
0 1795 R
0 -63 V
600 251 M
2817 0 V
0 1858 V
-2817 0 V
600 251 L
LT0
2000 1946 M
180 0 V
600 669 M
12 8 V
46 29 V
47 31 V
46 31 V
47 29 V
46 27 V
46 27 V
47 28 V
46 27 V
46 26 V
47 24 V
46 24 V
47 24 V
46 24 V
47 25 V
46 25 V
46 23 V
47 22 V
46 22 V
47 23 V
46 23 V
47 25 V
46 25 V
46 25 V
47 26 V
46 25 V
47 25 V
46 24 V
46 24 V
47 24 V
46 22 V
47 20 V
46 21 V
46 19 V
47 20 V
46 19 V
47 21 V
46 22 V
47 21 V
46 22 V
46 22 V
47 21 V
46 22 V
47 22 V
46 21 V
47 21 V
46 20 V
46 20 V
47 20 V
46 18 V
46 16 V
47 16 V
46 16 V
47 16 V
46 16 V
46 16 V
47 16 V
46 19 V
47 19 V
46 19 V
47 20 V
20 9 V
LT1
2000 1846 M
180 0 V
600 656 M
12 7 V
46 27 V
47 29 V
46 29 V
47 26 V
46 24 V
46 25 V
47 25 V
46 24 V
46 22 V
47 20 V
46 20 V
47 20 V
46 21 V
47 20 V
46 21 V
46 19 V
47 18 V
46 19 V
47 18 V
46 19 V
47 21 V
46 21 V
46 22 V
47 21 V
46 21 V
47 21 V
46 20 V
46 20 V
47 19 V
46 18 V
47 16 V
46 15 V
46 16 V
47 15 V
46 15 V
47 17 V
46 17 V
47 18 V
46 18 V
46 18 V
47 18 V
46 18 V
47 17 V
46 18 V
47 17 V
46 16 V
46 16 V
47 15 V
46 14 V
46 11 V
47 12 V
46 11 V
47 10 V
46 11 V
46 11 V
47 12 V
46 14 V
47 15 V
46 15 V
47 16 V
20 7 V
LT2
2000 1746 M
180 0 V
600 642 M
12 7 V
46 25 V
47 28 V
46 28 V
47 24 V
46 23 V
46 23 V
47 24 V
46 22 V
46 21 V
47 18 V
46 18 V
47 17 V
46 18 V
47 18 V
46 19 V
46 18 V
47 18 V
46 18 V
47 18 V
46 19 V
47 20 V
46 19 V
46 20 V
47 19 V
46 19 V
47 19 V
46 18 V
46 18 V
47 17 V
46 16 V
47 13 V
46 13 V
46 13 V
47 12 V
46 13 V
47 14 V
46 15 V
47 15 V
46 15 V
46 15 V
47 15 V
46 15 V
47 3 V
46 26 V
47 15 V
46 13 V
46 14 V
47 13 V
46 11 V
46 10 V
47 10 V
46 9 V
47 9 V
46 9 V
46 10 V
47 9 V
46 11 V
47 11 V
46 12 V
47 12 V
20 5 V
LT3
2000 1646 M
180 0 V
600 644 M
12 7 V
46 26 V
47 28 V
46 28 V
47 25 V
46 23 V
46 23 V
47 24 V
46 23 V
46 21 V
47 20 V
46 18 V
47 18 V
46 19 V
47 19 V
46 19 V
46 19 V
47 18 V
46 19 V
47 19 V
46 20 V
47 21 V
46 20 V
46 21 V
47 20 V
46 21 V
47 20 V
46 19 V
46 19 V
47 18 V
46 17 V
47 14 V
46 15 V
46 13 V
47 14 V
46 13 V
47 16 V
46 16 V
47 16 V
46 17 V
46 16 V
47 17 V
46 16 V
47 16 V
46 16 V
47 15 V
46 15 V
46 14 V
47 14 V
46 13 V
46 11 V
47 11 V
46 11 V
47 10 V
46 10 V
46 11 V
47 11 V
46 13 V
47 13 V
46 13 V
47 14 V
20 6 V
LT4
2000 1546 M
180 0 V
600 645 M
12 8 V
46 28 V
47 26 V
46 26 V
47 25 V
46 24 V
46 25 V
47 25 V
46 24 V
46 23 V
47 21 V
46 20 V
47 18 V
46 18 V
47 18 V
46 17 V
46 21 V
47 22 V
46 22 V
47 23 V
46 22 V
47 20 V
46 19 V
46 20 V
47 20 V
46 21 V
47 23 V
46 23 V
46 23 V
47 23 V
46 24 V
47 25 V
46 24 V
46 24 V
47 23 V
46 24 V
47 21 V
46 21 V
47 20 V
46 19 V
46 19 V
47 19 V
46 18 V
47 17 V
46 17 V
47 17 V
46 17 V
46 17 V
47 18 V
46 17 V
46 19 V
47 19 V
46 19 V
47 20 V
46 20 V
46 20 V
47 21 V
46 21 V
47 22 V
46 23 V
47 22 V
20 11 V
stroke
grestore
end
showpage
}
\put(1944,1546){\makebox(0,0)[r]{${\cal S}$ with Argonne $V_{18}$}}
\put(1944,1646){\makebox(0,0)[r]{${\cal S}$ with Reid93}}
\put(1944,1746){\makebox(0,0)[r]{${\cal S}$ with Nijm II}}
\put(1944,1846){\makebox(0,0)[r]{${\cal S}$ with Nijm I}}
\put(1944,1946){\makebox(0,0)[r]{${\cal S}$ with CD-Bonn}}
\put(2008,21){\makebox(0,0){$n$ (fm$^{-3}$)}}
\put(100,1180){%
\special{ps: gsave currentpoint currentpoint translate
270 rotate neg exch neg exch translate}%
\makebox(0,0)[b]{\shortstack{${\cal S}(n)$ (MeV)}}%
\special{ps: currentpoint grestore moveto}%
}
\put(3417,151){\makebox(0,0){1}}
\put(3104,151){\makebox(0,0){0.9}}
\put(2791,151){\makebox(0,0){0.8}}
\put(2478,151){\makebox(0,0){0.7}}
\put(2165,151){\makebox(0,0){0.6}}
\put(1852,151){\makebox(0,0){0.5}}
\put(1539,151){\makebox(0,0){0.4}}
\put(1226,151){\makebox(0,0){0.3}}
\put(913,151){\makebox(0,0){0.2}}
\put(600,151){\makebox(0,0){0.1}}
\put(540,2109){\makebox(0,0)[r]{100}}
\put(540,1737){\makebox(0,0)[r]{80}}
\put(540,1366){\makebox(0,0)[r]{60}}
\put(540,994){\makebox(0,0)[r]{40}}
\put(540,623){\makebox(0,0)[r]{20}}
\put(540,251){\makebox(0,0)[r]{0}}
\end{picture}
     \caption{Symmetry energy ${\cal S}$ as function of density $n$.}
     \label{fig:sec2fig5}
\end{center}\end{figure}
The differences seen in Figs.\ \ref{fig:sec2fig3} and \ref{fig:sec2fig4}
should also be reflected in the symmetry energy 
defined in Eq.\ (\ref{eq:symenergy}). This is seen 
in Fig.\ \ref{fig:sec2fig5} where we display the
symmetry energy for the above potentials
 as function of density $n$.
 From the differences in symmetry energies
one would then expect that properties like proton fractions in
$\beta$-stable matter will be influenced. This in turn has important 
consequences for 
the composition of matter in a neutron star and thereby eventual phases
present in dense matter. 
In Fig.\ \ref{fig:sec2fig6} we display the corresponding proton fractions 
obtained by
calculating the energy per particle through  the $G$-matrix of Eq.\ 
(\ref{eq:gnonrel})
and imposing the equilibrium conditions of Eq.\ (\ref{eq:npebetaequilibrium}) 
and including muons, see again Ref.\ \cite{ehobo96} for further details.
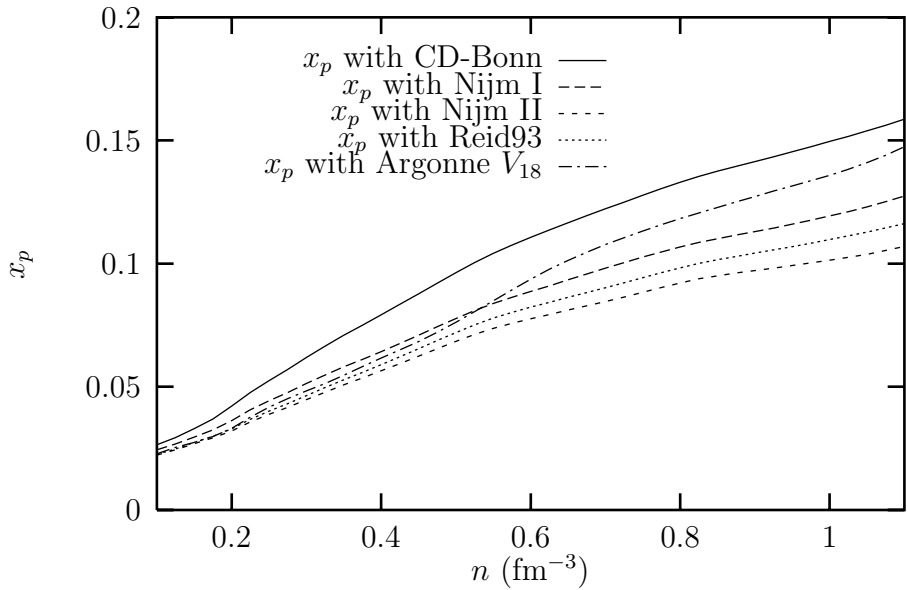
\begin{figure}\begin{center} 
     % GNUPLOT: LaTeX picture with Postscript
\setlength{\unitlength}{0.1bp}
\special{!
%!PS-Adobe-2.0
%%Creator: gnuplot
%%DocumentFonts: Helvetica
%%BoundingBox: 50 50 770 554
%%Pages: (atend)
%%EndComments
/gnudict 40 dict def
gnudict begin
/Color false def
/Solid false def
/gnulinewidth 5.000 def
/vshift -33 def
/dl {10 mul} def
/hpt 31.5 def
/vpt 31.5 def
/M {moveto} bind def
/L {lineto} bind def
/R {rmoveto} bind def
/V {rlineto} bind def
/vpt2 vpt 2 mul def
/hpt2 hpt 2 mul def
/Lshow { currentpoint stroke M
  0 vshift R show } def
/Rshow { currentpoint stroke M
  dup stringwidth pop neg vshift R show } def
/Cshow { currentpoint stroke M
  dup stringwidth pop -2 div vshift R show } def
/DL { Color {setrgbcolor Solid {pop []} if 0 setdash }
 {pop pop pop Solid {pop []} if 0 setdash} ifelse } def
/BL { stroke gnulinewidth 2 mul setlinewidth } def
/AL { stroke gnulinewidth 2 div setlinewidth } def
/PL { stroke gnulinewidth setlinewidth } def
/LTb { BL [] 0 0 0 DL } def
/LTa { AL [1 dl 2 dl] 0 setdash 0 0 0 setrgbcolor } def
/LT0 { PL [] 0 1 0 DL } def
/LT1 { PL [4 dl 2 dl] 0 0 1 DL } def
/LT2 { PL [2 dl 3 dl] 1 0 0 DL } def
/LT3 { PL [1 dl 1.5 dl] 1 0 1 DL } def
/LT4 { PL [5 dl 2 dl 1 dl 2 dl] 0 1 1 DL } def
/LT5 { PL [4 dl 3 dl 1 dl 3 dl] 1 1 0 DL } def
/LT6 { PL [2 dl 2 dl 2 dl 4 dl] 0 0 0 DL } def
/LT7 { PL [2 dl 2 dl 2 dl 2 dl 2 dl 4 dl] 1 0.3 0 DL } def
/LT8 { PL [2 dl 2 dl 2 dl 2 dl 2 dl 2 dl 2 dl 4 dl] 0.5 0.5 0.5 DL } def
/P { stroke [] 0 setdash
  currentlinewidth 2 div sub M
  0 currentlinewidth V stroke } def
/D { stroke [] 0 setdash 2 copy vpt add M
  hpt neg vpt neg V hpt vpt neg V
  hpt vpt V hpt neg vpt V closepath stroke
  P } def
/A { stroke [] 0 setdash vpt sub M 0 vpt2 V
  currentpoint stroke M
  hpt neg vpt neg R hpt2 0 V stroke
  } def
/B { stroke [] 0 setdash 2 copy exch hpt sub exch vpt add M
  0 vpt2 neg V hpt2 0 V 0 vpt2 V
  hpt2 neg 0 V closepath stroke
  P } def
/C { stroke [] 0 setdash exch hpt sub exch vpt add M
  hpt2 vpt2 neg V currentpoint stroke M
  hpt2 neg 0 R hpt2 vpt2 V stroke } def
/T { stroke [] 0 setdash 2 copy vpt 1.12 mul add M
  hpt neg vpt -1.62 mul V
  hpt 2 mul 0 V
  hpt neg vpt 1.62 mul V closepath stroke
  P  } def
/S { 2 copy A C} def
end
}
\begin{picture}(3600,2160)(0,0)
\special{"
gnudict begin
gsave
50 50 translate
0.100 0.100 scale
0 setgray
/Helvetica findfont 100 scalefont setfont
newpath
-500.000000 -500.000000 translate
LTa
600 251 M
2817 0 V
LTb
600 251 M
63 0 V
2754 0 R
-63 0 V
600 716 M
63 0 V
2754 0 R
-63 0 V
600 1180 M
63 0 V
2754 0 R
-63 0 V
600 1645 M
63 0 V
2754 0 R
-63 0 V
600 2109 M
63 0 V
2754 0 R
-63 0 V
882 251 M
0 63 V
0 1795 R
0 -63 V
1445 251 M
0 63 V
0 1795 R
0 -63 V
2009 251 M
0 63 V
0 1795 R
0 -63 V
2572 251 M
0 63 V
0 1795 R
0 -63 V
3135 251 M
0 63 V
0 1795 R
0 -63 V
600 251 M
2817 0 V
0 1858 V
-2817 0 V
600 251 L
LT0
2114 1946 M
180 0 V
600 497 M
70 27 V
71 34 V
70 36 V
71 49 V
70 51 V
71 45 V
70 42 V
70 45 V
71 43 V
70 41 V
71 38 V
70 39 V
71 40 V
70 39 V
70 40 V
71 40 V
70 38 V
71 35 V
70 31 V
70 29 V
71 28 V
70 27 V
71 27 V
70 26 V
71 25 V
70 26 V
70 25 V
71 24 V
70 22 V
71 20 V
70 18 V
71 18 V
70 18 V
70 19 V
71 19 V
70 20 V
71 19 V
70 21 V
71 21 V
70 23 V
LT1
2114 1846 M
180 0 V
600 478 M
70 22 V
71 26 V
70 27 V
71 35 V
70 41 V
71 34 V
70 31 V
70 34 V
71 32 V
70 31 V
71 28 V
70 29 V
71 31 V
70 31 V
70 32 V
71 31 V
70 30 V
71 27 V
70 23 V
70 22 V
71 21 V
70 23 V
71 22 V
70 22 V
71 21 V
70 21 V
70 19 V
71 19 V
70 17 V
71 14 V
70 14 V
71 13 V
70 13 V
70 15 V
71 15 V
70 16 V
71 16 V
70 18 V
71 20 V
70 21 V
LT2
2114 1746 M
180 0 V
600 458 M
70 20 V
71 23 V
70 22 V
71 26 V
70 34 V
71 29 V
70 26 V
70 29 V
71 28 V
70 28 V
71 26 V
70 27 V
71 27 V
70 28 V
70 28 V
71 28 V
70 26 V
71 23 V
70 19 V
70 17 V
71 16 V
70 17 V
71 16 V
70 16 V
71 17 V
70 18 V
70 18 V
71 17 V
70 16 V
71 11 V
70 11 V
71 9 V
70 9 V
70 10 V
71 10 V
70 10 V
71 11 V
70 11 V
71 15 V
70 15 V
LT3
2114 1646 M
180 0 V
600 461 M
70 20 V
71 25 V
70 23 V
71 28 V
70 36 V
71 31 V
70 28 V
70 30 V
71 31 V
70 29 V
71 28 V
70 29 V
71 30 V
70 30 V
70 31 V
71 30 V
70 29 V
71 26 V
70 21 V
70 20 V
71 18 V
70 19 V
71 18 V
70 18 V
71 19 V
70 19 V
70 19 V
71 18 V
70 17 V
71 14 V
70 12 V
71 13 V
70 12 V
70 12 V
71 13 V
70 14 V
71 14 V
70 14 V
71 16 V
70 16 V
LT4
2114 1546 M
180 0 V
600 463 M
70 24 V
71 19 V
70 22 V
71 31 V
70 42 V
71 38 V
70 32 V
70 29 V
71 28 V
70 31 V
71 33 V
70 32 V
71 31 V
70 31 V
70 35 V
71 38 V
70 40 V
71 41 V
70 41 V
70 40 V
71 38 V
70 33 V
71 31 V
70 29 V
71 26 V
70 25 V
70 24 V
71 23 V
70 21 V
71 21 V
70 21 V
71 20 V
70 21 V
70 19 V
71 20 V
70 20 V
71 23 V
70 25 V
71 28 V
70 32 V
stroke
grestore
end
showpage
}
\put(2054,1546){\makebox(0,0)[r]{$x_p$ with Argonne $V_{18}$}}
\put(2054,1646){\makebox(0,0)[r]{$x_p$ with Reid93}}
\put(2054,1746){\makebox(0,0)[r]{$x_p$ with Nijm II}}
\put(2054,1846){\makebox(0,0)[r]{$x_p$ with Nijm I}}
\put(2054,1946){\makebox(0,0)[r]{$x_p$ with CD-Bonn}}
\put(2008,21){\makebox(0,0){$n$ (fm$^{-3}$)}}
\put(100,1180){%
\special{ps: gsave currentpoint currentpoint translate
270 rotate neg exch neg exch translate}%
\makebox(0,0)[b]{\shortstack{$x_p$}}%
\special{ps: currentpoint grestore moveto}%
}
\put(3135,151){\makebox(0,0){1}}
\put(2572,151){\makebox(0,0){0.8}}
\put(2009,151){\makebox(0,0){0.6}}
\put(1445,151){\makebox(0,0){0.4}}
\put(882,151){\makebox(0,0){0.2}}
\put(540,2109){\makebox(0,0)[r]{0.2}}
\put(540,1645){\makebox(0,0)[r]{0.15}}
\put(540,1180){\makebox(0,0)[r]{0.1}}
\put(540,716){\makebox(0,0)[r]{0.05}}
\put(540,251){\makebox(0,0)[r]{0}}
\end{picture}
     \caption{Proton fraction $x_p$ for $\beta$-stable 
              matter as function of density $n$}
     \label{fig:sec2fig6}
\end{center}\end{figure}
 From Fig.\ \ref{fig:sec2fig6} one notices that the potential with the largest 
symmetry energy, the CD-Bonn interaction, is also the one which gives the 
largest
proton fractions. This means in turn that the so-called direct Urca process
can occcur at lower densities. For the CD-Bonn interaction this happens
at 0.88 fm$^{-3}$, for the Nijm I it starts at 1.25 fm$^{-3}$ while for the 
Reid93 interaction one reaches the critical density at 1.36 fm$^{-3}$.
The Argonne potential allows for the direct Urca process at a density of
1.05 fm$^{-3}$. For the Nijm II we were not able to get the direct Urca process
for densities below 1.5 fm$^{-3}$.

It is also interesting to notice that the symmetry energy increases rather
monotonously for all potentials. This means that the higher order derivatives 
in Eq.\ (\ref{eq:energyexpansion}) can be neglected and that we can, to a good
approximation,  associate 
the second derivative $d^2 {\cal E}/dx_p^2$ 
with the symmetry energy ${\cal S} (n)$ in the empirical
mass formula. With this mind, one can calculate the proton fraction employing 
the 
theoretically derived symmetry energy shown in Fig.\ \ref{fig:sec2fig5} using
the simple formula of Eq.\ (\ref{eq:crudeprotonfraction}). A good agreement
is in general obtained with the above simple formula, see e.g.\  Ref.\
\cite{ehobo96} and the parametrization in subsection \ref{subsec:paramsec2}. 

\begin{figure}\begin{center} 
     % GNUPLOT: LaTeX picture with Postscript
\setlength{\unitlength}{0.1bp}
\special{!
%!PS-Adobe-2.0
%%Creator: gnuplot
%%DocumentFonts: Helvetica
%%BoundingBox: 50 50 770 554
%%Pages: (atend)
%%EndComments
/gnudict 40 dict def
gnudict begin
/Color false def
/Solid false def
/gnulinewidth 5.000 def
/vshift -33 def
/dl {10 mul} def
/hpt 31.5 def
/vpt 31.5 def
/M {moveto} bind def
/L {lineto} bind def
/R {rmoveto} bind def
/V {rlineto} bind def
/vpt2 vpt 2 mul def
/hpt2 hpt 2 mul def
/Lshow { currentpoint stroke M
  0 vshift R show } def
/Rshow { currentpoint stroke M
  dup stringwidth pop neg vshift R show } def
/Cshow { currentpoint stroke M
  dup stringwidth pop -2 div vshift R show } def
/DL { Color {setrgbcolor Solid {pop []} if 0 setdash }
 {pop pop pop Solid {pop []} if 0 setdash} ifelse } def
/BL { stroke gnulinewidth 2 mul setlinewidth } def
/AL { stroke gnulinewidth 2 div setlinewidth } def
/PL { stroke gnulinewidth setlinewidth } def
/LTb { BL [] 0 0 0 DL } def
/LTa { AL [1 dl 2 dl] 0 setdash 0 0 0 setrgbcolor } def
/LT0 { PL [] 0 1 0 DL } def
/LT1 { PL [4 dl 2 dl] 0 0 1 DL } def
/LT2 { PL [2 dl 3 dl] 1 0 0 DL } def
/LT3 { PL [1 dl 1.5 dl] 1 0 1 DL } def
/LT4 { PL [5 dl 2 dl 1 dl 2 dl] 0 1 1 DL } def
/LT5 { PL [4 dl 3 dl 1 dl 3 dl] 1 1 0 DL } def
/LT6 { PL [2 dl 2 dl 2 dl 4 dl] 0 0 0 DL } def
/LT7 { PL [2 dl 2 dl 2 dl 2 dl 2 dl 4 dl] 1 0.3 0 DL } def
/LT8 { PL [2 dl 2 dl 2 dl 2 dl 2 dl 2 dl 2 dl 4 dl] 0.5 0.5 0.5 DL } def
/P { stroke [] 0 setdash
  currentlinewidth 2 div sub M
  0 currentlinewidth V stroke } def
/D { stroke [] 0 setdash 2 copy vpt add M
  hpt neg vpt neg V hpt vpt neg V
  hpt vpt V hpt neg vpt V closepath stroke
  P } def
/A { stroke [] 0 setdash vpt sub M 0 vpt2 V
  currentpoint stroke M
  hpt neg vpt neg R hpt2 0 V stroke
  } def
/B { stroke [] 0 setdash 2 copy exch hpt sub exch vpt add M
  0 vpt2 neg V hpt2 0 V 0 vpt2 V
  hpt2 neg 0 V closepath stroke
  P } def
/C { stroke [] 0 setdash exch hpt sub exch vpt add M
  hpt2 vpt2 neg V currentpoint stroke M
  hpt2 neg 0 R hpt2 vpt2 V stroke } def
/T { stroke [] 0 setdash 2 copy vpt 1.12 mul add M
  hpt neg vpt -1.62 mul V
  hpt 2 mul 0 V
  hpt neg vpt 1.62 mul V closepath stroke
  P  } def
/S { 2 copy A C} def
end
}
\begin{picture}(3600,2160)(0,0)
\special{"
gnudict begin
gsave
50 50 translate
0.100 0.100 scale
0 setgray
/Helvetica findfont 100 scalefont setfont
newpath
-500.000000 -500.000000 translate
LTa
600 251 M
2817 0 V
LTb
600 251 M
63 0 V
2754 0 R
-63 0 V
600 561 M
63 0 V
2754 0 R
-63 0 V
600 870 M
63 0 V
2754 0 R
-63 0 V
600 1180 M
63 0 V
2754 0 R
-63 0 V
600 1490 M
63 0 V
2754 0 R
-63 0 V
600 1799 M
63 0 V
2754 0 R
-63 0 V
600 2109 M
63 0 V
2754 0 R
-63 0 V
600 251 M
0 63 V
0 1795 R
0 -63 V
913 251 M
0 63 V
0 1795 R
0 -63 V
1226 251 M
0 63 V
0 1795 R
0 -63 V
1539 251 M
0 63 V
0 1795 R
0 -63 V
1852 251 M
0 63 V
0 1795 R
0 -63 V
2165 251 M
0 63 V
0 1795 R
0 -63 V
2478 251 M
0 63 V
0 1795 R
0 -63 V
2791 251 M
0 63 V
0 1795 R
0 -63 V
3104 251 M
0 63 V
0 1795 R
0 -63 V
3417 251 M
0 63 V
0 1795 R
0 -63 V
600 251 M
2817 0 V
0 1858 V
-2817 0 V
600 251 L
LT0
3114 1900 M
180 0 V
600 1130 M
78 67 V
79 59 V
78 51 V
78 46 V
78 43 V
79 39 V
78 37 V
78 34 V
78 32 V
79 30 V
78 28 V
78 27 V
78 25 V
79 25 V
78 24 V
78 23 V
78 22 V
79 23 V
78 21 V
78 21 V
78 21 V
79 19 V
78 17 V
78 18 V
78 16 V
79 17 V
78 15 V
78 15 V
78 15 V
79 16 V
78 15 V
78 15 V
78 15 V
79 15 V
78 15 V
78 15 V
LT1
3114 1750 M
180 0 V
600 536 M
78 31 V
79 26 V
78 33 V
78 32 V
78 31 V
79 24 V
78 26 V
78 25 V
78 25 V
79 26 V
78 26 V
78 26 V
78 25 V
79 22 V
78 19 V
78 22 V
78 18 V
79 16 V
78 16 V
78 13 V
78 13 V
79 18 V
78 20 V
78 19 V
78 19 V
79 19 V
78 21 V
78 19 V
78 16 V
79 13 V
78 17 V
78 13 V
78 13 V
79 11 V
78 12 V
78 8 V
LT2
3114 1600 M
180 0 V
600 535 M
78 29 V
79 23 V
78 28 V
78 26 V
78 21 V
79 19 V
78 18 V
78 18 V
78 18 V
79 18 V
78 19 V
78 18 V
78 17 V
79 17 V
78 15 V
78 15 V
78 15 V
79 12 V
78 11 V
78 10 V
78 10 V
79 13 V
78 15 V
78 14 V
78 15 V
79 15 V
78 15 V
78 13 V
78 13 V
79 11 V
78 11 V
78 10 V
78 11 V
79 9 V
78 8 V
78 7 V
LT3
3114 1450 M
180 0 V
600 251 M
78 0 V
79 52 V
78 99 V
78 54 V
78 38 V
79 30 V
78 28 V
78 26 V
78 26 V
79 24 V
78 24 V
78 24 V
78 21 V
79 21 V
78 19 V
78 18 V
78 17 V
79 15 V
78 12 V
78 12 V
78 13 V
79 14 V
78 16 V
78 16 V
78 17 V
79 17 V
78 16 V
78 16 V
78 14 V
79 13 V
78 11 V
78 12 V
78 11 V
79 10 V
78 9 V
78 8 V
stroke
grestore
end
showpage
}
\put(3054,1450){\makebox(0,0)[r]{$k_F^{\mu}$}}
\put(3054,1600){\makebox(0,0)[r]{$k_F^e$}}
\put(3054,1750){\makebox(0,0)[r]{$k_F^p$}}
\put(3054,1900){\makebox(0,0)[r]{$k_F^n$}}
\put(2008,20){\makebox(0,0){$n$ (fm$^{-3}$)}}
\put(100,1180){%
\special{ps: gsave currentpoint currentpoint translate
270 rotate neg exch neg exch translate}%
\makebox(0,0)[b]{\shortstack{$k_F$ (fm$^{-1}$)}}%
\special{ps: currentpoint grestore moveto}%
}
\put(3417,151){\makebox(0,0){1}}
\put(3104,151){\makebox(0,0){0.9}}
\put(2791,151){\makebox(0,0){0.8}}
\put(2478,151){\makebox(0,0){0.7}}
\put(2165,151){\makebox(0,0){0.6}}
\put(1852,151){\makebox(0,0){0.5}}
\put(1539,151){\makebox(0,0){0.4}}
\put(1226,151){\makebox(0,0){0.3}}
\put(913,151){\makebox(0,0){0.2}}
\put(600,151){\makebox(0,0){0.1}}
\put(540,2109){\makebox(0,0)[r]{3}}
\put(540,1799){\makebox(0,0)[r]{2.5}}
\put(540,1490){\makebox(0,0)[r]{2}}
\put(540,1180){\makebox(0,0)[r]{1.5}}
\put(540,870){\makebox(0,0)[r]{1}}
\put(540,561){\makebox(0,0)[r]{0.5}}
\put(540,251){\makebox(0,0)[r]{0}}
\end{picture}
     \caption{Fermi momenta for neutrons, protons, electrons and muons in 
              $\beta$-stable matter with the CD-Bonn interaction.}
     \label{fig:sec2fig7}
\end{center}\end{figure}
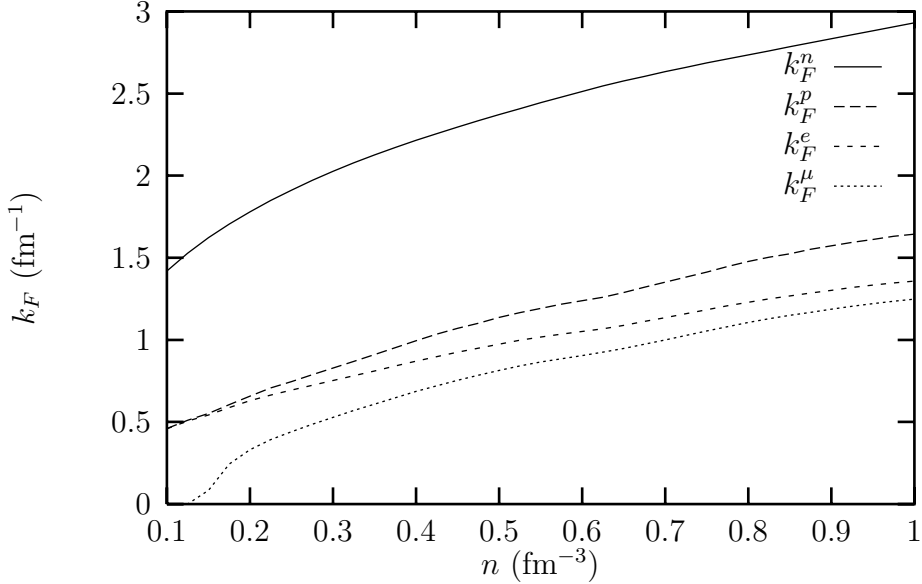

With the results for the proton fractions of Fig.\ \ref{fig:sec2fig6} in mind,
we plot in Fig.\ \ref{fig:sec2fig7} 
the Fermi momenta for  electrons, muons and nucleons obtained 
with the CD-Bonn interaction, the other 
potentials yield qualitatively similar results although the proton fraction
is slightly smaller. 
 From this figure one notices that muons appear at a density close to the 
saturation
density of nuclear matter, as expected. This can easily be seen if one were to
calculate $\beta$-stable matter with non-interacting particles only,
see e.g.\  Ref.\ \cite{st83}.

\begin{figure}\begin{center} 
     \input{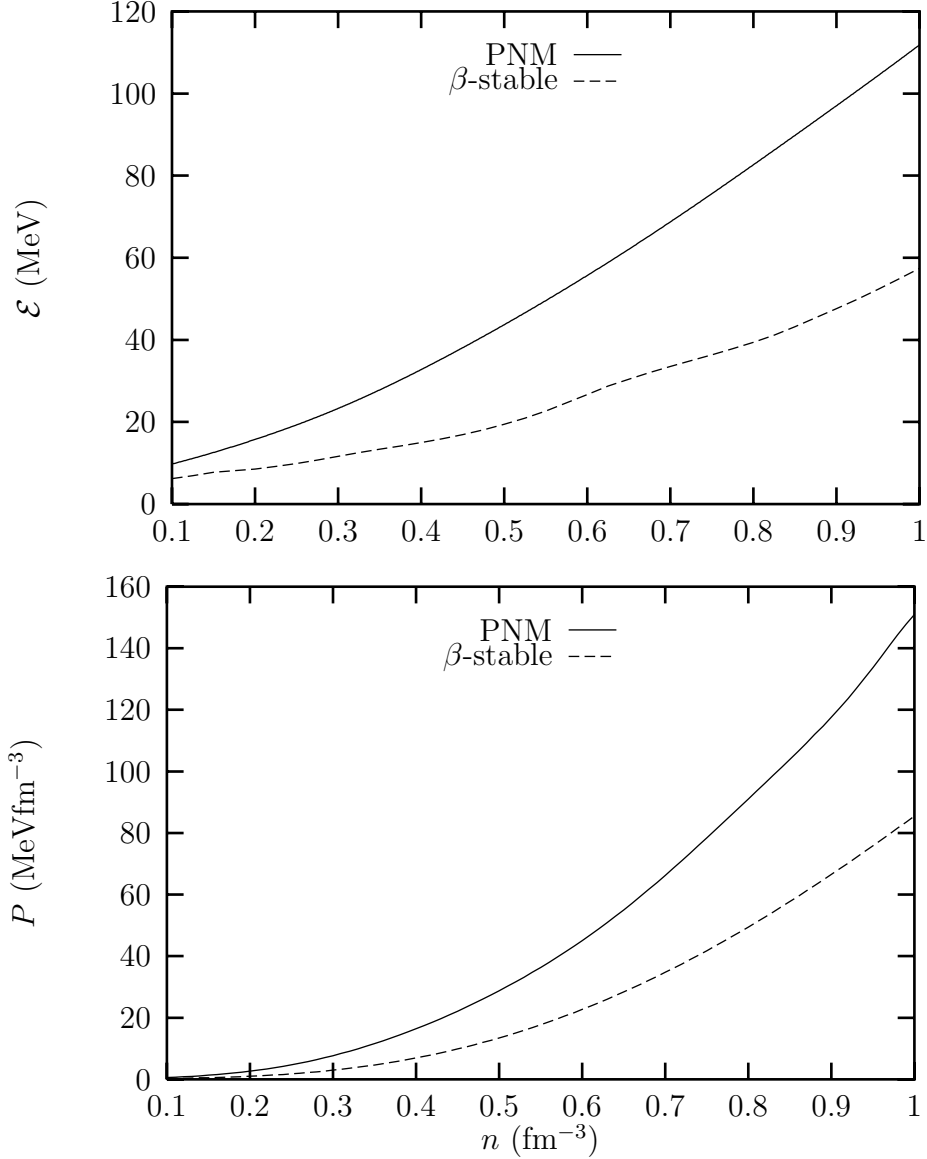}
     \caption{Upper panel: BHF energy per nucleon for pure neutron matter 
              and $\beta$-stable matter
              obtained with the CD-Bonn interaction.
              Lower panel : the corresponding pressure $P$.}
     \label{fig:sec2fig8}
\end{center}\end{figure}

Of further interest is the difference in energy per particle ${\cal E}(n,x_p)$
for pure neutron matter (PNM) and matter in $\beta$-equilibrium.
In Fig.\ \ref{fig:sec2fig8} 
we display  ${\cal E}(n,x_p)$ for PNM and $\beta$-stable matter
for results with the CD-Bonn interaction only since the other potentials
yield qualitatively similar results.
Obviously, as seen from Fig.\ \ref{fig:sec2fig8} the energy per particle
for $\beta$-stable matter yields a softer EoS, since the repulsive
energy per particle in PNM receives attractive contributions
from the $T_z=0$ channel. This is reflected in the corresponding 
pressure as well (lower
panel of the same figure) and 
will in turn result in neutron stars
with smaller total masses compared with the PNM case.

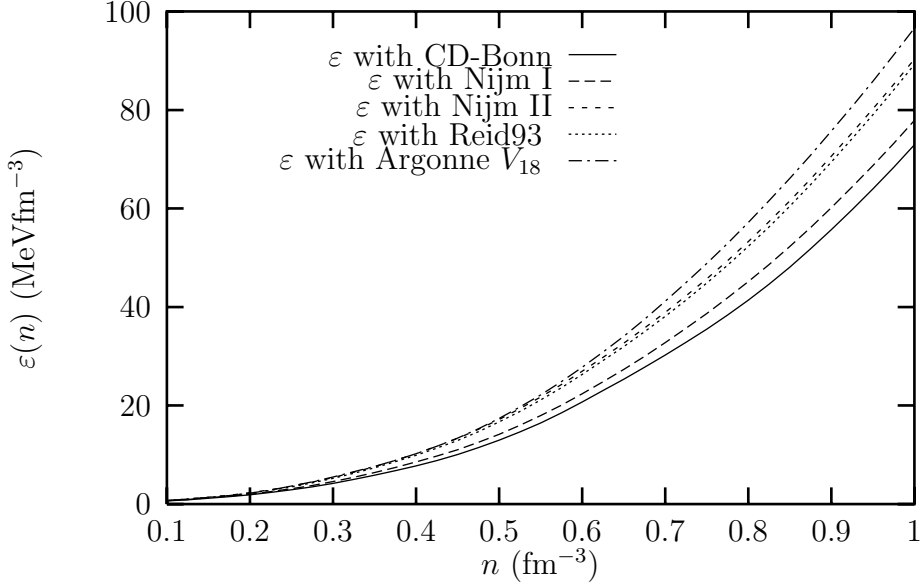
\begin{figure}\begin{center} 
     % GNUPLOT: LaTeX picture with Postscript
\setlength{\unitlength}{0.1bp}
\special{!
%!PS-Adobe-2.0
%%Creator: gnuplot
%%DocumentFonts: Helvetica
%%BoundingBox: 50 50 770 554
%%Pages: (atend)
%%EndComments
/gnudict 40 dict def
gnudict begin
/Color false def
/Solid false def
/gnulinewidth 5.000 def
/vshift -33 def
/dl {10 mul} def
/hpt 31.5 def
/vpt 31.5 def
/M {moveto} bind def
/L {lineto} bind def
/R {rmoveto} bind def
/V {rlineto} bind def
/vpt2 vpt 2 mul def
/hpt2 hpt 2 mul def
/Lshow { currentpoint stroke M
  0 vshift R show } def
/Rshow { currentpoint stroke M
  dup stringwidth pop neg vshift R show } def
/Cshow { currentpoint stroke M
  dup stringwidth pop -2 div vshift R show } def
/DL { Color {setrgbcolor Solid {pop []} if 0 setdash }
 {pop pop pop Solid {pop []} if 0 setdash} ifelse } def
/BL { stroke gnulinewidth 2 mul setlinewidth } def
/AL { stroke gnulinewidth 2 div setlinewidth } def
/PL { stroke gnulinewidth setlinewidth } def
/LTb { BL [] 0 0 0 DL } def
/LTa { AL [1 dl 2 dl] 0 setdash 0 0 0 setrgbcolor } def
/LT0 { PL [] 0 1 0 DL } def
/LT1 { PL [4 dl 2 dl] 0 0 1 DL } def
/LT2 { PL [2 dl 3 dl] 1 0 0 DL } def
/LT3 { PL [1 dl 1.5 dl] 1 0 1 DL } def
/LT4 { PL [5 dl 2 dl 1 dl 2 dl] 0 1 1 DL } def
/LT5 { PL [4 dl 3 dl 1 dl 3 dl] 1 1 0 DL } def
/LT6 { PL [2 dl 2 dl 2 dl 4 dl] 0 0 0 DL } def
/LT7 { PL [2 dl 2 dl 2 dl 2 dl 2 dl 4 dl] 1 0.3 0 DL } def
/LT8 { PL [2 dl 2 dl 2 dl 2 dl 2 dl 2 dl 2 dl 4 dl] 0.5 0.5 0.5 DL } def
/P { stroke [] 0 setdash
  currentlinewidth 2 div sub M
  0 currentlinewidth V stroke } def
/D { stroke [] 0 setdash 2 copy vpt add M
  hpt neg vpt neg V hpt vpt neg V
  hpt vpt V hpt neg vpt V closepath stroke
  P } def
/A { stroke [] 0 setdash vpt sub M 0 vpt2 V
  currentpoint stroke M
  hpt neg vpt neg R hpt2 0 V stroke
  } def
/B { stroke [] 0 setdash 2 copy exch hpt sub exch vpt add M
  0 vpt2 neg V hpt2 0 V 0 vpt2 V
  hpt2 neg 0 V closepath stroke
  P } def
/C { stroke [] 0 setdash exch hpt sub exch vpt add M
  hpt2 vpt2 neg V currentpoint stroke M
  hpt2 neg 0 R hpt2 vpt2 V stroke } def
/T { stroke [] 0 setdash 2 copy vpt 1.12 mul add M
  hpt neg vpt -1.62 mul V
  hpt 2 mul 0 V
  hpt neg vpt 1.62 mul V closepath stroke
  P  } def
/S { 2 copy A C} def
end
}
\begin{picture}(3600,2160)(0,0)
\special{"
gnudict begin
gsave
50 50 translate
0.100 0.100 scale
0 setgray
/Helvetica findfont 100 scalefont setfont
newpath
-500.000000 -500.000000 translate
LTa
600 251 M
2817 0 V
LTb
600 251 M
63 0 V
2754 0 R
-63 0 V
600 623 M
63 0 V
2754 0 R
-63 0 V
600 994 M
63 0 V
2754 0 R
-63 0 V
600 1366 M
63 0 V
2754 0 R
-63 0 V
600 1737 M
63 0 V
2754 0 R
-63 0 V
600 2109 M
63 0 V
2754 0 R
-63 0 V
600 251 M
0 63 V
0 1795 R
0 -63 V
913 251 M
0 63 V
0 1795 R
0 -63 V
1226 251 M
0 63 V
0 1795 R
0 -63 V
1539 251 M
0 63 V
0 1795 R
0 -63 V
1852 251 M
0 63 V
0 1795 R
0 -63 V
2165 251 M
0 63 V
0 1795 R
0 -63 V
2478 251 M
0 63 V
0 1795 R
0 -63 V
2791 251 M
0 63 V
0 1795 R
0 -63 V
3104 251 M
0 63 V
0 1795 R
0 -63 V
3417 251 M
0 63 V
0 1795 R
0 -63 V
600 251 M
2817 0 V
0 1858 V
-2817 0 V
600 251 L
LT0
2114 1946 M
180 0 V
600 263 M
78 4 V
79 6 V
78 6 V
78 7 V
78 9 V
79 10 V
78 11 V
78 13 V
78 15 V
79 16 V
78 17 V
78 18 V
78 20 V
79 23 V
78 26 V
78 28 V
78 31 V
79 34 V
78 38 V
78 41 V
78 43 V
79 43 V
78 45 V
78 46 V
78 48 V
79 50 V
78 53 V
78 56 V
78 60 V
79 65 V
78 69 V
78 72 V
78 75 V
79 79 V
78 81 V
78 85 V
LT1
2114 1846 M
180 0 V
600 264 M
78 4 V
79 7 V
78 6 V
78 7 V
78 10 V
79 11 V
78 12 V
78 15 V
78 16 V
79 18 V
78 19 V
78 21 V
78 22 V
79 24 V
78 28 V
78 30 V
78 34 V
79 37 V
78 39 V
78 43 V
78 45 V
79 47 V
78 49 V
78 52 V
78 54 V
79 56 V
78 59 V
78 61 V
78 64 V
79 69 V
78 71 V
78 75 V
78 78 V
79 81 V
78 83 V
78 87 V
LT2
2114 1746 M
180 0 V
600 264 M
78 6 V
79 7 V
78 7 V
78 10 V
78 11 V
79 14 V
78 15 V
78 17 V
78 20 V
79 21 V
78 24 V
78 25 V
78 28 V
79 30 V
78 34 V
78 37 V
78 41 V
79 44 V
78 47 V
78 50 V
78 52 V
79 54 V
78 57 V
78 59 V
78 62 V
79 65 V
78 68 V
78 72 V
78 75 V
79 79 V
78 82 V
78 86 V
78 88 V
79 91 V
78 92 V
78 95 V
LT3
2114 1646 M
180 0 V
600 264 M
78 6 V
79 6 V
78 7 V
78 9 V
78 11 V
79 13 V
78 15 V
78 17 V
78 19 V
79 21 V
78 23 V
78 24 V
78 27 V
79 29 V
78 33 V
78 37 V
78 39 V
79 44 V
78 46 V
78 50 V
78 52 V
79 53 V
78 56 V
78 58 V
78 61 V
79 65 V
78 67 V
78 71 V
78 73 V
79 79 V
78 81 V
78 85 V
78 88 V
79 91 V
78 93 V
78 96 V
LT4
2114 1546 M
180 0 V
600 264 M
78 6 V
79 6 V
78 7 V
78 9 V
78 12 V
79 15 V
78 16 V
78 18 V
78 20 V
79 20 V
78 23 V
78 25 V
78 28 V
79 31 V
78 35 V
78 39 V
78 43 V
79 46 V
78 50 V
78 54 V
78 57 V
79 60 V
78 64 V
78 68 V
78 70 V
79 73 V
78 75 V
78 79 V
78 82 V
79 85 V
78 87 V
78 91 V
78 92 V
79 96 V
78 99 V
78 101 V
stroke
grestore
end
showpage
}
\put(2054,1546){\makebox(0,0)[r]{${\varepsilon}$ with Argonne $V_{18}$ }}
\put(2054,1646){\makebox(0,0)[r]{${\varepsilon}$ with Reid93 }}
\put(2054,1746){\makebox(0,0)[r]{${\varepsilon}$ with Nijm II}}
\put(2054,1846){\makebox(0,0)[r]{${\varepsilon}$ with Nijm I}}
\put(2054,1946){\makebox(0,0)[r]{${\varepsilon}$ with CD-Bonn}}
\put(2008,21){\makebox(0,0){$n$ (fm$^{-3}$)}}
\put(100,1180){%
\special{ps: gsave currentpoint currentpoint translate
270 rotate neg exch neg exch translate}%
\makebox(0,0)[b]{\shortstack{${\varepsilon}(n)$ (MeVfm$^{-3}$)}}%
\special{ps: currentpoint grestore moveto}%
}
\put(3417,151){\makebox(0,0){1}}
\put(3104,151){\makebox(0,0){0.9}}
\put(2791,151){\makebox(0,0){0.8}}
\put(2478,151){\makebox(0,0){0.7}}
\put(2165,151){\makebox(0,0){0.6}}
\put(1852,151){\makebox(0,0){0.5}}
\put(1539,151){\makebox(0,0){0.4}}
\put(1226,151){\makebox(0,0){0.3}}
\put(913,151){\makebox(0,0){0.2}}
\put(600,151){\makebox(0,0){0.1}}
\put(540,2109){\makebox(0,0)[r]{100}}
\put(540,1737){\makebox(0,0)[r]{80}}
\put(540,1366){\makebox(0,0)[r]{60}}
\put(540,994){\makebox(0,0)[r]{40}}
\put(540,623){\makebox(0,0)[r]{20}}
\put(540,251){\makebox(0,0)[r]{0}}
\end{picture}
     \caption{The total energy in $\beta$-stable 
              matter as function of density $n$.}
     \label{fig:sec2fig10}
\end{center}\end{figure}

We end this subsection by plotting
in Fig.\ \ref{fig:sec2fig10} the energy density per baryon
${\cal E}$ (including the contribution from leptons)
for  $\beta$-stable matter. 
Since the proton fractions are not too large,
see Fig.\ \ref{fig:sec2fig6}, 
the most important contribution to $\varepsilon$ and ${\cal E}$ 
stems from the $T_z=-1$
channel
and  the contribution from the nuclear tensor force, especially via the 
$^3S_1$ and $^3S_1$-$^3D_1$ contributions with $T_z=0$, plays 
a less significant role than what seen for the potential energy in symmetric
nuclear matter in Fig.\ \ref{fig:sec2fig3}. Thus,
the main contribution to the differences between the various
potentials arises from the $T_z=1$ channel.
For the $T_Z=1$ channel, the new potentials yield also results for 
$\beta$-stable 
matter in close agreement.

It is thus gratifying that the new NN interactions yield similar 
energies per particle in neutron star matter. As we will see in the 
next subsection also, more sophisticated many-body calculations 
at the two-body level yield rather similar results for neutron matter,
since the strong nuclear tensor force in the $T_z=0$ channel
is not present in pure neutron matter or in a less important
way in $\beta$-stable matter. However, contributions coming from
real three-body interaction or relativistic effects, may alter this picture.
This is the topic of the next two subsections.

\subsubsection{Higher-order many-body calculations }\label{subsubsec:manybody}

The previous subsection served to establish the connection between
the nucleon-nucleon interaction and the simplest many-body approach 
possible, namely the summation of so-called ladder diagrams by the 
Brueckner-Hartree-Fock (BHF) method. We will label these calculations
as lowest-order Brueckner (LOB) theory.
This allowed us to see how the free
NN interaction gets modified in a nuclear medium. Eventual differences
in e.g.\  the EoS for neutron star matter could then be retraced
to properties of the various NN interactions. The new 
high-quality NN interactions
have also narrowed the differences  at higher densities observed
in the literature for older 
interaction models, see e.g.\  the discussion in Ref.\ \cite{ehmmp97}.
However, it is well-known that the BHF method in its simplest
form is not fully appropriate
for a  description of dense matter. More complicated many-body terms
arising from core-polarization effects, effective three-body 
and many-body diagrams and eventually the inclusion of three-body 
forces are expected to be important at densities above $n_0$.
Moreover, LOB theory suffers from other pathologies like
the lack of conservation of number of particles \cite{ms92}.
The need to include e.g.\  three-body interactions is seen already at the
level of the triton since all the above potentials underbind the triton,
albeit the discrepancies which existed earlier have been reduced.
To give an example, the old Reid potential \cite{reid68} gave a binding
energy of $-7.35$ MeV while a precursor to the CD-Bonn interaction, the Bonn A
interaction \cite{mac89} resulted in $-8.35$ MeV. The modern potentials
yield results in a range from $-7.6$ to $-8.0$ MeV. 

Here we will therefore review three possible improvements 
to LOB theory and discuss the resulting equations of state
in detail.
The first improvement will be to consider the summation to infinite
order of the chain of particle-particle hole-hole diagrams (PPHH)
\cite{angels,rpd89,shk86,shk87}.
Thereafter, we will discuss the recent three-hole line results of 
Baldo and co-workers \cite{baldo97,baldo98} and finally the calculations
with three-body forces as well by Akmal et al.\  \cite{ap97,apr98}.

In brief, the summation of PPHH diagrams means that the Pauli
operator in Eq.\ (\ref{eq:pauli}) is extended in order to prevent scattering
into intermediate hole-hole states as well. 
This means that in addition to summing up to infinite order diagrams
with particle-particle intermediate states, we will now sum, still
to infinite order, a larger class of diagrams containing 
hole-hole intermediate 
states as well.

\begin{figure}\begin{center}
       \setlength{\unitlength}{1mm}
       \begin{picture}(80,20)
       \put(25,5){\epsfxsize=5cm \epsfbox{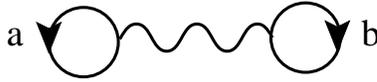}}
       \end{picture}
       \caption{First order contribution 
                to the ground-state energy shift 
                $ \Delta E_0$ in BHF theory.}
       \label{fig:gbhf}
\end{center}\end{figure}

For LOB, as discussed in the previous subsection,
the  effective two-body interaction in nuclear matter is
given by 
the $G$-matrix, which includes all ladder-type diagrams with 
particle-particle intermediate states to infinite order. 
The ground-state energy shift
$\Delta E_0$ in terms of the $G$-matrix is represented
by the first-order diagram  of Fig.\ \ref{fig:gbhf} and reads
\begin{equation}
        \Delta E_0^{LOB}=
	\sum _{ab} n_a n_b\langle ab\vert G(\omega =
        \varepsilon _a + \varepsilon _b )\vert ab\rangle.
        \label{eq:g_lob}
\end{equation}
In Eq.\ (\ref{eq:g_lob}) the n's are the unperturbed Fermi-Dirac 
distribution functions, namely $n_k$ =1 if $k\leq k_F$ and =0 
if $k>k_F$ where  $k_F$ is the Fermi momentum.

\begin{figure}
       \setlength{\unitlength}{1mm}
       \begin{picture}(80,60)
       \put(25,3){\epsfxsize=10cm \epsfbox{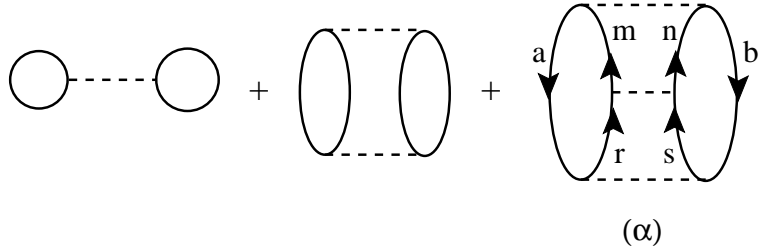}}
       \end{picture}
        \caption{Goldstone diagrams contained in the 
         Brueckner-Hartree-Fock  $G$-matrix.}
       \label{fig:vdiagr}
\end{figure}

The Brueckner-Hartree-Fock  
$G$-matrix  contains  repeated interactions
between a pair of "particle" lines, as illustrated by the diagrams  of 
Fig.\ \ref{fig:vdiagr}.
Note that they are so-called Goldstone diagrams, with
an explicit time ordering.
The third-order diagram $(\alpha)$ of 
Fig.\ \ref{fig:vdiagr} is given by
\begin{equation}
      Diag.(\alpha)=(\frac{1}{2})^3\frac{V_{abmn}V_{mnrs}V_{rsab}}
      {(\epsilon_a+\epsilon_b-\epsilon_m-\epsilon_n)
      (\epsilon_a+\epsilon_b-\epsilon_r-\epsilon_s)}.
\end{equation}
Here m, n, r, s are all particle lines, and $V_{ijkl}$ represents
the anti-symmetrized matrix elements of the NN interaction V.

\begin{figure}
       \setlength{\unitlength}{1mm}
       \begin{picture}(80,50)
       \put(40,3){\epsfxsize=3.3cm \epsfbox{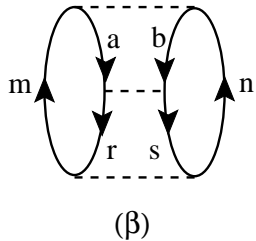}}
       \end{picture}
       \caption{Goldstone diagram with repeated 
                interactions between hole 
                lines.}
       \label{fig:vdiagrb}
\end{figure}

To the same order, there is also a diagram
with hole-hole interactions, as  shown by diagram $(\beta)$ in 
Fig.\ \ref{fig:vdiagrb}, where a, b, r, s are all hole lines.
This diagram is not included in standard LOB calculations of 
nuclear matter, for the following reason.
In earlier times, nuclear-matter calculations were based on, 
by and large, the so-called hole-line-expansion. 
The essence of the hole-line approach is that diagrams with (n+1) 
hole lines are generally much smaller than those with n hole lines.
With this criterion,  diagram ($\beta$) which has 3 (independent) 
hole lines would be negligible compared with diagram $(\alpha)$ 
which has 2 hole lines. Thus the former could be neglected. 
To investigate the validity of this criterion,
it may be useful to actually calculate diagrams like ($\beta$).

A motivation behind the PPHH-diagram method of nuclear matter 
is thus to include
diagrams with hole-hole correlations like diagram $(\beta)$
to infinite order
For further details on how to obtain an effective
interaction and the energy per particle in neutron star matter,
see e.g.\  Refs.\ \cite{rpd89,shk87,engvik97}.

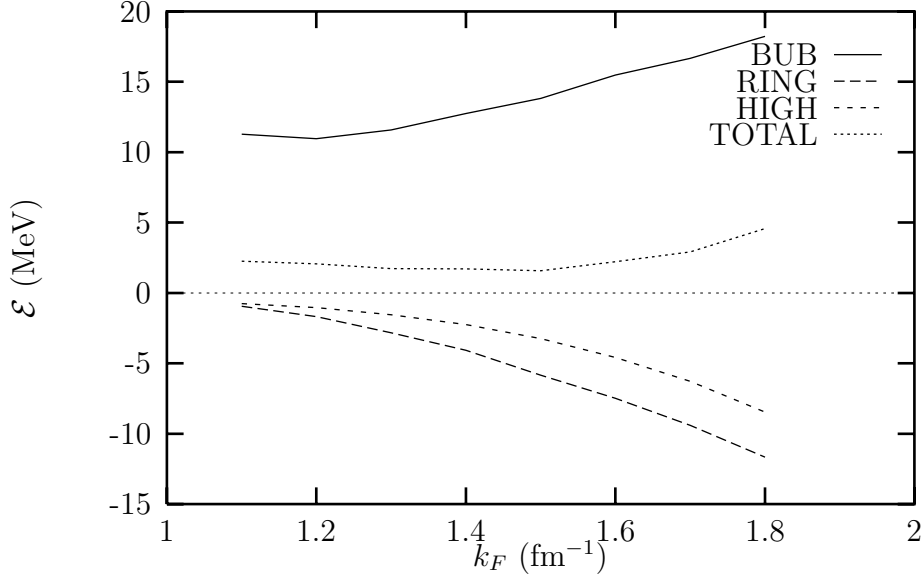
\begin{figure}\begin{center}
     % GNUPLOT: LaTeX picture with Postscript
\setlength{\unitlength}{0.1bp}
\special{!
%!PS-Adobe-2.0
%%Creator: gnuplot
%%DocumentFonts: Helvetica
%%BoundingBox: 50 50 770 554
%%Pages: (atend)
%%EndComments
/gnudict 40 dict def
gnudict begin
/Color false def
/Solid false def
/gnulinewidth 5.000 def
/vshift -33 def
/dl {10 mul} def
/hpt 31.5 def
/vpt 31.5 def
/M {moveto} bind def
/L {lineto} bind def
/R {rmoveto} bind def
/V {rlineto} bind def
/vpt2 vpt 2 mul def
/hpt2 hpt 2 mul def
/Lshow { currentpoint stroke M
  0 vshift R show } def
/Rshow { currentpoint stroke M
  dup stringwidth pop neg vshift R show } def
/Cshow { currentpoint stroke M
  dup stringwidth pop -2 div vshift R show } def
/DL { Color {setrgbcolor Solid {pop []} if 0 setdash }
 {pop pop pop Solid {pop []} if 0 setdash} ifelse } def
/BL { stroke gnulinewidth 2 mul setlinewidth } def
/AL { stroke gnulinewidth 2 div setlinewidth } def
/PL { stroke gnulinewidth setlinewidth } def
/LTb { BL [] 0 0 0 DL } def
/LTa { AL [1 dl 2 dl] 0 setdash 0 0 0 setrgbcolor } def
/LT0 { PL [] 0 1 0 DL } def
/LT1 { PL [4 dl 2 dl] 0 0 1 DL } def
/LT2 { PL [2 dl 3 dl] 1 0 0 DL } def
/LT3 { PL [1 dl 1.5 dl] 1 0 1 DL } def
/LT4 { PL [5 dl 2 dl 1 dl 2 dl] 0 1 1 DL } def
/LT5 { PL [4 dl 3 dl 1 dl 3 dl] 1 1 0 DL } def
/LT6 { PL [2 dl 2 dl 2 dl 4 dl] 0 0 0 DL } def
/LT7 { PL [2 dl 2 dl 2 dl 2 dl 2 dl 4 dl] 1 0.3 0 DL } def
/LT8 { PL [2 dl 2 dl 2 dl 2 dl 2 dl 2 dl 2 dl 4 dl] 0.5 0.5 0.5 DL } def
/P { stroke [] 0 setdash
  currentlinewidth 2 div sub M
  0 currentlinewidth V stroke } def
/D { stroke [] 0 setdash 2 copy vpt add M
  hpt neg vpt neg V hpt vpt neg V
  hpt vpt V hpt neg vpt V closepath stroke
  P } def
/A { stroke [] 0 setdash vpt sub M 0 vpt2 V
  currentpoint stroke M
  hpt neg vpt neg R hpt2 0 V stroke
  } def
/B { stroke [] 0 setdash 2 copy exch hpt sub exch vpt add M
  0 vpt2 neg V hpt2 0 V 0 vpt2 V
  hpt2 neg 0 V closepath stroke
  P } def
/C { stroke [] 0 setdash exch hpt sub exch vpt add M
  hpt2 vpt2 neg V currentpoint stroke M
  hpt2 neg 0 R hpt2 vpt2 V stroke } def
/T { stroke [] 0 setdash 2 copy vpt 1.12 mul add M
  hpt neg vpt -1.62 mul V
  hpt 2 mul 0 V
  hpt neg vpt 1.62 mul V closepath stroke
  P  } def
/S { 2 copy A C} def
end
}
\begin{picture}(3600,2160)(0,0)
\special{"
gnudict begin
gsave
50 50 translate
0.100 0.100 scale
0 setgray
/Helvetica findfont 100 scalefont setfont
newpath
-500.000000 -500.000000 translate
LTa
600 1047 M
2817 0 V
LTb
600 251 M
63 0 V
2754 0 R
-63 0 V
600 516 M
63 0 V
2754 0 R
-63 0 V
600 782 M
63 0 V
2754 0 R
-63 0 V
600 1047 M
63 0 V
2754 0 R
-63 0 V
600 1313 M
63 0 V
2754 0 R
-63 0 V
600 1578 M
63 0 V
2754 0 R
-63 0 V
600 1844 M
63 0 V
2754 0 R
-63 0 V
600 2109 M
63 0 V
2754 0 R
-63 0 V
600 251 M
0 63 V
0 1795 R
0 -63 V
1163 251 M
0 63 V
0 1795 R
0 -63 V
1727 251 M
0 63 V
0 1795 R
0 -63 V
2290 251 M
0 63 V
0 1795 R
0 -63 V
2854 251 M
0 63 V
0 1795 R
0 -63 V
3417 251 M
0 63 V
0 1795 R
0 -63 V
600 251 M
2817 0 V
0 1858 V
-2817 0 V
600 251 L
LT0
3114 1946 M
180 0 V
882 1646 M
281 -17 V
282 33 V
282 62 V
282 57 V
281 88 V
282 63 V
282 83 V
LT1
3114 1846 M
180 0 V
882 997 M
281 -39 V
282 -61 V
282 -66 V
282 -94 V
281 -87 V
2572 548 L
2854 428 L
LT2
3114 1746 M
180 0 V
882 1007 M
281 -15 V
282 -27 V
282 -37 V
282 -53 V
281 -71 V
282 -90 V
2854 598 L
LT3
3114 1646 M
180 0 V
882 1167 M
281 -10 V
282 -18 V
282 -1 V
282 -7 V
281 34 V
282 37 V
282 88 V
stroke
grestore
end
showpage
}
\put(3054,1646){\makebox(0,0)[r]{TOTAL}}
\put(3054,1746){\makebox(0,0)[r]{HIGH}}
\put(3054,1846){\makebox(0,0)[r]{RING}}
\put(3054,1946){\makebox(0,0)[r]{BUB}}
\put(2008,51){\makebox(0,0){$k_F$ (fm$^{-1}$)}}
\put(100,1180){%
\special{ps: gsave currentpoint currentpoint translate
270 rotate neg exch neg exch translate}%
\makebox(0,0)[b]{\shortstack{${\cal E}$ (MeV)}}%
\special{ps: currentpoint grestore moveto}%
}
\put(3417,151){\makebox(0,0){2}}
\put(2854,151){\makebox(0,0){1.8}}
\put(2290,151){\makebox(0,0){1.6}}
\put(1727,151){\makebox(0,0){1.4}}
\put(1163,151){\makebox(0,0){1.2}}
\put(600,151){\makebox(0,0){1}}
\put(540,2109){\makebox(0,0)[r]{20}}
\put(540,1844){\makebox(0,0)[r]{15}}
\put(540,1578){\makebox(0,0)[r]{10}}
\put(540,1313){\makebox(0,0)[r]{5}}
\put(540,1047){\makebox(0,0)[r]{0}}
\put(540,782){\makebox(0,0)[r]{-5}}
\put(540,516){\makebox(0,0)[r]{-10}}
\put(540,251){\makebox(0,0)[r]{-15}}
\end{picture}
       \caption{ The contributions from the bubble (BUB), ring (RING), 
                  and higher-order three hole-line diagrams (HIGH) employing the 
                  continuous choice. Results are for symmetric nuclear matter,
                  employing the Argonne $V_{14}$ model for the NN interaction.}
       \label{fig:baldo3holeline}
\end{center}\end{figure}

The next set of diagrams which can be included is the summation of
so-called three-hole line diagrams through the solution of the 
Bethe-Fadeev equations, originally pionereed by Day \cite{day81} and 
recently taken up again by Baldo and co-workers \cite{baldo97,baldo98}.
The whole set of three hole-line diagrams can be
grouped into three main sets of diagrams. The so-called ring diagrams,
the bubble diagram,  
and the so-called higher-order diagrams, with an arbitrary 
number of particle lines. 
Each of these three-hole line contributions are quite large, see Refs.\   
\cite{baldo97,baldo98} and Fig.\ \ref{fig:baldo3holeline}, 
but there is a strong degree 
of cancellation among the various terms. Thus,  the total 
three-hole line contribution turns out to be substantially smaller than the 
two-body (i.e.\  two hole-line)  contribution. 
This can be seen in Fig.\ 
\ref{fig:baldo3holeline} where we plot the various 
three hole-line contributions and 
their total contribution, all with the continuous 
choice for the single particle energies. 
The interesting feature of the calculations of Baldo et al.\  \cite{baldo98},
is that the results with the continuous single-particle choice lead to 
three hole-line results which are rather close to the total two hole-line 
results. Diagrams of the PPHH type are however not included. In Ref.\
\cite{day83} these were estimated to be of the same sign in both
the standard and the continuum choice. 
In spite of this methodological progress in perturbative approaches, there
are still classes of diagrams which need to be summed up. 
As discussed by Jackson
in Ref.\ \cite{jackson83}, one can prove that 
there is a minimal set of diagrams
which need to be summed up in order get the physics of a many-body system
right. This set of diagrams is the so-called Parquet class of diagrams
\cite{parquet} where ring diagrams and ladder diagrams are summed up to all 
orders in a
self-consistent way. Since it is rather hard to sum this set of diagrams in a 
practical
way, other many-body methods like the coupled-cluster ansatz \cite{ccm} and 
optimized
hypernetted-chain theory \cite{fhnc} provide a systematic approximation method
to sets of Feynman diagrams that cannot be calculated exactly.
In the remainder of this section will therefore focus on 
a variational method based on the hypernetted chain summation
techniques developed by Pandharipande, Wiringa and co-workers.
The approach was developed
in the 1970s \cite{pw}, particularly to include
the effects of many-body correlations, presumably important in
dense neutron star matter. Calculations performed since then have
confirmed that many-body clusters make significant contributions
to the binding energies of equilibrium nuclear matter
and  light nuclei \cite{day81,pw,dw,forest,pscppr}.
We will base our discussion of the EoS on the recent
results of Akmal, Pandharipande and Ravenhall \cite{ap97,apr98,aryathesis98}.
We will refer the reader to the latter references and Ref.\ 
\cite{pw} for more details.
In brief, 
the variational wavefunction has the form:
\begin{equation}
\Psi_v = (S \prod_{i<j} F_{ij}) \Phi ,
\end{equation}
consisting of a symmetrized product of pair correlation operators
$F_{ij}$ operating on the Fermi gas wavefunction $\Phi$. In symmetric
nuclear matter,
the function $F_{ij}$ includes eight terms:
\begin{equation}
F_{ij}=\sum_{p=1,8} f^p(r_{ij})O^p_{ij} ,
\end{equation}
representing central, spin-spin, 
tensor and spin-orbit
correlations with and without isospin factors. In pure neutron matter, 
the $F_{ij}$ reduce to a sum of four terms with only odd $p\leq$7.
The correlation operators $F_{ij}$ are determined from Euler-Lagrange 
equations \cite{lp3} that minimize the two-body cluster contribution
of an interaction $(V-\lambda)$, where:
\begin{eqnarray}
    V_{ij} & = & \sum_{p=1,14}\alpha^p V^p(r_{ij}) O^p_{ij}, \\
    \lambda_{ij} & = & \sum_{p=1,8}\lambda^p(r_{ij})O^p_{ij}.
\end{eqnarray}
The variational parameters $\alpha^p$ are meant to simulate the
quenching of the spin-isospin interaction between particles i and j,
due to flipping of the spin and/or isospin of particle i or j via
interaction with other particles in matter.
The NN interaction used in Refs.\  \cite{ap97,apr98,aryathesis98}
is the recent parametrization of the 
Argonne group, the so-called
Argonne $V_{18}$ two-nucleon interaction discussed above. It has the form
\begin{equation}
    V_{18,ij}=\sum_{p=1,18}V^p(r_{ij})O^p_{ij}+V_{em} .
\end{equation}
The electromagnetic part $V_{em}$ consists of Coulomb and
magnetic interactions in the $nn$, $np$ and $pp$ pairs, and it is omitted
from all nuclear matter studies. The strong interaction
part of the potential includes fourteen isoscalar terms with operators
among these the central, spin-spin, spin-orbit and tensor operators.
In addition, a phenomenological three-body force $V_{ijk}$ is included,
represented by the
Urbana models of $V_{ijk}$ containing two isoscalar terms:
\begin{equation}
    V_{ijk}=V^{2\pi}_{ijk} + V^{R}_{ijk} \ .
     \label{eq:threebodyforce}
\end{equation}
The first term represents the Fujita-Miyazawa two-pion exchange interaction:
\begin{eqnarray}
     V^{2\pi}_{ijk}=&\sum_{cyc} A_{2\pi} \left( \left\{{\bf \tau}_i \cdot {\bf 
\tau}_j,
     {\bf \tau}_i \cdot {\bf \tau}_k \right\}
     \left\{ X_{ij},X_{ik} \right\}
     + \frac{1}{4} [{\bf\tau}_i \cdot {\bf\tau}_j, {\bf\tau}_i \cdot 
{\bf\tau}_k]
    [X_{ij},X_{ik}] \right),  \\
     &X_{ij} = S_{ij}T_\pi(r_{ij})+{\bf\sigma}_i \cdot {\bf\sigma}_j 
Y_\pi(r_{ij}),
\end{eqnarray}
with strength $A_{2\pi}$ and where the matrices $\sigma$ and $\tau$ are
the Pauli matrices for spin and isospin, respectively. The functions 
$T_{\pi}(r_{ij})$ and $Y_{\pi}(r_{ij})$
describe the radial shapes of the one-pion exchange tensor and Yukawa
potentials. These functions are calculated using the average value
of the pion mass and include the short-range cutoffs used in the
Argonne $V_{18}$ NN interaction. The term denoted by $V^R_{ijk}$ is purely
phenomenological, and has the form:
\begin{equation}
    V^{R}_{ijk}=U_0\sum_{cyc}T_\pi^2(r_{ij})T_\pi^2(r_{ik}).
\end{equation}
This term is meant to represent the modification of N$\Delta$- and 
$\Delta\Delta$-contributions in the two-body interaction 
by other particles in the medium, and also 
accounts for relativistic effects. The spin-isospin 
dependence of these effects is neglected.
The two parameters $A_{2\pi}$ and $U_0$ are chosen to yield the observed
energy of $^3$H and the equilibrium density of nuclear matter,
$\rho_0=0.16$ fm$^{-3}$. Obviously, this fitting procedure will yield
different parameters if another NN interaction is employed, e.g.\ 
the CD-Bonn interaction since, see discussion above, the various
potentials yield slightly different binding energies for  $^3$H
at the two-body level.
The inclusion of many-body clusters is described in Ref.\ \cite{pw}.
Relativistic boost corrections were also evaluated in Refs.\ 
\cite{ap97,apr98,aryathesis98} but the latter will be included
in our discussion of relativistic effects in subsection 
\ref{subsubsec:relativistic}
below.

In the remainder of this subsection we will henceforth discuss the 
consequences for the EoS from the above many-body corrections.

\begin{figure}\begin{center} 
     \input{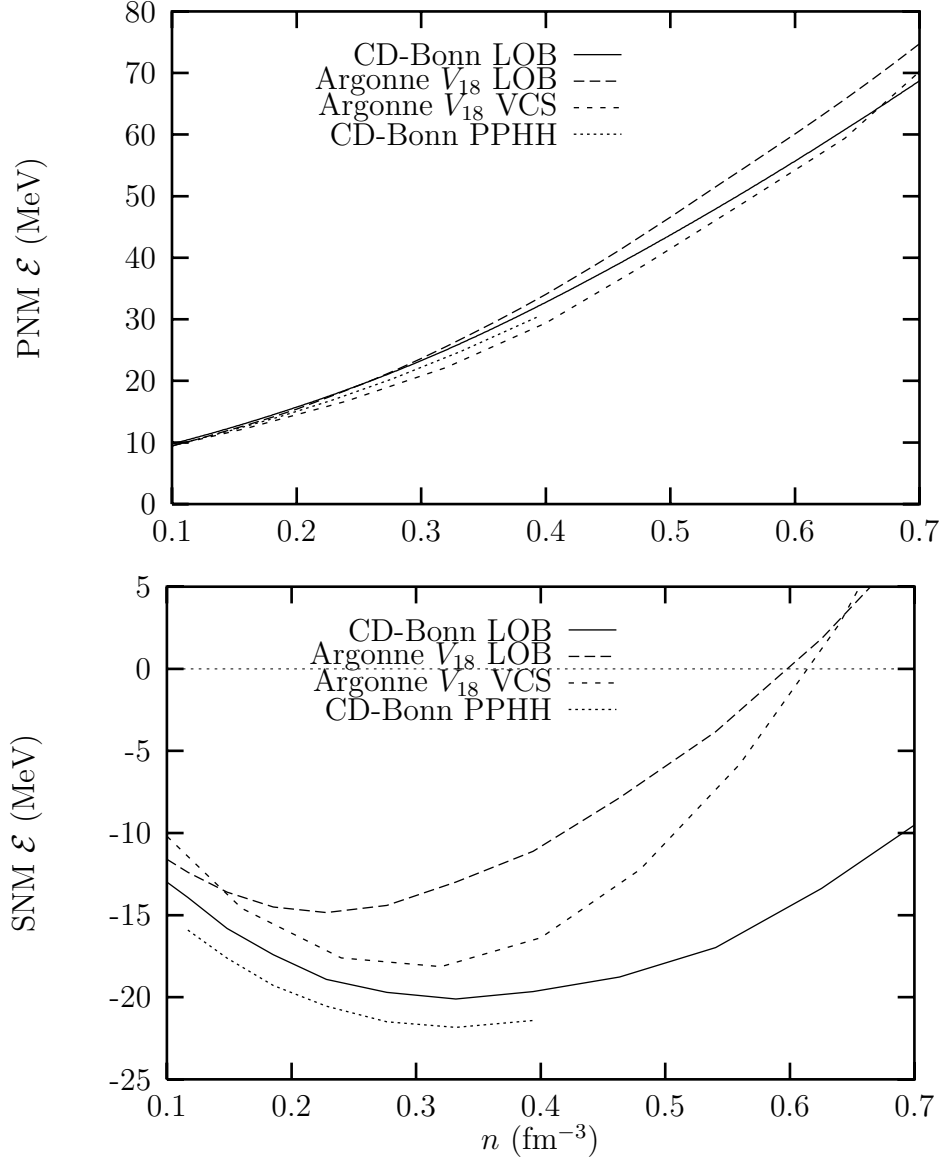}
     \caption{Upper panel: Energy per particle in PNM for various many-body
              approaches with two-body interactions only, i.e.\  
              PPHH results with the CD-Bonn interaction (stable results were
              obtained up to densities $0.4$ fm$^{-3}$ only),
              LOB with the Argonne $V_{18}$ interaction and
              variational cluster (VCS) results 
               for the Argonne $V_{18}$ interaction model.
              Lower panel: The corresponding results for SNM.}
     \label{fig:sec2fig11}
\end{center}\end{figure}

In Fig.\ \ref{fig:sec2fig11} we plot the results for PNM and SNM 
obtained with two-body interactions
including only PPHH diagrams or higher-order
diagrams stemming from the variational cluster approach. 
It is noteworthy to observe that in PNM 
the energy per particle up to 2-3 times $n_0$ is rather similar for all
calculations at the two-body level.
To a certain extent this is expected since the strong nuclear tensor
force contribution 
from the $T=0$ channel is absent. This means in turn that more complicated
many-body terms are not so important in PNM. We will see this also in connection
with the three-hole line discussion below. 
For symmetric nuclear matter the situation is however different due to the 
strong 
tensor force in the $np$ channel, leading to larger higher-order corrections.
This is clearly seen in the lower panel of Fig.\ \ref{fig:sec2fig11}. 
In e.g.\  the PPHH calculation, where the correlations are due to hole-hole
and particle-particle ladders, the tensor force plays the main role.
Again this is similar to the situation for the effective interaction
in finite nuclei \cite{hko95}. There,
to e.g.\  second order in the interaction, diagrams with hole-hole and 
particle-particle
intermediate states tend to be bigger than screening diagrams. 
In general however, none of the calculations at the two-body level reproduce
properly the saturation properties of SNM.

\begin{figure}\begin{center} 
     % GNUPLOT: LaTeX picture with Postscript
\setlength{\unitlength}{0.1bp}
\special{!
%!PS-Adobe-2.0
%%Creator: gnuplot
%%DocumentFonts: Helvetica
%%BoundingBox: 50 50 770 554
%%Pages: (atend)
%%EndComments
/gnudict 40 dict def
gnudict begin
/Color false def
/Solid false def
/gnulinewidth 5.000 def
/vshift -33 def
/dl {10 mul} def
/hpt 31.5 def
/vpt 31.5 def
/M {moveto} bind def
/L {lineto} bind def
/R {rmoveto} bind def
/V {rlineto} bind def
/vpt2 vpt 2 mul def
/hpt2 hpt 2 mul def
/Lshow { currentpoint stroke M
  0 vshift R show } def
/Rshow { currentpoint stroke M
  dup stringwidth pop neg vshift R show } def
/Cshow { currentpoint stroke M
  dup stringwidth pop -2 div vshift R show } def
/DL { Color {setrgbcolor Solid {pop []} if 0 setdash }
 {pop pop pop Solid {pop []} if 0 setdash} ifelse } def
/BL { stroke gnulinewidth 2 mul setlinewidth } def
/AL { stroke gnulinewidth 2 div setlinewidth } def
/PL { stroke gnulinewidth setlinewidth } def
/LTb { BL [] 0 0 0 DL } def
/LTa { AL [1 dl 2 dl] 0 setdash 0 0 0 setrgbcolor } def
/LT0 { PL [] 0 1 0 DL } def
/LT1 { PL [4 dl 2 dl] 0 0 1 DL } def
/LT2 { PL [2 dl 3 dl] 1 0 0 DL } def
/LT3 { PL [1 dl 1.5 dl] 1 0 1 DL } def
/LT4 { PL [5 dl 2 dl 1 dl 2 dl] 0 1 1 DL } def
/LT5 { PL [4 dl 3 dl 1 dl 3 dl] 1 1 0 DL } def
/LT6 { PL [2 dl 2 dl 2 dl 4 dl] 0 0 0 DL } def
/LT7 { PL [2 dl 2 dl 2 dl 2 dl 2 dl 4 dl] 1 0.3 0 DL } def
/LT8 { PL [2 dl 2 dl 2 dl 2 dl 2 dl 2 dl 2 dl 4 dl] 0.5 0.5 0.5 DL } def
/P { stroke [] 0 setdash
  currentlinewidth 2 div sub M
  0 currentlinewidth V stroke } def
/D { stroke [] 0 setdash 2 copy vpt add M
  hpt neg vpt neg V hpt vpt neg V
  hpt vpt V hpt neg vpt V closepath stroke
  P } def
/A { stroke [] 0 setdash vpt sub M 0 vpt2 V
  currentpoint stroke M
  hpt neg vpt neg R hpt2 0 V stroke
  } def
/B { stroke [] 0 setdash 2 copy exch hpt sub exch vpt add M
  0 vpt2 neg V hpt2 0 V 0 vpt2 V
  hpt2 neg 0 V closepath stroke
  P } def
/C { stroke [] 0 setdash exch hpt sub exch vpt add M
  hpt2 vpt2 neg V currentpoint stroke M
  hpt2 neg 0 R hpt2 vpt2 V stroke } def
/T { stroke [] 0 setdash 2 copy vpt 1.12 mul add M
  hpt neg vpt -1.62 mul V
  hpt 2 mul 0 V
  hpt neg vpt 1.62 mul V closepath stroke
  P  } def
/S { 2 copy A C} def
end
}
\begin{picture}(3600,2160)(0,0)
\special{"
gnudict begin
gsave
50 50 translate
0.100 0.100 scale
0 setgray
/Helvetica findfont 100 scalefont setfont
newpath
-500.000000 -500.000000 translate
LTa
600 251 M
2817 0 V
LTb
600 251 M
63 0 V
2754 0 R
-63 0 V
600 457 M
63 0 V
2754 0 R
-63 0 V
600 664 M
63 0 V
2754 0 R
-63 0 V
600 870 M
63 0 V
2754 0 R
-63 0 V
600 1077 M
63 0 V
2754 0 R
-63 0 V
600 1283 M
63 0 V
2754 0 R
-63 0 V
600 1490 M
63 0 V
2754 0 R
-63 0 V
600 1696 M
63 0 V
2754 0 R
-63 0 V
600 1903 M
63 0 V
2754 0 R
-63 0 V
600 2109 M
63 0 V
2754 0 R
-63 0 V
600 251 M
0 63 V
0 1795 R
0 -63 V
1070 251 M
0 63 V
0 1795 R
0 -63 V
1539 251 M
0 63 V
0 1795 R
0 -63 V
2009 251 M
0 63 V
0 1795 R
0 -63 V
2478 251 M
0 63 V
0 1795 R
0 -63 V
2948 251 M
0 63 V
0 1795 R
0 -63 V
3417 251 M
0 63 V
0 1795 R
0 -63 V
600 251 M
2817 0 V
0 1858 V
-2817 0 V
600 251 L
LT0
2114 1946 M
180 0 V
600 349 M
282 29 V
375 45 V
376 57 V
376 75 V
375 98 V
376 105 V
375 106 V
282 112 V
LT1
2114 1846 M
180 0 V
600 364 M
94 18 V
188 48 V
188 63 V
187 56 V
376 149 V
376 211 V
375 276 V
376 345 V
375 428 V
102 151 V
LT2
2114 1746 M
180 0 V
600 351 M
178 21 V
277 33 V
343 49 V
423 81 V
502 125 V
596 203 V
498 229 V
stroke
grestore
end
showpage
}
\put(2054,1746){\makebox(0,0)[r]{3-hole line}}
\put(2054,1846){\makebox(0,0)[r]{Argonne $V_{18}+$UIX}}
\put(2054,1946){\makebox(0,0)[r]{Argonne $V_{18}$ VCS}}
\put(100,1180){%
\special{ps: gsave currentpoint currentpoint translate
270 rotate neg exch neg exch translate}%
\makebox(0,0)[b]{\shortstack{PNM ${\cal E}$ (MeV)}}%
\special{ps: currentpoint grestore moveto}%
}
\put(3417,151){\makebox(0,0){0.7}}
\put(2948,151){\makebox(0,0){0.6}}
\put(2478,151){\makebox(0,0){0.5}}
\put(2009,151){\makebox(0,0){0.4}}
\put(1539,151){\makebox(0,0){0.3}}
\put(1070,151){\makebox(0,0){0.2}}
\put(600,151){\makebox(0,0){0.1}}
\put(540,2109){\makebox(0,0)[r]{180}}
\put(540,1903){\makebox(0,0)[r]{160}}
\put(540,1696){\makebox(0,0)[r]{140}}
\put(540,1490){\makebox(0,0)[r]{120}}
\put(540,1283){\makebox(0,0)[r]{100}}
\put(540,1077){\makebox(0,0)[r]{80}}
\put(540,870){\makebox(0,0)[r]{60}}
\put(540,664){\makebox(0,0)[r]{40}}
\put(540,457){\makebox(0,0)[r]{20}}
\put(540,251){\makebox(0,0)[r]{0}}
\end{picture}
\begin{picture}(3600,2160)(0,0)
\special{"
gnudict begin
gsave
50 50 translate
0.100 0.100 scale
0 setgray
/Helvetica findfont 100 scalefont setfont
newpath
-500.000000 -500.000000 translate
LTa
600 561 M
2817 0 V
LTb
600 251 M
63 0 V
2754 0 R
-63 0 V
600 561 M
63 0 V
2754 0 R
-63 0 V
600 870 M
63 0 V
2754 0 R
-63 0 V
600 1180 M
63 0 V
2754 0 R
-63 0 V
600 1490 M
63 0 V
2754 0 R
-63 0 V
600 1799 M
63 0 V
2754 0 R
-63 0 V
600 2109 M
63 0 V
2754 0 R
-63 0 V
600 251 M
0 63 V
0 1795 R
0 -63 V
1070 251 M
0 63 V
0 1795 R
0 -63 V
1539 251 M
0 63 V
0 1795 R
0 -63 V
2009 251 M
0 63 V
0 1795 R
0 -63 V
2478 251 M
0 63 V
0 1795 R
0 -63 V
2948 251 M
0 63 V
0 1795 R
0 -63 V
3417 251 M
0 63 V
0 1795 R
0 -63 V
600 251 M
2817 0 V
0 1858 V
-2817 0 V
600 251 L
LT0
2114 1946 M
180 0 V
600 403 M
882 335 L
375 -47 V
376 -8 V
376 27 V
375 65 V
376 99 V
375 132 V
282 130 V
LT1
2114 1846 M
180 0 V
600 417 M
94 -19 V
882 377 L
188 9 V
187 35 V
376 153 V
376 176 V
375 309 V
376 431 V
375 542 V
40 77 V
LT2
2114 1746 M
180 0 V
600 392 M
78 -15 V
825 350 L
999 328 L
202 -16 V
230 -2 V
258 17 V
291 32 V
324 85 V
362 111 V
404 165 V
347 180 V
stroke
grestore
end
showpage
}
\put(2054,1746){\makebox(0,0)[r]{3-hole line}}
\put(2054,1846){\makebox(0,0)[r]{Argonne $V_{18}+$UIX}}
\put(2054,1946){\makebox(0,0)[r]{Argonne $V_{18}$ VCS}}
\put(2008,21){\makebox(0,0){$n$ (fm$^{-3}$)}}
\put(100,1180){%
\special{ps: gsave currentpoint currentpoint translate
270 rotate neg exch neg exch translate}%
\makebox(0,0)[b]{\shortstack{SNM ${\cal E}$ (MeV)}}%
\special{ps: currentpoint grestore moveto}%
}
\put(3417,151){\makebox(0,0){0.7}}
\put(2948,151){\makebox(0,0){0.6}}
\put(2478,151){\makebox(0,0){0.5}}
\put(2009,151){\makebox(0,0){0.4}}
\put(1539,151){\makebox(0,0){0.3}}
\put(1070,151){\makebox(0,0){0.2}}
\put(600,151){\makebox(0,0){0.1}}
\put(540,2109){\makebox(0,0)[r]{100}}
\put(540,1799){\makebox(0,0)[r]{80}}
\put(540,1490){\makebox(0,0)[r]{60}}
\put(540,1180){\makebox(0,0)[r]{40}}
\put(540,870){\makebox(0,0)[r]{20}}
\put(540,561){\makebox(0,0)[r]{0}}
\put(540,251){\makebox(0,0)[r]{-20}}
\end{picture}
     \caption{Upper panel: Energy per particle in PNM for variational 
              calculations with 
              the Argonne $V_{18}$ interaction without (VCS) and with   
              three-body interactions ($V_{18}+$UIX).
              The result from the three-hole line
              expansion of Baldo and co-workers employing the $V_{14}$ 
interaction
              is
              included as well.
              Lower panel : The corresponding results for SNM.}
     \label{fig:sec2fig13}
\end{center}\end{figure}
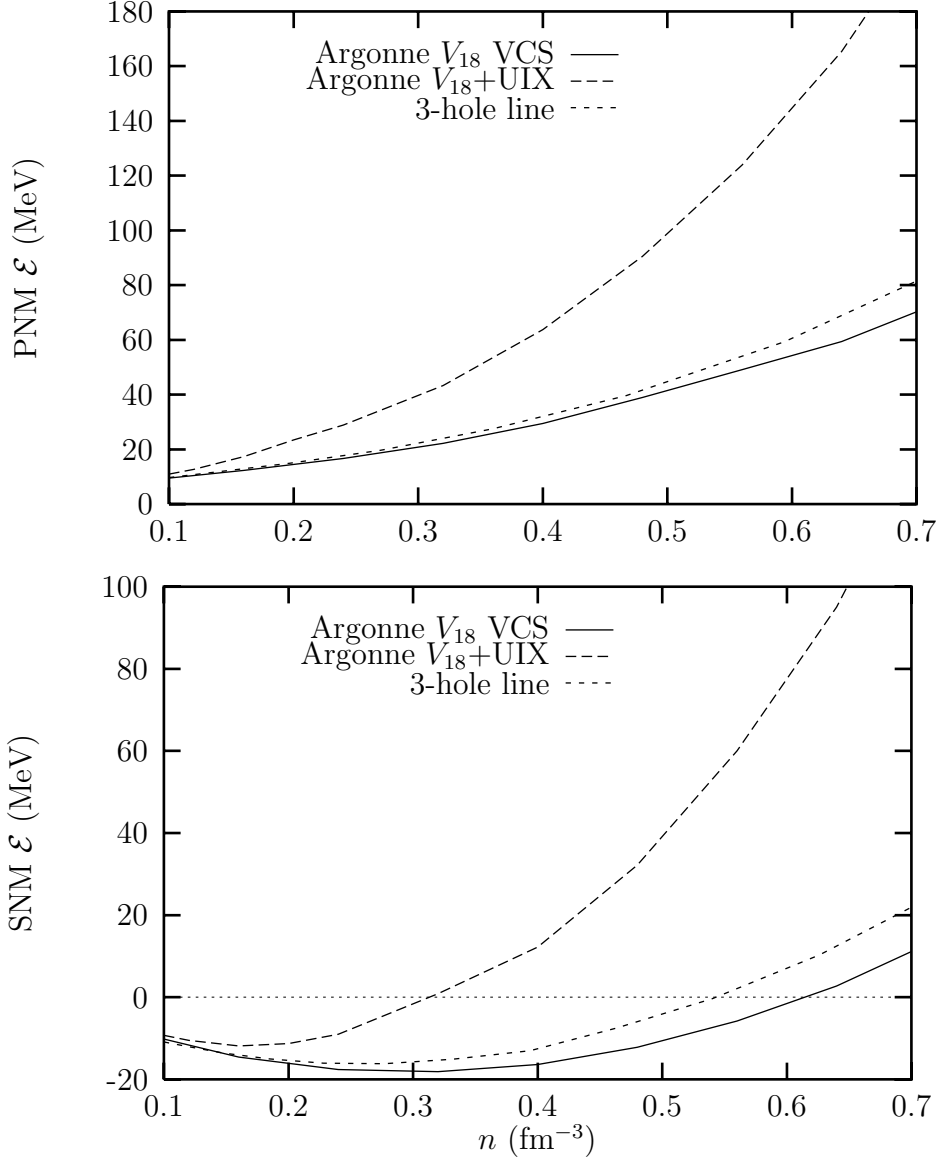

In Fig.\ \ref{fig:sec2fig13} we have then included results from calculations
with effective three-body terms and the phenomenological three-body forces.
The results are compared with those from LOB with the Argonne $V_{18}$ 
interaction.
With the inclusion of the phenomenological three-body force
described in Eq.\ (\ref{eq:threebodyforce}), which was fitted to reproduce
the binding energy of the triton and
the alpha particle, a clear change is seen  both in PNM and SNM.
The energy per particle  gets more repulsive at higher densities. 
Three hole-line diagrams however, with the continuous choice are rather
close to LOB with the continuous choice in SNM, while they
are almost negligible in PNM, in line with our observation above about the 
tensor force component in the $T=0$ channel.

Thus, as a summary, the inclusion of phenomenological three-body
forces in non-relativistic calculations are needed in order to
improve the saturation properties of the microscopically 
calculated EoS, whereas the inclusion of additional many-body
effects, either three-hole line diagrams or through VCS calculations
yield results at low densities similar to those from LOB. 
However, even with the three-body force, one is not able to reproduce
properly the binding energy. In the next subsection we will thence
discuss further corrections stemming from relativistic effects.

\subsubsection{Relativistic effects}\label{subsubsec:relativistic}

The properties of neutron stars depend on the equation of state
at densities up to an order of magnitude higher than those observed
in ordinary nuclei. At such densities, relativistic effects
certainly prevail. Among relativistic approaches to the nuclear
many-body problem, the so-called Dirac-Hartree and Dirac-Hartree-Fock
approaches have received much interest 
(Serot and  Walecka \cite{sw86}; Serot 1992 \cite{serot92} 
Horowitz and Serot \cite{hs87}).
One of the early successes of these
approaches was the quantitative reproduction of spin observables,
which are only poorly described by the non-relativistic theory.
Important to these methods
was the introduction of a strongly attractive scalar component
and a repulsive vector component 
in the nucleon self-energy (Brockmann \cite{brockmann78} and
Serot and Walecka \cite{sw86}). 
Inspired
by the successes of the Dirac-Hartree-Fock method, a relativistic
extension of Brueckner theory was proposed by Celenza and Shakin \cite{cs87}, 
known as the Dirac-Brueckner theory.
One of the appealing features of the Dirac-Brueckner approach
is the self-consistent determination of the relativistic sp
energies and wave functions.
The Dirac-Brueckner approach differs from the Dirac-Hartree-Fock one
in the sense that in the former one starts from the free NN interaction
which is only constrained by a fit to the NN data, whereas the
Dirac-Hartree-Fock method pursues a line where the parameters of the
theory are determined so as to reproduce the bulk properties
of nuclear matter. It ought, however, to be stressed that the Dirac-Brueckner
approach, 
which starts from NN interactions based on meson exchange,
is a non-renormalizable theory where the short-range part
of the potential depends on additional parameters like
vertex cut-offs, clearly minimizing the
sensitivity of calculated results to short-distance inputs, see e.g.\ 
Refs.\ \cite{cs87,bm90,tm87}.
The description presented here for the Dirac-Brueckner approach follows
closely that of Brockmann and Machleidt\cite{bm90}. We will thus
use the meson-exchange models of the Bonn group, defined
in table A.2 of Ref.\ \cite{mac89}. There the three-dimensional
reduction of the Bethe-Salpeter equation as given by the Thompson equation
is used to solve the equation for the scattering
matrix \cite{thompsson70}. 
Hence, including the necessary medium effects like the
Pauli operators discussed in the previous subsection and the starting
energy, we shall rewrite Eq. (\ref{eq:gnonrel}) departing from
the Thompson equation. Then, in a self-consistent way, we
determine the above-mentioned scalar and vector components  which
define the nucleon self-energy. 
Note that negative energy solutions are not included.
An account of these can be found in the recent work of de Jong
and Lenske \cite{fredhorst98}.
In order to introduce the relativistic nomenclature, we consider first the
Dirac equation for a free nucleon, i.e.\ 
\[
  (i\not \partial -m )\psi (x)=0,
\]
where $m$ is the free nucleon mass 
and $\psi (x)$ is  the nucleon field operator
($x$ is a four-point)
which is conventionally expanded in terms of plane wave states
and the Dirac spinors $u(p,s)$, and $ v(p,s)$, where
 $p=(p^0 ,{\bf p})$ is
a four momentum\footnote{Further notation is  as
given  in Itzykson  and Zuber \cite{iz80}.}
 and  $s$ is the spin projection.  

The positive energy Dirac spinors are
(with $\overline{u}u=1$)
\begin{equation}
   u(p,s)=\sqrt{\frac{E(p)+m}{2m}}
	  \left(\begin{array}{c} \chi_s\\ \\
	  \frac{\mbox{\boldmath $\sigma$}\cdot{\bf p}}{E(p)+m}\chi_s
	  \end{array}\right),
   \label{eq:freespinor}
\end{equation}
where $\chi_s$ is the Pauli spinor and $E(p) =\sqrt{m^2 +|{\bf p}|^2}$.
To account for medium modifications to the free Dirac equation,
we introduce the notion of the self-energy $\Sigma (p)$.
As we assume parity to be a good quantum number, the self-energy of a
nucleon can be formally written as
\[
       \Sigma(p) =
       \Sigma_S(p) -\gamma_0 \Sigma^0(p)
       +\mbox{\boldmath $\gamma$}{\bf p}\Sigma^V(p).
\]
The momentum dependence of $\Sigma^0$ 
and $\Sigma_S$ is rather weak (Serot and Walecka 1986).
Moreover, $\Sigma^V << 1$, such 
that the features of the Dirac-Brueckner-Hartree-Fock
procedure can be discussed within the framework of the phenomenological
Dirac-Hartree ansatz, i.e. we  approximate
\[      
\Sigma \approx \Sigma_S -\gamma_0 \Sigma^0 = U_S + U_V,
\]
where $U_S$ is an attractive 
scalar field, and $U_V$ is the time-like component
of a repulsive vector field. 
The finite self-energy modifies the
free Dirac spinors of Eq. (\ref{eq:freespinor}) as
\[
   \tilde{u}(p,s)=\sqrt{\frac{\tilde{E}(p)+\tilde{m}}{2\tilde{m}}}
	  \left(\begin{array}{c} \chi_s\\ \\
	  \frac{\mbox{\boldmath $\sigma$}\cdot{\bf p}}
          {\tilde{E}(p)+\tilde{m}}\chi_s
	  \end{array}\right),
\]
where we let the terms with tilde 
represent the medium modified quantities. 
Here we have defined 
\[      
   \tilde{m}=m+{U_S},
\]                   
and
\begin{equation}
      \tilde{E}_{i}=
      \tilde{E}(p_{i})=\sqrt{\tilde{m}^2_i+{\bf p}_{i}^2}.
\label{eq:eirel}
\end{equation}
As in the previous subsection, the subscripts $i$, and $h$ below,
represent the quantum numbers of the single-particle states, such as
isospin projections $\tau_i$ and $\tau_h$, momenta $k_i$ and $k_h$, etc.

The sp energy is 
\begin{equation}
   \tilde{\varepsilon}_{i} =\tilde{E}_{i} +U_V^i,
   \label{eq:sprelen}
\end{equation}
and the sp potential is given by the G-matrix as
\begin{equation}
   u_{i} =\sum_{h\leq k_F} 
         \frac{ \tilde{m}_i\tilde{m}_h }{ \tilde{E}_i \tilde{E}_h }
	\left\langle i h \right| \tilde{G}(\tilde{E}=
        \tilde{E}_i +\tilde{E}_h)\left | i h\right\rangle_{\mathrm{AS}},
	\label{eq:urel}
\end{equation}
or, if we wish to express it in terms of the constants $U_S$ and
$U_V$,
we have
\begin{equation}
   u_{i} = \frac{\tilde{m}_i}{\tilde{E}_{i}}{U_S}^{i} +U_V^{i}.
   \label{eq:sppotrel}
\end{equation}
In Eq. (\ref{eq:urel}), we have introduced the relativistic 
$\tilde{G}$-matrix. If the two interacting particles,
with isospins $\tau_1$ and $\tau_2$, give a total isospin projection
$T_z$, the relativistic $\tilde{G}$-matrix
in a partial wave representation is given by
\begin{eqnarray}
   \tilde{G}_{ll'}^{\alpha T_z}(kk'K\tilde{E})&=&
   \tilde{V}_{ll'}^{\alpha T_z}(kk')
   +\sum_{l''}\int \frac{d^3 q}{(2\pi )^3} \label{eq:grel} \\
   &&\times\tilde{V}_{ll''}^{\alpha T_z}(kq)
   \frac{\tilde{m}_1 \tilde{m}_2} { \tilde{E}_1^{q} \tilde{E}_2^{q} }
   \frac{Q^{T_z}(q,K)} {( \tilde{E}-\tilde{E}_1^{q}-\tilde{E}_2^{q})}
   \tilde{G}_{l''l'}^{\alpha T_z}(qk'K\tilde{E}), \nonumber
\end{eqnarray}
where the relativistic starting energy 
is defined according to Eq. (\ref{eq:eirel}) as
\[\tilde{E}=\tilde{E}(\sqrt{k^2+K^2/4},\tau_1)+
\tilde{E}(\sqrt{k^2+K^2/4},\tau_2).\]
and 
\[\tilde{E}_{1 (2)}^{q}=\tilde{E}(\sqrt{q^2+K^2/4},\tau_{1(2)}).\]

Equations (\ref{eq:sprelen})-(\ref{eq:grel}) are solved self-consistently, 
starting
with adequate values for the scalar and vector components
$U_S$ and $U_V$. This iterative scheme is continued until these
parameters show little variation.
Finally, the relativistic version of Eq. (\ref{eq:enrel}) reads
\begin{eqnarray}               
   {\cal E}/A &=&
   \frac{1}{A}\sum_{h\leq k_F}
   \frac{ \tilde{m_h}m+ k_h^2} {\tilde{E}_h}+  
  \label{eq:erel} \\
& &   \frac{1}{2A}\sum_{h \leq k_F,h'\leq k_F}
   \frac{\tilde{m}_h\tilde{m}_{h'}}{\tilde{E}_h\tilde{E}_{h'}}
   \left\langle hh'\right |\tilde{G}
   (\tilde{E}=
    \tilde{E}_h +\tilde{E}_{h'})
    \left | hh' \right\rangle_{\mathrm{AS}}
    -m.  \nonumber
\end{eqnarray}

An alternative approach which we will also discuss is the 
inclusion of relativistic boost corrections to non-relativistic
NN interactions in the VCS calculations of Akmal et al.\  
\cite{apr98,aryathesis98}.
In all analyses, the NN scattering data are reduced to 
the center-of-mass  frame and 
fitted using phase shifts calculated from the nucleon-nucleon
interaction, $V$, 
in that frame.  The interaction obtained by this procedure 
describes the NN interaction in the c.m.\  frame, in which the total momentum 
${\bf P}_{ij} = {\bf p}_i + {\bf p}_j$, is zero.  In general, the 
interaction between particles depends upon their total momentum,  
and can be written as
\begin{equation}
v({\bf P}_{ij}) = v_{ij} + {\delta}v({\bf P}_{ij}),
\label{eq:vofp}
\end{equation}
where $v_{ij}$ is the interaction for ${\bf P}_{ij} = 0$, and 
${\delta}v({\bf P}_{ij})$ is the boost interaction \cite{FPF95} which
is zero when ${\bf P}_{ij} = 0$.

Following the work of Krajcik and Foldy \cite{KF74}, Friar \cite{Fri75} 
obtained the following equation relating the boost interaction of order
$P^2$ to the interaction in the center of mass frame:
\begin{equation}
{\delta}v({\bf P}) = -\frac{P^2}{8m^2} v 
+\frac{1}{8m^2}[{\bf P \cdot r \; P \cdot {\nabla}},v] 
+\frac{1}{8m^2}[({\bf \sigma}_i - {\bf \sigma}_j) \times 
{\bf P \cdot \nabla}, v] .
\label{eq:friar}
\end{equation}
The general validity of this equation in relativistic mechanics and field 
theory was recently discussed \cite{FPF95}. 
Incorporating the boost into the interaction yields a non-relativistic 
Hamiltonian of the form:
\begin{equation}
H^\ast_{NR} = \sum \frac{p^2_i}{2m} + \sum (v_{ij} + {\delta}v({\bf P}_{ij})) 
+ \sum V^*_{ijk} + \cdots ,
\label{eq:nrhwb}
\end{equation}
where the ellipsis denotes the three-body boost, and four and higher body 
interactions. This $H^\ast_{NR}$ contains all terms quadratic in the particle 
velocities, and is therefore suitable for complete studies in the 
non-relativistic limit.

The authors of Refs.\ \cite{forest,CPS93} find that 
the contribution of the two-body boost interaction to the energy is
repulsive, with a magnitude which is 37\% of the $V^R_{ijk}$ contribution.
The boost interaction thus accounts for a significant part of
the $V^R_{ijk}$ in Hamiltonians which fit nuclear energies
neglecting ${\delta}v$.

In this work we will follow Refs.\ \cite{ap97,apr98,aryathesis98}
and only keep the terms of the boost 
interaction associated with the static part of $V$, and neglect the
last term in Eq.~(\ref{eq:friar}).  That term is responsible for Thomas 
precession and quantum contributions that are negligibly 
small here \cite{FPC95}.
The correction $\delta v$ is then given by
\begin{equation}
{\delta}v({\bf P}) = -\frac{P^2}{8m^2}v^s +\frac{1}{8m^2}
{\bf P \cdot r}\ {\bf P \cdot \nabla} v^s.
\label{eq:afriar}
\end{equation}
The two terms are due to the relativistic energy expression and Lorentz 
contraction, and are denoted $\delta v^{RE}$ and $\delta v^{LC}$,
respectively.  The three-nucleon 
interaction used in the $H^\ast_{NR}$ of Eq.~(\ref{eq:nrhwb})
is denoted by $V^\ast_{ijk}$.  Its parameters are obtained by fitting the 
binding energies of $^3$H and $^4$He, and the equilibrium density of 
SNM, including $\delta v$. The strength of $V^{R\ast}_{ijk}$ is 0.63 times that
of $V^R_{ijk}$ in UIX, while that of $V^{2 \pi}_{ijk}$ is unchanged.  The 
resulting model of $V_{ijk}$ is called UIX$^\ast$.

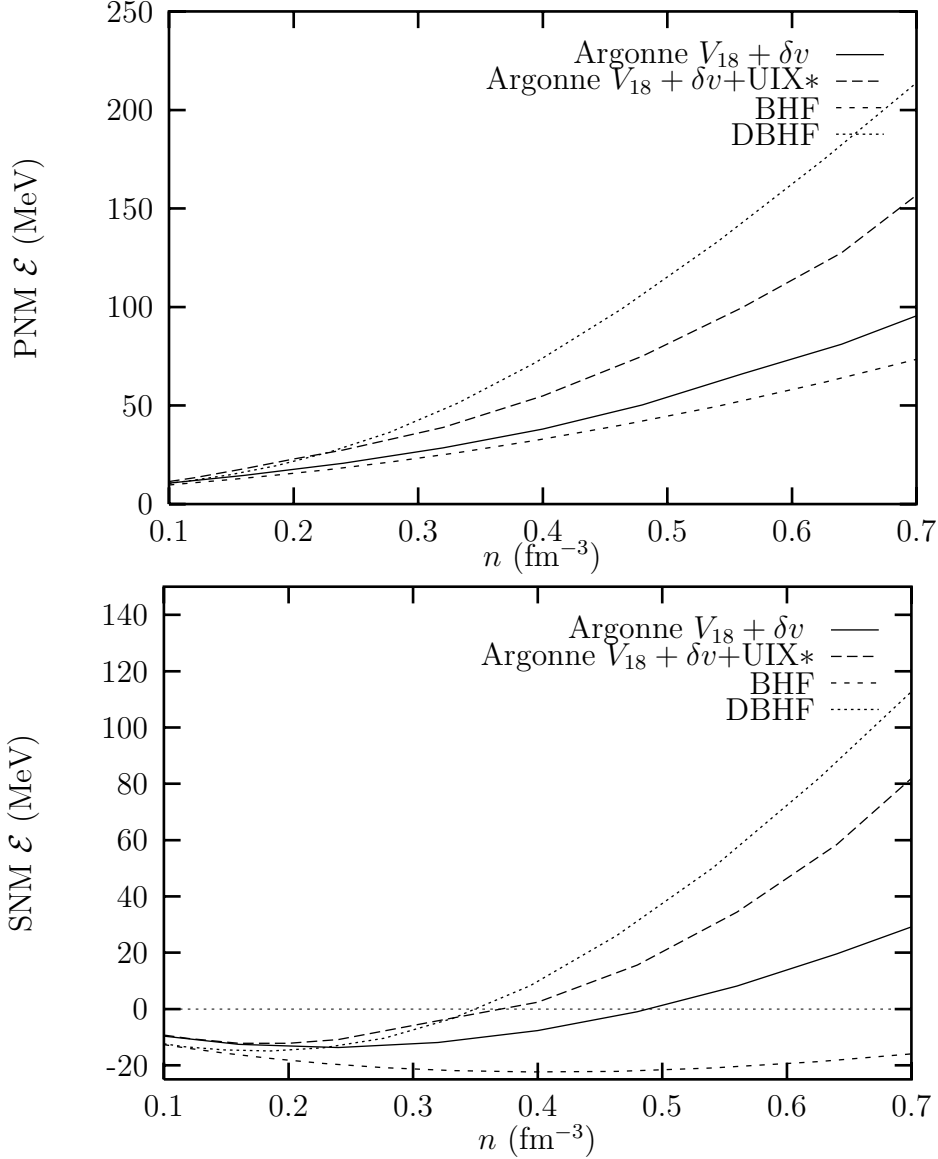
\begin{figure}\begin{center} 
     % GNUPLOT: LaTeX picture with Postscript
\setlength{\unitlength}{0.1bp}
\special{!
%!PS-Adobe-2.0
%%Creator: gnuplot
%%DocumentFonts: Helvetica
%%BoundingBox: 50 50 770 554
%%Pages: (atend)
%%EndComments
/gnudict 40 dict def
gnudict begin
/Color false def
/Solid false def
/gnulinewidth 5.000 def
/vshift -33 def
/dl {10 mul} def
/hpt 31.5 def
/vpt 31.5 def
/M {moveto} bind def
/L {lineto} bind def
/R {rmoveto} bind def
/V {rlineto} bind def
/vpt2 vpt 2 mul def
/hpt2 hpt 2 mul def
/Lshow { currentpoint stroke M
  0 vshift R show } def
/Rshow { currentpoint stroke M
  dup stringwidth pop neg vshift R show } def
/Cshow { currentpoint stroke M
  dup stringwidth pop -2 div vshift R show } def
/DL { Color {setrgbcolor Solid {pop []} if 0 setdash }
 {pop pop pop Solid {pop []} if 0 setdash} ifelse } def
/BL { stroke gnulinewidth 2 mul setlinewidth } def
/AL { stroke gnulinewidth 2 div setlinewidth } def
/PL { stroke gnulinewidth setlinewidth } def
/LTb { BL [] 0 0 0 DL } def
/LTa { AL [1 dl 2 dl] 0 setdash 0 0 0 setrgbcolor } def
/LT0 { PL [] 0 1 0 DL } def
/LT1 { PL [4 dl 2 dl] 0 0 1 DL } def
/LT2 { PL [2 dl 3 dl] 1 0 0 DL } def
/LT3 { PL [1 dl 1.5 dl] 1 0 1 DL } def
/LT4 { PL [5 dl 2 dl 1 dl 2 dl] 0 1 1 DL } def
/LT5 { PL [4 dl 3 dl 1 dl 3 dl] 1 1 0 DL } def
/LT6 { PL [2 dl 2 dl 2 dl 4 dl] 0 0 0 DL } def
/LT7 { PL [2 dl 2 dl 2 dl 2 dl 2 dl 4 dl] 1 0.3 0 DL } def
/LT8 { PL [2 dl 2 dl 2 dl 2 dl 2 dl 2 dl 2 dl 4 dl] 0.5 0.5 0.5 DL } def
/P { stroke [] 0 setdash
  currentlinewidth 2 div sub M
  0 currentlinewidth V stroke } def
/D { stroke [] 0 setdash 2 copy vpt add M
  hpt neg vpt neg V hpt vpt neg V
  hpt vpt V hpt neg vpt V closepath stroke
  P } def
/A { stroke [] 0 setdash vpt sub M 0 vpt2 V
  currentpoint stroke M
  hpt neg vpt neg R hpt2 0 V stroke
  } def
/B { stroke [] 0 setdash 2 copy exch hpt sub exch vpt add M
  0 vpt2 neg V hpt2 0 V 0 vpt2 V
  hpt2 neg 0 V closepath stroke
  P } def
/C { stroke [] 0 setdash exch hpt sub exch vpt add M
  hpt2 vpt2 neg V currentpoint stroke M
  hpt2 neg 0 R hpt2 vpt2 V stroke } def
/T { stroke [] 0 setdash 2 copy vpt 1.12 mul add M
  hpt neg vpt -1.62 mul V
  hpt 2 mul 0 V
  hpt neg vpt 1.62 mul V closepath stroke
  P  } def
/S { 2 copy A C} def
end
}
\begin{picture}(3600,2160)(0,0)
\special{"
gnudict begin
gsave
50 50 translate
0.100 0.100 scale
0 setgray
/Helvetica findfont 100 scalefont setfont
newpath
-500.000000 -500.000000 translate
LTa
600 251 M
2817 0 V
LTb
600 251 M
63 0 V
2754 0 R
-63 0 V
600 623 M
63 0 V
2754 0 R
-63 0 V
600 994 M
63 0 V
2754 0 R
-63 0 V
600 1366 M
63 0 V
2754 0 R
-63 0 V
600 1737 M
63 0 V
2754 0 R
-63 0 V
600 2109 M
63 0 V
2754 0 R
-63 0 V
600 251 M
0 63 V
0 1795 R
0 -63 V
1070 251 M
0 63 V
0 1795 R
0 -63 V
1539 251 M
0 63 V
0 1795 R
0 -63 V
2009 251 M
0 63 V
0 1795 R
0 -63 V
2478 251 M
0 63 V
0 1795 R
0 -63 V
2948 251 M
0 63 V
0 1795 R
0 -63 V
3417 251 M
0 63 V
0 1795 R
0 -63 V
600 251 M
2817 0 V
0 1858 V
-2817 0 V
600 251 L
LT0
3114 1946 M
180 0 V
600 330 M
282 29 V
375 46 V
376 58 V
376 71 V
375 91 V
376 117 V
375 112 V
282 107 V
LT1
3114 1846 M
180 0 V
600 336 M
94 14 V
188 34 V
188 37 V
187 34 V
376 85 V
376 119 V
375 150 V
376 183 V
375 207 V
282 217 V
LT2
3114 1746 M
180 0 V
600 323 M
78 7 V
149 13 V
174 17 V
200 21 V
229 28 V
259 36 V
291 46 V
326 58 V
362 71 V
400 90 V
349 87 V
LT3
3114 1646 M
180 0 V
600 329 M
78 10 V
149 22 V
174 34 V
200 51 V
229 75 V
259 111 V
291 152 V
326 203 V
362 254 V
400 310 V
349 289 V
stroke
grestore
end
showpage
}
\put(3054,1646){\makebox(0,0)[r]{DBHF}}
\put(3054,1746){\makebox(0,0)[r]{BHF}}
\put(3054,1846){\makebox(0,0)[r]{Argonne $V_{18}+\delta v$+UIX$\ast$}}
\put(3054,1946){\makebox(0,0)[r]{Argonne $V_{18}+\delta v$ }}
\put(2008,51){\makebox(0,0){$n$ (fm$^{-3}$)}}
\put(100,1180){%
\special{ps: gsave currentpoint currentpoint translate
270 rotate neg exch neg exch translate}%
\makebox(0,0)[b]{\shortstack{PNM ${\cal E}$ (MeV)}}%
\special{ps: currentpoint grestore moveto}%
}
\put(3417,151){\makebox(0,0){0.7}}
\put(2948,151){\makebox(0,0){0.6}}
\put(2478,151){\makebox(0,0){0.5}}
\put(2009,151){\makebox(0,0){0.4}}
\put(1539,151){\makebox(0,0){0.3}}
\put(1070,151){\makebox(0,0){0.2}}
\put(600,151){\makebox(0,0){0.1}}
\put(540,2109){\makebox(0,0)[r]{250}}
\put(540,1737){\makebox(0,0)[r]{200}}
\put(540,1366){\makebox(0,0)[r]{150}}
\put(540,994){\makebox(0,0)[r]{100}}
\put(540,623){\makebox(0,0)[r]{50}}
\put(540,251){\makebox(0,0)[r]{0}}
\end{picture}
\begin{picture}(3600,2160)(0,0)
\special{"
gnudict begin
gsave
50 50 translate
0.100 0.100 scale
0 setgray
/Helvetica findfont 100 scalefont setfont
newpath
-500.000000 -500.000000 translate
LTa
600 516 M
2817 0 V
LTb
600 304 M
63 0 V
2754 0 R
-63 0 V
600 516 M
63 0 V
2754 0 R
-63 0 V
600 729 M
63 0 V
2754 0 R
-63 0 V
600 941 M
63 0 V
2754 0 R
-63 0 V
600 1153 M
63 0 V
2754 0 R
-63 0 V
600 1366 M
63 0 V
2754 0 R
-63 0 V
600 1578 M
63 0 V
2754 0 R
-63 0 V
600 1790 M
63 0 V
2754 0 R
-63 0 V
600 2003 M
63 0 V
2754 0 R
-63 0 V
600 251 M
0 63 V
0 1795 R
0 -63 V
1070 251 M
0 63 V
0 1795 R
0 -63 V
1539 251 M
0 63 V
0 1795 R
0 -63 V
2009 251 M
0 63 V
0 1795 R
0 -63 V
2478 251 M
0 63 V
0 1795 R
0 -63 V
2948 251 M
0 63 V
0 1795 R
0 -63 V
3417 251 M
0 63 V
0 1795 R
0 -63 V
600 251 M
2817 0 V
0 1858 V
-2817 0 V
600 251 L
LT0
3114 1946 M
180 0 V
600 414 M
882 383 L
375 -12 V
376 19 V
376 45 V
375 71 V
376 97 V
375 121 V
282 102 V
LT1
3114 1846 M
180 0 V
600 418 M
94 -13 V
882 387 L
188 0 V
187 14 V
376 71 V
376 70 V
375 140 V
376 200 V
375 254 V
282 250 V
LT2
3114 1746 M
180 0 V
600 382 M
78 -12 V
827 350 L
174 -20 V
200 -19 V
229 -16 V
259 -11 V
291 -5 V
326 2 V
362 13 V
400 24 V
349 29 V
LT3
3114 1646 M
180 0 V
600 385 M
78 -11 V
827 362 L
174 -4 V
200 12 V
229 36 V
259 75 V
291 123 V
326 189 V
362 255 V
400 338 V
349 327 V
stroke
grestore
end
showpage
}
\put(3054,1646){\makebox(0,0)[r]{DBHF}}
\put(3054,1746){\makebox(0,0)[r]{BHF}}
\put(3054,1846){\makebox(0,0)[r]{Argonne $V_{18}+\delta v+$UIX$\ast$}}
\put(3054,1946){\makebox(0,0)[r]{Argonne $V_{18}+\delta v$ }}
\put(2008,21){\makebox(0,0){$n$ (fm$^{-3}$)}}
\put(100,1180){%
\special{ps: gsave currentpoint currentpoint translate
270 rotate neg exch neg exch translate}%
\makebox(0,0)[b]{\shortstack{SNM ${\cal E}$ (MeV)}}%
\special{ps: currentpoint grestore moveto}%
}
\put(3417,151){\makebox(0,0){0.7}}
\put(2948,151){\makebox(0,0){0.6}}
\put(2478,151){\makebox(0,0){0.5}}
\put(2009,151){\makebox(0,0){0.4}}
\put(1539,151){\makebox(0,0){0.3}}
\put(1070,151){\makebox(0,0){0.2}}
\put(600,151){\makebox(0,0){0.1}}
\put(540,2003){\makebox(0,0)[r]{140}}
\put(540,1790){\makebox(0,0)[r]{120}}
\put(540,1578){\makebox(0,0)[r]{100}}
\put(540,1366){\makebox(0,0)[r]{80}}
\put(540,1153){\makebox(0,0)[r]{60}}
\put(540,941){\makebox(0,0)[r]{40}}
\put(540,729){\makebox(0,0)[r]{20}}
\put(540,516){\makebox(0,0)[r]{0}}
\put(540,304){\makebox(0,0)[r]{-20}}
\end{picture}
     \caption{Upper panel: DBHF and BHF energy per nucleon for PNM. The 
       results with boost corrections of
       Akmal et al.\  \protect\cite{apr98} with
       ($V_{18}+\delta v$+UIX$\ast$) and 
       without ($V_{18}+\delta v$) three-body forces.
       Lower panel: the corresponding results for SNM.}
     \label{fig:sec2fig15}
\end{center}\end{figure}

\begin{figure}\begin{center} 
     \input{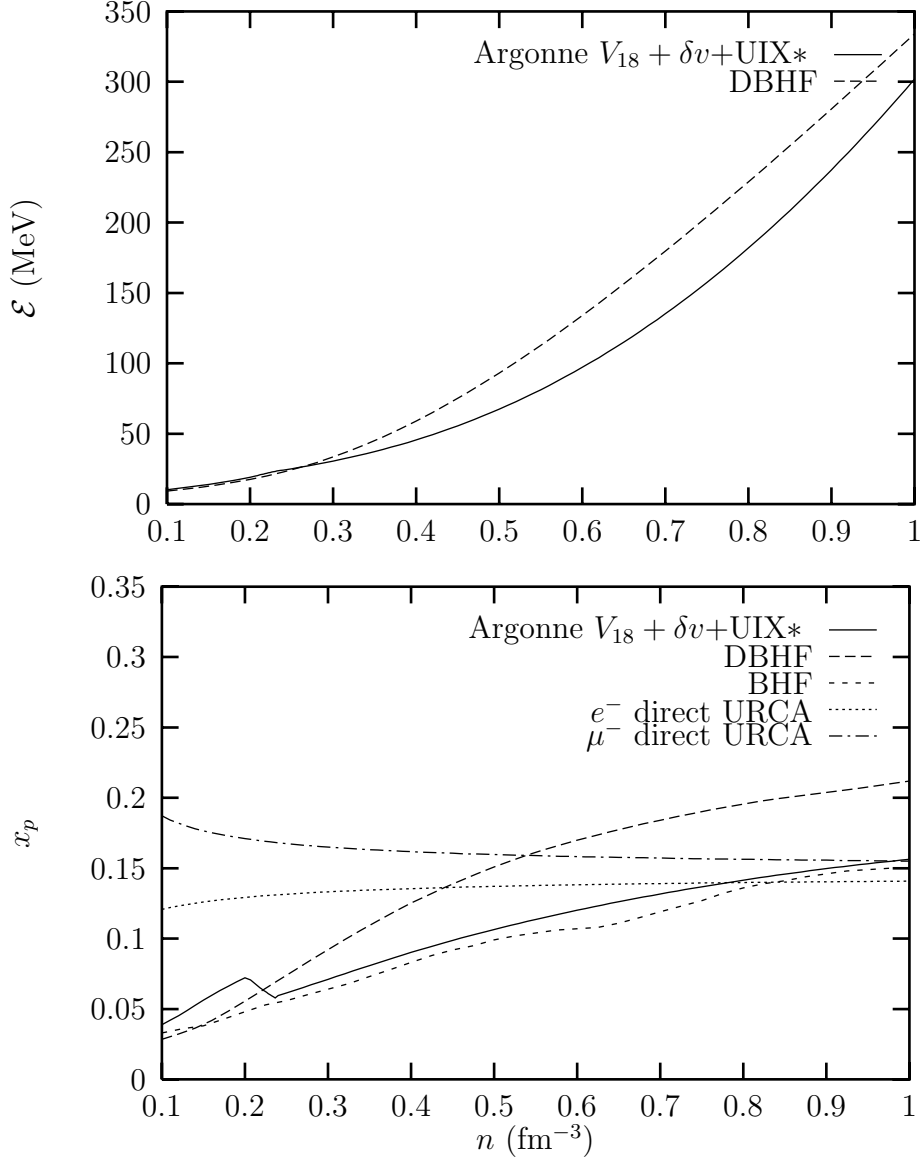}
     \caption{Upper panel: Energy per nucleon in $\beta-stable$ matter for the 
DBHF 
      approach and the results of
       Akmal et al.\ with boost
       corrections and three-body forces ($V_{18}+\delta v$+UIX$\ast$).
       Lower panel: the corresponding proton fraction $x_p$.
       For the DBHF calculation both electrons and muons are included.}
     \label{fig:sec2fig16}
\end{center}\end{figure}

In Fig.\ \ref{fig:sec2fig15} we plot the non-relativistic BHF 
and DBHF energies per particle for both
SNM and PNM. These results were discussed in Ref.\ \cite{ehobo96}
and were obtained with an older version of the Bonn interactions, namely the
so-called Bonn A potential. The relativistic corrections include in an
effective way  a certain
class of three-body corrections, the so-called Z-graphs \cite{brown87}, although
the calculations are done only at the two-body level.
The non-relativistic results are similar to those obtained with the 
CD-Bonn interaction discussed in the previous subsection. One notices also
that the DBHF results yield saturation properties close to the empirical data,
a binding energy of $-15.1$ MeV and saturation density of $0.19$ fm$^{-3}$,
while the non-relativistic results saturate at $0.37$   fm$^{-3}$ with a binding
energy of $-22.4$ MeV.
For comparison we have also included the results of Akmal et al.\  
\cite{ap97,apr98} with
three-body diagrams and boost corrections. 
Again, one notices that at densities below $n_0$ all many-body schemes tend
to give similar results, although in this case, the relativistic results
are in better agreement with the results of the full calculation of 
Akmal et al.\  \cite{ap97,apr98}. 
For completeness, we display the energy per particle for 
$\beta$-stable matter obtained 
with the DBHF 
 and the results from the calculations of Akmal et al.\ 
\cite{apr98,aryathesis98} with boost corrections and three-body
forces in Fig.\  \ref{fig:sec2fig16}. The corresponding proton fractions 
are also plotted. The dip in the proton fraction of Akmal et al.\
at $\sim 0.2$ fm$^{-3}$ is due to the formation of a pion condensate.
For higher densities we note that BHF and the calculations 
of Akmal et al.\ yield rather similar proton fractions.
The direct Urca starts only at densities $\sim 1$ fm$^{-3}$, while the 
DBHF calculations allow the direct Urca at much lower densities.

\subsection{A causal parametrization of the nuclear matter EoS}
\label{subsec:paramsec2}. 

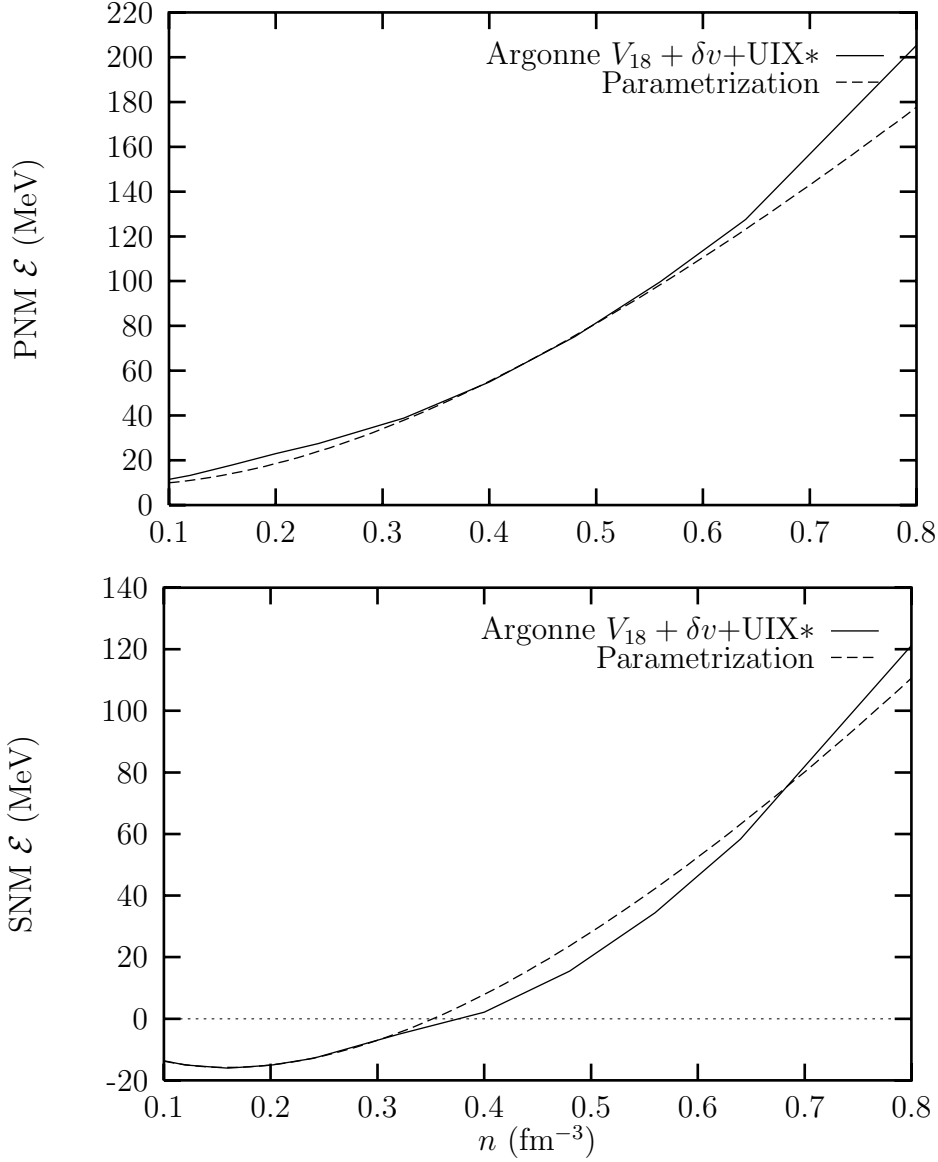
\begin{figure}\begin{center} 
     % GNUPLOT: LaTeX picture with Postscript
\setlength{\unitlength}{0.1bp}
\special{!
%!PS-Adobe-2.0
%%Creator: gnuplot
%%DocumentFonts: Helvetica
%%BoundingBox: 50 50 770 554
%%Pages: (atend)
%%EndComments
/gnudict 40 dict def
gnudict begin
/Color false def
/Solid false def
/gnulinewidth 5.000 def
/vshift -33 def
/dl {10 mul} def
/hpt 31.5 def
/vpt 31.5 def
/M {moveto} bind def
/L {lineto} bind def
/R {rmoveto} bind def
/V {rlineto} bind def
/vpt2 vpt 2 mul def
/hpt2 hpt 2 mul def
/Lshow { currentpoint stroke M
  0 vshift R show } def
/Rshow { currentpoint stroke M
  dup stringwidth pop neg vshift R show } def
/Cshow { currentpoint stroke M
  dup stringwidth pop -2 div vshift R show } def
/DL { Color {setrgbcolor Solid {pop []} if 0 setdash }
 {pop pop pop Solid {pop []} if 0 setdash} ifelse } def
/BL { stroke gnulinewidth 2 mul setlinewidth } def
/AL { stroke gnulinewidth 2 div setlinewidth } def
/PL { stroke gnulinewidth setlinewidth } def
/LTb { BL [] 0 0 0 DL } def
/LTa { AL [1 dl 2 dl] 0 setdash 0 0 0 setrgbcolor } def
/LT0 { PL [] 0 1 0 DL } def
/LT1 { PL [4 dl 2 dl] 0 0 1 DL } def
/LT2 { PL [2 dl 3 dl] 1 0 0 DL } def
/LT3 { PL [1 dl 1.5 dl] 1 0 1 DL } def
/LT4 { PL [5 dl 2 dl 1 dl 2 dl] 0 1 1 DL } def
/LT5 { PL [4 dl 3 dl 1 dl 3 dl] 1 1 0 DL } def
/LT6 { PL [2 dl 2 dl 2 dl 4 dl] 0 0 0 DL } def
/LT7 { PL [2 dl 2 dl 2 dl 2 dl 2 dl 4 dl] 1 0.3 0 DL } def
/LT8 { PL [2 dl 2 dl 2 dl 2 dl 2 dl 2 dl 2 dl 4 dl] 0.5 0.5 0.5 DL } def
/P { stroke [] 0 setdash
  currentlinewidth 2 div sub M
  0 currentlinewidth V stroke } def
/D { stroke [] 0 setdash 2 copy vpt add M
  hpt neg vpt neg V hpt vpt neg V
  hpt vpt V hpt neg vpt V closepath stroke
  P } def
/A { stroke [] 0 setdash vpt sub M 0 vpt2 V
  currentpoint stroke M
  hpt neg vpt neg R hpt2 0 V stroke
  } def
/B { stroke [] 0 setdash 2 copy exch hpt sub exch vpt add M
  0 vpt2 neg V hpt2 0 V 0 vpt2 V
  hpt2 neg 0 V closepath stroke
  P } def
/C { stroke [] 0 setdash exch hpt sub exch vpt add M
  hpt2 vpt2 neg V currentpoint stroke M
  hpt2 neg 0 R hpt2 vpt2 V stroke } def
/T { stroke [] 0 setdash 2 copy vpt 1.12 mul add M
  hpt neg vpt -1.62 mul V
  hpt 2 mul 0 V
  hpt neg vpt 1.62 mul V closepath stroke
  P  } def
/S { 2 copy A C} def
end
}
\begin{picture}(3600,2160)(0,0)
\special{"
gnudict begin
gsave
50 50 translate
0.100 0.100 scale
0 setgray
/Helvetica findfont 100 scalefont setfont
newpath
-500.000000 -500.000000 translate
LTa
600 251 M
2817 0 V
LTb
600 251 M
63 0 V
2754 0 R
-63 0 V
600 420 M
63 0 V
2754 0 R
-63 0 V
600 589 M
63 0 V
2754 0 R
-63 0 V
600 758 M
63 0 V
2754 0 R
-63 0 V
600 927 M
63 0 V
2754 0 R
-63 0 V
600 1096 M
63 0 V
2754 0 R
-63 0 V
600 1264 M
63 0 V
2754 0 R
-63 0 V
600 1433 M
63 0 V
2754 0 R
-63 0 V
600 1602 M
63 0 V
2754 0 R
-63 0 V
600 1771 M
63 0 V
2754 0 R
-63 0 V
600 1940 M
63 0 V
2754 0 R
-63 0 V
600 2109 M
63 0 V
2754 0 R
-63 0 V
600 251 M
0 63 V
0 1795 R
0 -63 V
1002 251 M
0 63 V
0 1795 R
0 -63 V
1405 251 M
0 63 V
0 1795 R
0 -63 V
1807 251 M
0 63 V
0 1795 R
0 -63 V
2210 251 M
0 63 V
0 1795 R
0 -63 V
2612 251 M
0 63 V
0 1795 R
0 -63 V
3015 251 M
0 63 V
0 1795 R
0 -63 V
3417 251 M
0 63 V
0 1795 R
0 -63 V
600 251 M
2817 0 V
0 1858 V
-2817 0 V
600 251 L
LT0
3114 1946 M
180 0 V
600 348 M
80 15 V
161 40 V
161 42 V
161 38 V
322 96 V
322 136 V
322 171 V
322 207 V
322 235 V
644 657 V
LT1
3114 1846 M
180 0 V
600 335 M
40 4 V
40 5 V
41 6 V
40 6 V
40 7 V
40 8 V
41 8 V
40 9 V
40 9 V
40 11 V
41 10 V
40 11 V
40 12 V
40 13 V
41 12 V
40 14 V
40 14 V
40 14 V
41 15 V
40 16 V
40 16 V
40 16 V
41 17 V
40 17 V
40 18 V
40 18 V
41 19 V
40 19 V
40 19 V
40 20 V
41 21 V
40 20 V
40 21 V
40 21 V
40 22 V
41 22 V
40 22 V
40 23 V
40 23 V
41 23 V
40 24 V
40 24 V
40 24 V
41 24 V
40 25 V
40 25 V
40 25 V
41 26 V
40 26 V
40 26 V
40 26 V
41 27 V
40 26 V
40 27 V
40 28 V
41 27 V
40 28 V
40 28 V
40 28 V
41 28 V
40 29 V
40 28 V
40 29 V
41 29 V
40 30 V
40 29 V
40 30 V
41 29 V
40 30 V
40 31 V
stroke
grestore
end
showpage
}
\put(3054,1846){\makebox(0,0)[r]{Parametrization}}
\put(3054,1946){\makebox(0,0)[r]{Argonne $V_{18}+\delta v+$UIX$\ast$}}
\put(100,1180){%
\special{ps: gsave currentpoint currentpoint translate
270 rotate neg exch neg exch translate}%
\makebox(0,0)[b]{\shortstack{PNM ${\cal E}$ (MeV)}}%
\special{ps: currentpoint grestore moveto}%
}
\put(3417,151){\makebox(0,0){0.8}}
\put(3015,151){\makebox(0,0){0.7}}
\put(2612,151){\makebox(0,0){0.6}}
\put(2210,151){\makebox(0,0){0.5}}
\put(1807,151){\makebox(0,0){0.4}}
\put(1405,151){\makebox(0,0){0.3}}
\put(1002,151){\makebox(0,0){0.2}}
\put(600,151){\makebox(0,0){0.1}}
\put(540,2109){\makebox(0,0)[r]{220}}
\put(540,1940){\makebox(0,0)[r]{200}}
\put(540,1771){\makebox(0,0)[r]{180}}
\put(540,1602){\makebox(0,0)[r]{160}}
\put(540,1433){\makebox(0,0)[r]{140}}
\put(540,1264){\makebox(0,0)[r]{120}}
\put(540,1096){\makebox(0,0)[r]{100}}
\put(540,927){\makebox(0,0)[r]{80}}
\put(540,758){\makebox(0,0)[r]{60}}
\put(540,589){\makebox(0,0)[r]{40}}
\put(540,420){\makebox(0,0)[r]{20}}
\put(540,251){\makebox(0,0)[r]{0}}
\end{picture}
\begin{picture}(3600,2160)(0,0)
\special{"
gnudict begin
gsave
50 50 translate
0.100 0.100 scale
0 setgray
/Helvetica findfont 100 scalefont setfont
newpath
-500.000000 -500.000000 translate
LTa
600 483 M
2817 0 V
LTb
600 251 M
63 0 V
2754 0 R
-63 0 V
600 483 M
63 0 V
2754 0 R
-63 0 V
600 716 M
63 0 V
2754 0 R
-63 0 V
600 948 M
63 0 V
2754 0 R
-63 0 V
600 1180 M
63 0 V
2754 0 R
-63 0 V
600 1412 M
63 0 V
2754 0 R
-63 0 V
600 1645 M
63 0 V
2754 0 R
-63 0 V
600 1877 M
63 0 V
2754 0 R
-63 0 V
600 2109 M
63 0 V
2754 0 R
-63 0 V
600 251 M
0 63 V
0 1795 R
0 -63 V
1002 251 M
0 63 V
0 1795 R
0 -63 V
1405 251 M
0 63 V
0 1795 R
0 -63 V
1807 251 M
0 63 V
0 1795 R
0 -63 V
2210 251 M
0 63 V
0 1795 R
0 -63 V
2612 251 M
0 63 V
0 1795 R
0 -63 V
3015 251 M
0 63 V
0 1795 R
0 -63 V
3417 251 M
0 63 V
0 1795 R
0 -63 V
600 251 M
2817 0 V
0 1858 V
-2817 0 V
600 251 L
LT0
3114 1946 M
180 0 V
600 325 M
80 -16 V
841 297 L
161 11 V
161 26 V
322 91 V
322 83 V
322 155 V
322 220 V
322 278 V
644 730 V
LT1
3114 1846 M
180 0 V
600 323 M
40 -7 V
40 -6 V
41 -5 V
40 -3 V
40 -2 V
40 0 V
41 0 V
40 2 V
40 3 V
40 4 V
41 5 V
40 6 V
40 7 V
40 8 V
41 9 V
40 10 V
40 11 V
40 11 V
41 13 V
40 13 V
40 14 V
40 15 V
41 15 V
40 17 V
40 17 V
40 17 V
41 19 V
40 19 V
40 20 V
40 20 V
41 21 V
40 21 V
40 23 V
40 22 V
40 24 V
41 23 V
40 25 V
40 24 V
40 26 V
41 26 V
40 26 V
40 27 V
40 27 V
41 28 V
40 28 V
40 28 V
40 29 V
41 30 V
40 29 V
40 31 V
40 30 V
41 31 V
40 31 V
40 32 V
40 32 V
41 33 V
40 32 V
40 33 V
40 34 V
41 33 V
40 34 V
40 35 V
40 34 V
41 35 V
40 35 V
40 36 V
40 36 V
41 36 V
40 36 V
40 37 V
stroke
grestore
end
showpage
}
\put(3054,1846){\makebox(0,0)[r]{Parametrization}}
\put(3054,1946){\makebox(0,0)[r]{Argonne $V_{18}+\delta v+$UIX$\ast$}}
\put(2008,21){\makebox(0,0){$n$ (fm$^{-3}$)}}
\put(100,1180){%
\special{ps: gsave currentpoint currentpoint translate
270 rotate neg exch neg exch translate}%
\makebox(0,0)[b]{\shortstack{SNM ${\cal E}$ (MeV)}}%
\special{ps: currentpoint grestore moveto}%
}
\put(3417,151){\makebox(0,0){0.8}}
\put(3015,151){\makebox(0,0){0.7}}
\put(2612,151){\makebox(0,0){0.6}}
\put(2210,151){\makebox(0,0){0.5}}
\put(1807,151){\makebox(0,0){0.4}}
\put(1405,151){\makebox(0,0){0.3}}
\put(1002,151){\makebox(0,0){0.2}}
\put(600,151){\makebox(0,0){0.1}}
\put(540,2109){\makebox(0,0)[r]{140}}
\put(540,1877){\makebox(0,0)[r]{120}}
\put(540,1645){\makebox(0,0)[r]{100}}
\put(540,1412){\makebox(0,0)[r]{80}}
\put(540,1180){\makebox(0,0)[r]{60}}
\put(540,948){\makebox(0,0)[r]{40}}
\put(540,716){\makebox(0,0)[r]{20}}
\put(540,483){\makebox(0,0)[r]{0}}
\put(540,251){\makebox(0,0)[r]{-20}}
\end{picture}
     \caption{Upper panel: Comparison of the parametrized EoS of 
       Eq.\ (\protect\ref{eq:EA})
       and the results of 
       Akmal et al.\ \protect\cite{apr98}  with boost
       corrections and three-body forces ($V_{18}+\delta v+$UIX$\ast$)
       for PNM.
       Lower panel: the corresponding results for SNM.}
     \label{fig:sec2fig18}
\end{center}\end{figure}

\begin{figure}\begin{center} 
     \input{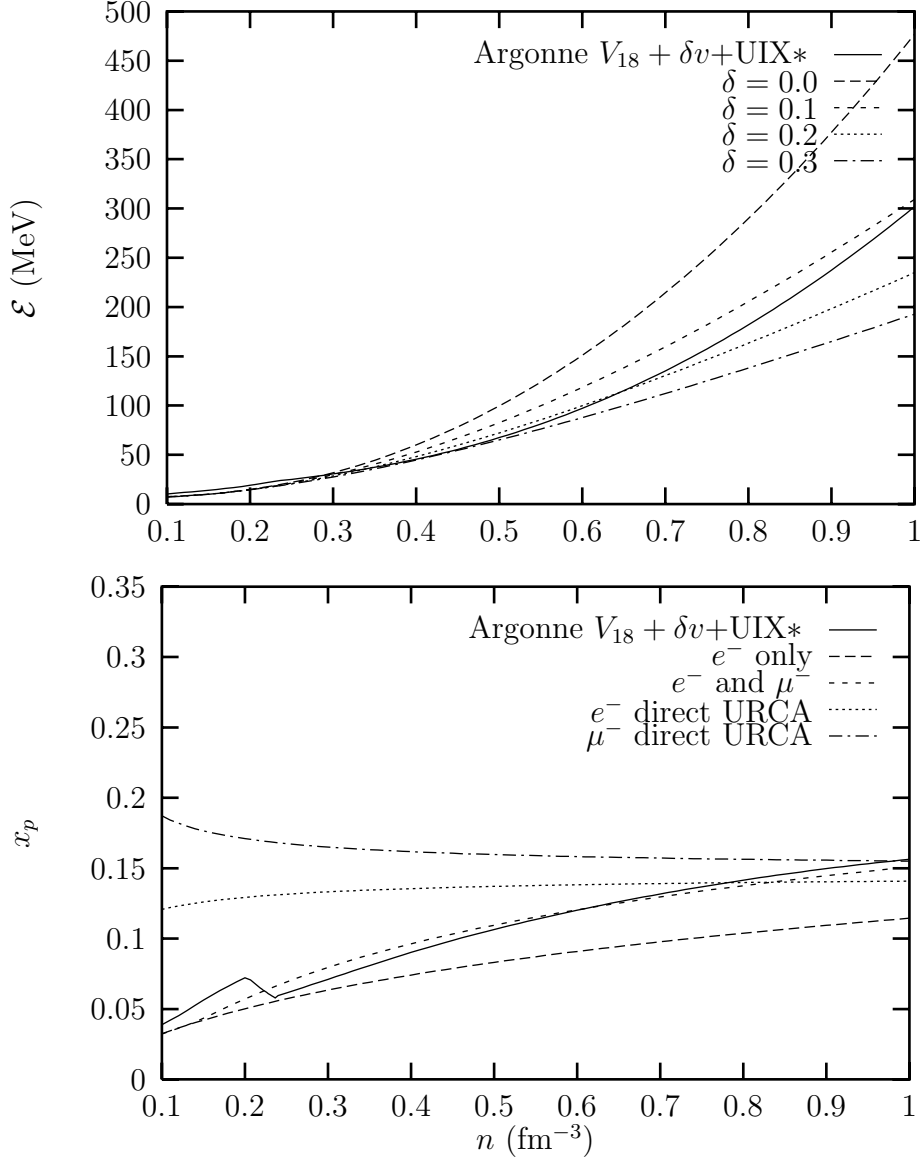}
     \caption{Upper panel: Energy per nucleon in $\beta$-stable matter for the
      parametrized EoS of Eq.\ \protect{(\ref{eq:EA})} 
       for $\delta=0,0.13,0.2,0.3$ and the results of
       Akmal et al.\  \protect\cite{apr98} with boost
       corrections and three-body forces ($V_{18}+\delta v+$UIX$\ast$).
       Lower panel: the corresponding proton fraction $x_p$.}
     \label{fig:sec2fig19}
\end{center}\end{figure}

\begin{figure}\begin{center}
     \input{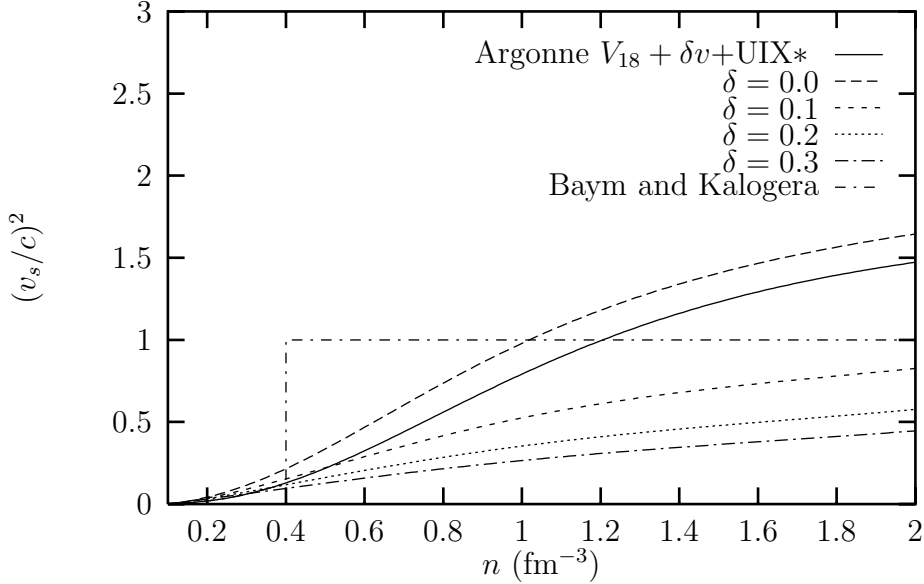}
       \caption{$(v_s/c)^2$ for $\delta=0,0.1,0.2,0.3$, 
                the results of Akmal et al. \protect\cite{apr98}, 
                and for Baym and Kalogera's \protect\cite{kalogera}
                patched EoS which shows 
                a discontinuous $(v_s/c)^2$. }
       \label{fig:sec2fig20}
\end{center}\end{figure}

Since three-body forces are expected to be important (it suffices
to mention studies of the Triton \cite{nhkg97}), 
we will in our discussions of the mixed phase in the next section and 
in connection with the structure of a neutron star, employ the 
recent EoS of Akmal et al.\  \cite{apr98}. 
A non-relativistic EoS (although with the inclusion of boost corrections)
is preferred here.
The EoS for nuclear matter is thus known to some accuracy for 
densities up to a few times nuclear saturation density, $n_0=0.16$ fm$^{-3}$.
Detailed knowledge of the EoS is crucial for the existence of, e.g.\  pion
condensates \cite{ap97,apr98} or the delicate structures in the inner crust
of neutron stars \cite{pr95}. However, for the gross properties
and our discussion of properties of neutron stars
we will adopt a simple form for the binding energy per nucleon in nuclear
matter consisting of a compressional term and a symmetry term
\begin{equation}
    {\cal E}=E_{comp}(n) + S(n)(1-2x_p)^2
    ={\cal E}_0 u\frac{u-2-\delta}{1+\delta u}
           +S_0 u^\gamma (1-2x_p)^2.   
    \label{eq:EA} 
\end{equation}
Here $u=n/n_0$ is the ratio of the baryon density to nuclear saturation density
and we have defined the proton fraction $x_p=n_p/n$.
The compressional term is in (\ref{eq:EA}) parametrized by a simple form which
reproduces the saturation density, binding energy and compressibility. 
The binding
energy per nucleon at saturation density excluding Coulomb energies is
${\cal E}_0=-15.8$MeV and the parameter $\delta= 0.2$ is determined by
fitting the energy per nucleon at high density to the EoS of
Akmal et al.\ \cite{apr98} with three-body forces and boost corrections, 
but taking the corrected values from  Table 6 of Ref.\ \cite{apr98}. 
The reason behind the choice of $\delta=0.2$ will explained below
in connection with the discussion of Figs.\ \ref{fig:sec2fig19} and
\ref{fig:sec2fig20}. Further discussions will be presented
in section \ref{sec:starproperties}.
The corresponding
compressibility is $K_0=18{\cal E}_0/(1+\delta)\simeq 200$MeV in agreement with
the experimental value. 
For the symmetry term we obtain  $S_0=32$ MeV and $\gamma=0.6$
for the best fit.
The quality of this simple functional exhibits a $\chi^2$ per datum close to 1
and is compared with the results of Akmal et al.\ \cite{apr98}
for both PNM and SNM in Fig.\ \ref{fig:sec2fig18}.
As can been from this figure, the agreement is rather good except at the
very high densities where the EoS of Akmal et al.\ \cite{apr98}
becomes superluminal
and therefore anyway must be wrong.
A much more sophisticated fit which reproduces the data in terms of
Skyrme functional approach is given by the Akmal et al.\ \cite{apr98}. However,
it is amazing that such a simple quadratic formula fits so well the
data coming from a microscopic calculation. In view of the
uncertainties which pertain to the EoS at higher densities, we feel that 
our parametrization is within present error margins. The agreement
between the microscopic calculation of Akmal et al.\  \cite{apr98} and 
the simple parametrization of Eq.\ (\ref{eq:EA})  may imply that 
the essential many-body physics close to the equilibrium density 
arises 
from two-body and three-body terms to ${\cal E}$ only.  The reason 
being\footnote{This argument is for the energy density, i.e.\ 
$\varepsilon={\cal E}n$.} 
that three-body terms
are proportional with $n^3$ while the two-body terms are proportional with
$n^2$. With three-body terms we obviously intend both effective interactions
and contributions from real three-body forces. The evaluation of the latter is,
as discussed above, still an unsettled problem. 

If we now restrict the attention to matter with electrons only, one can easily
obtain an analytic equation for the proton fraction through the asymmetry
parameter $x$. Recalling the equilibrium conditions for $\beta$-stable matter 
of Eqs.\  (\ref{eq:npebetaequilibrium})-(\ref{eq:chargeconserv})
and using the definitions of the chemical potentials for particle species $i$
of Eq.\ (\ref{eq:chemicalpotdef}) one finds that
\begin{equation}
   \mu_e=\frac{1}{n}\frac{\partial \varepsilon}{\partial x_p},
\end{equation}
and from the latter it is rather  
easy to show that the asymmetry parameter $x$ is given by
(assuming ultra-relativistic electrons) 
\begin{equation}
     n x_p=\frac{\left(4S_0u^{\gamma}(1-2x_p)\right)^3}{3\pi^2}.
     \label{eq:chieq}
\end{equation}
Defining
\begin{equation} 
     a=\frac{2\left(4S_0u^{\gamma}\right)^3}{\pi^2n},
\end{equation}
Eq.\ (\ref{eq:chieq}) reduces to
\begin{equation} 
     3x^3+ax-a=0 \,,
     \label{eq:cubic}
\end{equation}
where $x=1-2x_p$.
Since we will always look at solutions
for densities greater than zero, the cubic equation for $x$
has actually an analytical solution which is real and given by
\begin{equation}
     x= -\frac{2\sqrt{a}}{\tan(2\psi)},
     \label{eq:chi_analytical}
\end{equation}
with $tan\psi =\left(\tan\frac{\phi}{2}\right)^{\frac{1}{3}}$
and $\tan\phi=-2\sqrt{a}/3$.
Note well that $x$ depends on the total baryon density $n$ only.
This means in turn that our parametrization of the EoS can now be rewritten
for $\beta$-stable matter as
\begin{equation}
    {\cal E}={\cal E}_0 u\frac{u-2-\delta}{1+\delta u}
           +S_0 u^\gamma \left(\frac{2\sqrt{a}}{\tan(2\psi)}\right)^2.   
    \label{eq:EAnew} 
\end{equation}
and is an analytical function of density only. 

The quality of our approximation to the EoS of Akmal et al.\ 
\cite{apr98} for other observables than the energy per 
particle is shown in Fig.\ \ref{fig:sec2fig19} for the proton fractions 
derived from the simple expression in Eq.\ (\ref{eq:chi_analytical}). 
In the same figure we display also the resulting 
energy per nucleon in $\beta$-stable matter and compare it with the 
results of  Akmal et al.\  
\cite{apr98} for various values of $\delta$. Note well that the proton fraction
does not depend on the value of $\delta$, see Eq.\ (\ref{eq:chieq}). 
As can be seen from this picture, the EoS with $\delta=0$ yields the stiffest
EoS, and as a consequence it results in a superluminal behavior
at densities greater than $ n\approx 1.0$ fm$^{-3}$. This is seen in 
Fig.\ \ref{fig:sec2fig20} where we plot the 
sound speed $(v_s/c)^2$ for various $\delta$ values and that resulting
from the microscopic calculation of Akmal et al.\  \cite{apr98} 
The form of (\ref{eq:EA}), with the inclusion of the parameter $\delta$,
provides therefore a smooth extrapolation from
small and large densities with the correct behavior in both limits,
i.e.\  the binding energy per nucleon
$E/A={\cal E}$ is linear in number density. In the dilute limit this
is the Lenz (optical) potential. 
At high densities the linearity is 
required by the condition that the sound speed 
$c_s^2=\partial P/\partial\varepsilon$
does not exceed the speed of light. 
This justifies  the introduction of the parameter
$\delta$ in our parametrization and 
explains also our deviation from the results
of Akmal et al.\  at densities greater than $0.6\sim 0.7$ fm$^{-3}$, see
Figs.\ \ref{fig:sec2fig18} and \ref{fig:sec2fig19}. 
For $\delta=0.1$ the EoS becomes superluminal at densities of the order of
6 fm$^{-3}$.
 From the definition of $(v_s/c)^2$ in
Eq.\ (\ref{eq:speedofsound}) and the EoS  of Eq.\ (\ref{eq:EA}) that 
for 
\begin{eqnarray}
 \delta \ga \sqrt{\frac{{\cal E}_0}{m_n}} \simeq 0.13 \,,\label{causal}
\end{eqnarray}
the EoS remains causal for all densities.  The EoS of Akmal et al.\
becomes superluminous at $n\approx 1.1$ fm$^{-3}$.  With this caveat
we have an EoS that reproduces the data of Akmal et al.\ at densities
up to $0.6\sim 0.7$ fm$^{-3}$ and has the right causal behavior at
higher densities.  Furthermore, the differences at higher densities
will also not be of importance in our analysis of the dynamics and
structure of neutrons stars, since the mixed baryon-quark phase, with
realistic values for the bag parameter and the coupling constant
$\alpha_s$ starts at densities around $0.5-0.8$ fm$^{-3}$.

Finally, in Fig.\ \ref{fig:sec2fig20} we have also plotted the sound speed
following the approach of Baym and Kalogera \cite{kalogera}, where 
the sound speed is allowed to jump discontinuously at a chosen density
in order to keep the EoS causal. With this prescription, Baym and Kalogera
were also able to obtain an upper bound for neutron star masses
of $2.9 M_{\odot}$.
The approach of Baym and Kalogera differs substantially from ours since
their EoS is discontinuously stiffened by taking
$v_s=c$ at densities above a certain value $n_c$ which, however, is
lower than $n_{s}=5n_0$ where their nuclear EoS becomes superluminous.
This stiffens the nuclear EoS for densities $n_c<n<n_s$ but softens it
at higher densities. Their resulting maximum masses lie in the range
$2.2M_\odot<M<2.9M_\odot$. Our approach incorporates causality by
reducing the sound speed smoothly towards the speed of light at high
densities. Therefore our maximum mass never exceeds that the of nuclear
EoS of Akmal et al.\ \cite{apr98}. In fact one may argue that at very high
densities particles become relativistic and the sound speed should be
even lower, $v_s^2\simeq c^2/3$, and therefore the softening we get
from incorporating causality is even on the low side.

\subsection{Hyperonic matter}\label{subsec:hyper}

\begin{figure}\begin{center} 
       \setlength{\unitlength}{1mm}
       \begin{picture}(140,120)
       \put(25,5){\epsfxsize=12cm \epsfbox{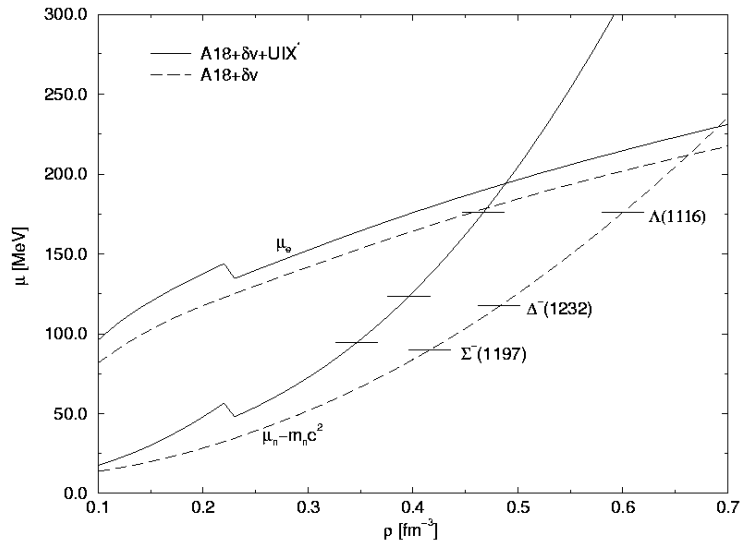}}
       \end{picture}
      \caption{
      The neutron and electron chemical potentials in beta stable matter 
      according to models $V_{18}+\delta v$+UIX$^*$ (full line) 
      and $V_{18}+\delta v$ 
      (dashed line).  Threshold densities for the 
       appearance of non-interacting hyperons are marked by 
      horizontal line segments. Taken from Ref.\ \protect\cite{apr98}.}
     \label{fig:chempots}
\end{center}\end{figure}

At nuclear matter density the electron chemical potential is
$\sim 110$ MeV, see e.g.\  Fig.\ \ref{fig:sec2fig7}. 
Once the rest mass of the muon is exceeded, it becomes
energetically favorable for an electron at the top
of the $e^-$ Fermi surface to decay into a
$\mu^-$. We then develope a Fermi sea of degenerate negative muons,
see again Fig.\ \ref{fig:sec2fig7}. 
In a similar way, as soon as the chemical potential
of the neutron becomes sufficiently large, energetic neutrons
can decay via weak strangeness non-conserving interactions
into $\Lambda$ hyperons leading to a $\Lambda$ Fermi sea
with $\mu_{\Lambda}=\mu_n$. 
However, if we neglect interactions, or assume that their effects
are small, one would expect the $\Sigma^-$ to appear via
\begin{equation}
    e^-+n \rightarrow \Sigma^- +\nu_e,
\end{equation}
at lower densities than the $\Lambda$, even though $\Sigma^-$ is more
massive the reason being that the above process removes both an
energetic neutron and an energetic electron, whereas the decay
to a $\Lambda$, being neutral, removes only an energetic
neutron. Stated differently, the negatively charged hyperons
appear in the ground state of matter when their masses
equal $\mu_e+\mu_n$, while the neutral hyperon $\Lambda$ 
appears when $\mu_n$ equals its mass. Since the 
electron chemical potential in matter is larger than 
the mass difference $m_{\Sigma^-}-m_{\Lambda}= 81.76$ MeV,
the $\Sigma^-$ will appear at lower densities than the $\Lambda$.
We show this in Fig.\ \ref{fig:chempots} where we plot 
the chemical potentials for electrons and neutrons in $\beta$-stable
matter. The threshold densities for $\Sigma^-$, $\Lambda$ and the isobar
$\Delta^-$ are indicated by the horizontal lines.

Since this work has an emphasis on many-body approaches,
we will try to delineate here as well how to obtain properties
of hyperons in dense nuclear matter within the framework
of microscopic theories. 
The main problem we have to face in our case is 
that the hyperon-nucleon interaction and especially the hyperon-hyperon
interaction are less constrained by the data as is the case in 
the nucleonic sector. For a recent version of the Nijmegen model
for the baryon-baryon interaction and applications to nucler matter,
see Stoks et al.\ in Refs.\ \cite{stoks99a,stoks99b}.

In spite of these problems, we will attempt at describing how to 
approach hyperonic matter starting with realistic hyperon-nucleon
and hyperon-hyperon interactions.
We will start with the simplest possible many-body scheme again,
namely lowest order Brueckener theory (LOB). 
The nuclear matter $YN$ $G$-matrix is solved in momentum space and the 
two-particle $YN$ states are defined in terms of 
relative and the center-of-mass
momenta, {\bf k} and {\bf K}, given by

\begin{eqnarray}
   {\bf k} &=&\frac{M_N{\bf k}_Y-M_Y{\bf k}_N}{M_N+M_Y}, \nonumber \\   
   {\bf K} &=&{\bf k}_N+{\bf k}_Y \ , \nonumber   
\end{eqnarray}
where ${\bf k}_N$ and ${\bf k}_Y$ are the nucleon and hyperon momenta,
respectively.
Using an angle-averaged Pauli operator, we perform a partial wave
decomposition of the Bethe-Goldstone equation, which, in terms of the 
quantum numbers of the relative and center-of-mass motion (RCM), is 
written as
\begin{eqnarray}
     &\left\langle (Y''N)k''l''KS''\right |
      G^{\alpha T_z}(\omega)\left | (YN)klKS\right\rangle = &\nonumber\\
      & \left\langle (Y''N)k''l''KS''\right |
      V^{\alpha T_z}\left | (YN)klKS\right\rangle & \nonumber \\
      & +{\displaystyle
      \sum_{l'S'}\sum_{Y'=\Lambda\Sigma}\int k'^{2}dk'}
      \left\langle (Y''N)k''l''KS''\right |
      V^{\alpha T_z}\left | (Y'N)k'l'KS'\right\rangle &\nonumber \\
      & \times \frac{Q_{YN}(k',K)}{\omega -\frac{K^2}{2(M_N+M_{Y^{'}})} -
      \frac{k'^2(M_N+M_{Y^{'}})}{2M_NM_{Y^{'}}}-M_{Y^{'}}+M_Y} &
      \nonumber \\
      & \times\left\langle (Y'N)k'l'KS'\right |
      G^{\alpha T_z}(\omega)\left | (YN)klKS\right\rangle & ,
   \label{eq:gmat}
\end{eqnarray}
where $Q_{YN}$ is the hyperon-nucleon nuclear 
Pauli operator, $V$ is the $YN$ potential
and $\omega$
is the starting energy which corresponds to the sum of 
non-relativistic single-particle energies of the interacting nucleon and 
hyperon. 
Note that kinetic energies are used in the intermediate $Y^\prime N$ states
and $M_{Y}-M_{Y^\prime}$ accounts for the mass difference of the initial 
and intermediate hyperon. The label $\alpha$ represents in this the angular
momentum ${\cal J}$ and the orbital angular momentum $L$ of the 
center-of-mass (c.m.) motion. 
The
variables $k$, $k'$, $k''$ and $l$, $l'$, $l''$ denote relative momenta
and angular momenta, respectively, while $K$ is the momentum 
of the center of mass motion. 

In the Brueckner-Hartree-Fock approach 
the hyperon single-particle potential $U_{Y}$ is obtained
self-consistently by the following sum of
diagonal $G$-matrix elements:
\begin{equation}
     U_{Y}^N(k_{Y})=
   \int_{k_{N}\leq k_{F}}d^3k_{N}\left\langle Y{\bf k}_{Y},N{\bf k}_{N}\right
| G(\varepsilon_N(k_N)+\varepsilon_{Y}(k_Y))\left | Y{\bf
k}_{Y},N{\bf k}_{N}\right\rangle \ ,  
    \label{eq:spenergyhyp}
\end{equation} 
where $\varepsilon_{N(Y)} (k_{N(Y)}) = k_{N(Y)}^2/(2 M_{N(Y)}) +
U_{N(Y)}(k_{N(Y)})$ is
the single-particle energy of the nucleon (hyperon). 
The superscript $N$ refers to the fact that we have 
only nucleon holes.
Using the
partial wave decomposition of the $G$-matrix, the single-particle 
potential $U_Y$  can be rewritten as
\begin{eqnarray}
   U_{Y}(k_{Y})&=&
\frac{(1+\xi_{Y})^3}{2(2t_{Y}+1)}\sum_{{\cal J},l,S,T}(2{\cal J}+1)(2T+1) 
\nonumber \\
& & \times \int_{0}^{k_{max}}k^2dk f(k,k_{Y})
   \left\langle YNklS\right | G^{\alpha T_z}(\omega)\left | YNklS\right\rangle ,
\label{eq:uy}
\end{eqnarray}
where an average over the hyperon spin 
and isospin ($t_Y$) has been performed and the weak center-of-mass
dependence of the $G$-matrix has been neglected.
In Eq. (\ref{eq:uy}), $k$ is the relative momentum of the $YN$ pair, 
$\xi_Y=M_N/M_Y$, $k_{max}$ is given by
\begin{equation}
   k_{max} = \frac{k_{F}+\xi_{Y}k_{Y}}{1+\xi_{Y}} \nonumber  
\end{equation}
and the weight function $f(k,k_Y)$ by 
\begin{equation}
f(k,k_{Y})= \left\{ \begin{array}{cl} 1 & \mbox{for $ k\leq
\frac{k_{F}-\xi_{F}k_{Y}}{1+\xi_{Y}} $ }, \\ 0 & \mbox{for $
|\xi_{Y}k_{Y}-(1+\xi_{Y})k| > k_{F} $ }, \\
\frac{k_{F}^2-[\xi_{Y}k_{Y}-(1+\xi_{Y})k]^2}{4\xi_{Y}(1+\xi_{Y})k_{Y}k
} & \mbox{otherwise} 

\end{array} \right. \nonumber 
\end{equation} 

\begin{figure}
%    \begin{center}
       \setlength{\unitlength}{1mm}
       \begin{picture}(100,100)
       \put(25,0){\epsfxsize=16cm \epsfbox{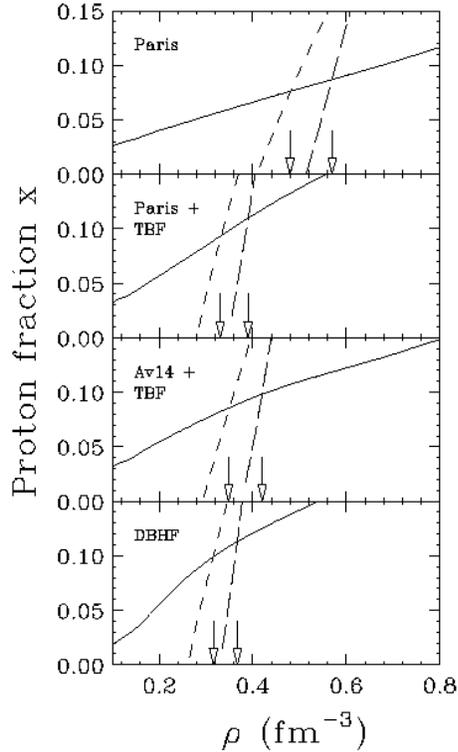}}
       \end{picture}
     \caption{The onset of $\Sigma^-$ and $\Lambda$ formation
            employing the Paris interaction in an BHF calculation,
            the Paris interaction added three-body forces (TBF),
            the Argonne $V_{14}$  plus TBF from Ref.\ \protect\cite{WFF},
            and the DBHF calculations discussed in subsection 
            \ref{subsubsec:relativistic}. The short dashes indicate the 
            occurence of chemical equilibrium for $\Sigma^-$ and
            the long dashes the corresponding one for $\Lambda$.}
   \label{fig:hyperonform}
%   \end{center}
\end{figure}

The binding energy of the
hyperon to its single-particle energy is then
\begin{equation}
    B_Y(k_Y)\equiv \varepsilon(k_Y)= \frac{k_Y^2}{2 M_Y} + U^N_Y(k_Y)
\label{eq:binding}
\end{equation}
However, we see from Eq.\ (\ref{eq:spenergyhyp}) that we have only
nucleons as intermediate hole states. This approach is viable only if we
are interested in studying properties of hyperons on top 
of $\beta$-stable matter. Such studies were recently performed by
Schulze et al.\ \cite{schulze97}. These calculations could be
viewed as way of establishing the possibility of hyperon
formation in matter.
The results for $\beta$-stable matter with hyperon formation
from Ref.\ \cite{schulze97} is shown in Fig.\ \ref{fig:hyperonform}
employing various models for the EoS. 
The soft-core nucleon-hyperon interaction of the Nijmegen group
is included here \cite{vynnim} in order to evaluate the hyperon single-particle
energy. Since the EoS which includes three-body interactions or relativistic
effects is stiffer than the non-relativistic EoS from BHF calculations
(the results labelled Paris), one expects the threshold for 
formation to be lower. We note from the above figure that the relativistic
and three-body results seem to agree  fairly well and yield 
an onset of hyperon formation at densities $\sim 2-3 n_0$.

However, with a finite hyperon concentration,
this entails also a summation over hyperon holes in
Eq.\ (\ref{eq:spenergyhyp}). Eq.\ (\ref{eq:binding})
gets an additional term $U^Y_Y(k_Y)$ where the latter is
obtained from an interaction between hyperons only.
Furthermore, in order to obtain the correct value for the Fermi
surface in $\beta$-stable matter with hyperons, one needs 
to solve e.g.\  Eqs.\ (\ref{eq:generalcharge}) and
(\ref{eq:generalbeta}) as well. 
Results of such calculations are in progress \cite{isaac99}
with realistic hyperon-hyperon interactions.

In this work, we will make life somewhat easier by just including
quark degrees of freedom through a simplified model, namely
the bag model, in order to account for degrees of freedom beyond the 
nucleonic ones. This is discussed in subsection \ref{subsec:qmeos}.

\subsection{Kaon condensation}

 Kaon condensation in dense matter was suggested by Kaplan and Nelson
\cite{kn87}, and has been discussed in many recent publications
\cite{BLRT,Weise}. Due to the attraction between $K^-$ and
nucleons its energy decreases with increasing density, and eventually
if it drops below the electron chemical potential in neutron star matter in
$\beta$-equilibrium, a Bose condensate of $K^-$ will appear.
It is found that $K^-$'s condense at densities above
$\sim 3-4\rho_0$, where $\rho_0=0.16$ fm$^{-3}$ is normal nuclear matter
density. This is to be compared to the central density of
$\sim4\rho_0$ for a neutron star of mass 1.4$M_\odot$ according to the
estimates of Wiringa, Fiks and Fabrocini \cite{WFF} using realistic
models of nuclear forces. 

In neutron matter at low densities, when the interparticle spacing is much
larger than the range of the interaction, $r_0\gg R$,
the kaon interacts strongly many times with the same nucleon before it
encounters and interacts with another nucleon. 
Thus one can use the scattering length as the ``effective'' kaon-nucleon
interaction, $a_{K^-N}\simeq -0.41$fm. The kaon energy deviates from its 
rest mass by the Lenz potential
\begin{equation}
   \omega_{Lenz} = m_K + \frac{2\pi}{m_R}\, a_{K^-N}\,\rho ,
               \label{Lenz}
\end{equation}
which is the optical potential obtained in the impulse 
approximation. If hadron masses furthermore 
decrease with density the condensation will occur at lower densities
\cite{BLRT}.

\begin{figure}
\begin{center}
{\centering
\mbox{\psfig{file=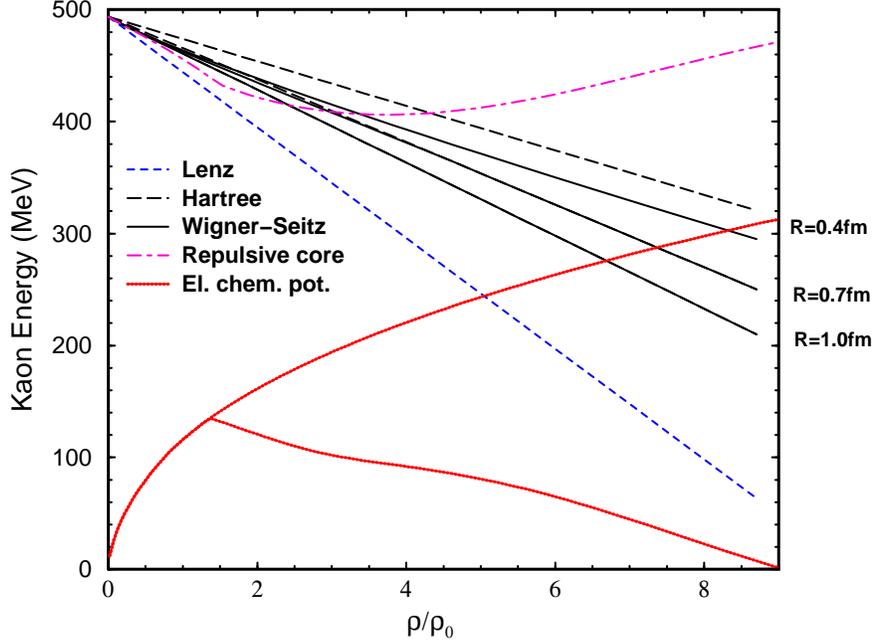,height=100mm,angle=-90}}}
\caption{
Kaon energy as function of neutron density. Including nuclear correlations
in the Wigner-Seitz cell approzimation is shown by
full curves for varios ranges of the $K^-n$ potentials
$R=0.4$ fm, $R=0.7$ fm and
$R=1.0$ fm. At low densities they approach the Lenz result
(Eq. (\ref{Lenz}), dotted curve) and at high densities they
approach the Hartree result (Eq.(\ref{Hartree}), dashed curves).
The electron chemical potential $\mu_e$ of our EoS (Eq. \ref{eq:EA}) with
$\delta=0.2$ is
shown with (lowest dotted curve) and without a transition to a
mixed phase of quark matter for a Bag constant of 
$B^{1/4}=100$ MeVfm$^{-3}$.
}
\end{center}
\label{fig:kaoncondens}
\end{figure}

At high densities when the interparticle spacing is much
less than the range of the interaction, $r_0\ll R$, the kaon
will interact with many nucleons on a distance scale much less the range
of the interaction. 
The kaon thus experiences the field from many nucleons
and the kaon energy deviates from its rest mass 
by the Hartree potential:
\begin{equation}
   \omega_{Hartree} = m_K + \rho \int V_{K^-N}(r)
           d^3r          \,,\label{Hartree}
\end{equation}
As shown in Ref.\ \cite{PPT}, the Hartree potential is considerably less
attractive than the Lenz potential.  Already at rather low densities,
when the interparticle distance is comparable to the range of the $KN$
interaction, the kaon-nucleon and nucleon-nucleon correlations
conspire to reduce the $K^-N$ attraction significantly \cite{kaon}.
This is also evident from Fig.\ \ref{fig:kaoncondens} 
where the transition from the low
density Lenz potential to the high density Hartree potential is
calculated by solving the Klein-Gordon equation for kaons in neutron
matter in the Wigner-Seitz cell approximation. Results are for square
well $K^-N$-potentials of various ranges $R$.

Kaon-nucleon correlations reduce the $K^-N$ interaction significantly
when its range is comparable to or larger than the nucleon-nucleon
interparticle spacing. The transition from the Lenz potential at low
densities to the Hartree potential at high densities occurs already
well below nuclear matter densities. For the measured $K^-n$
scattering lengths and reasonable ranges of interactions the
attraction is reduced by about a factor of 2-3 in cores of neutron
stars. Relativistic effects further reduce the attraction at high
densities. Consequently, a kaon condensate is less likely in neutron
stars due to nuclear correlations. However, if kaon masses drop with
densities \cite{BLRT} condensation will set in at lower densities.

If the kaon condensate occurs a mixed phase of kaon condensates and
ordinary nuclear matter may coexist in a mixed phase \cite{Schaffner}
depending on surface and Coulomb energies involved. The structures
would be much like the quark and nuclear matter mixed phases described
above.

\subsection{Pion condensation}

Pion condensation is like kaon condensation possible
in dense neutron star matter. For an in depth survey see e.g.\ Refs.\
\cite{migdal90,toki} and references therein.
If we first neglect the effect of strong correlations of pions
with the matter in modifying the pion self-energy, one finds
it is favorable for a neutron on the top of the Fermi sea to 
turn into a proton and a $\pi^-$ when 
\begin{equation}
    \mu_n=\mu_p+\mu_e > m_{\pi},
\end{equation}
where $m_{\pi}=139.6$ MeV is the $\pi^{-}$ rest mass. As discussed 
in the previous subsection, at nuclear matter 
saturation density the electron chemical potential is
$\sim 100$ MeV and one might therefore expect the appeareance
of $\pi^{-}$ at a slightly higher density.
One can however not neglect the interaction of the pion with the 
background matter. Such interactions can enhance the pion self-energy
and thereby the pion threshold density, and depending on the chosen
parameters , see again Ref.\ \cite{migdal90}, the critical density 
for pion condensation may vary from $n_0$ to $4n_0$.  
These matters are however
not yet settled in a satisfying way, and models with strong nucleon-nucleon
correlations tend to suppress both the $\pi NN$ and $\pi\Delta N$ 
interaction vertices so that a pion condensation in neutron star matter
does not occur. 
 However, in addition
to a charged pion condensate, one may also form a $\pi^0$ condensate
through the reaction $n\rightarrow n + \pi^0$ if the $\pi^0$ effective
mass in the medium is zero. The recent analysis, based
on the $V_{18}$ interaction model \cite{v18} 
of Akmal et al.\
\cite{ap97,apr98,aryathesis98} suggests such a pion condensate.
The effects were partly discussed in section \ref{subsec:nucdeg} and
the impact on the proton fraction was shown in Fig.\ \ref{fig:sec2fig16}.
The $\pi^0$ condensation of Akmal et al.\ \cite{apr98} for pure
neutron matter appears at a density of $\sim 0.2$ fm$^{-3}$ when 
the three-body interaction is included, whereas without $V_{ijk}$ 
it appears at much higher densities, i.e.\  $\sim 0.5$ fm$^{-3}$.
Although it is a robust mechanism in the variational
calculation of Ref.\ \cite{apr98}, the conclusion
relies on both the model of the NN and three-body interactions adopted
in the calculations. As noted in Ref.\ \cite{ehmmp97} for 
pure neutron matter, the $V_{18}$ interaction resulted in a slightly 
different energy per particle at densities greater than $0.4$ fm$^{-3}$
when compared with the CD-Bonn and Nijmegen interactions. 
This topic was also discussed in 
in section \ref{subsec:nucdeg} and shown in Fig.\ \ref{fig:sec2fig4}. 
Thus, before a firm conclusion can be reached about $\pi^0$ condensation,
it is our belief that it should also be obtained 
at the two-body level with the other phase-shift
equivalent NN interactions. That would lend strong support 
to the conclusions reached in Ref.\ \cite{apr98}.
The inclusion of three-body interactions
introduces a further model dependence.

Due to these uncertainties, we will refrain in this work from presenting
a  thourough
discussion of pion condensation. Rather, we will take the liberty to
refer to, e.g.\  Refs.\
\cite{apr98,tsuruta98,migdal90,toki}.

\subsection{Superfluidity in baryonic matter}\label{subsec:superf}

The presence of neutron superfluidity in 
the crust and the inner part 
of neutron stars 
are considered well established 
in the physics of these compact stellar objects. 
In the low density outer part of a neutron star, 
the neutron superfluidity is expected 
mainly in the attractive $^1S_0$ channel. 
At higher density, the nuclei in the crust dissolve, and one 
expects a region consisting of a quantum liquid of neutrons and 
protons in beta equilibrium. 
The proton contaminant should be superfluid 
in the $^1S_0$ channel, while neutron superfluidity is expected to  
occur mainly in the coupled $^3P_2$-$^3F_2$ two-neutron channel. 
In the core of the star any superfluid 
phase should finally disappear.
 
The presence of two different superfluid regimes 
is suggested by the known trend of the 
nucleon-nucleon (NN) phase shifts 
in each scattering channel. 
In both the $^1S_0$ and $^3P_2$-$^3F_2$ channels the
phase shifts indicate that the NN interaction is attractive. 
In particular for the $^1S_0$ channel, the occurrence of 
the well known virtual state in the neutron-neutron channel
strongly suggests the possibility of a 
pairing condensate at low density, 
while for the $^3P_2$-$^3F_2$ channel the 
interaction becomes strongly attractive only
at higher energy, which therefore suggests a possible 
pairing condensate
in this channel at higher densities. 
In recent years the BCS gap equation
has actually been solved with realistic interactions, 
and the results confirm
these expectations. 

The $^1S_0$ neutron superfluid is relevant for phenomena
that can occur in the inner crust of neutron stars, like the 
formation of glitches, which may to be related to vortex pinning  
of the superfluid phase in the solid crust \cite{glitch}. 
The results of different groups are in close agreement
on the $^1S_0$ pairing gap values and on 
its density dependence, which
shows a peak value of about 3 MeV at a Fermi momentum close to
$k_F \approx 0.8\; {\rm fm}^{-1}$ \cite{bcll90,kkc96,eh98,sclbl96}. 
All these calculations adopt the bare
NN interaction as the pairing force, and it has been pointed out
that the screening by the medium of the interaction 
could strongly reduce
the pairing strength in this channel \cite{sclbl96,chen86,ains89}. 
However, the issue of the 
many-body calculation of the pairing 
effective interaction is a complex
one and still far from a satisfactory solution.

The precise knowledge of the $^3P_2$-$^3F_2$ pairing gap is of 
paramount relevance for, e.g.\  the cooling of neutron stars, 
and different values correspond to drastically
different scenarios for the cooling process \cite{tsuruta98}.
Generally, the gap suppresses the cooling by a factor
$\sim\exp(-\Delta/T)$, see e.g.\   Ref.\ \cite{st83}, 
which is severe for
temperatures well below the gap energy.
Unfortunately, only few and partly
contradictory calculations of the pairing gap exist in the literature, 
even at the level of the bare NN interaction 
\cite{amu85,bcll92,taka93,elga96,khodel97}. 
However, when comparing the results, one should note that the  
NN interactions used in these calculations are not phase-shift 
equivalent, i.e.\  they do not 
predict exactly the same NN phase shifts.  
Furthermore, for the interactions used in 
Refs.~\cite{amu85,bcll92,taka93,elga96} the predicted 
phase shifts do not agree accurately with modern phase shift 
analyses, and the fit of the NN data has typically 
$\chi^2/{\rm datum}\approx 3$.  
As we discussed in subsection \ref{subsec:nucdeg}, 
progress has 
been made not only in the accuracy and the consistency of the 
phase-shift analysis, but also in the fit of realistic NN interactions 
to these data.  As a result, several new NN interactions have 
been constructed which fit the world data for $pp$ and $np$ scattering 
below 350 MeV with high precision.  Potentials like the recent 
Argonne $V_{18}$ \cite{v18}, the CD-Bonn \cite{cdbonn} 
or the new Nijmegen potentials \cite{nim} yield a 
$\chi^2/{\rm datum}$ of about 1 and may be called phase-shift 
equivalent.  
In Table \ref{tab:pgaps} we show the recent non-relativistic
pairing gaps for the $^3P_2$-$^3F_2$ partial waves, where
effective nucleon masses from the lowest-order Brueckner-Hartree-Fock
calculation of subsection \ref{subsec:nucdeg} have been 
employed, see Ref.\ \cite{beehs98} for more details.
These results are for pure neutron matter and we observe that
up to $k_F\sim 2$ fm$^{-1}$, the various potentials
give more
or less the same pairing gap. Above this Fermi momentum, which
corresponds to a lab energy of $\sim 350$ MeV, the results start
to differ. This is simply due to the fact that the potentials
are basically fit to reproduce scattering data up to this
lab energy. Beyond this energy, the potentials predict rather
different phase shifts for the 
$^3P_2$-$^3F_2$ partial waves, see e.g.\  Ref.\ \cite{beehs98}.
\begin{table}[hbtp]
\begin{center}
\caption{Collection of $^3P_2$-$^3F_2$ energy gaps (in MeV) for the 
various potentials discussed in subsection \ref{subsec:nucdeg}.  
BHF single-particle energies have been used. In case of no results,
a vanishing gap was found.}
\begin{tabular}{ccccc}\hline 
\multicolumn{1}{c}{$k_F\;({\rm fm}^{-1})$}& 
\multicolumn{1}{c}{CD-Bonn}&\multicolumn{1}{c}{$V_{18}$}&
\multicolumn{1}{c}{Nijm I}&\multicolumn{1}{c}{Nijm II} \\ \hline  
     1.2  & 0.04 & 0.04 & 0.04  & 0.04  \\
     1.4  & 0.10 & 0.10 & 0.10  & 0.10  \\
     1.6  & 0.18 & 0.17 & 0.18  & 0.18  \\
     1.8  & 0.25 & 0.23 & 0.26  & 0.26  \\
     2.0  & 0.29 & 0.22 & 0.34  & 0.36  \\
     2.2  & 0.29 & 0.16 & 0.40  & 0.47  \\
     2.4  & 0.27 & 0.07 & 0.46  & 0.67  \\
     2.6  & 0.21 &      & 0.47  & 0.99  \\
     2.8  & 0.17 &      & 0.49  & 1.74  \\
     3.0  & 0.11 &      & 0.43  & 3.14  \\ \hline
\end{tabular}
\label{tab:pgaps}
\end{center}
\end{table} 
Thus, before a precise calculation of $^3P_2$-$^3F_2$ energy gaps
can be made, one needs NN interactions that fit the scattering
data up to lab energies of $\sim 1$ GeV. This means 
in turn that the interaction models have to 
account for the opening
of inelasticities above $350$ MeV due to the
$N\Delta$ channel.

The reader should however note that the above results are
for pure neutron matter. We end therefore this subsection
with a discussion of the pairing gap for $\beta$-stable
matter of relevance for the neutron star cooling discussed
in section \ref{sec:starproperties}. 
We will also omit a discussion on neutron pairing gaps in the
$^1S_0$ channel, since these appear at densities corresponding 
to the crust of the neutron star. The gap in the crustal material 
is unlikely
to have any significant effect on cooling processes \cite{pr95}, 
though
it is expected to be important in the explanation 
of glitch phenomena.
Therefore, the relevant pairing gaps for neutron star cooling
should stem from the 
the proton contaminant 
in the $^1S_0$ channel, and superfluid neutrons yielding energy gaps 
in the coupled $^3P_2$-$^3F_2$ two-neutron channel. 
If in addition one studies closely the phase shifts for
various higher partial waves of the NN interaction, one notices
that at the densities which will correspond to the  
core of the star, any superfluid 
phase should eventually disappear. This is due to the fact that
an attractive NN interaction is needed in order to
obtain a positive energy gap.

Since the relevant total baryonic densities for these types of
pairing will be higher than the saturation
density of nuclear matter, we will account for relativistic
effects as well in the calculation of the pairing gaps.
To do so, we resort to the Dirac-Brueckner-Hartree-Fock (DBHF)
formalism discussed in subsection \ref{subsubsec:relativistic}. 

As an example, consider the evaluation of the proton
$^1S_0$ pairing gap using the DBHF approach.
To evaluate the pairing gap we follow 
the scheme
of Baldo et al.\  \cite{bcll90}.
These authors introduced an
effective interaction $\tilde{V}_{k,k'}$.
This effective interaction
sums up all two-particle excitations 
above a cutoff momentum $k_M$, $k_M=3$ fm$^{-1}$ in this work. 
It is defined according to
\begin{equation}
       \tilde{V}_{k,k'}=V_{k,k'}-\sum_{k''>k_M}V_{k,k''}
       \frac{1}{2{\cal E}_{k''}}
       \tilde{V}_{k'',k'},
       \label{eq:gap1}
\end{equation}
where the energy ${\cal E}_k$ is given by
${\cal E}_k =\sqrt{\left(\tilde{\varepsilon}_k-
\tilde{\varepsilon}_F\right)^2+\Delta_k^2}$,
$\tilde{\varepsilon}_F$ being the single-particle 
energy at the Fermi surface,
$V_{k,k'}$ is the free nucleon-nucleon potential 
in momentum space, defined
by the three-momenta $k,k'$. 
The renormalized potential $\tilde{V}_{k,k'}$ and 
the free NN interaction
$V_{k,k'}$ carry a factor 
$\tilde{m}^2/\tilde{E}_k\tilde{E}_{k'}$,
due to the normalization chosen for the 
Dirac spinors in nuclear matter.
These
constants are also included in the evaluation of the $G$-matrix,
as discussed in \cite{hko95,bm90}. 
For the $^1S_0$ channel, the pairing gap
$\Delta_k$ is \cite{bcll90,am61,kr90} 
\begin{equation}
       \Delta_k=-\sum_{k'\leq k_M}\tilde{V}_{k,k'}
       \frac{\Delta_{k'}}{2{\cal E}_{k'}}.
       \label{eq:gap3}
\end{equation}
These equations are solved self-consistently
in order to obtain the pairing gap
$\Delta$ for protons and neutrons for different partial waves.

\begin{figure}\begin{center}
       \setlength{\unitlength}{1mm}
       \begin{picture}(140,100)
       \put(25,5){\epsfxsize=7cm \epsfbox{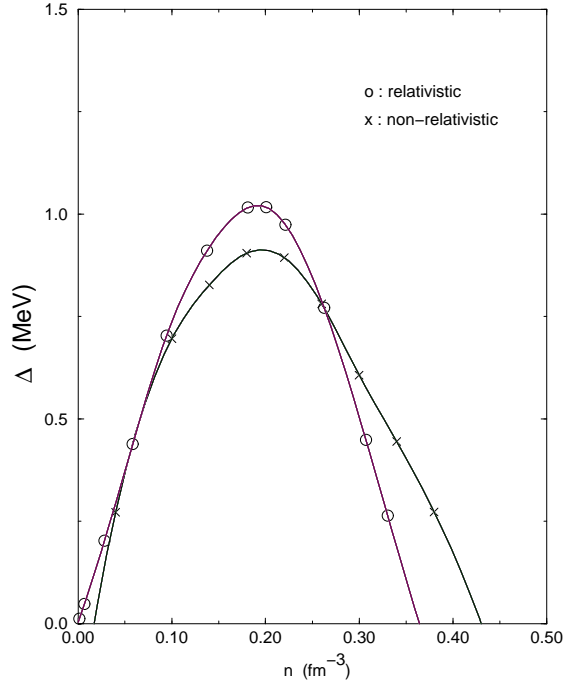}}
       \end{picture}
      \caption{Proton pairing in $\beta$-stable matter for 
          the $^1S_0$ partial wave. The non-relativistic results are taken from
          Ref. \protect\cite{eeho96a}.}
     \label{fig:figgap1s0}
\end{center}\end{figure}

\begin{figure}\begin{center}
       \setlength{\unitlength}{1mm}
       \begin{picture}(140,100)
       \put(25,5){\epsfxsize=7cm \epsfbox{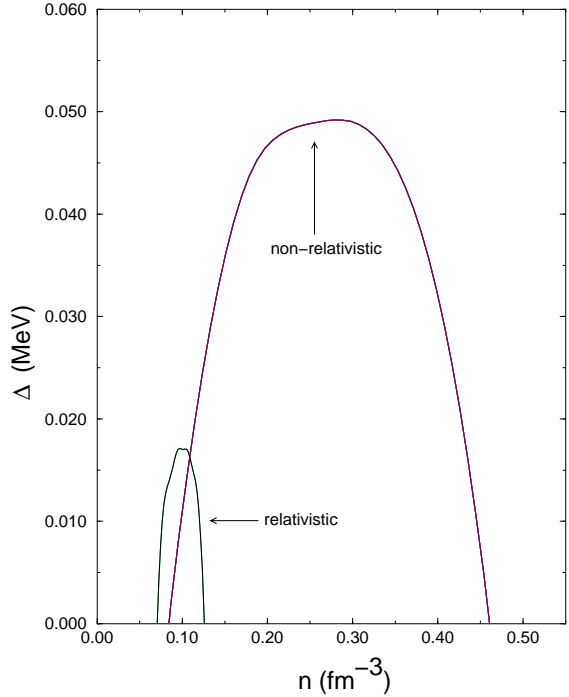}}
       \end{picture}
    \caption{Neutron pairing in $\beta$-stable matter for the $^3P_2$
     partial wave. The non-relativistic 
     results are taken from Ref. \protect\cite{eeho96a}. }
    \label{fig:figgap3p2}
\end{center}\end{figure}

In Fig.\ \ref{fig:figgap1s0} we plot as function of the total baryonic 
density the pairing gap for protons in the $^1S_0$
state, together with the results from the non-relativistic 
approach discussed in  Refs.\
\cite{elga96,eeho96c}. The results in the latter
references were  also obtained with the Bonn A potential of
Ref.\ \cite{mac89}. These results are all 
for matter in $\beta$-equilibrium. In Fig.\ \ref{fig:figgap3p2} 
we plot the 
corresponding relativistic 
results for the neutron energy gap in the $^3P_2$ channel. 
For the 
$^1D_2$ channel we found both 
the non-relativistic and the relativistic
energy gaps to vanish. 
The non-relativistic
results for  the Bonn A potential are 
taken from Ref.\ \cite{eeho96a}.

As can be seen from Fig.\ \ref{fig:figgap1s0}, there are only small
differences (except for higher densities) between the non-relativistic
and relativistic proton gaps in the $^1S_0$ wave\footnote{Even smaller
differences are obtained for neutrons in the $^1S_0$ channel.}.  This
is expected since the proton fractions (and their respective Fermi
momenta) are rather small, see Fig.\ \ref{fig:sec2fig7}.

For neutrons however, 
the Fermi momenta are larger, and we would 
expect relativistic effects to be important. At Fermi momenta
which correspond to the
saturation point of nuclear matter, $k_F=1.36$ fm$^{-1}$,
the lowest relativistic correction to the kinetic energy per 
particle is of the order of 2 MeV. 
At densities higher than the saturation
point, relativistic effects should be even 
more important, as can clearly
be seen in the calculations of Ref.\ \cite{bm90}. 
Since we are dealing with
very small proton fractions, a Fermi momentum
of $k_F=1.36$ fm$^{-1}$, would correspond to a total baryonic 
density $\sim 0.09$  fm$^{-3}$. Thus, at larger densities 
relativistic effects for neutrons should
be important.
This is also reflected in Fig.\ \ref{fig:figgap3p2} for the pairing
gap in the $^3P_2$ channel.
The relativistic $^3P_2$ gap is less  than half
the corresponding non-relativistic one, and the 
density region is also much smaller. This is mainly due to the 
inclusion of relativistic single-particle energies in the 
energy denominator of Eq.\ (\ref{eq:gap3}) and the normalization
factors for the Dirac spinors in the NN interaction. As an example,
at a neutron Fermi momentum $k_F=1.5$ fm$^{-1}$, the gap has 
a value of 0.17 MeV when one uses free single-particle energies 
and a bare NN  potential. Including the normalization
factors in the NN interaction, but employing free single-particle 
energies reduces the gap to 0.08 MeV.
If we employ only DBHF single-particle energies and the bare
NN interaction, the gap drops from 0.17 MeV to 0.04 MeV. 
Thus,
the largest effect stems from the 
change in the single-particle energies, although, 
the combined action 
of both mechanisms reduce the gap from 0.17 MeV to 0.015 at
$k_F=1.5$ fm$^{-1}$. The NN interaction in the 
$^3P_2$ channel depends also strongly on the 
spin-orbit force, see e.g.\ Fig.\ 3.3 in Ref.\ \cite{mac89}, 
and relativistic
effects tend to make the NN spin-orbit interaction from the 
$\omega$-meson in $P$-waves more repulsive \cite{mac89}. 
This leads
to a less attractive NN interaction in the  $^3P_2$ channel 
and a smaller
pairing gap.

The present results can be summarized as follows.
\begin{itemize}
      \item The $^1S_0$ proton gap in $\beta$-stable matter
            is $ \le 1$ MeV, and if polarization
            effects were taken into account \cite{sclbl96},
            it could be further reduced by a factor 2-3.
      \item The $^3P_2$ gap is also small, of the order
            of $\sim 0.1$ MeV in $\beta$-stable matter.
            If relativistic effects are taken into account,
            it is almost vanishing. However, there is
            quite some uncertainty with the value for this
            pairing gap for densities above $\sim 0.3$
            fm$^{-3}$ due to the fact that the NN interactions
            are not fitted for the corresponding lab energies. 
      \item Higher partial waves give essentially vanishing
            pairing gaps in $\beta$-stable matter.
      \item We have omitted a discussion of hyperon pairing,
            due to the uncertainties in the determination
            of the hyperon-hyperon interaction. We refer the reader
            here to Ref.\ \cite{nir98} for a discussion of these
            gaps.
\end{itemize}

Consequences for cooling histories will be discussed in section
\ref{sec:starproperties}.

\subsection{Quark matter} \label{subsec:qmeos}

When nuclear matter is compressed to densities so high that the
nucleon cores substantially overlap, one expects the nucleons to merge
and undergo a phase transition to chiral symmetric and/or
deconfined quark matter.  Rephrased in terms of the relevant field
excitations, we expect a transition from hadronic to quark degrees of
freedom at high densities.  Knowledge of the EOS of both hadronic and
quark matter is necessary to estimate the possible effects of this
transition in neutron stars.

Recent advances in the QCD phase diagram 
include improved lattice QCD calculations, random matrix models
\cite{Jackson}, and models addressing the possibility of color
superconductivity at fintie density \cite{RW}. 
Lattice QCD can only treat the case of
zero baryon chemical potential and is therefore not useful for
neutron stars.
Lattice calculations suggest that QCD has a first order transition at 
finite temperature and
zero chemical potential, provided that the strange quark is sufficiently 
light \cite{Lattice}. The transition weakens and might change to second
order for large strange quark masses in the limit of QCD with two massless
flavors.

Recent work using chiral random matrix models (chRMM) \cite{Jackson} suggests 
an effective thermodynamic potential of the form
\begin{equation}
\Omega(\phi;\mu,T) /N_f = \Omega_0(\mu,T) +
\phi^2 - \frac{1}{2}{\rm ln} \{ [ \phi^2 - 
(\mu + iT)^2] \cdot [\phi^2 - (\mu - iT)^2] \} \ .  \label{Omega}
\end{equation}
Here $\Omega_0(\mu,T)$ is unspecified and independent of the chiral mean
field $\phi$.
The scale of dimensional quantities cannot be determined within the chRMM,
and must be estimated by relating the variance of the Gaussian random matrix 
ensemble to the vacuum expectation value of the $\phi$ field via the 
Casher-Banks relation \cite{Jackson}.  The value of $\phi$ at the minima of 
$\Omega (\phi;\mu,T)$ is related to the quark condensate, 
$\langle {\bar \psi} \psi \rangle$, which is the order parameter for 
chiral symmetry breaking.  
Minimization of Eq.\ (\ref{Omega}) leads to a fifth order polynomial equation 
for $\phi$ which is identical in form to the results of Landau-Ginzberg 
theory using a $\phi^6$ potential.  One solution to this equation corresponds 
to the restored symmetry phase with $\phi = 0$.  This model predicts 
a second order transition for $\mu = 0$ at a temperature, $T_c$, which 
is generally agreed to be in the range 140-170 MeV.  For $T=0$, a first 
order transition occurs at some $\mu_0$.  
Since the phases in 
which chiral symmetry is broken and restored must be separated in the 
$( \mu , T )$ plane by an unbroken line of phase transitions, 
this implies the existence of a tricritical point in the theory of massless
quarks.

  There has recently been speculation regarding color
superconductivity at medium densities resulting from non-perturbative
attraction between quarks.  At finite chemical potential, this invariably
leads to the possibility of a diquark condensate which breaks global 
color invariance \cite{RW}. The associated color gap is $\sim 100$ MeV
and may become the thermodynamically favorable phase at high baryon
densities. When the strange quark is taken into account many different 
phases may exist \cite{SW98} and such effects require more analysis.

\subsubsection{Bag Models}

Since we do not have a fully reliable theory for the quark matter
phase, we will for simplicity employ the simple Bag model in our
actual studies of the mixed phase and neutron start properties.  In
the bag model the quarks in the hadrons are assumed to be confined to
a finite region of space, the so-called 'bag', by a vacuum pressure
$B$.  The pressure from the quarks inside the bag is provided by the
Fermi pressure and interactions computed to order $\alpha_s=g^2/4\pi$
where $g$ is the QCD coupling constant.
The pressure for quarks of flavor $f$, with $f=u,d$  or $s$ is 
\cite{kapusta}
\begin{eqnarray}
    P_f&=&{1\over 4\pi^2}\left[\mu_f k_f(\mu_f^2-2.5m_f^2)+
    1.5m_f^4 ln\left({\mu_f+k_f\over m_f}\right)\right]\nonumber\\
    &-&{\alpha_s\over \pi^3}\left[ {3\over 2}\left(\mu_f k_f-m_f^2
    ln \left({\mu_f+k_f\over m_f}\right)\right)^2-k_f^4\right].
\end{eqnarray}
The Fermi momentum is $k_f=(\mu_f^2-m_f^2)^{1/2}$. The total pressure,
including the bag constant B simulating confinement is
\begin{equation}
    P=P_e+\sum_f P_f-B.
     \label{pquark}
\end{equation}
The electron pressure is
\begin{equation}
    P_e={\mu_e^4\over 12\pi^2}.
\end{equation}

A Fermi gas of quarks of flavor {\em i} has density $n_i =
k_{Fi}^3/\pi^2$, due to the three color states.  There is no one-gluon
exchange interaction energy between quarks of different flavor, while
that between quarks of flavour {\em i} is given by $(2 \alpha/3 \pi)
E_i$ per quark {\em i} \cite{BC76}.  Here $E_i$ is the average kinetic
energy per quark, and $\alpha$ is the strong interaction coupling
constant, assumed to have a value of 0.5.  The $u$ and $d$ quarks are
taken to be massless, and $s$ quarks to have a mass of
150~MeV. Typical quark chemical potentials $\mu_q\ga m_N/3$ are
generally much larger. The value of the bag constant {\em B} is
poorly known, and we present results using three representative
values, $B^{1/4}=100$~MeVfm$^{-3}$ \cite{CLS86}, $B^{1/4}=150$~MeVfm$^{-3}$ and
$B^{1/4}=200$~MeVfm$^{-3}$ \cite{Sat82}.

Another possible model which has been applied to neutron star studies,
and which differs from the Bag-model is a massive quark model, the
so-called Color-Dielectric model (CDM) \cite{pirner92,birse90,dfb95}.
The CDM is a confinement model which has been used with success to
study properties of single nucleons, such as structure functions
\cite{barone} and form factors \cite{ff}, or to describe the
interaction potential between two nucleons \cite{kurt}, or to
investigate quark matter \cite{dfb95,mitja}.  In particular, it is
possible, using the same set of parameters, both to describe the
single nucleon properties and to obtain meaningful results for the
deconfinement phase transition \cite{dfb95}.  The latter happens at a
density of the order of 2-3 times $n_0$ when symmetric nuclear matter
is considered, and at even smaller densities for matter in
$\beta$-equilibrium, as discussed below in this work.  Another
important feature is that effective quark masses in the CDM are always
larger than a value of the order of 100 MeV, hence chiral symmetry is
broken and the Goldstone bosons are relevant degrees of freedom. This
is to be contrasted with models like the MIT bag, where quarks have
masses of a few MeV.  We therefore expect the CDM to be relevant for
computing the cooling rate of neutron stars {\it via} the Urca
mechanism, as suggested by Iwamoto \cite{iwamoto82}.

We will however stick to the Bag-model in our discussion of properties
of neutron stars.

\section{THERMODYNAMICS OF MULTI-COMPONENT PHASE TRANSITIONS} \label{sec:phases}

Numerous phase transitions may occur in neutron stars, e.g.\
the nuclear liquid-gas transition in the inner crust and in
the interior quark matter and/or condensates of kaons,
pions, hyperons, etc.  may be present.
Even if the transition between two such phases is first order microscopically,
the transition macroscopically (in bulk)  may be second order when there are 
several components present. We shall in this section briefly
describe this curious phenomenon, the thermodynamics of multi-component systems
and the corresponding mixed phases. It will be employed for neutron stars 
in the subsequent section.

\begin{figure}[htbp]
\begin{center}
{\centering
\mbox{\psfig{figure=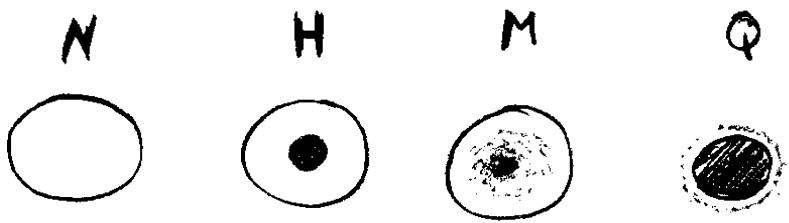,height=6cm,width=14cm,angle=0}}
}
\caption{Sketch of a neutron, hybrid, mixed and quark star.}
\label{fig:stars}
\end{center}
\end{figure}

\subsection{Maxwell construction for one component systems} 
\label{subsec:maxwell}

In the usual picture the transition between two phases occurs
at a unique pressure, temperature and chemical potential. 
Consequently the density is
expected to jump discontinuously at the boundary between the two phases.
This is not only true for systems of one component as in the
everyday example of water freezing
or evaporating. It is also the case for some two component systems as,
e.g.\  electrically neutral nuclear matter in $\beta$-equilibrium.   
Electric neutrality requires that the proton and electron densities
are the same in bulk
\begin{equation}
   n_p = n_e \,. \label{cnNM}
\end{equation}
$\beta$-equilibrium requires that the chemical potentials of neutrons and 
protons
only differ by that of the electrons
\begin{equation}
   \mu_n = \mu_p + \mu_e^{NM} \,, \label{beNM}
\end{equation}
in nuclear matter.
These two conditions restrict two of the three components
leaving only one independent variable, the baryon density.

Likewise, in quark matter charge neutrality implies that
\begin{equation}
   \frac{2}{3}n_u-\frac{1}{3}n_d-\frac{1}{3}n_s = n_e \,. \label{cnQM}
\end{equation}
$\beta$-equilibrium requires analogously
\begin{equation}
   \mu_d=\mu_s = \mu_u + \mu_e^{QM} \,. \label{beQM}
\end{equation}

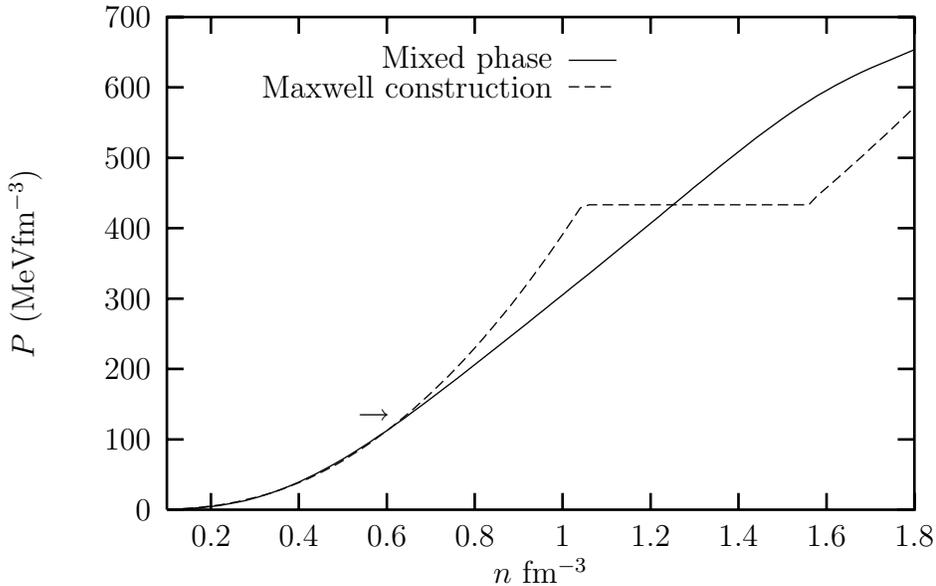
\begin{figure}[htb]
       \begin{center}
       % GNUPLOT: LaTeX picture with Postscript
\setlength{\unitlength}{0.1bp}
\special{!
%!PS-Adobe-2.0
%%Creator: gnuplot
%%DocumentFonts: Helvetica
%%BoundingBox: 50 50 770 554
%%Pages: (atend)
%%EndComments
/gnudict 40 dict def
gnudict begin
/Color false def
/Solid false def
/gnulinewidth 5.000 def
/vshift -33 def
/dl {10 mul} def
/hpt 31.5 def
/vpt 31.5 def
/M {moveto} bind def
/L {lineto} bind def
/R {rmoveto} bind def
/V {rlineto} bind def
/vpt2 vpt 2 mul def
/hpt2 hpt 2 mul def
/Lshow { currentpoint stroke M
  0 vshift R show } def
/Rshow { currentpoint stroke M
  dup stringwidth pop neg vshift R show } def
/Cshow { currentpoint stroke M
  dup stringwidth pop -2 div vshift R show } def
/DL { Color {setrgbcolor Solid {pop []} if 0 setdash }
 {pop pop pop Solid {pop []} if 0 setdash} ifelse } def
/BL { stroke gnulinewidth 2 mul setlinewidth } def
/AL { stroke gnulinewidth 2 div setlinewidth } def
/PL { stroke gnulinewidth setlinewidth } def
/LTb { BL [] 0 0 0 DL } def
/LTa { AL [1 dl 2 dl] 0 setdash 0 0 0 setrgbcolor } def
/LT0 { PL [] 0 1 0 DL } def
/LT1 { PL [4 dl 2 dl] 0 0 1 DL } def
/LT2 { PL [2 dl 3 dl] 1 0 0 DL } def
/LT3 { PL [1 dl 1.5 dl] 1 0 1 DL } def
/LT4 { PL [5 dl 2 dl 1 dl 2 dl] 0 1 1 DL } def
/LT5 { PL [4 dl 3 dl 1 dl 3 dl] 1 1 0 DL } def
/LT6 { PL [2 dl 2 dl 2 dl 4 dl] 0 0 0 DL } def
/LT7 { PL [2 dl 2 dl 2 dl 2 dl 2 dl 4 dl] 1 0.3 0 DL } def
/LT8 { PL [2 dl 2 dl 2 dl 2 dl 2 dl 2 dl 2 dl 4 dl] 0.5 0.5 0.5 DL } def
/P { stroke [] 0 setdash
  currentlinewidth 2 div sub M
  0 currentlinewidth V stroke } def
/D { stroke [] 0 setdash 2 copy vpt add M
  hpt neg vpt neg V hpt vpt neg V
  hpt vpt V hpt neg vpt V closepath stroke
  P } def
/A { stroke [] 0 setdash vpt sub M 0 vpt2 V
  currentpoint stroke M
  hpt neg vpt neg R hpt2 0 V stroke
  } def
/B { stroke [] 0 setdash 2 copy exch hpt sub exch vpt add M
  0 vpt2 neg V hpt2 0 V 0 vpt2 V
  hpt2 neg 0 V closepath stroke
  P } def
/C { stroke [] 0 setdash exch hpt sub exch vpt add M
  hpt2 vpt2 neg V currentpoint stroke M
  hpt2 neg 0 R hpt2 vpt2 V stroke } def
/T { stroke [] 0 setdash 2 copy vpt 1.12 mul add M
  hpt neg vpt -1.62 mul V
  hpt 2 mul 0 V
  hpt neg vpt 1.62 mul V closepath stroke
  P  } def
/S { 2 copy A C} def
end
}
\begin{picture}(3600,2160)(0,0)
\special{"
gnudict begin
gsave
50 50 translate
0.100 0.100 scale
0 setgray
/Helvetica findfont 100 scalefont setfont
newpath
-500.000000 -500.000000 translate
LTa
600 251 M
2817 0 V
LTb
600 251 M
63 0 V
2754 0 R
-63 0 V
600 516 M
63 0 V
2754 0 R
-63 0 V
600 782 M
63 0 V
2754 0 R
-63 0 V
600 1047 M
63 0 V
2754 0 R
-63 0 V
600 1313 M
63 0 V
2754 0 R
-63 0 V
600 1578 M
63 0 V
2754 0 R
-63 0 V
600 1844 M
63 0 V
2754 0 R
-63 0 V
600 2109 M
63 0 V
2754 0 R
-63 0 V
766 251 M
0 63 V
0 1795 R
0 -63 V
1097 251 M
0 63 V
0 1795 R
0 -63 V
1429 251 M
0 63 V
0 1795 R
0 -63 V
1760 251 M
0 63 V
0 1795 R
0 -63 V
2091 251 M
0 63 V
0 1795 R
0 -63 V
2423 251 M
0 63 V
0 1795 R
0 -63 V
2754 251 M
0 63 V
0 1795 R
0 -63 V
3086 251 M
0 63 V
0 1795 R
0 -63 V
3417 251 M
0 63 V
0 1795 R
0 -63 V
600 251 M
2817 0 V
0 1858 V
-2817 0 V
600 251 L
LT0
2114 1946 M
180 0 V
600 252 M
33 1 V
33 2 V
33 2 V
34 2 V
33 4 V
33 4 V
33 6 V
33 6 V
33 7 V
33 9 V
34 10 V
33 11 V
33 12 V
33 13 V
33 14 V
33 16 V
33 16 V
34 18 V
33 18 V
33 19 V
33 20 V
33 21 V
33 21 V
33 23 V
34 22 V
33 24 V
33 23 V
33 25 V
33 24 V
33 25 V
33 25 V
34 26 V
33 25 V
33 26 V
33 26 V
33 26 V
33 26 V
33 26 V
34 27 V
33 26 V
33 26 V
33 27 V
33 26 V
33 27 V
33 26 V
33 27 V
34 27 V
33 26 V
33 27 V
33 27 V
33 27 V
33 27 V
33 27 V
34 28 V
33 27 V
33 27 V
33 28 V
33 27 V
33 27 V
33 28 V
34 27 V
33 27 V
33 27 V
33 26 V
33 26 V
33 26 V
33 26 V
34 25 V
33 24 V
33 24 V
33 23 V
33 22 V
33 21 V
33 21 V
34 19 V
33 19 V
33 17 V
33 17 V
33 16 V
33 15 V
33 14 V
34 14 V
33 14 V
33 14 V
33 14 V
LT1
2114 1846 M
180 0 V
600 252 M
33 1 V
33 2 V
33 2 V
34 4 V
33 4 V
33 4 V
33 6 V
33 6 V
33 8 V
33 8 V
34 9 V
33 10 V
33 11 V
33 13 V
33 13 V
33 14 V
33 16 V
34 17 V
33 17 V
33 19 V
33 21 V
33 21 V
33 22 V
33 24 V
34 25 V
33 26 V
33 27 V
33 28 V
33 30 V
33 30 V
33 32 V
34 33 V
33 34 V
33 35 V
33 37 V
33 37 V
33 39 V
33 40 V
34 41 V
33 43 V
33 43 V
33 45 V
33 45 V
33 47 V
33 48 V
33 49 V
34 51 V
33 12 V
33 0 V
33 0 V
33 0 V
33 0 V
33 0 V
34 0 V
33 0 V
33 0 V
33 0 V
33 0 V
33 0 V
33 0 V
34 0 V
33 0 V
33 0 V
33 0 V
33 0 V
33 0 V
33 0 V
34 0 V
33 0 V
33 0 V
33 0 V
33 0 V
33 0 V
33 35 V
34 29 V
33 29 V
33 29 V
33 30 V
33 29 V
33 31 V
33 30 V
34 31 V
33 32 V
33 31 V
33 32 V
stroke
grestore
end
showpage
}
\put(2054,1846){\makebox(0,0)[r]{Maxwell construction}}
\put(2054,1946){\makebox(0,0)[r]{Mixed phase}}
\put(2008,21){\makebox(0,0){$n$ fm$^{-3}$}}
\put(100,1180){%
\special{ps: gsave currentpoint currentpoint translate
270 rotate neg exch neg exch translate}%
\makebox(0,0)[b]{\shortstack{$P$ (MeVfm$^{-3}$})}%
\special{ps: currentpoint grestore moveto}%
}
\put(3417,151){\makebox(0,0){1.8}}
\put(3086,151){\makebox(0,0){1.6}}
\put(2754,151){\makebox(0,0){1.4}}
\put(2423,151){\makebox(0,0){1.2}}
\put(2091,151){\makebox(0,0){1}}
\put(1760,151){\makebox(0,0){0.8}}
\put(1429,151){\makebox(0,0){0.6}}
\put(1097,151){\makebox(0,0){0.4}}
\put(766,151){\makebox(0,0){0.2}}
\put(540,2109){\makebox(0,0)[r]{700}}
\put(540,1844){\makebox(0,0)[r]{600}}
\put(540,1578){\makebox(0,0)[r]{500}}
\put(540,1313){\makebox(0,0)[r]{400}}
\put(540,1047){\makebox(0,0)[r]{300}}
\put(540,782){\makebox(0,0)[r]{200}}
\put(540,516){\makebox(0,0)[r]{100}}
\put(540,251){\makebox(0,0)[r]{0}}
\put(1440,600){\makebox(0,0)[r]{$\rightarrow$}}
\end{picture}
       \caption{Pressure as function of density. 
                Maxwell construction and mixed phase for bag parameter
                 $B^{1/4}=$ 200 MeVfm$^{-3}$. The arrow indicates where the
               Maxwell construction starts. }
       \end{center}
       \label{fig:sec3maxwell}
\end{figure}

Over the past two decades many authors have considered the properties
of neutron stars with a core of quark matter \cite{qmref}.
In such ``hybrid'' stars (see Fig. \ref{fig:stars})
it is assumed that each of the two phases are
electrically neutral separately as in Eqs. (\ref{cnNM}) and (\ref{cnQM})
and in $\beta$-equilibrium separately as in Eqs. (\ref{beNM}) and (\ref{beQM}).
Gibb's conditions $P_{NM}=P_{QM}$ and $\mu_n^{NM}=\mu_n^{QM}$ (the
temperatures are vanishing in both phases) then determine a unique
density at which the two bulk neutral phases coexist. This is the
standard Maxwell construction and is seen in Fig. \ref{fig:sec3maxwell} as the
double-tangent. In a gravitational field the denser phase (QM) will
sink to the center whereas the lighter phase (NM) will float on top as
a mantle as icebergs in the sea. At the phase transition there is a sharp 
density
discontinuity and generally $\mu_e^{NM}\ne\mu_e^{QM}$ so that the
electron densities $n_e=\mu_e^3/3\pi^2$ are {\it different} in the two
phases. This assumes that the sizes of QM structures are larger than
electron screening lengths which, as discussed in \cite{hps93}, is {\it
not} the case.

For small values of the bag constants 
the phase transition occurs at densities lower than $n_0$ and the
whole neutron star is a quark star except possibly for a hadronic
crust \cite{wg91}.

\subsection{Two component systems in a mixed phase} \label{subsec:mixed}

For several coexisting components as, e.g. dissolved chemicals, there
may exist a mixed phase where the various chemical potentials of the
solvents vary continuously - as does the pressure and densities.  A
similar phenomenon was predicted for the nuclear liquid-gas phase
transitions in the inner crust of neutron stars confirmed by recent
detailed numerical calculations \cite{lrp93,prl95}.  In the inner
crust the nuclei are surrounded by a neutron gas with an
interpenetrating constant background of electrons, i.e.\ the nuclear
matter and neutron gas form a mixed phase.  Going a few hundred meters
down in the neutron star crust the density of nuclei and thus also the
average density increase continuously. Thus there is no sharp density
discontinuity in bulk, i.e.\ over macroscopical distances of more than
hundreds of Fermi's and up to several meters (see Fig. \ref{fig:stars}). 
However, on microscopical distances of a few
Fermi's the density varies rapidly.  Another example first considered
by Glendenning \cite{glendenning92} is that of a mixed phase of
nuclear and quark matter.

Contrary to the Maxwell construction described in the previous
subsection, where the condition of charge neutrality
applies to both phases, Eqs.\  (\ref{cnNM}) and (\ref{cnQM}), 
it is relaxed to {\it overall} charge
neutrality only. Thus two conditions are relaxed to one - allowing for one
new variable quantity, which is usually taken as the {\it filling
fraction} $f$ of one of the phases in their coexisting mixture.
For the nuclear and quark matter mixed phase the
filling fraction is defined as the fraction of the volume which is
in the quark phase
\begin{equation}
   f \equiv \frac{V_{QM}}{V_{QM}+V_{NM}} \,, \label{f}
\end{equation}
and so $(1-f)$ is the filling fraction of nuclear matter.
The overall charge neutrality requirement is
\begin{equation}
   f n_p +(1-f)(\frac{2}{3}n_u-\frac{1}{3}n_d-\frac{1}{3}n_s)= n_e \,. 
\label{cn}
\end{equation}

A number of requirements must be met in order to 
form such a mixed phase as addressed in \cite{hps93}. These will be
discussed in the following section for the mixed phase of
nuclear and quark matter.

\section{STRUCTURE OF NEUTRON STARS}  \label{sec:structure}

The structure of the neutron star is seriously affected by
phase transitions, the order of the phase transition and whether
mixed phases can occur over a significant part of the star.
The important questions to be addressed are numerous.
When is it legitimate to regard the electron
density as uniform, what is the spatial structure of the new phase, and is
it energetically favorable?
In order to answer these questions one must first investigate screening
lengths of the various charged particles and compare to typical size
scales of structures. For that matter Coulomb and
surface energies must be calculated.

We will mainly discuss the mixed phase of nuclear and quark matter
in cores of neutron stars.
The mixed phase of nuclei and a neutron gas is in many ways similar
and has been calculated in detail in Refs. \cite{lrp93,prl95,Lasagna}

\subsection{Screening lengths}  \label{sec:screen}

As described in \cite{hps93} the mixed phase of quark and nuclear matter
may be regarded as droplets of quark matter immersed in nuclear matter
at lower densities, usually referred to as the {\it droplet
phase}, even though at higher densities its structure is more
complicated.  If droplet sizes and separations are
small compared with Debye screening lengths, the electron density will
be uniform to a good approximation.

The Debye screening length, $\lambda_ D$ is given by
\begin{equation}
    1/\lambda_D^2 = 4\pi \sum_i Q_i^2
    \left(\frac{\partial n_i}{\partial\mu_i} \right)_{n_j,j\neq i}
     \, , \label{qD}
\end{equation}
where $n_i$, $\mu_i$, and $Q_i$ are the number density,
chemical potential, and charge of particle species $i$. Considering only
electrons gives a screening length
\begin{equation}
    \lambda^{(e)}_{D}\, =\,
          \,  \frac{\sqrt{\pi /4\alpha}}{k_{F,e}}   \, ,     \label{le}
\end{equation}
where $\alpha\simeq 1/137$ and
the Fermi momentum $k_{F,e}=\mu_e$ since the electrons are always
relativistic at these densities.
For $\mu_e \raisebox{-.5ex}{$\stackrel{<}{\sim}$} 150$ MeV
we thus obtain $\lambda^{(e)}_{D}
\raisebox{-.5ex}{$\stackrel{>}{\sim}$} 13$ fm.
The screening length for protons alone
$\lambda^{(p)}_{D}$, is given by
$(\pi v_{F,p}/c 4\alpha(1+F_0) )^{1/2}/ k_{F,p}$, where
$F_0$ is the Landau parameter which gives the energy for proton density
variations.
 At the saturation density for symmetric nuclear
matter, $F_0\simeq 0$,
whereas at higher densities
$F_0\sim 1$ \cite{PR88}.
Since $\mu_p\sim m$, the nucleon mass, we find $\lambda^{(p)}_{D}
\raisebox{-.5ex}{$\stackrel{>}{\sim}
$} 10 $ fm, somewhat
shorter than the electron screening length, and therefore in the nuclear
matter phase, protons are the particles most effective at screening.
The screening length for quarks is $\lambda_{D}^{(q)}\simeq 7/k_{F,q}$ where
{\it q=u, d,} and $s$ refer to up, down and strange quarks.
It depends only slightly on whether or not $s$-quarks are present, so for
$\mu_q\simeq m/3$ we find $\lambda_{D}^{(q)}\simeq 5$ fm.

In a composite system, such as the one we consider, screening cannot be
described using a single screening length, but it is clear from our estimates,
that if the characteristic spatial scales of structures are less than about
10 fm for the nuclear phase, and less than about 5 fm for the quark phase,
screening effects will be unimportant, and the electron density will be
essentially uniform.  In the opposite case, when screening lengths are
short compared with spatial scales, the total charge densities in bulk nuclear
matter and quark matter will both vanish.

\subsection{Surface and Coulomb energies of the mixed phase} \label{sec:SC}

    When screening lengths are much larger than the
spatial scale of structures.  This condition implies that the electron
density is uniform everywhere, and all other particle densities are uniform
within a given phase.  The problem is essentially identical to that of matter
at subnuclear densities \cite{Lasagna}, and the structure is determined by
competition between Coulomb and interface energies.  When quark matter
occupies a small fraction, $f$, of the total volume, it will form spherical
droplets immersed in nuclear matter.  For higher filling fractions, the quark
matter will adopt shapes more like rods (``spaghetti") and plates
(``lasagna"), rather than spheres.  For $f\ge 0.5$, the structures expected
are the same as for a filling factor $1-f$, but with the roles of nuclear
matter and quark matter reversed.  Thus one expects for increasing $f$ that
there will be regions with nuclear matter in rod-like structures, and roughly
spherical droplets.

To estimate characteristic dimensions, some special cases can be considered.
The intricate structures in the
general case will be discussed in Sec. \ref{sec:FP}.
When $f$ is small or close to unity, the minority phase will form spherical
droplets.  The surface energy per droplet is given by
\begin{equation}
    {\cal E}_S=\sigma 4\pi R^2  \, ,   \label{ESurf}
\end{equation}
where $\sigma$ is the surface tension, and the Coulomb energy is
\begin{equation}
    {\cal E}_C=\frac{3}{5}\frac{Z^2e^2}{R}
            =\frac{16\pi^2}{15} \left(\rho_{QM}-\rho_{NM}\right)^2 R^5.
\label{ECoul}
\end{equation}
Here $Z$ is the excess charge of the droplet compared with the
surrounding medium,
$Ze=(\rho_{QM}-\rho_{NM})V_D$ where
$V_D=(4\pi/3)R^3$ is the droplet volume and $\rho_{QM}$ and $\rho_{NM}$
are the total charge densities in bulk quark
and nuclear matter, respectively.
Minimizing the energy density with respect to $R$ one
obtain the usual result that ${\cal E}_S=2{\cal E}_C$  and find a droplet radius
\begin{eqnarray}
    R&&=\left(\frac{15}{8\pi}\frac{\sigma}{
       (\rho_{QM}-\rho_{NM})^2}\right)^{1/3} \nonumber \\
      &&\simeq 5.0\, {\rm fm} \, \left(\frac{\sigma}{\sigma_0}
\right)^{1/3}
         \left( \frac{\rho_{QM}-\rho_{NM}}{\rho_0}\right) ^{-2/3}.    \label{R}
\end{eqnarray}
In the second formula we have introduced the quantities $\rho_0=e~ 0.4$
fm$^{-3}$ and $\sigma_0 = 50$ MeV$\cdot$fm$^{-2}$ which, as we shall
argue below, are typical scales for the quantities. (A droplet of
symmetric nuclear matter in vacuum has a surface tension $\sigma = 1$
MeV$\cdot$fm$^{-2}$ for which (\ref{R}) gives $R\simeq$ 4 fm, which
agrees with the fact that nuclei like $^{56}$Fe are the most stable form
of matter for roughly symmetric nuclear matter at low density.)  The
form of Eq. (\ref{R}) reflects the fact that on dimensional grounds, the
characteristic length scale is $(\sigma /(\rho_{QM}-\rho_{NM})^2)^{1/3}$ times
a function of $f$.  The total Coulomb and surface energy per unit volume
is given for small $f$ by
\begin{eqnarray}
     \varepsilon_{S+C}&=& f\, 9\left(\frac{\pi}{15}\sigma^2
(\rho_{QM}-\rho_{NM})^2\right)^{1/3}\nonumber \\
&\simeq& 44\, {\rm MeV~fm^{-3}} \, f \,
      \left(\frac{\sigma}{\sigma_0}
 \frac{\rho_{QM}-\rho_{NM}}{\rho_0}\right) ^{2/3}.    \label{energy}
\end{eqnarray}
The result for $f$ close to unity is given by replacing $f$ by $1-f$.
In the case when the volumes of quark and nuclear matter
are equal, $f=1/2$,  the structure can be approximated as alternating layers,
of quark and nuclear matter and was considered in \cite{hps93}.

To estimate length scales and energy densities, one needs the surface
tension of quark matter and the charge densities in the two phases.  A
rough estimate of the surface tension is the bag constant, $B$, times a
typical hadronic length scale $\sim$1 fm.  Estimates of the bag constant
range from 50 to 450 MeV$\cdot$fm$^{-3}$ \cite{MITSh83}.  The kinetic
contribution to the  surface tension at zero temperature has been
calculated in the bag model in Ref. \cite{BJ87}.   Only massive quarks
contribute because relativistic particles, unlike non-relativistic ones,
are not excluded near the surface due to the boundary conditions. The
kinetic contribution to $\sigma$ from a quark species depends strongly
on its mass and chemical potential. For $m_s\ll \mu_s$ it behaves as $
(3/4\pi^2) \mu_s^2 m_s$, and it vanishes as $m_s$ approaches $\mu_s$. 
If we adopt for the strange quark mass the value $m_s\simeq 150$ MeV,
and for the quark chemical potentials one third of the baryon chemical
potential, which  generally  is slightly larger than the nucleon mass,
$\mu_s\, \simeq\, \mu_B /3 \, \raisebox{-.5ex}{$\stackrel{>}{\sim}$} \,
m/3$,  one obtains from  Ref. \cite{BJ87} $\sigma\simeq$ 10
MeV$\cdot$fm$^{-2}$, which is close to the maximum value it can attain
for any choice of $m_s$.  We conclude that the surface tension for quark
matter is poorly known, but lies most probably in  the range $10~-~100$
MeV$\cdot$fm$^{-2}$. (Lattice gauge theory estimates of $\sigma$ at high
temperatures and zero quark chemical potentials lie in the range
$\sigma\simeq 0.14-0.28 T_c^3\sim$ 10-60 MeV$\cdot$fm$^{-2}$ \cite{Latt}
for $T_c\sim 150-200$ MeV, comparable to our estimates for cold quark
matter, but it is unclear to what extent this agreement is accidental.)

To estimate charge densities, we consider quark matter immersed in a
uniform background of electrons.  $\beta$-equilibrium insures that
$\mu_d=\mu_s=\mu_u+\mu_e$, and therefore in the absence of quark-quark
interactions, one finds the total electric charge density in the quark
matter phase is given for $\mu_e\ll \mu_u\sim\mu_d\equiv\mu_q$ and
$m_s\ll\mu_q$  by
\begin{equation}
 \rho_{QM} = \frac{e}{3}(2n_u-n_d-n_s-3n_e)
         \simeq \frac{e}{\pi^2}
         \left(\frac{1}{2}m_s^2\mu_q-2\mu_e\mu_q^2\right)
    \,  . \label{rhoq}
\end{equation}
Assuming $m_s\simeq 150$ MeV and $\mu_q\simeq m/3$ the
second term dominates except for small $\mu_e$ and so the droplet is negatively
charged and for $\mu_e \simeq 170$ MeV the density is about $-0.4e$ fm$^{-3
}$, the characteristic scale of
densities adopted in making estimates above.

Due to the high quark density, $\rho_{NM}$ is small compared with
$\rho_{QM}$ in Eq. (\ref{R}) when quark matter occupies a small fraction
of the volume. The electron chemical potential in neutron stars
depends strongly on the model for the nuclear equation of state, but
generally one finds $\mu_e\raisebox{-.5ex}{$\stackrel{<}{\sim}$} 170$
MeV. Consequently, for $\sigma \simeq 10$ MeV$\cdot$fm$^{-2}$ we find
from Eq. (\ref{R}) a radius of
$R\raisebox{-.5ex}{$\stackrel{>}{\sim}$}$ 3.1 fm, whereas $\sigma
\simeq 100$ MeV$\cdot$fm$^{-2}$ gives
$R\raisebox{-.5ex}{$\stackrel{>}{\sim}$}$ 6.6 fm. For $f$ close to
unity one finds nuclear bubble radii which are comparable with those
for quark droplets, and for the layer-like structures expected for $f
\simeq 0.5$, half the layer thickness is of comparable size. Estimates
of characteristic scales for rod-like structures give similar values.

Detailed calculations show that the effects of nonuniformity of the charge
distribution affect estimates of Coulomb energies significantly if the
characteristic lengths, $R$ and $a$, exceed the Debye screening
length.  The estimates of screening lengths made above show that screening will
be not be dominant for surface tensions below about 100 MeV, if the charge
density difference is $\rho_0$, but for higher values the simple picture of
coexisting uniform bulk phases would become invalid, and the droplet phase
would resemble increasingly two electrically neutral phases in
equilibrium.

For the smallest droplets of size 
$R\sim 5$fm the charge $Z=(4\pi/3)R^3\rho_{QM}$ and baryon
number $A=(4\pi/3)R^3n$ are typically a few hundreds.

\subsection{Is the mixed phase energetically favored?} \label{sec:Why}

If the bulk energy gained by going to the mixed phase is larger
than the costs of the associated Coulomb and surface energies, then
the mixed phase is energetically favored. Before going on to
estimate these crucial energies we point out the basic physical
reason why the bulk energy is lower in the mixed phase.
As mentioned in connection with Eq. (\ref{rhoq}) the QM droplets are 
{\it negatively} charged.
By immersing QM in
the positively charged NM we can either remove some of the electrons
from the top of the Fermi levels with energy $\mu_e$, or we can
increase the proton fraction in NM by which the symmetry energy is
lowered. In equilibrium a combination of both will occur and in both
cases {\it bulk energy is saved} and a lower energy density is
achieved as seen in Fig. \ref{fig:sec3maxwell}. 

To calculate the bulk energy
we adopt a simple form for the energy density of nuclear
matter consisting of a compressional term, a symmetry term,
and an electron energy density as discussed in Eq.\ (\ref{eq:EA})
\begin{eqnarray}
    \varepsilon_{NM}&=& nE \,
    = n[m+E_{comp}(n) + S(n)x^2] +\varepsilon_e \nonumber \\
%          &=&n[m+\frac{{\cal E}_0}{18}(\frac{n}{n_0}-1)^2
          &=&n[m+{\cal E}_0\frac{n}{n_0}(\frac{n}{n_0}-2-\delta)
    (1+\delta\frac{n}{n_0})^{-1}
           +S_0 (\frac{n}{n_0})^\gamma x^2 ] 
    + \, \frac{\mu_e^4}{12\pi^2}
          \label{ENM} .
\end{eqnarray}
The
electron chemical potential is never much above the muon mass and
therefore muons may be ignored. For quark matter we assume the bag model
equation of state
\begin{equation}
    \varepsilon_{QM}=
    (1-\frac{2\alpha_s}{\pi})
    \left(\sum_{q=u,d,s} \frac{3\mu_q^4}{4\pi^2}\right)  + B
    +\frac{\mu_e^4}{12\pi^2}   \, ,
    \label{EQM}
\end{equation}
with the QCD fine structure constant $\alpha_s\simeq 0.4$ and bag
constant $B^{1/4}\simeq 120$ MeV$\cdot$fm$^{-3}$. 
We have taken all quark masses to be zero. 
In the absence of surface and Coulomb effects the
equilibrium conditions for the droplet phase are that the quark and
nuclear matter should have equal pressures, and that it should cost no
energy to convert a neutron or a proton in nuclear matter into quarks
in quark matter.  The last condition amounts to $\mu_n=2\mu_d+\mu_u$
and $\mu_p=\mu_d+2\mu_u$.  The electron density is the same in quark
and nuclear matter, and we assume that matter is electrically neutral
and in $\beta$-equilibrium, that is  $\mu_n=\mu_p + \mu_e$ and
$\mu_d=\mu_u+\mu_e$. The chemical potentials are related to
the Fermi momenta by $\mu_q=p_{F,q}(1-2\alpha_s/\pi)^{-1/3}$.
Electrons contribute little to pressures, but they play an important
role through the $\beta$-equilibrium and charge neutrality conditions.

\begin{figure}[htbp]
\begin{center}
{\centering
\mbox{\psfig{figure=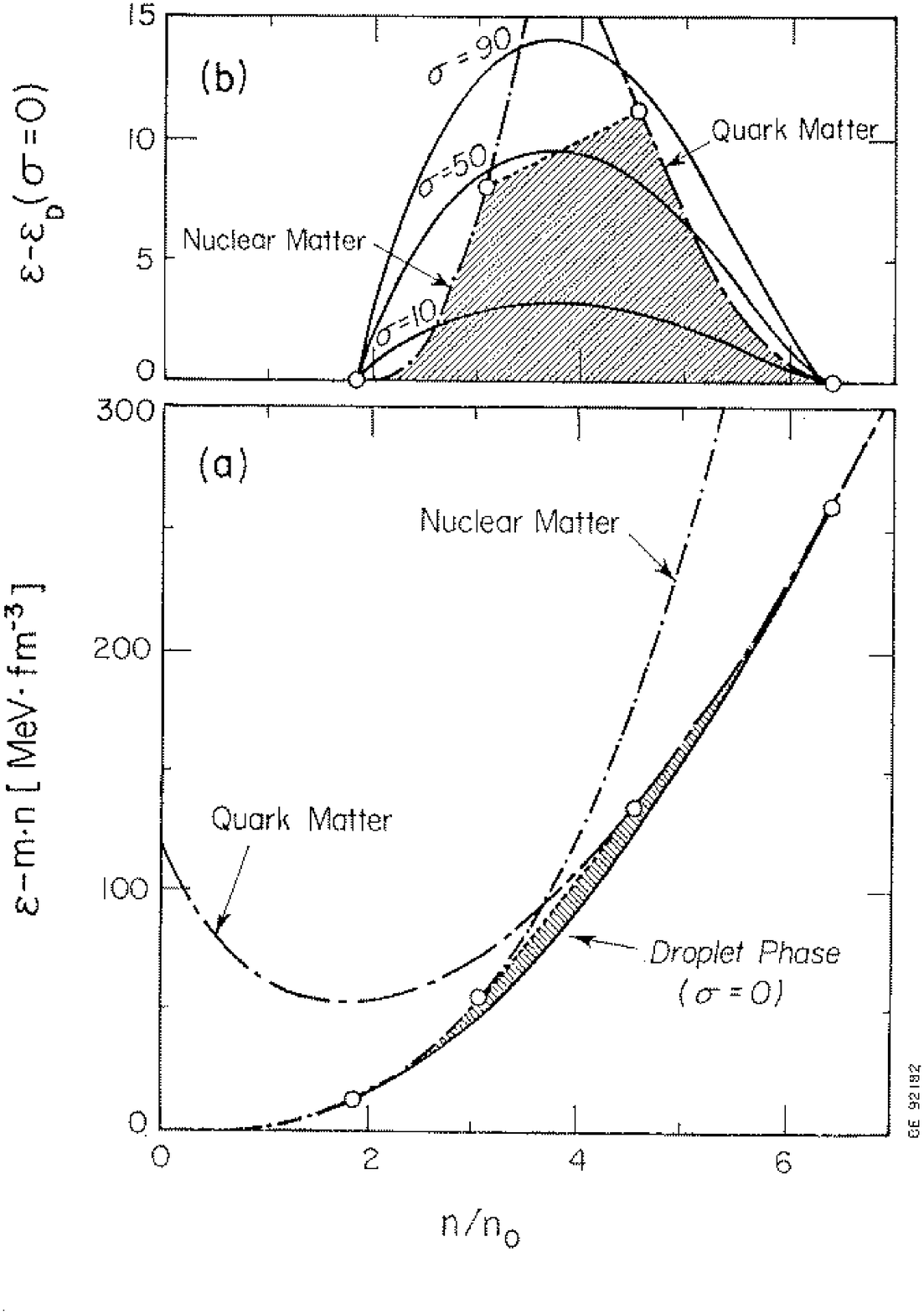,height=14cm,width=12cm,angle=0}}
}
\caption{
(a) The full line gives the energy density of the droplet phase without 
surface and Coulomb energies ($\sigma$ = 0).  Also shown are the
energy densities of
electrically neutral bulk nuclear matter,
quark matter in $\beta$-equilibrium, and 
the double tangent construction (dashed line)
corresponding to the coexistence of bulk electrically neutral phases.
(b) Energy densities of the droplet phase relative to its value for
$\sigma$= 0 for $\sigma$ = 10, 50, and 90 MeV$\cdot$fm$^{-2}$.
When the energy density of the droplet phase falls within the
hatched area it is energetically favored.  From \protect\cite{hps93}. }
       \label{fig:e}
\end{center}
\end{figure}

Figure \ref{fig:e}(a) shows the density dependence of the energy density of the
droplet phase calculated neglecting surface and Coulomb energies
($\sigma = 0$) for a simple quadratic EoS for nuclear matter \cite{hps93}.  
The energy of uniform, electrically neutral, bulk
nuclear matter in $\beta$-equilibrium is also shown, together with the
corresponding result for quark matter. The double-tangent construction
gives the energy density for densities at which the two bulk neutral
phases coexist. This corresponds to the standard treatment of the
phase transition between nuclear matter and quark matter, in which the
pressure remains constant throughout the transition, and consequently
neutron stars have a core of quark matter and a mantle of nuclear
matter, with a sharp density discontinuity at the phase transition. 
As one sees, if surface and Coulomb effects may be ignored, the
transition from nuclear matter to the droplet phase occurs at a lower
density than the transition to two bulk neutral phases, a  feature
also apparent in Ref. \cite{glendenning92}. In addition, droplets of nuclear
matter survive up to densities above those at which bulk neutral
phases can coexist.  We also observe that bulk contributions to the
energy density of the droplet phase are always lower than those for
coexisting bulk neutral phases.  While detailed properties of the
droplet phase depend strongly on the bulk energies, the qualitative
picture we find persists over a wide range of possible bulk matter
properties.

We now estimate surface and Coulomb energies. When quark matter
occupies a small fraction of space, $f$, one can show that the
difference in energy between the droplet phase and bulk neutral
nuclear matter varies as $f^2$. In contrast to this, the contributions
to the energy density from surface and Coulomb energies are linear in
$f$, see Eq. (\ref{energy}).  Similar results apply for $f$ close to unity. 
This shows that the transitions to the droplet phase must occur via a
first-order transition.  However, if the surface and Coulomb energies
are sufficiently large, the droplet phase may never be favorable. The
energy-density difference between the droplet phase, neglecting
surface and Coulomb effects, and two coexisting neutral phases is at
most 10 MeV$\cdot$fm$^{-3}$, as may be seen from Fig. \ref{fig:e}. 
This is very
small compared with characteristic energy densities which are of order
1000 MeV$\cdot$fm$^{-3}$.  In Fig. \ref{fig:e}(b) we show the energy density of
the droplet phase for various values of the surface tension, relative
to the value for $\sigma = 0$.  In these calculations the geometry of
the droplets was characterized by a continuous dimensionality, $d$, as
described in Ref. \cite{Lasagna}, with $d$ = 3, 2 and 1 corresponding to
spheres, rods and plates, respectively.  For the droplet phase to be
favorable, its energy density must lie below those of nuclear matter,
quark matter, and coexisting electrically neutral phases of nuclear 
and quark matter.  That is the droplet phase will be favored if its
energy lies within the hatched region in Fig. \ref{fig:e}(a+b).  We see that
whether or  not the droplet phase is energetically favorable depends
crucially on properties of quark matter and nuclear matter.  For our
model the droplet phase is energetically favorable at some densities
provided  $\sigma \raisebox{-.5ex}{$\stackrel{<}{\sim}$} 70$
MeV$\cdot$fm$^{-2}$.  However, given the large uncertainties in
estimates of bulk and surface properties one cannot at present claim
that the droplet phase is definitely favored energetically.

\subsection{Melting temperatures} \label{sec:Melt}

Should the quark-droplet phase exist in neutron stars, it could have
important observational consequences.  First, as Glendenning showed,
the pressure difference across the droplet phase can be large, of
order 250 MeV$\cdot$fm$^{-3}$. This is also seen from Fig. \ref{fig:e}(a), 
since
the pressure is the negative intercept of the tangent to the curve.
Consequently, a large portion of a neutron star could consist of matter
in the droplet phase.  Secondly, phases with isolated droplets would be
expected to be solid.  The melting temperature is  $\sim Z^2 e^2
f^{1/3}/(170 R)$ \cite{melt}  typically some hundreds of MeV, while
spaghetti- and lasagna-like structures would exhibit anisotropic
elastic properties, being rigid to some shear strains but not others
in much the same way as liquid crystals.  This could be important for
quake phenomena, which have been invoked to explain observations in a
number of different contexts. Third, neutrino generation, and hence
cooling of  neutron stars could be influenced.  This could come about
because nuclear matter in the droplet phase has a higher proton
concentration than bulk, neutral nuclear matter, and this could make
it easier to attain the threshold condition for the nucleon direct
Urca process \cite{Pe91}.  Another is that the presence of the spatial
structure of the  droplet phase might allow processes to occur which
would be forbidden in a translationally invariant system. Finally, one
should bear in mind the possibility that even if the droplet phase
were favored energetically, it would not be realized in practice if 
the time required to nucleate is too long.

\subsection{Funny phases} \label{sec:FP}

Surface and Coulomb energies
determine the topology and length scales of the structures. Denoting the
dimensionality of the structures by $d$ ($d=3$ for droplets and bubbles, $d=2$
for rods and $d=1$ for plates) the surface and Coulomb energies are
generally \cite{pr95,lrp93} 
\begin{equation}  
    {\cal E}_S= d\sigma \frac{4\pi}{3} R^2  \, 
\end{equation}
\begin{equation}
    {\cal E}_C=\frac{8\pi^2}{3(d+2)} (\rho_{QM}-\rho_{NM})^2R^5
   \left[\frac{2}{d-2}(1-\frac{d}{2}f^{1-2/d})+f\right] \, , 
\end{equation}  
where $\sigma$ is the surface tension, $R$ the size of the structure, and 
$\rho_{QM}$ and $\rho_{NM}$ are the total charge densities in bulk QM and NM,
respectively.  For droplets ($f\simeq0$) or bubbles ($f\simeq1$) $d=3$  and the
Coulomb energies reduce to the usual term ${\cal E}_C=(3/5)Z^2e^2/R$ where $Z$
is the excess charge of the droplet compared with the surrounding medium,
$Ze=(4\pi/3)(\rho_{QM}-\rho_{NM})R^3$. 
Minimizing the energy density with respect to
$R$ we obtain the usual result that ${\cal E}_S=2{\cal E}_C$. \footnote{
The condition for fission instability 
is contrarily: $2{\cal E}_S\le {\cal E}_C$ .}. 
Minimizing with respect to the continuous dimensionality as well thus 
determines both $R$ and $d$

  For the Walecka model the
droplet phase is energetically favorable at some densities provided  $\sigma
\raisebox{-.5ex}{$\stackrel{<}{\sim}$} 20$ MeV/fm$^2$ \cite{HH}. 
For comparison, using a quadratic EoS for NM\cite{hps93} one finds in
stead the more favorable condition
$\sigma\raisebox{-.5ex}{$\stackrel{<}{\sim}$} 70$ MeV/fm$^2$.  Given
the large uncertainties in estimates of bulk and surface properties
one cannot at present claim that the droplet phase is definitely
favored energetically.

\begin{figure}[htbp]
\begin{center}
{\centering
\mbox{\psfig{figure=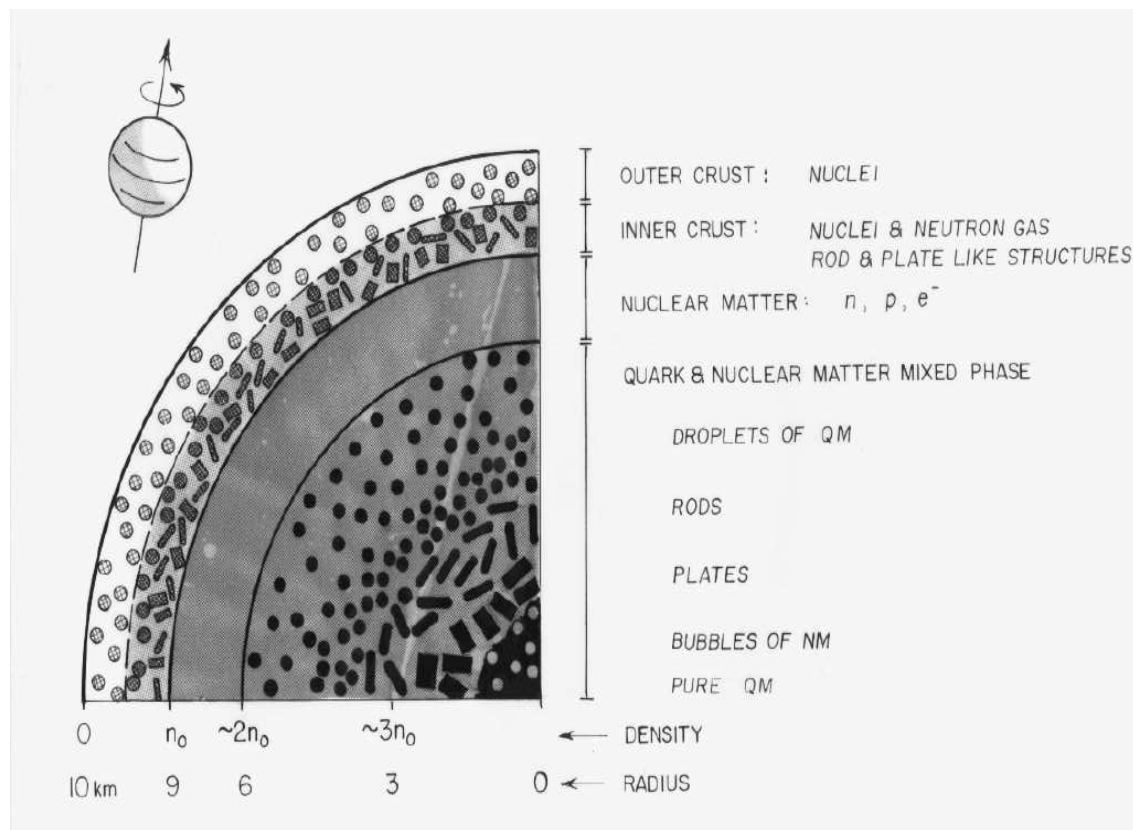,height=11cm,width=15cm,angle=90}}
}
\caption{Nuclear and quark matter structures in a 
$\sim1.4M_\odot$ neutron star.
Typical sizes of structures are $\sim10^{-14}$ m but have been scaled up
to be seen.}
\label{fig:qms}
\end{center}
\end{figure}

The mixed phase in the inner crust of neutron stars consists of
nuclear matter and a neutron gas in $\beta$-equilibrium with a
background of electrons such that the matter is overall electrically
neutral \cite{lrp93,prl95}. Likewise, quark and nuclear matter can have a
mixed phase \cite{glendenning92} and possible also nuclear matter with
and without condensate of any negatively charged particles such as
$K^-$ \cite{Schaffner}, $\pi^-$, $\Sigma^-$, etc.  The quarks are
confined in droplet, rod- and plate-like structures \cite{hps93}
(see Fig. \ref{fig:qms})
analogous to the nuclear matter and neutron gas structures in the
inner crust of neutron stars \cite{lrp93,prl95}.  Depending on the equation of
state, normal nuclear matter exists only at moderate densities,
$\rho\sim 1-2\rho_0$.  With increasing density, droplets of quark
matter form in nuclear matter and may merge into rod- and later
plate-like structures.  At even higher densities the structures invert
forming plates, rods and droplets of nuclear matter in quark matter.
Finally, pure quark matter is formed at very high densities unless the
star already has exceeded its maximum mass.

A necessary condition for forming these structures and the mixed phase
is that the additional surface and Coulomb energies of these
structures are sufficiently small. Excluding them makes the mixed
phase energetically favored \cite{glendenning92}. That is also the
case when surface energies are small (see \cite{hps93} for a quantitative
condition). If they are
too large the neutron star will 
have a core of pure quark matter with a mantle of nuclear matter
surrounding and the two phases are coexisting by an ordinary first
order phase transition.

The quark and nuclear matter mixed phase has continuous pressures and
densities \cite{glendenning92} when surface and Coulomb energies are
excluded.  There are at most two second order phase transitions. Namely,
at a lower density, where quark matter first appears in nuclear
matter, and at a very high density, where all nucleons are finally
dissolved into quark matter, if the star is gravitationally stable at
such high central densities.  However, due to the finite Coulomb and
surface energies associated with forming these structures, the
transitions change from second to first order at each topological change in
structure \cite{hps93}. If the surface and Coulomb energies are very
small the transitions will be only weakly first order but there may be
several of them.

\subsection{Summary of neutron star structures} \label{subsec:phases_summary}

To summarize, we have shown that whether or not a droplet phase consisting of
quark matter and nuclear matter can exist in neutron stars depends not only on
bulk properties, but also on the surface tension. In order to make better
estimates it is important to improve our understanding of the transition
between bulk nuclear matter and bulk quark matter. For the droplet phase to be
possible, this must be first-order. If the transition is indeed first-order,
better estimates of the surface tension are needed to determine whether the
droplet phase is favored energetically.

In these analyses several restrictions were made: the interfaces were
sharp, the charge densities constant in both NM and QM and the
background electron density was also assumed constant.  Relaxing these
restrictions generally allow the system to minimize its energy
further.  Constant charge densities may be a good approximation when
screening lengths are much larger than spatial length scales of
structures but since they are only slightly larger \cite{hps93} the
system may save significant energy by rearranging the charges.

\section{OBSERVATIONAL CONSEQUENCES FOR NEUTRON STARS} 
\label{sec:starproperties}

In this section we first briefly review the observational status of
neutron star masses from binary pulsars and X-ray
binaries. Subsequently, we revisit the equations for calculating the
structure of rotating neutron stars and calculate masses, radii,
moments of inertia, etc.\  for rotating neutron stars with the equations
of state described in the previous sections with and without phase
transitions.  Glitches are then discussed in subsections
\ref{subsec:glitches} and 5.7 while a discussion on neutron star cooling is
given in subsection \ref{subsec:cooling}.
In the last two subsections we discuss supernovae and Gamma-Ray-Bursters.

\subsection{Masses from radio pulsars, X-ray binaries and QPO's}

\begin{figure}
       \begin{center}
       \setlength{\unitlength}{1mm}
       \begin{picture}(140,100)
       \put(25,5){\epsfxsize=8cm \epsfbox{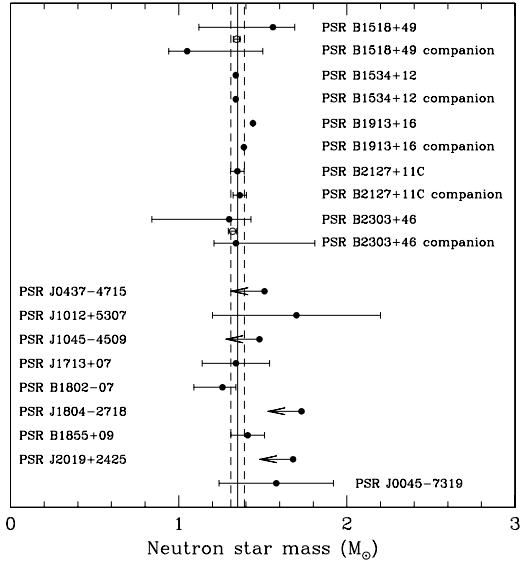}}
       \end{picture}
        \caption{Neutron star masses from
         observations of radio pulsar systems. Error bars indicate
         central 68\% confidence limits, except upper limits are one-sided
         95\% confidence limits. 
         The vertical lines are drawn at a mass $1.35\pm0.04M_\odot$.
         Taken from Ref.\ \protect\cite{tc98}.}
       \end{center}
        \label{fig:starmassdata}
\end{figure}

The measurements of  masses and radii of neutron stars 
(as well as detailed study of
their cooling histories and rotational instabilities) may provide a unique
window on the behavior of matter at densities well above that found in
atomic nuclei. The most precisely 
measured physical parameter of any rotating neutron star, or pulsar, is its
spin frequency. The frequencies of the fastest observed pulsars
(PSR~B1937+21 at 641.9~Hz and B1957+20 at 622.1~Hz) have already been
used to set constraints on the nuclear equation of state at high
densities under the assumption that these
pulsars are near their maximum (breakup) spin frequency.  However, the
fastest observed spin frequencies may be limited by complex accretion
physics rather than fundamental nuclear and gravitational physics. A
quantity more directly useful for comparison with physical theories is
the neutron star mass.  
In Fig.\ \ref{fig:starmassdata} we show the latest compilation
of Thorsett and Chakrabarty \cite{tc98} of neutron star
masses in binary radio pulsar systems.
As can be seen from this figure, most of observed binary pulsars
exhibit masses around $1.4M_{\odot}$. 
One exception may be PSR J1012+5307 with mass
$M=(2.1\pm 0.4)M_\odot$ \cite{Paradijs}. 
However, its mass is less well determined
as the system is non-relativistic.
The  masses are determined from pulse delays of the millisecond pulsar
as well as radial-velocity curves and
spectral lines of the white dwarf companion

The recent discovery of high-frequency brightness oscillations from
11  neutron stars in low-mass X-ray binaries may provide us with
a new promising method for determining masses and radii of neutron
stars. These quasi-periodic oscillations (QPO) 
are observed in both the persistent X-ray emission and in bursts.
According to the most successful model of
Miller et al.\ \cite{mlp98}
the QPO's are most likely the orbital frequencies of accreting gas
in Keplerian orbits around neutron stars.
The orbital frequency of the gas at distance $r$ from the neutron star is
\begin{equation}
   \nu_{\mathrm{QPO}} = 
   \frac{1}{2\pi} \left(\frac{M}{r}\right)^{1/2}.
\end{equation}
As the QPO last many periods, the gas has to be in a stable orbit.
The innermost stable orbit $R_{ms}$ for a slowly rotating neutron
star is related to its mass as
\begin{equation}
   M = \frac{c^2}{G} R_{ms}.
\end{equation}
Thus we obtain the limits on the non-rotating neutron star masses and radii:
\begin{eqnarray}
   M &\le& 2.2 M_\odot \frac{{\mathrm{kHz}}}{\nu_{\mathrm{QPO}}} \\
   R &\le&  20\mathrm{km} \frac{\mathrm{kHz}}{\nu_{\mathrm{QPO}}}.
\end{eqnarray}
For example, the 1220 Hz QPO observed in the atoll source 4U 1636-536
limits the neutron star mass to $M\le 1.8M_\odot$ and its radius to
$R\le 16$km.
It was predicted by
Miller et al.\ \cite{mlp98} that as the accretion flux increases towards the
innermost stable orbit the QPO frequency should stop increasing. 
This was subsequently observed by 
Zhang et al.  \cite{zss97} and Kaaret et al. \cite{kfc97}  for
4U 1820-30 which has $\nu_{QPO}=0.8-0.9$ kHz. The resulting neutron star
mass is $M\simeq2.3M_\odot$ when rotation is included. 
The deduced mass is consistent with the hypothesis that these
neutron stars were born with $M\simeq 1.4 M_\odot$
and have been accreting matter at a fraction of the Eddington limit
for $10^8$ yr.
The  QPO's provide an important tool for determining
neutron star masses or at least restricting them 
and for limiting radii. Due to accretion they are expected to be heavier
than X-ray binaries and therefore potentially more interesting for
restricting maximum masses and the EoS.

A new determination of the mass of the accretion powered X-ray
pulsar Vela X-1 gives the mass (95\% confidence limit)
\begin{equation}
   M = 1.93^{+0.23}_{-0.17} \, M_\odot \,. \label{Vela}
\end{equation}
The mass could be determined from the velocities of the binaries
from pulse timing of the neutron star and doppler shifts in the spectral
lines of its companion as well as the inclination from orbital flux
variations.

Recently, also the mass of the low mass X-ray binary Cygnus X-2
has been estimated from U-B-V light curves \cite{Orosz}
\begin{equation}
   M = (1.78 \pm 0.23) \, M_\odot \,. \label{Cygnus}
\end{equation}

The existence of such large neutron star masses will require a rather
stiff EoS for nuclear matter and restricts the softening due to phase
transitions severely, as will be discussed in the following subsection.

\subsection{TOV and Hartle's equations} \label{subsec:generalprop}

The theoretical description of a neutron star is governed by 
conditions imposed by general relativity.
General relativity has to be
taken into account for the determination of the gross properties of a star
with approximately one solar mass $M_\odot$ and a radius $R$ of approximately
10\,km, since relativistic effects 
are of the order \cite{st83}
\begin{equation}
    \frac{M}{R} \sim 0.1-0.2 .
\end{equation}
The starting point for such studies is
how to determine Einstein's
curvature tensor $G_{\mu\nu}$ for a massive star ($R_{\mu\nu}, g_{\mu\nu}$,
and $R$ denote the Ricci tensor, metric tensor, and Ricci scalar,
respectively).
\begin{equation}
     G_{\mu\nu} \equiv R_{\mu\nu} - \frac{1}{2} g_{\mu\nu} R =
     8\pi\,T_{\mu\nu}\left(\varepsilon, P(\varepsilon)\right)~.
\end{equation}
A necessary ingredient for solving this equation is the energy-momentum tensor
density $T_{\mu\nu}$, for which knowledge of the EoS, i.e.\ 
pressure $P$ as function of the energy density $\varepsilon$ is necessary.

For a spherically symmetric and static star, the metric has the 
Schwarzschild form
\begin{equation}
   ds^2 = -\,e^{2\phi(r)} dt^2 + e^{2\Lambda(r)} dr^2 + r^2 
   (d \theta^2 + sin^2 \theta d\phi^2)~,
\end{equation}
where the metric functions are given by:
\begin{equation}
     e^{2\Lambda(r)} = (1 - \gamma(r))^{-1} ~,
\end{equation}
\begin{equation}
    e^{2\phi(r)} = e^{-2\Lambda(r)} = (1 - \gamma(r)) \quad \rm{for} \quad r >
    R_{star}~,
\end{equation}
with
\begin{equation}
    \gamma(r) = \left\{
\begin{array}{ccc}
     2M(r)/r &,& r\le R_s \\[2mm]  2M_s/r &,& r\ge R_s
\end{array} \right.
\end{equation}
Einstein's equations for a static star reduce then to the familiar
Tolman-Oppenheimer-Volkoff equation (TOV) \cite{st83,hartle67,ov39}:
\begin{equation}
      \frac{dP(r)}{dr} = -\,\frac{1}{r^2} \left(\varepsilon(r) + P(r)\right)
      \left(M(r) + 4\pi r^3P(r)\right)  e^{2\Lambda(r)} ~,
       \label{eq:tov}
\end{equation}
where the gravitational mass $M(r)$ contained in a sphere with radius $r$
is determined via the energy-density $\varepsilon(r)$ by:
\begin{equation}
     M(r) = 4\pi \int^r_0 \varepsilon(r) r^2 dr ~.
    \label{eq:massofstar}
\end{equation}
The metric function $\phi(r)$ obeys the differential equation
\begin{equation}
     \frac{d\phi}{dr} = -\, \frac{1}{\varepsilon(r)+P(r)} \, \frac{dP}{dr} ~,
     \label{eq:metrics}
\end{equation}
with the boundary condition
\begin{equation}
     \phi(r = R_s) = \frac{1}{2} \ln (1-\gamma(R_s)) ~.
\end{equation}
For a given EoS, i.e. $P(\varepsilon)$, one can now solve the TOV equation by
integrating them for a given central energy density $\varepsilon_c$
from the star's centre to the star's radius, defined by $P(R_s) = 0$.

More complicated is the case of rotating stars, where due to the rotation
changes occur in the pressure, energy density, etc. The energy-momentum
density tensor $T_{\mu\nu}$ takes the form $(g^{\mu\nu}u_\mu u_\nu = -\,1)$
\cite{wg91,46,47}:
\begin{equation}
    T_{\mu\nu} = T^0_{\mu 0} + \Delta T_{\mu\nu} ~,
\end{equation}
with
\begin{equation}
     T_{\mu\nu}^0 = (\varepsilon + P) u_\mu u_\nu + P g_{\mu\nu} ~,
\end{equation}
\begin{equation}
    \Delta T_{\mu\nu} = (\Delta \varepsilon + \Delta P) u_\mu u_\nu + \Delta P
    g_{\mu\nu} ~.
\end{equation}
$P, \varepsilon$, and $\rho$ are quantities in a local inertial frame comoving
with the fluid at the instant of measurement. For the rotationally deformed, 
axially-symmetric
configurations one assumes a multipole expansion up to second order ($P_2$
denotes the Legrendre polynomial):
\begin{eqnarray}
   \Delta P & = & (\varepsilon + P) (p_0 + p_2 P_2 (\cos \theta)) ~, \\
   \Delta \varepsilon & = & \Delta P \frac{\partial \varepsilon}{\partial P} ~, 
\\
   \Delta \rho & = & \Delta P \frac{\partial \rho}{\partial P} ~.
\end{eqnarray}
For the rotating and deformed star with the rotational frequency $\Omega$ one
has now to deal with a generalized Schwarzschild metric, given by
\cite{48,49}
\begin{eqnarray}
    ds^2 = & -\,e^{2\nu(r,\theta,\phi)} dt^2 + e^{2\psi(r,\theta,\Omega)}
     (d\phi - \omega(r,\Omega)dt)^2 + e^{2\mu(r,\theta,\phi)} d\theta^2 
   \nonumber \\
   & + e^{2\lambda(r,\theta,\phi)} dr^2 + {\cal O}(\Omega^3) ~.
\label{eq:metric}
\end{eqnarray}
Here, $\omega(r)$ denotes the angular velocity of the local inertial
frame, which, due to the dragging of the local system is
proportional to $\Omega$.

The metric functions of Eq.\ (\ref{eq:metric}) which correspond to
stationary rotation and axial symmetry with respect to the axis of
rotation are expanded up to second order as (independent of $\phi$ and
$t$)
\begin{eqnarray}
   e^{2\nu(r,\theta,\Omega)} & = & e^{2\phi(r)} \left[ 1 + 2\left(h_0(r,\Omega)
     + h_2(r,\Omega) P_2(\cos \phi)\right)\right] ~,\\
e^{2\psi(r,\phi,\Omega)} & = & r^2 \sin^2 \theta \left[ 1 + 2
\left(v_2(r,\Omega) - h_2(r,\Omega)\right) P_2 (\cos\theta)\right] ~, \\
e^{2\mu(r,\theta,\Omega)} & = & r^2 \left[ 1 + 2 \left(v_2(r,\Omega) - h_2
(r,\Omega)\right) P_2(\cos\theta) \right] \\
e^{2\lambda(r,\theta,\Omega)} & = & e^{2\wedge(r)} \left[ 1 + \frac{2}{r}\,
   \frac{m_0(r,\Omega)G + m_2(r,\Omega) P_2(\cos\theta)}{1 - \gamma (r)}
  \right] ~.
\end{eqnarray}
The angular velocity in the local inertial frame is determined by the
differential equation
\begin{equation}
\frac{d}{dr} \left(r^4 j(r) \frac{d\omega}{dr}\right)
  + 4r^3 \frac{dj(r)}{dr} \omega(r) = 0 ~, \quad r < R_s ~,
\label{eq:ddr}
\end{equation}
where $\omega(r)$ is regular for $r = 0$ with $\frac{d\omega}{dr} = 0$.
 $j(r)$ abbreviates
\begin{equation}
     j(r) \equiv e^{-\phi(r)} \sqrt{1-\gamma(r)} ~.
\end{equation}
Outside the star $\omega(r,\Omega)$ is given by:
\begin{equation}
       \omega(r,\Omega) = \Omega - \,\frac{2}{r^3} J(\Omega) \quad , \quad
          r > R_s ~.
\end{equation}
The total angular momentum is defined by:
\begin{equation}
      J(\Omega) = \frac{R_s^4}{6} \left(\frac{d\omega}{dr}\right)_{r=R_s} ~.
\end{equation}
 From the last two equations one obtains then an angular frequency
$\Omega$ as a function of central angular velocity $\omega_c =
\omega(r=0)$ (starting value for the iteration):
\begin{equation}
\Omega(\omega_c) = \omega(R_s) + \frac{2}{R^3_s} J(\Omega) ~.
\end{equation}
Due to the linearity of Eq.\ (\ref{eq:ddr}) for $\omega(r)$ new values for
$\omega(r)$ emerge simply by rescaling of $\omega_c$. The momentum of
inertia, defined by $I = \frac{J}{\Omega}$, is given by
\begin{equation}
    I = \frac{J(\Omega)}{\Omega} = \frac{8\pi}{3} \int^{R_s}_0 dv\,r^4 
   \frac{\varepsilon + P}{\sqrt{1-\gamma(r)}} \frac{\omega-\Omega}{\Omega} 
   e^{-\phi} ~.
\end{equation}
Relativistic changes from the Newtonian value are caused by the
dragging of the local systems, i.e. $\bar\omega/\Omega$, the redshift
($e^{-\phi}$), and the space-curvature $\left(
\left(1-\gamma(r)\right)^{-1/2} \right)$. For slowly rotating stars
with low masses, one can neglect the dragging $(\frac{\omega}{\Omega} \to
1)$ and rotational deformations, but we would like to emphasize that
the described treatment is not restricted by low masses and/or slow
rotations.

With $\omega(r)$, one can also solve the 
coupled mass monopole equations $(\ell = 0)$ for $m_0, p_0$, where the 
latter represents the 
monopole pressure perturbation,  and $h_0$,
Refs.\ \cite{wg91,46}). The quadrupole distortions $h_2$ and $v_2$
$(\ell = 2)$ determine the star's shape (see
Refs.\,\cite{wg91,46}). We will here just state the equations for the 
monopole functions $m_0$ and $p_0$ in order to obtain the 
corrections to the mass due to rotation. We will not deal with
quadrupole corrections in this work. 
The equations read
\begin{equation}
  \frac{dm_0}{dr}=4\pi^2\frac{\partial \varepsilon}{\partial P}
  \left(\varepsilon + P\right)p_0+\frac{1}{12}j^2r^4
   \left(\frac{d\omega}{dr}\right)^2+\frac{8\pi}{3}r^4j^2
   \frac{\varepsilon + P}{1-\gamma}\omega^2,
  \label{eq:monopolem}
\end{equation}
and
\begin{equation}
   \frac{dp_0}{dr}=-\frac{1+8\pi^2P}{r^2(1-\gamma)^2}m_0-
                     4\pi\frac{(\varepsilon + P)}{1-\gamma}p_0+
                   \frac{1}{12}\frac{j^2r^3}{1-\gamma}
                    \left(\frac{d\omega}{dr}\right)^2+
                    \frac{1}{3}\frac{d}{dr}
                    \left(\frac{r^2j^2\omega^2}{1-\gamma}\right).
   \label{eq:monopolep}
\end{equation}
The boundary conditions are $m_0\rightarrow 0$ and 
$p_0\rightarrow 0$ when $r\rightarrow 0$. Outside the star one has
\[
  m_0 =\Delta M - \frac{1}{r^3}J(\Omega)^2 \hspace{1cm}
          r > R_s ~.
\]
with $\Delta M$ being the rotational  correction to the 
gravitational mass. This corrections is given by
\begin{equation} 
     \Delta M=m_0+\frac{1}{R_s^3}J(\Omega)^2,
     \label{eq:rotcorr}
\end{equation}
at the surface of the star. Thus, when we solve the monopole
equations we know also the correction to the gravitational
mass. 
These two equations, together with Eqs.\ (\ref{eq:tov}),
(\ref{eq:massofstar}), (\ref{eq:metrics}) and (\ref{eq:ddr}) 
form the starting point for our numerical procedure for obtaining
the total mass, radius, moment of Inertia and rotational mass.
Results for various approaches to the EoS are discussed in the next
subsection.

\subsection{Neutron star properties from various equations of state}
\label{subsec:mass-radius}

For a given EoS we obtain the mass and radii of the neutron star by solving
the equations for a weakly rotating neutron star as given by Hartle 
\cite{hartle67} and discussed in 
Eqs.\ (\ref{eq:tov}),
(\ref{eq:massofstar}), (\ref{eq:metrics}), (\ref{eq:ddr}), (\ref{eq:monopolem})
and (\ref{eq:monopolep}). Various results with and without
rotational corrections are displayed in Figs.\ 
\ref{fig:sec5fig1}-\ref{fig:sec5fig9}, where we show total masses,
mass-radius relations and moments of inertia for various approximations
to the EoS. 
The following possible properties pertain to the various approximations
to the EoS.
\begin{itemize}
\item For the EoS parametrization of Akmal et al.\ \cite{apr98} with just $pn$
      degrees of freedom,
      the EoS with $\delta=0.13$ gives the stiffest EoS and thereby the 
      largest neutron star mass. For $\delta < 0.13$ the EoS is superluminal.
      See also the discussion in connection with Figs.\ \ref{fig:sec2fig19}
      and \ref{fig:sec2fig20}. For $\delta=0.3$ or $\delta=0.4$ the EoS
      differs from that with $\delta=0.2$ or $\delta=0.13$ at densities
      below $n=0.3$ fm$^{-3}$. This explains the differences in masses 
      seen at low central densities in Fig.\ \ref{fig:sec5fig1}.
      The reader should also recall that in section 2, the best fit to the 
      results of Ref.\ \cite{apr98} was obtained with $\delta=0.2$.  
\item We have selected three representative values for the Bag-model
      parameter $B$, namely, 100, 150 and 200 MeVfm$^{-3}$. 
      For $B^{1/4}=100$ MeVfm$^{-3}$, the mixed phase starts already at
      $0.22$ fm$^{-3}$ and the pure quark phase starts at 1.54 fm$^{-3}$.
      For $B^{1/4}=150$ MeVfm$^{-3}$, the mixed phase begins at 0.51 fm$^{-3}$
      and the pure quark matter phase begins at $1.89$ fm$^{-3}$.
      Finally, for $B^{1/4}=200$ MeVfm$^{-3}$, the mixed phase starts
      at $0.72$ MeVfm$^{-3}$ while the pure quark phase starts at 
      $2.11$ fm$^{-3}$.
\item In case of a Maxwell construction, in order to link the 
      hadronic and the quark matter EoS, we obtain for $B^{1/4}=100$ 
MeVfm$^{-3}$
      that the pure $pn$ phase ends  at $0.58$ fm$^{-3}$
      and that the pure quark phase starts  
      at $0.67$ fm$^{-3}$. For $B^{1/4}=150$ MeVfm$^{-3}$, 
      the numbers are $0.92$ fm$^{-3}$
      and $1.215$ fm$^{-3}$ , while the corresponding numbers for
      $B^{1/4}=200$ MeVfm$^{-3}$ are $1.04$ and $1.57$ fm$^{-3}$.
\end{itemize}
As can be
seen from Figs.\ \ref{fig:sec5fig1}, \ref{fig:sec5fig3}, 
\ref{fig:sec5fig5}, \ref{fig:sec5fig7} and \ref{fig:sec5fig9},
none of the equations of state from either the pure $pn$ phase
or with a mixed phase construction with quark degrees of freedom, result in
stable configurations for densities above $\approx 10 n_0$, implying
thereby, see e.g.\  Figs.\ \ref{fig:sec5fig3} and \ref{fig:sec5fig7}, 
that none of the stars have cores with
a pure quark phase.  The EoS with $pn$ degrees of freedom only
results in the largest mass $\approx 2.2M_{\odot}$ when the rotational
correction of Eq.\ (\ref{eq:rotcorr}) 
is accounted for, see Fig.\ \ref{fig:sec5fig7}. 
With the inclusion of the mixed phase,
the total mass is reduced since the EoS is softer.

Several interesting conclusions can be inferred from the results
displayed in Figs.\ \ref{fig:sec5fig1}-\ref{fig:sec5fig9}.  
Firstly, to obtain neutron
star masses of the order $M\sim 2.2M_\odot$ as may now have been observed
in QPO's, we need the stiffest EoS allowed by causality (i.e.\ 
$\delta\simeq 0.13-0.2$) and to include rotation, see Figs.\ 
\ref{fig:sec5fig1} and \ref{fig:sec5fig7}. Furthermore, a phase
transition to quark matter below densities of order $\sim 5 n_0$ can be
excluded, corresponding to restricting the Bag constant to $B^{1/4}\ga
200$ MeVfm$^{-3}$. This can be seen in Fig.\ \ref{fig:sec5fig3}
where we plot star masses as function of the central density $n_c$ and
bag-model parameter $B$. 
These results differ significantly from those of 
Akmal et al.\ and Kalogera and Baym \cite{apr98,kalogera} 
due to the very different recipes we use to
incorporate causality at high densities. In Refs.\ 
\cite{apr98,kalogera}  the EoS is discontinuously stiffened by taking
$v_s=c$ at densities above a certain value $n_c$ which, however, is
lower than $n_{s}=5n_0$ where their nuclear EoS becomes superluminal.
This stiffens the nuclear EoS for densities $n_c<n<n_s$ but softens it
at higher densities. Their resulting maximum masses are in the range
$2.2M_\odot<M<3M_\odot$. Our approach incorporates causality by
reducing the sound speed smoothly towards the speed of light at high
densities. Therefore, our maximum mass never exceeds that of the nuclear
EoS of Akmal et al.\ \cite{apr98}. In fact one may argue that at very high
densities particles become relativistic and the sound speed should be
even lower, $v_s^2\simeq c^2/3$.
On the other hand, if it turns out that the QPOs are not from the innermost
stable orbits and that even accreting neutron stars have small masses,
say like the binary pulsars $M\la 1.5M_\odot$, this may indicate that
heavier neutron stars are not stable. Therefore, the EoS is
soft at high densities $\delta\ga 0.4$ or that a phase transition
occurs at a few times nuclear matter densities. For the nuclear to
quark matter transition this would require $B^{1/4}\la 80$MeVfm$^{-3}$  for
$\delta=0.2$. For such small Bag parameters there is an appreciable
quark and nuclear matter mixed phase in the neutron star interior
but even in these extreme cases a pure quark matter core is not obtained
for stable neutron star configurations.  
\begin{figure}\begin{center}
   % GNUPLOT: LaTeX picture with Postscript
\setlength{\unitlength}{0.1bp}
\special{!
%!PS-Adobe-2.0
%%Creator: gnuplot
%%DocumentFonts: Helvetica
%%BoundingBox: 50 50 770 554
%%Pages: (atend)
%%EndComments
/gnudict 40 dict def
gnudict begin
/Color false def
/Solid false def
/gnulinewidth 5.000 def
/vshift -33 def
/dl {10 mul} def
/hpt 31.5 def
/vpt 31.5 def
/M {moveto} bind def
/L {lineto} bind def
/R {rmoveto} bind def
/V {rlineto} bind def
/vpt2 vpt 2 mul def
/hpt2 hpt 2 mul def
/Lshow { currentpoint stroke M
  0 vshift R show } def
/Rshow { currentpoint stroke M
  dup stringwidth pop neg vshift R show } def
/Cshow { currentpoint stroke M
  dup stringwidth pop -2 div vshift R show } def
/DL { Color {setrgbcolor Solid {pop []} if 0 setdash }
 {pop pop pop Solid {pop []} if 0 setdash} ifelse } def
/BL { stroke gnulinewidth 2 mul setlinewidth } def
/AL { stroke gnulinewidth 2 div setlinewidth } def
/PL { stroke gnulinewidth setlinewidth } def
/LTb { BL [] 0 0 0 DL } def
/LTa { AL [1 dl 2 dl] 0 setdash 0 0 0 setrgbcolor } def
/LT0 { PL [] 0 1 0 DL } def
/LT1 { PL [4 dl 2 dl] 0 0 1 DL } def
/LT2 { PL [2 dl 3 dl] 1 0 0 DL } def
/LT3 { PL [1 dl 1.5 dl] 1 0 1 DL } def
/LT4 { PL [5 dl 2 dl 1 dl 2 dl] 0 1 1 DL } def
/LT5 { PL [4 dl 3 dl 1 dl 3 dl] 1 1 0 DL } def
/LT6 { PL [2 dl 2 dl 2 dl 4 dl] 0 0 0 DL } def
/LT7 { PL [2 dl 2 dl 2 dl 2 dl 2 dl 4 dl] 1 0.3 0 DL } def
/LT8 { PL [2 dl 2 dl 2 dl 2 dl 2 dl 2 dl 2 dl 4 dl] 0.5 0.5 0.5 DL } def
/P { stroke [] 0 setdash
  currentlinewidth 2 div sub M
  0 currentlinewidth V stroke } def
/D { stroke [] 0 setdash 2 copy vpt add M
  hpt neg vpt neg V hpt vpt neg V
  hpt vpt V hpt neg vpt V closepath stroke
  P } def
/A { stroke [] 0 setdash vpt sub M 0 vpt2 V
  currentpoint stroke M
  hpt neg vpt neg R hpt2 0 V stroke
  } def
/B { stroke [] 0 setdash 2 copy exch hpt sub exch vpt add M
  0 vpt2 neg V hpt2 0 V 0 vpt2 V
  hpt2 neg 0 V closepath stroke
  P } def
/C { stroke [] 0 setdash exch hpt sub exch vpt add M
  hpt2 vpt2 neg V currentpoint stroke M
  hpt2 neg 0 R hpt2 vpt2 V stroke } def
/T { stroke [] 0 setdash 2 copy vpt 1.12 mul add M
  hpt neg vpt -1.62 mul V
  hpt 2 mul 0 V
  hpt neg vpt 1.62 mul V closepath stroke
  P  } def
/S { 2 copy A C} def
end
}
\begin{picture}(3600,2160)(0,0)
\special{"
gnudict begin
gsave
50 50 translate
0.100 0.100 scale
0 setgray
/Helvetica findfont 100 scalefont setfont
newpath
-500.000000 -500.000000 translate
LTa
LTb
600 251 M
63 0 V
2754 0 R
-63 0 V
600 623 M
63 0 V
2754 0 R
-63 0 V
600 994 M
63 0 V
2754 0 R
-63 0 V
600 1366 M
63 0 V
2754 0 R
-63 0 V
600 1737 M
63 0 V
2754 0 R
-63 0 V
600 2109 M
63 0 V
2754 0 R
-63 0 V
788 251 M
0 63 V
0 1795 R
0 -63 V
1163 251 M
0 63 V
0 1795 R
0 -63 V
1539 251 M
0 63 V
0 1795 R
0 -63 V
1915 251 M
0 63 V
0 1795 R
0 -63 V
2290 251 M
0 63 V
0 1795 R
0 -63 V
2666 251 M
0 63 V
0 1795 R
0 -63 V
3041 251 M
0 63 V
0 1795 R
0 -63 V
3417 251 M
0 63 V
0 1795 R
0 -63 V
600 251 M
2817 0 V
0 1858 V
-2817 0 V
600 251 L
LT0
3114 1946 M
180 0 V
656 474 M
57 95 V
56 92 V
56 86 V
57 81 V
56 75 V
56 67 V
57 61 V
56 55 V
56 49 V
57 44 V
56 37 V
56 34 V
57 29 V
56 26 V
56 21 V
57 19 V
56 16 V
56 14 V
57 11 V
56 9 V
56 7 V
57 6 V
56 5 V
57 3 V
56 3 V
56 1 V
57 0 V
56 0 V
56 -1 V
57 -1 V
56 -2 V
56 -3 V
57 -3 V
56 -3 V
56 -3 V
57 -4 V
56 -4 V
56 -4 V
57 -5 V
56 -4 V
56 -5 V
57 -5 V
56 -5 V
56 -4 V
57 -6 V
56 -4 V
56 -5 V
57 -5 V
56 -6 V
LT1
3114 1846 M
180 0 V
656 491 M
57 83 V
56 79 V
56 74 V
57 68 V
56 63 V
56 57 V
57 52 V
56 47 V
56 41 V
57 37 V
56 34 V
56 29 V
57 26 V
56 22 V
56 20 V
57 17 V
56 16 V
56 12 V
57 12 V
56 9 V
56 8 V
57 6 V
56 6 V
57 4 V
56 4 V
56 2 V
57 2 V
56 1 V
56 1 V
57 0 V
56 -1 V
56 -2 V
57 0 V
56 -3 V
56 -2 V
57 -2 V
56 -2 V
56 -3 V
57 -3 V
56 -4 V
56 -3 V
57 -4 V
56 -3 V
56 -4 V
57 -4 V
56 -3 V
56 -5 V
57 -4 V
56 -4 V
LT2
3114 1746 M
180 0 V
656 369 M
57 65 V
56 63 V
56 60 V
57 57 V
56 53 V
56 49 V
57 46 V
56 41 V
56 39 V
57 35 V
56 32 V
56 29 V
57 26 V
56 24 V
56 21 V
57 19 V
56 17 V
56 16 V
57 14 V
56 12 V
56 11 V
57 10 V
56 9 V
57 7 V
56 7 V
56 6 V
57 4 V
56 5 V
56 4 V
57 3 V
56 2 V
56 2 V
57 1 V
56 1 V
56 1 V
57 1 V
56 0 V
56 -1 V
57 0 V
56 -1 V
56 -1 V
57 -1 V
56 -2 V
56 -1 V
57 -2 V
56 -2 V
56 -2 V
57 -1 V
56 -3 V
LT3
3114 1646 M
180 0 V
656 327 M
57 54 V
56 53 V
56 50 V
57 48 V
56 44 V
56 42 V
57 40 V
56 35 V
56 34 V
57 31 V
56 28 V
56 26 V
57 24 V
56 22 V
56 20 V
57 19 V
56 16 V
56 15 V
57 14 V
56 13 V
56 11 V
57 10 V
56 10 V
57 8 V
56 8 V
56 6 V
57 6 V
56 6 V
56 5 V
57 4 V
56 4 V
56 3 V
57 3 V
56 3 V
56 1 V
57 2 V
56 2 V
56 1 V
57 0 V
56 1 V
56 1 V
57 0 V
56 0 V
56 -1 V
57 0 V
56 -1 V
56 0 V
57 -1 V
56 -1 V
stroke
grestore
end
showpage
}
\put(3054,1646){\makebox(0,0)[r]{$\delta=0.4$}}
\put(3054,1746){\makebox(0,0)[r]{$\delta=0.3$}}
\put(3054,1846){\makebox(0,0)[r]{$\delta=0.2$}}
\put(3054,1946){\makebox(0,0)[r]{$\delta=0.13$}}
\put(2008,21){\makebox(0,0){$n_c$ (fm$^{-3}$)}}
\put(100,1180){%
\special{ps: gsave currentpoint currentpoint translate
270 rotate neg exch neg exch translate}%
\makebox(0,0)[b]{\shortstack{$M/M_{\odot}$}}%
\special{ps: currentpoint grestore moveto}%
}
\put(3417,151){\makebox(0,0){1.8}}
\put(3041,151){\makebox(0,0){1.6}}
\put(2666,151){\makebox(0,0){1.4}}
\put(2290,151){\makebox(0,0){1.2}}
\put(1915,151){\makebox(0,0){1}}
\put(1539,151){\makebox(0,0){0.8}}
\put(1163,151){\makebox(0,0){0.6}}
\put(788,151){\makebox(0,0){0.4}}
\put(540,2109){\makebox(0,0)[r]{3}}
\put(540,1737){\makebox(0,0)[r]{2.5}}
\put(540,1366){\makebox(0,0)[r]{2}}
\put(540,994){\makebox(0,0)[r]{1.5}}
\put(540,623){\makebox(0,0)[r]{1}}
\put(540,251){\makebox(0,0)[r]{0.5}}
\end{picture}
   \caption{Total mass $M$ for various values of $\delta$. See text for 
            further details.}
   \label{fig:sec5fig1}
\end{center}\end{figure}
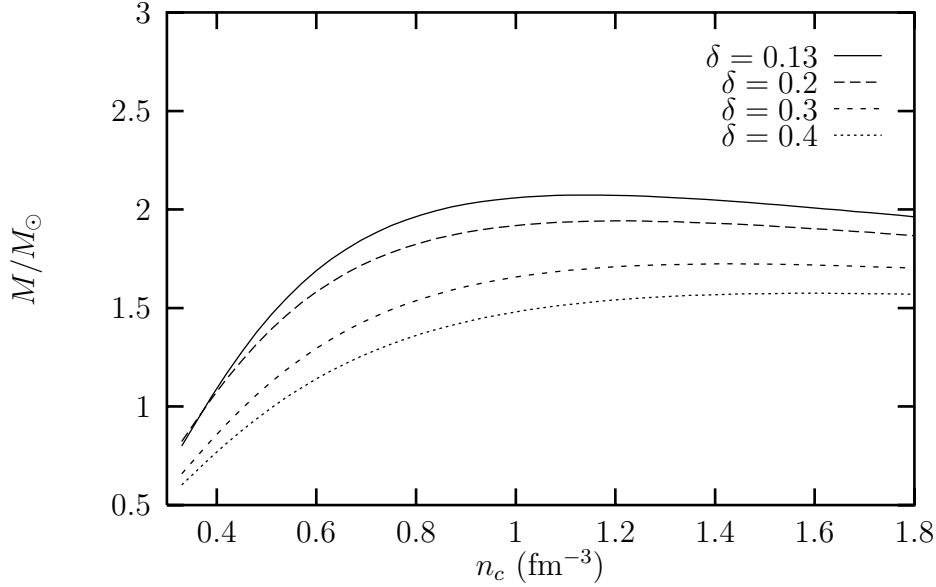
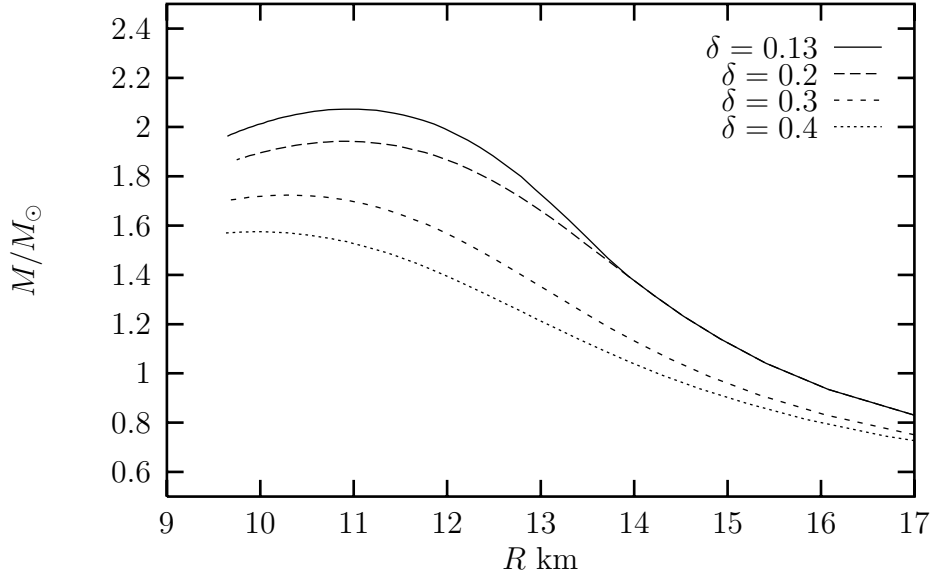
\begin{figure}\begin{center}
   % GNUPLOT: LaTeX picture with Postscript
\setlength{\unitlength}{0.1bp}
\special{!
%!PS-Adobe-2.0
%%Creator: gnuplot
%%DocumentFonts: Helvetica
%%BoundingBox: 50 50 770 554
%%Pages: (atend)
%%EndComments
/gnudict 40 dict def
gnudict begin
/Color false def
/Solid false def
/gnulinewidth 5.000 def
/vshift -33 def
/dl {10 mul} def
/hpt 31.5 def
/vpt 31.5 def
/M {moveto} bind def
/L {lineto} bind def
/R {rmoveto} bind def
/V {rlineto} bind def
/vpt2 vpt 2 mul def
/hpt2 hpt 2 mul def
/Lshow { currentpoint stroke M
  0 vshift R show } def
/Rshow { currentpoint stroke M
  dup stringwidth pop neg vshift R show } def
/Cshow { currentpoint stroke M
  dup stringwidth pop -2 div vshift R show } def
/DL { Color {setrgbcolor Solid {pop []} if 0 setdash }
 {pop pop pop Solid {pop []} if 0 setdash} ifelse } def
/BL { stroke gnulinewidth 2 mul setlinewidth } def
/AL { stroke gnulinewidth 2 div setlinewidth } def
/PL { stroke gnulinewidth setlinewidth } def
/LTb { BL [] 0 0 0 DL } def
/LTa { AL [1 dl 2 dl] 0 setdash 0 0 0 setrgbcolor } def
/LT0 { PL [] 0 1 0 DL } def
/LT1 { PL [4 dl 2 dl] 0 0 1 DL } def
/LT2 { PL [2 dl 3 dl] 1 0 0 DL } def
/LT3 { PL [1 dl 1.5 dl] 1 0 1 DL } def
/LT4 { PL [5 dl 2 dl 1 dl 2 dl] 0 1 1 DL } def
/LT5 { PL [4 dl 3 dl 1 dl 3 dl] 1 1 0 DL } def
/LT6 { PL [2 dl 2 dl 2 dl 4 dl] 0 0 0 DL } def
/LT7 { PL [2 dl 2 dl 2 dl 2 dl 2 dl 4 dl] 1 0.3 0 DL } def
/LT8 { PL [2 dl 2 dl 2 dl 2 dl 2 dl 2 dl 2 dl 4 dl] 0.5 0.5 0.5 DL } def
/P { stroke [] 0 setdash
  currentlinewidth 2 div sub M
  0 currentlinewidth V stroke } def
/D { stroke [] 0 setdash 2 copy vpt add M
  hpt neg vpt neg V hpt vpt neg V
  hpt vpt V hpt neg vpt V closepath stroke
  P } def
/A { stroke [] 0 setdash vpt sub M 0 vpt2 V
  currentpoint stroke M
  hpt neg vpt neg R hpt2 0 V stroke
  } def
/B { stroke [] 0 setdash 2 copy exch hpt sub exch vpt add M
  0 vpt2 neg V hpt2 0 V 0 vpt2 V
  hpt2 neg 0 V closepath stroke
  P } def
/C { stroke [] 0 setdash exch hpt sub exch vpt add M
  hpt2 vpt2 neg V currentpoint stroke M
  hpt2 neg 0 R hpt2 vpt2 V stroke } def
/T { stroke [] 0 setdash 2 copy vpt 1.12 mul add M
  hpt neg vpt -1.62 mul V
  hpt 2 mul 0 V
  hpt neg vpt 1.62 mul V closepath stroke
  P  } def
/S { 2 copy A C} def
end
}
\begin{picture}(3600,2160)(0,0)
\special{"
gnudict begin
gsave
50 50 translate
0.100 0.100 scale
0 setgray
/Helvetica findfont 100 scalefont setfont
newpath
-500.000000 -500.000000 translate
LTa
LTb
600 344 M
63 0 V
2754 0 R
-63 0 V
600 530 M
63 0 V
2754 0 R
-63 0 V
600 716 M
63 0 V
2754 0 R
-63 0 V
600 901 M
63 0 V
2754 0 R
-63 0 V
600 1087 M
63 0 V
2754 0 R
-63 0 V
600 1273 M
63 0 V
2754 0 R
-63 0 V
600 1459 M
63 0 V
2754 0 R
-63 0 V
600 1645 M
63 0 V
2754 0 R
-63 0 V
600 1830 M
63 0 V
2754 0 R
-63 0 V
600 2016 M
63 0 V
2754 0 R
-63 0 V
600 251 M
0 63 V
0 1795 R
0 -63 V
952 251 M
0 63 V
0 1795 R
0 -63 V
1304 251 M
0 63 V
0 1795 R
0 -63 V
1656 251 M
0 63 V
0 1795 R
0 -63 V
2009 251 M
0 63 V
0 1795 R
0 -63 V
2361 251 M
0 63 V
0 1795 R
0 -63 V
2713 251 M
0 63 V
0 1795 R
0 -63 V
3065 251 M
0 63 V
0 1795 R
0 -63 V
3417 251 M
0 63 V
0 1795 R
0 -63 V
600 251 M
2817 0 V
0 1858 V
-2817 0 V
600 251 L
LT0
3114 1946 M
180 0 V
3417 558 M
-324 97 V
-236 99 V
-172 92 V
-138 85 V
-112 79 V
-96 72 V
-77 68 V
-77 76 V
-71 69 V
-67 62 V
-60 53 V
-52 48 V
-57 42 V
-49 36 V
-49 32 V
-46 27 V
-46 23 V
-42 21 V
-39 17 V
-42 14 V
-39 11 V
-35 9 V
-35 7 V
-36 6 V
-31 5 V
-35 2 V
-32 2 V
-28 0 V
-28 0 V
-32 -1 V
-25 -1 V
-24 -3 V
-29 -3 V
-24 -4 V
-21 -4 V
-25 -4 V
-21 -5 V
-25 -5 V
-21 -5 V
-18 -6 V
-21 -5 V
-17 -7 V
-21 -5 V
-18 -7 V
-17 -5 V
-17 -7 V
-16 -5 V
-15 -7 V
-17 -6 V
-14 -7 V
LT1
3114 1846 M
180 0 V
3417 558 M
-324 97 V
-236 99 V
-172 92 V
-138 85 V
-112 79 V
-96 72 V
-84 64 V
-77 58 V
-67 52 V
-64 47 V
-60 41 V
-56 37 V
-53 32 V
-49 28 V
-49 25 V
-43 21 V
-45 20 V
-43 16 V
-38 14 V
-39 12 V
-39 9 V
-32 8 V
-35 8 V
-35 5 V
-32 4 V
-31 3 V
-29 3 V
-28 1 V
-31 1 V
-25 0 V
-28 -1 V
-25 -2 V
-25 -1 V
-24 -3 V
-25 -3 V
-21 -2 V
-21 -3 V
-21 -4 V
-21 -4 V
-21 -4 V
-22 -4 V
-17 -5 V
-18 -4 V
-20 -5 V
-16 -5 V
-18 -4 V
-17 -6 V
-16 -4 V
-16 -6 V
LT2
3114 1746 M
180 0 V
3417 484 M
-335 75 V
-250 75 V
-190 71 V
-148 66 V
-123 62 V
-105 56 V
-88 52 V
-81 49 V
-71 43 V
-67 40 V
-60 36 V
-56 33 V
-53 30 V
-49 27 V
-46 24 V
-42 21 V
-42 20 V
-39 17 V
-42 15 V
-36 14 V
-35 12 V
-32 11 V
-35 10 V
-31 8 V
-29 7 V
-31 6 V
-28 6 V
-29 4 V
-28 4 V
-24 3 V
-25 2 V
-25 2 V
-24 1 V
-25 1 V
-21 1 V
-21 0 V
-21 -1 V
-22 0 V
-21 -1 V
-21 -2 V
-18 -1 V
-19 -1 V
-19 -2 V
-18 -3 V
-18 -2 V
-16 -3 V
-17 -2 V
-16 -2 V
LT3
3114 1646 M
180 0 V
3417 462 M
-109 17 V
-299 63 V
-222 60 V
-180 56 V
-144 52 V
-120 49 V
-102 45 V
-92 42 V
-81 39 V
-74 35 V
-63 32 V
-63 30 V
-57 28 V
-53 25 V
-49 23 V
-46 21 V
-42 18 V
-42 18 V
-39 16 V
-39 14 V
-35 13 V
-35 12 V
-35 10 V
-32 9 V
-32 9 V
-28 7 V
-28 8 V
-32 5 V
-24 6 V
-28 4 V
-25 4 V
-25 4 V
-24 3 V
-22 2 V
-24 3 V
-21 2 V
-25 1 V
-21 1 V
-21 1 V
-21 1 V
-19 0 V
-18 0 V
-20 -1 V
-18 0 V
-18 -1 V
-17 -1 V
-17 -1 V
-17 -1 V
stroke
grestore
end
showpage
}
\put(3054,1646){\makebox(0,0)[r]{$\delta=0.4$}}
\put(3054,1746){\makebox(0,0)[r]{$\delta=0.3$}}
\put(3054,1846){\makebox(0,0)[r]{$\delta=0.2$}}
\put(3054,1946){\makebox(0,0)[r]{$\delta=0.13$}}
\put(2008,21){\makebox(0,0){$R$ km}}
\put(100,1180){%
\special{ps: gsave currentpoint currentpoint translate
270 rotate neg exch neg exch translate}%
\makebox(0,0)[b]{\shortstack{$M/M_{\odot}$}}%
\special{ps: currentpoint grestore moveto}%
}
\put(3417,151){\makebox(0,0){17}}
\put(3065,151){\makebox(0,0){16}}
\put(2713,151){\makebox(0,0){15}}
\put(2361,151){\makebox(0,0){14}}
\put(2009,151){\makebox(0,0){13}}
\put(1656,151){\makebox(0,0){12}}
\put(1304,151){\makebox(0,0){11}}
\put(952,151){\makebox(0,0){10}}
\put(600,151){\makebox(0,0){9}}
\put(540,2016){\makebox(0,0)[r]{2.4}}
\put(540,1830){\makebox(0,0)[r]{2.2}}
\put(540,1645){\makebox(0,0)[r]{2}}
\put(540,1459){\makebox(0,0)[r]{1.8}}
\put(540,1273){\makebox(0,0)[r]{1.6}}
\put(540,1087){\makebox(0,0)[r]{1.4}}
\put(540,901){\makebox(0,0)[r]{1.2}}
\put(540,716){\makebox(0,0)[r]{1}}
\put(540,530){\makebox(0,0)[r]{0.8}}
\put(540,344){\makebox(0,0)[r]{0.6}}
\end{picture}
   \caption{Mass-radius relation for various values of $\delta$.}
   \label{fig:sec5fig2}
\end{center}\end{figure}
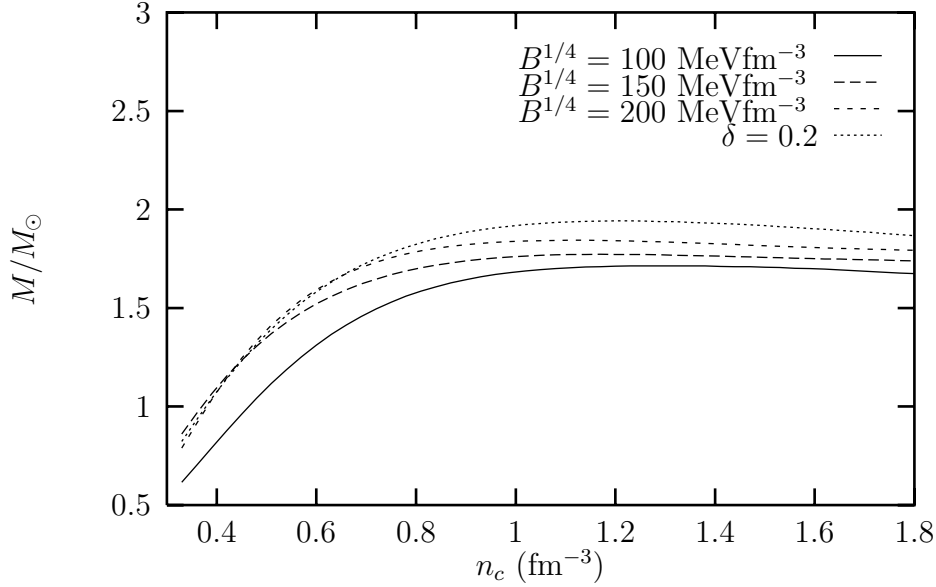
\begin{figure}\begin{center}
   % GNUPLOT: LaTeX picture with Postscript
\setlength{\unitlength}{0.1bp}
\special{!
%!PS-Adobe-2.0
%%Creator: gnuplot
%%DocumentFonts: Helvetica
%%BoundingBox: 50 50 770 554
%%Pages: (atend)
%%EndComments
/gnudict 40 dict def
gnudict begin
/Color false def
/Solid false def
/gnulinewidth 5.000 def
/vshift -33 def
/dl {10 mul} def
/hpt 31.5 def
/vpt 31.5 def
/M {moveto} bind def
/L {lineto} bind def
/R {rmoveto} bind def
/V {rlineto} bind def
/vpt2 vpt 2 mul def
/hpt2 hpt 2 mul def
/Lshow { currentpoint stroke M
  0 vshift R show } def
/Rshow { currentpoint stroke M
  dup stringwidth pop neg vshift R show } def
/Cshow { currentpoint stroke M
  dup stringwidth pop -2 div vshift R show } def
/DL { Color {setrgbcolor Solid {pop []} if 0 setdash }
 {pop pop pop Solid {pop []} if 0 setdash} ifelse } def
/BL { stroke gnulinewidth 2 mul setlinewidth } def
/AL { stroke gnulinewidth 2 div setlinewidth } def
/PL { stroke gnulinewidth setlinewidth } def
/LTb { BL [] 0 0 0 DL } def
/LTa { AL [1 dl 2 dl] 0 setdash 0 0 0 setrgbcolor } def
/LT0 { PL [] 0 1 0 DL } def
/LT1 { PL [4 dl 2 dl] 0 0 1 DL } def
/LT2 { PL [2 dl 3 dl] 1 0 0 DL } def
/LT3 { PL [1 dl 1.5 dl] 1 0 1 DL } def
/LT4 { PL [5 dl 2 dl 1 dl 2 dl] 0 1 1 DL } def
/LT5 { PL [4 dl 3 dl 1 dl 3 dl] 1 1 0 DL } def
/LT6 { PL [2 dl 2 dl 2 dl 4 dl] 0 0 0 DL } def
/LT7 { PL [2 dl 2 dl 2 dl 2 dl 2 dl 4 dl] 1 0.3 0 DL } def
/LT8 { PL [2 dl 2 dl 2 dl 2 dl 2 dl 2 dl 2 dl 4 dl] 0.5 0.5 0.5 DL } def
/P { stroke [] 0 setdash
  currentlinewidth 2 div sub M
  0 currentlinewidth V stroke } def
/D { stroke [] 0 setdash 2 copy vpt add M
  hpt neg vpt neg V hpt vpt neg V
  hpt vpt V hpt neg vpt V closepath stroke
  P } def
/A { stroke [] 0 setdash vpt sub M 0 vpt2 V
  currentpoint stroke M
  hpt neg vpt neg R hpt2 0 V stroke
  } def
/B { stroke [] 0 setdash 2 copy exch hpt sub exch vpt add M
  0 vpt2 neg V hpt2 0 V 0 vpt2 V
  hpt2 neg 0 V closepath stroke
  P } def
/C { stroke [] 0 setdash exch hpt sub exch vpt add M
  hpt2 vpt2 neg V currentpoint stroke M
  hpt2 neg 0 R hpt2 vpt2 V stroke } def
/T { stroke [] 0 setdash 2 copy vpt 1.12 mul add M
  hpt neg vpt -1.62 mul V
  hpt 2 mul 0 V
  hpt neg vpt 1.62 mul V closepath stroke
  P  } def
/S { 2 copy A C} def
end
}
\begin{picture}(3600,2160)(0,0)
\special{"
gnudict begin
gsave
50 50 translate
0.100 0.100 scale
0 setgray
/Helvetica findfont 100 scalefont setfont
newpath
-500.000000 -500.000000 translate
LTa
LTb
600 251 M
63 0 V
2754 0 R
-63 0 V
600 623 M
63 0 V
2754 0 R
-63 0 V
600 994 M
63 0 V
2754 0 R
-63 0 V
600 1366 M
63 0 V
2754 0 R
-63 0 V
600 1737 M
63 0 V
2754 0 R
-63 0 V
600 2109 M
63 0 V
2754 0 R
-63 0 V
788 251 M
0 63 V
0 1795 R
0 -63 V
1163 251 M
0 63 V
0 1795 R
0 -63 V
1539 251 M
0 63 V
0 1795 R
0 -63 V
1915 251 M
0 63 V
0 1795 R
0 -63 V
2290 251 M
0 63 V
0 1795 R
0 -63 V
2666 251 M
0 63 V
0 1795 R
0 -63 V
3041 251 M
0 63 V
0 1795 R
0 -63 V
3417 251 M
0 63 V
0 1795 R
0 -63 V
600 251 M
2817 0 V
0 1858 V
-2817 0 V
600 251 L
LT0
3114 1946 M
180 0 V
656 337 M
57 65 V
56 65 V
56 64 V
57 63 V
56 60 V
56 56 V
57 52 V
56 48 V
56 44 V
57 41 V
56 35 V
56 32 V
57 29 V
56 25 V
56 23 V
57 19 V
56 16 V
56 15 V
57 12 V
56 11 V
56 9 V
57 7 V
56 6 V
57 5 V
56 4 V
56 3 V
57 3 V
56 1 V
56 2 V
57 1 V
56 0 V
56 0 V
57 0 V
56 0 V
56 0 V
57 -2 V
56 -1 V
56 0 V
57 -2 V
56 -1 V
56 -3 V
57 -1 V
56 -2 V
56 -3 V
57 -3 V
56 -3 V
56 -3 V
57 -3 V
56 -2 V
LT1
3114 1846 M
180 0 V
656 519 M
57 79 V
56 73 V
56 66 V
57 60 V
56 53 V
56 48 V
57 42 V
56 37 V
56 33 V
57 28 V
56 25 V
56 22 V
57 18 V
56 17 V
56 14 V
57 12 V
56 10 V
56 8 V
57 8 V
56 6 V
56 5 V
57 4 V
56 3 V
57 3 V
56 2 V
56 1 V
57 0 V
56 1 V
56 -1 V
57 0 V
56 0 V
56 -1 V
57 -2 V
56 -1 V
56 -1 V
57 -1 V
56 -2 V
56 -1 V
57 -2 V
56 -1 V
56 -2 V
57 -1 V
56 -2 V
56 0 V
57 -2 V
56 -1 V
56 -1 V
57 -2 V
56 -1 V
LT2
3114 1746 M
180 0 V
656 466 M
57 94 V
56 89 V
56 81 V
57 75 V
56 66 V
56 59 V
57 50 V
56 45 V
56 38 V
57 32 V
56 29 V
56 23 V
57 20 V
56 17 V
56 14 V
57 12 V
56 9 V
56 8 V
57 6 V
56 5 V
56 4 V
57 3 V
56 2 V
57 1 V
56 1 V
56 1 V
57 0 V
56 -1 V
56 -1 V
57 -1 V
56 -2 V
56 -1 V
57 -2 V
56 -2 V
56 -2 V
57 -2 V
56 -2 V
56 -3 V
57 -2 V
56 -2 V
56 -2 V
57 -2 V
56 -2 V
56 -2 V
57 -1 V
56 -2 V
56 -1 V
57 -2 V
56 -1 V
LT3
3114 1646 M
180 0 V
656 491 M
57 83 V
56 79 V
56 74 V
57 68 V
56 63 V
56 57 V
57 52 V
56 47 V
56 41 V
57 37 V
56 34 V
56 29 V
57 26 V
56 22 V
56 20 V
57 17 V
56 16 V
56 12 V
57 12 V
56 9 V
56 8 V
57 6 V
56 6 V
57 4 V
56 4 V
56 2 V
57 2 V
56 1 V
56 1 V
57 0 V
56 -1 V
56 -2 V
57 0 V
56 -3 V
56 -2 V
57 -2 V
56 -2 V
56 -3 V
57 -3 V
56 -4 V
56 -3 V
57 -4 V
56 -3 V
56 -4 V
57 -4 V
56 -3 V
56 -5 V
57 -4 V
56 -4 V
stroke
grestore
end
showpage
}
\put(3054,1646){\makebox(0,0)[r]{$\delta=0.2$}}
\put(3054,1746){\makebox(0,0)[r]{$B^{1/4}=200$ MeVfm$^{-3}$ }}
\put(3054,1846){\makebox(0,0)[r]{$B^{1/4}=150$ MeVfm$^{-3}$ }}
\put(3054,1946){\makebox(0,0)[r]{$B^{1/4}=100$ MeVfm$^{-3}$ }}
\put(2008,21){\makebox(0,0){$n_c$ (fm$^{-3}$)}}
\put(100,1180){%
\special{ps: gsave currentpoint currentpoint translate
270 rotate neg exch neg exch translate}%
\makebox(0,0)[b]{\shortstack{$M/M_{\odot}$}}%
\special{ps: currentpoint grestore moveto}%
}
\put(3417,151){\makebox(0,0){1.8}}
\put(3041,151){\makebox(0,0){1.6}}
\put(2666,151){\makebox(0,0){1.4}}
\put(2290,151){\makebox(0,0){1.2}}
\put(1915,151){\makebox(0,0){1}}
\put(1539,151){\makebox(0,0){0.8}}
\put(1163,151){\makebox(0,0){0.6}}
\put(788,151){\makebox(0,0){0.4}}
\put(540,2109){\makebox(0,0)[r]{3}}
\put(540,1737){\makebox(0,0)[r]{2.5}}
\put(540,1366){\makebox(0,0)[r]{2}}
\put(540,994){\makebox(0,0)[r]{1.5}}
\put(540,623){\makebox(0,0)[r]{1}}
\put(540,251){\makebox(0,0)[r]{0.5}}
\end{picture}
   \caption{Total mass $M$ for various values of the bag parameter $B$ for
            the mixed phase EoS. For comparison we include also the results
            from the $pn$-matter EoS for $\beta$-stable with $\delta=0.2$.}
   \label{fig:sec5fig3}
\end{center}\end{figure}
\begin{figure}\begin{center}
    % GNUPLOT: LaTeX picture with Postscript
\setlength{\unitlength}{0.1bp}
\special{!
%!PS-Adobe-2.0
%%Creator: gnuplot
%%DocumentFonts: Helvetica
%%BoundingBox: 50 50 770 554
%%Pages: (atend)
%%EndComments
/gnudict 40 dict def
gnudict begin
/Color false def
/Solid false def
/gnulinewidth 5.000 def
/vshift -33 def
/dl {10 mul} def
/hpt 31.5 def
/vpt 31.5 def
/M {moveto} bind def
/L {lineto} bind def
/R {rmoveto} bind def
/V {rlineto} bind def
/vpt2 vpt 2 mul def
/hpt2 hpt 2 mul def
/Lshow { currentpoint stroke M
  0 vshift R show } def
/Rshow { currentpoint stroke M
  dup stringwidth pop neg vshift R show } def
/Cshow { currentpoint stroke M
  dup stringwidth pop -2 div vshift R show } def
/DL { Color {setrgbcolor Solid {pop []} if 0 setdash }
 {pop pop pop Solid {pop []} if 0 setdash} ifelse } def
/BL { stroke gnulinewidth 2 mul setlinewidth } def
/AL { stroke gnulinewidth 2 div setlinewidth } def
/PL { stroke gnulinewidth setlinewidth } def
/LTb { BL [] 0 0 0 DL } def
/LTa { AL [1 dl 2 dl] 0 setdash 0 0 0 setrgbcolor } def
/LT0 { PL [] 0 1 0 DL } def
/LT1 { PL [4 dl 2 dl] 0 0 1 DL } def
/LT2 { PL [2 dl 3 dl] 1 0 0 DL } def
/LT3 { PL [1 dl 1.5 dl] 1 0 1 DL } def
/LT4 { PL [5 dl 2 dl 1 dl 2 dl] 0 1 1 DL } def
/LT5 { PL [4 dl 3 dl 1 dl 3 dl] 1 1 0 DL } def
/LT6 { PL [2 dl 2 dl 2 dl 4 dl] 0 0 0 DL } def
/LT7 { PL [2 dl 2 dl 2 dl 2 dl 2 dl 4 dl] 1 0.3 0 DL } def
/LT8 { PL [2 dl 2 dl 2 dl 2 dl 2 dl 2 dl 2 dl 4 dl] 0.5 0.5 0.5 DL } def
/P { stroke [] 0 setdash
  currentlinewidth 2 div sub M
  0 currentlinewidth V stroke } def
/D { stroke [] 0 setdash 2 copy vpt add M
  hpt neg vpt neg V hpt vpt neg V
  hpt vpt V hpt neg vpt V closepath stroke
  P } def
/A { stroke [] 0 setdash vpt sub M 0 vpt2 V
  currentpoint stroke M
  hpt neg vpt neg R hpt2 0 V stroke
  } def
/B { stroke [] 0 setdash 2 copy exch hpt sub exch vpt add M
  0 vpt2 neg V hpt2 0 V 0 vpt2 V
  hpt2 neg 0 V closepath stroke
  P } def
/C { stroke [] 0 setdash exch hpt sub exch vpt add M
  hpt2 vpt2 neg V currentpoint stroke M
  hpt2 neg 0 R hpt2 vpt2 V stroke } def
/T { stroke [] 0 setdash 2 copy vpt 1.12 mul add M
  hpt neg vpt -1.62 mul V
  hpt 2 mul 0 V
  hpt neg vpt 1.62 mul V closepath stroke
  P  } def
/S { 2 copy A C} def
end
}
\begin{picture}(3600,2160)(0,0)
\special{"
gnudict begin
gsave
50 50 translate
0.100 0.100 scale
0 setgray
/Helvetica findfont 100 scalefont setfont
newpath
-500.000000 -500.000000 translate
LTa
LTb
600 344 M
63 0 V
2754 0 R
-63 0 V
600 530 M
63 0 V
2754 0 R
-63 0 V
600 716 M
63 0 V
2754 0 R
-63 0 V
600 901 M
63 0 V
2754 0 R
-63 0 V
600 1087 M
63 0 V
2754 0 R
-63 0 V
600 1273 M
63 0 V
2754 0 R
-63 0 V
600 1459 M
63 0 V
2754 0 R
-63 0 V
600 1644 M
63 0 V
2754 0 R
-63 0 V
600 1830 M
63 0 V
2754 0 R
-63 0 V
600 2016 M
63 0 V
2754 0 R
-63 0 V
600 251 M
0 63 V
0 1795 R
0 -63 V
952 251 M
0 63 V
0 1795 R
0 -63 V
1304 251 M
0 63 V
0 1795 R
0 -63 V
1656 251 M
0 63 V
0 1795 R
0 -63 V
2009 251 M
0 63 V
0 1795 R
0 -63 V
2361 251 M
0 63 V
0 1795 R
0 -63 V
2713 251 M
0 63 V
0 1795 R
0 -63 V
3065 251 M
0 63 V
0 1795 R
0 -63 V
3417 251 M
0 63 V
0 1795 R
0 -63 V
600 251 M
2817 0 V
0 1858 V
-2817 0 V
600 251 L
LT0
3114 1946 M
180 0 V
3417 470 M
-275 51 V
-299 81 V
-222 78 V
-169 75 V
-137 69 V
-109 65 V
-95 61 V
-81 55 V
-74 51 V
-64 44 V
-56 40 V
-53 36 V
-49 32 V
-46 28 V
-39 24 V
-38 20 V
-36 19 V
-35 15 V
-28 13 V
-32 11 V
-24 9 V
-25 8 V
-25 6 V
-21 5 V
-21 4 V
-17 3 V
-15 2 V
-17 2 V
-14 1 V
-14 1 V
-14 0 V
-15 0 V
-10 -1 V
-11 0 V
-14 -2 V
-10 -1 V
-11 -1 V
-10 -2 V
-11 -2 V
-11 -2 V
-14 -2 V
-10 -3 V
-14 -4 V
-14 -2 V
-14 -4 V
-11 -4 V
-14 -4 V
-14 -3 V
LT1
3114 1846 M
180 0 V
3417 558 M
-324 97 V
-236 99 V
-172 92 V
-138 85 V
-112 79 V
-96 72 V
-84 64 V
-4 13 V
-63 41 V
-60 35 V
-56 31 V
-53 27 V
-46 24 V
-45 20 V
-43 18 V
-42 15 V
-39 13 V
-35 10 V
-35 9 V
-32 7 V
-32 7 V
-31 6 V
-28 3 V
-25 3 V
-28 3 V
-25 1 V
-21 1 V
-25 1 V
-21 -1 V
-17 0 V
-22 -1 V
-17 -1 V
-14 -2 V
-14 -2 V
-18 -1 V
-10 -2 V
-15 -2 V
-10 -1 V
-7 -2 V
-11 -2 V
-7 -2 V
-7 -2 V
-7 -2 V
-7 -1 V
-3 -1 V
-7 -1 V
-4 -2 V
-3 -2 V
-8 -2 V
LT2
3114 1746 M
180 0 V
3417 558 M
-324 97 V
-236 99 V
-172 92 V
-138 85 V
-112 79 V
-96 72 V
-84 64 V
-77 58 V
-67 52 V
-64 47 V
-67 39 V
-45 29 V
-39 25 V
-42 21 V
-36 18 V
-35 14 V
-32 12 V
-31 10 V
-28 7 V
-29 7 V
-24 4 V
-25 4 V
-25 3 V
-21 2 V
-21 1 V
-21 1 V
-21 0 V
-18 -1 V
-21 -1 V
-17 -2 V
-14 -2 V
-18 -2 V
-18 -2 V
-14 -3 V
-14 -2 V
-14 -3 V
-10 -3 V
-15 -3 V
-10 -2 V
-11 -3 V
-7 -3 V
-10 -2 V
-7 -3 V
-7 -2 V
-7 -1 V
-4 -2 V
-3 -2 V
-8 -2 V
-3 -2 V
LT3
3114 1646 M
180 0 V
3417 558 M
-324 97 V
-236 99 V
-172 92 V
-138 85 V
-112 79 V
-96 72 V
-84 64 V
-77 58 V
-67 52 V
-64 47 V
-60 41 V
-56 37 V
-53 32 V
-49 28 V
-49 25 V
-43 21 V
-45 20 V
-43 16 V
-38 14 V
-39 12 V
-39 9 V
-32 8 V
-35 8 V
-35 5 V
-32 4 V
-31 3 V
-29 3 V
-28 1 V
-31 1 V
-25 0 V
-28 -1 V
-25 -2 V
-25 -1 V
-24 -3 V
-25 -3 V
-21 -2 V
-21 -3 V
-21 -4 V
-21 -4 V
-21 -4 V
-22 -4 V
-17 -5 V
-18 -4 V
-20 -5 V
-16 -5 V
-18 -4 V
-17 -6 V
-16 -4 V
-16 -6 V
stroke
grestore
end
showpage
}
\put(3054,1646){\makebox(0,0)[r]{$\delta=0.2$}}
\put(3054,1746){\makebox(0,0)[r]{$B^{1/4}=200$ MeVfm$^{-3}$ }}
\put(3054,1846){\makebox(0,0)[r]{$B^{1/4}=150$ MeVfm$^{-3}$ }}
\put(3054,1946){\makebox(0,0)[r]{$B^{1/4}=100$ MeVfm$^{-3}$ }}
\put(2008,21){\makebox(0,0){$R$ km}}
\put(100,1180){%
\special{ps: gsave currentpoint currentpoint translate
270 rotate neg exch neg exch translate}%
\makebox(0,0)[b]{\shortstack{$M/M_{\odot}$}}%
\special{ps: currentpoint grestore moveto}%
}
\put(3417,151){\makebox(0,0){17}}
\put(3065,151){\makebox(0,0){16}}
\put(2713,151){\makebox(0,0){15}}
\put(2361,151){\makebox(0,0){14}}
\put(2009,151){\makebox(0,0){13}}
\put(1656,151){\makebox(0,0){12}}
\put(1304,151){\makebox(0,0){11}}
\put(952,151){\makebox(0,0){10}}
\put(600,151){\makebox(0,0){9}}
\put(540,2016){\makebox(0,0)[r]{2.4}}
\put(540,1830){\makebox(0,0)[r]{2.2}}
\put(540,1644){\makebox(0,0)[r]{2}}
\put(540,1459){\makebox(0,0)[r]{1.8}}
\put(540,1273){\makebox(0,0)[r]{1.6}}
\put(540,1087){\makebox(0,0)[r]{1.4}}
\put(540,901){\makebox(0,0)[r]{1.2}}
\put(540,716){\makebox(0,0)[r]{1}}
\put(540,530){\makebox(0,0)[r]{0.8}}
\put(540,344){\makebox(0,0)[r]{0.6}}
\end{picture}
    \caption{Mass-radius relation 
            for various values of the bag parameter $B$ for
            the mixed phase EoS. For comparison we include also the results
            from the $pn$-matter EoS for $\beta$-stable with $\delta=0.2$.}
    \label{fig:sec5fig4}
\end{center}\end{figure}
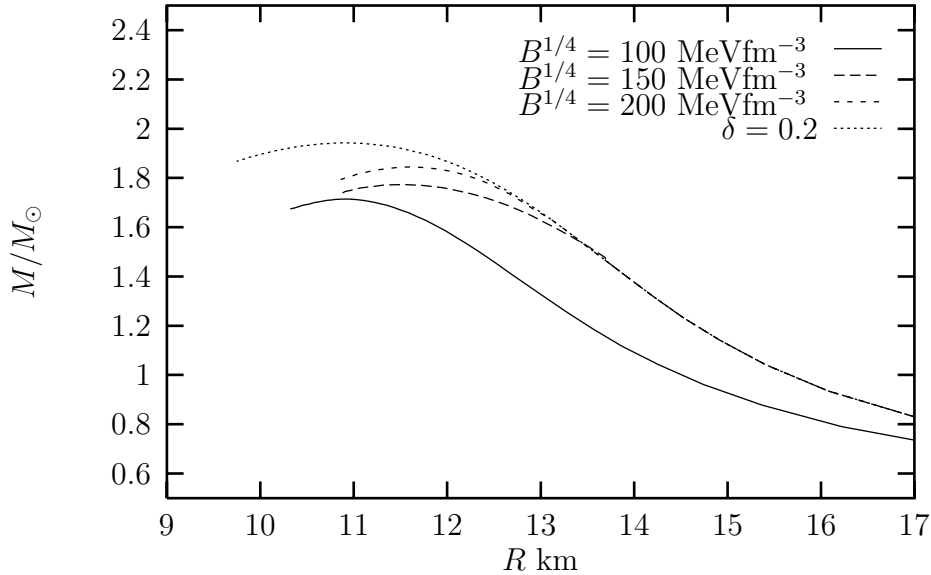
\begin{figure}\begin{center}
    % GNUPLOT: LaTeX picture with Postscript
\setlength{\unitlength}{0.1bp}
\special{!
%!PS-Adobe-2.0
%%Creator: gnuplot
%%DocumentFonts: Helvetica
%%BoundingBox: 50 50 770 554
%%Pages: (atend)
%%EndComments
/gnudict 40 dict def
gnudict begin
/Color false def
/Solid false def
/gnulinewidth 5.000 def
/vshift -33 def
/dl {10 mul} def
/hpt 31.5 def
/vpt 31.5 def
/M {moveto} bind def
/L {lineto} bind def
/R {rmoveto} bind def
/V {rlineto} bind def
/vpt2 vpt 2 mul def
/hpt2 hpt 2 mul def
/Lshow { currentpoint stroke M
  0 vshift R show } def
/Rshow { currentpoint stroke M
  dup stringwidth pop neg vshift R show } def
/Cshow { currentpoint stroke M
  dup stringwidth pop -2 div vshift R show } def
/DL { Color {setrgbcolor Solid {pop []} if 0 setdash }
 {pop pop pop Solid {pop []} if 0 setdash} ifelse } def
/BL { stroke gnulinewidth 2 mul setlinewidth } def
/AL { stroke gnulinewidth 2 div setlinewidth } def
/PL { stroke gnulinewidth setlinewidth } def
/LTb { BL [] 0 0 0 DL } def
/LTa { AL [1 dl 2 dl] 0 setdash 0 0 0 setrgbcolor } def
/LT0 { PL [] 0 1 0 DL } def
/LT1 { PL [4 dl 2 dl] 0 0 1 DL } def
/LT2 { PL [2 dl 3 dl] 1 0 0 DL } def
/LT3 { PL [1 dl 1.5 dl] 1 0 1 DL } def
/LT4 { PL [5 dl 2 dl 1 dl 2 dl] 0 1 1 DL } def
/LT5 { PL [4 dl 3 dl 1 dl 3 dl] 1 1 0 DL } def
/LT6 { PL [2 dl 2 dl 2 dl 4 dl] 0 0 0 DL } def
/LT7 { PL [2 dl 2 dl 2 dl 2 dl 2 dl 4 dl] 1 0.3 0 DL } def
/LT8 { PL [2 dl 2 dl 2 dl 2 dl 2 dl 2 dl 2 dl 4 dl] 0.5 0.5 0.5 DL } def
/P { stroke [] 0 setdash
  currentlinewidth 2 div sub M
  0 currentlinewidth V stroke } def
/D { stroke [] 0 setdash 2 copy vpt add M
  hpt neg vpt neg V hpt vpt neg V
  hpt vpt V hpt neg vpt V closepath stroke
  P } def
/A { stroke [] 0 setdash vpt sub M 0 vpt2 V
  currentpoint stroke M
  hpt neg vpt neg R hpt2 0 V stroke
  } def
/B { stroke [] 0 setdash 2 copy exch hpt sub exch vpt add M
  0 vpt2 neg V hpt2 0 V 0 vpt2 V
  hpt2 neg 0 V closepath stroke
  P } def
/C { stroke [] 0 setdash exch hpt sub exch vpt add M
  hpt2 vpt2 neg V currentpoint stroke M
  hpt2 neg 0 R hpt2 vpt2 V stroke } def
/T { stroke [] 0 setdash 2 copy vpt 1.12 mul add M
  hpt neg vpt -1.62 mul V
  hpt 2 mul 0 V
  hpt neg vpt 1.62 mul V closepath stroke
  P  } def
/S { 2 copy A C} def
end
}
\begin{picture}(3600,2160)(0,0)
\special{"
gnudict begin
gsave
50 50 translate
0.100 0.100 scale
0 setgray
/Helvetica findfont 100 scalefont setfont
newpath
-500.000000 -500.000000 translate
LTa
LTb
600 251 M
63 0 V
2754 0 R
-63 0 V
600 623 M
63 0 V
2754 0 R
-63 0 V
600 994 M
63 0 V
2754 0 R
-63 0 V
600 1366 M
63 0 V
2754 0 R
-63 0 V
600 1737 M
63 0 V
2754 0 R
-63 0 V
600 2109 M
63 0 V
2754 0 R
-63 0 V
788 251 M
0 63 V
0 1795 R
0 -63 V
1163 251 M
0 63 V
0 1795 R
0 -63 V
1539 251 M
0 63 V
0 1795 R
0 -63 V
1915 251 M
0 63 V
0 1795 R
0 -63 V
2290 251 M
0 63 V
0 1795 R
0 -63 V
2666 251 M
0 63 V
0 1795 R
0 -63 V
3041 251 M
0 63 V
0 1795 R
0 -63 V
3417 251 M
0 63 V
0 1795 R
0 -63 V
600 251 M
2817 0 V
0 1858 V
-2817 0 V
600 251 L
LT0
3114 1946 M
180 0 V
656 491 M
769 653 L
882 795 L
994 915 L
113 112 V
113 0 V
75 0 V
37 16 V
113 26 V
113 20 V
112 14 V
113 11 V
113 8 V
113 4 V
112 3 V
113 1 V
113 0 V
112 -1 V
113 -2 V
113 -3 V
112 -3 V
113 -4 V
113 -5 V
112 -4 V
113 -5 V
113 -5 V
LT1
3114 1846 M
180 0 V
656 491 M
769 653 L
882 795 L
994 915 L
113 99 V
113 78 V
112 63 V
113 48 V
113 37 V
112 28 V
113 17 V
113 0 V
113 0 V
112 0 V
113 0 V
113 0 V
112 -2 V
113 -7 V
113 -9 V
112 -10 V
113 -9 V
113 -10 V
112 -10 V
113 -10 V
113 -10 V
LT2
3114 1746 M
180 0 V
656 491 M
769 653 L
882 795 L
994 915 L
113 99 V
113 78 V
112 63 V
113 48 V
113 37 V
112 28 V
113 21 V
113 14 V
113 10 V
112 0 V
113 0 V
113 0 V
112 0 V
113 0 V
113 0 V
112 0 V
113 0 V
113 -2 V
112 -6 V
113 -8 V
113 -9 V
stroke
grestore
end
showpage
}
\put(3054,1746){\makebox(0,0)[r]{$B^{1/4}=200$ MeVfm$^{-3}$ }}
\put(3054,1846){\makebox(0,0)[r]{$B^{1/4}=150$ MeVfm$^{-3}$ }}
\put(3054,1946){\makebox(0,0)[r]{$B^{1/4}=100$ MeVfm$^{-3}$ }}
\put(2008,21){\makebox(0,0){$n_c$ (fm$^{-3}$)}}
\put(100,1180){%
\special{ps: gsave currentpoint currentpoint translate
270 rotate neg exch neg exch translate}%
\makebox(0,0)[b]{\shortstack{$M/M_{\odot}$}}%
\special{ps: currentpoint grestore moveto}%
}
\put(3417,151){\makebox(0,0){1.8}}
\put(3041,151){\makebox(0,0){1.6}}
\put(2666,151){\makebox(0,0){1.4}}
\put(2290,151){\makebox(0,0){1.2}}
\put(1915,151){\makebox(0,0){1}}
\put(1539,151){\makebox(0,0){0.8}}
\put(1163,151){\makebox(0,0){0.6}}
\put(788,151){\makebox(0,0){0.4}}
\put(540,2109){\makebox(0,0)[r]{3}}
\put(540,1737){\makebox(0,0)[r]{2.5}}
\put(540,1366){\makebox(0,0)[r]{2}}
\put(540,994){\makebox(0,0)[r]{1.5}}
\put(540,623){\makebox(0,0)[r]{1}}
\put(540,251){\makebox(0,0)[r]{0.5}}
\end{picture}
   \caption{Total mass $M$ for various values of the bag parameter $B$ for
            the Maxwell contructed EoS. For $B^{1/4}=100$ MeVfm$^{-3}$, 
            the pure $pn$ phase ends  at $0.58$ fm$^{-3}$
            and the pure quark phase starts  
            at $0.67$ fm$^{-3}$. For $B^{1/4}=150$ MeVfm$^{-3}$, 
            the numbers are $0.92$ fm$^{-3}$
            and $1.215$ fm$^{-3}$ , while the corresponding numbers for
            $B^{1/4}=200$ MeVfm$^{-3}$ are $1.04$ and $1.57$ fm$^{-3}$. In the 
            density regions where the two phases coexist, 
            the pressure is constant, a fact reflected
            in the constant value of the neutron star mass. }
    \label{fig:sec5fig5}
\end{center}\end{figure}
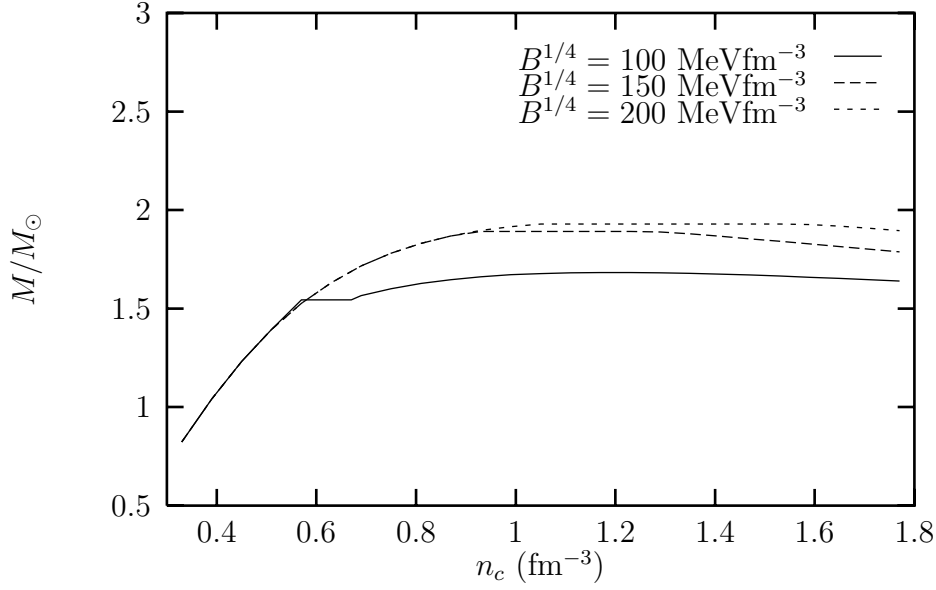
\begin{figure}\begin{center}
    % GNUPLOT: LaTeX picture with Postscript
\setlength{\unitlength}{0.1bp}
\special{!
%!PS-Adobe-2.0
%%Creator: gnuplot
%%DocumentFonts: Helvetica
%%BoundingBox: 50 50 770 554
%%Pages: (atend)
%%EndComments
/gnudict 40 dict def
gnudict begin
/Color false def
/Solid false def
/gnulinewidth 5.000 def
/vshift -33 def
/dl {10 mul} def
/hpt 31.5 def
/vpt 31.5 def
/M {moveto} bind def
/L {lineto} bind def
/R {rmoveto} bind def
/V {rlineto} bind def
/vpt2 vpt 2 mul def
/hpt2 hpt 2 mul def
/Lshow { currentpoint stroke M
  0 vshift R show } def
/Rshow { currentpoint stroke M
  dup stringwidth pop neg vshift R show } def
/Cshow { currentpoint stroke M
  dup stringwidth pop -2 div vshift R show } def
/DL { Color {setrgbcolor Solid {pop []} if 0 setdash }
 {pop pop pop Solid {pop []} if 0 setdash} ifelse } def
/BL { stroke gnulinewidth 2 mul setlinewidth } def
/AL { stroke gnulinewidth 2 div setlinewidth } def
/PL { stroke gnulinewidth setlinewidth } def
/LTb { BL [] 0 0 0 DL } def
/LTa { AL [1 dl 2 dl] 0 setdash 0 0 0 setrgbcolor } def
/LT0 { PL [] 0 1 0 DL } def
/LT1 { PL [4 dl 2 dl] 0 0 1 DL } def
/LT2 { PL [2 dl 3 dl] 1 0 0 DL } def
/LT3 { PL [1 dl 1.5 dl] 1 0 1 DL } def
/LT4 { PL [5 dl 2 dl 1 dl 2 dl] 0 1 1 DL } def
/LT5 { PL [4 dl 3 dl 1 dl 3 dl] 1 1 0 DL } def
/LT6 { PL [2 dl 2 dl 2 dl 4 dl] 0 0 0 DL } def
/LT7 { PL [2 dl 2 dl 2 dl 2 dl 2 dl 4 dl] 1 0.3 0 DL } def
/LT8 { PL [2 dl 2 dl 2 dl 2 dl 2 dl 2 dl 2 dl 4 dl] 0.5 0.5 0.5 DL } def
/P { stroke [] 0 setdash
  currentlinewidth 2 div sub M
  0 currentlinewidth V stroke } def
/D { stroke [] 0 setdash 2 copy vpt add M
  hpt neg vpt neg V hpt vpt neg V
  hpt vpt V hpt neg vpt V closepath stroke
  P } def
/A { stroke [] 0 setdash vpt sub M 0 vpt2 V
  currentpoint stroke M
  hpt neg vpt neg R hpt2 0 V stroke
  } def
/B { stroke [] 0 setdash 2 copy exch hpt sub exch vpt add M
  0 vpt2 neg V hpt2 0 V 0 vpt2 V
  hpt2 neg 0 V closepath stroke
  P } def
/C { stroke [] 0 setdash exch hpt sub exch vpt add M
  hpt2 vpt2 neg V currentpoint stroke M
  hpt2 neg 0 R hpt2 vpt2 V stroke } def
/T { stroke [] 0 setdash 2 copy vpt 1.12 mul add M
  hpt neg vpt -1.62 mul V
  hpt 2 mul 0 V
  hpt neg vpt 1.62 mul V closepath stroke
  P  } def
/S { 2 copy A C} def
end
}
\begin{picture}(3600,2160)(0,0)
\special{"
gnudict begin
gsave
50 50 translate
0.100 0.100 scale
0 setgray
/Helvetica findfont 100 scalefont setfont
newpath
-500.000000 -500.000000 translate
LTa
LTb
600 420 M
63 0 V
2754 0 R
-63 0 V
600 758 M
63 0 V
2754 0 R
-63 0 V
600 1096 M
63 0 V
2754 0 R
-63 0 V
600 1433 M
63 0 V
2754 0 R
-63 0 V
600 1771 M
63 0 V
2754 0 R
-63 0 V
600 2109 M
63 0 V
2754 0 R
-63 0 V
788 251 M
0 63 V
0 1795 R
0 -63 V
1163 251 M
0 63 V
0 1795 R
0 -63 V
1539 251 M
0 63 V
0 1795 R
0 -63 V
1915 251 M
0 63 V
0 1795 R
0 -63 V
2290 251 M
0 63 V
0 1795 R
0 -63 V
2666 251 M
0 63 V
0 1795 R
0 -63 V
3041 251 M
0 63 V
0 1795 R
0 -63 V
3417 251 M
0 63 V
0 1795 R
0 -63 V
600 251 M
2817 0 V
0 1858 V
-2817 0 V
600 251 L
LT0
3114 1946 M
180 0 V
656 1612 M
57 -165 V
56 -113 V
56 -83 V
57 -66 V
56 -54 V
56 -46 V
57 -40 V
56 -37 V
56 -10 V
57 0 V
56 0 V
56 -19 V
57 -22 V
56 -24 V
56 -23 V
57 -21 V
56 -21 V
56 -19 V
57 -19 V
56 -18 V
56 -17 V
57 -15 V
56 -16 V
57 -15 V
56 -13 V
56 -14 V
57 -12 V
56 -11 V
56 -14 V
57 -10 V
56 -10 V
56 -12 V
57 -8 V
56 -11 V
56 -10 V
57 -8 V
56 -9 V
56 -8 V
57 -8 V
56 -7 V
56 -9 V
57 -6 V
56 -9 V
56 -7 V
57 -6 V
56 -7 V
56 -7 V
57 -7 V
56 -5 V
LT1
3114 1846 M
180 0 V
656 1612 M
57 -165 V
56 -113 V
56 -83 V
57 -66 V
56 -54 V
56 -46 V
57 -40 V
56 -37 V
56 -31 V
57 -32 V
56 -28 V
56 -28 V
57 -25 V
56 -24 V
56 -21 V
57 -22 V
56 -22 V
56 -19 V
57 -20 V
56 -10 V
56 0 V
57 0 V
56 0 V
57 0 V
56 0 V
56 0 V
57 0 V
56 0 V
56 0 V
57 0 V
56 -2 V
56 -5 V
57 -5 V
56 -7 V
56 -7 V
57 -6 V
56 -7 V
56 -7 V
57 -7 V
56 -5 V
56 -7 V
57 -6 V
56 -7 V
56 -7 V
57 -5 V
56 -7 V
56 -5 V
57 -6 V
56 -7 V
LT2
3114 1746 M
180 0 V
656 1612 M
57 -165 V
56 -113 V
56 -83 V
57 -66 V
56 -54 V
56 -46 V
57 -40 V
56 -37 V
56 -32 V
57 -31 V
56 -28 V
56 -28 V
57 -25 V
56 -24 V
56 -23 V
57 -20 V
56 -22 V
56 -21 V
57 -18 V
56 -19 V
56 -18 V
57 -16 V
56 -16 V
57 -14 V
56 0 V
56 0 V
57 0 V
56 0 V
56 0 V
57 0 V
56 0 V
56 0 V
57 0 V
56 0 V
56 0 V
57 0 V
56 0 V
56 0 V
57 0 V
56 0 V
56 0 V
57 -3 V
56 -4 V
56 -5 V
57 -3 V
56 -5 V
56 -5 V
57 -5 V
56 -4 V
LT3
3114 1646 M
180 0 V
656 1612 M
57 -165 V
56 -113 V
56 -83 V
57 -66 V
56 -54 V
56 -46 V
57 -40 V
56 -37 V
56 -32 V
57 -31 V
56 -28 V
56 -28 V
57 -25 V
56 -24 V
56 -23 V
57 -20 V
56 -22 V
56 -21 V
57 -18 V
56 -19 V
56 -18 V
57 -16 V
56 -16 V
57 -17 V
56 -16 V
56 -15 V
57 -13 V
56 -14 V
56 -15 V
57 -12 V
56 -13 V
56 -12 V
57 -12 V
56 -12 V
56 -12 V
57 -10 V
56 -10 V
56 -10 V
57 -10 V
56 -10 V
56 -10 V
57 -9 V
56 -8 V
56 -10 V
57 -8 V
56 -8 V
56 -8 V
57 -8 V
56 -8 V
stroke
grestore
end
showpage
}
\put(3054,1646){\makebox(0,0)[r]{$\delta=0.2$}}
\put(3054,1746){\makebox(0,0)[r]{$B^{1/4}=200$ MeVfm$^{-3}$ }}
\put(3054,1846){\makebox(0,0)[r]{$B^{1/4}=150$ MeVfm$^{-3}$}}
\put(3054,1946){\makebox(0,0)[r]{$B^{1/4}=100$ MeVfm$^{-3}$}}
\put(2008,21){\makebox(0,0){$n_c$ (fm$^{-3}$)}}
\put(100,1180){%
\special{ps: gsave currentpoint currentpoint translate
270 rotate neg exch neg exch translate}%
\makebox(0,0)[b]{\shortstack{$R$ km}}%
\special{ps: currentpoint grestore moveto}%
}
\put(3417,151){\makebox(0,0){1.8}}
\put(3041,151){\makebox(0,0){1.6}}
\put(2666,151){\makebox(0,0){1.4}}
\put(2290,151){\makebox(0,0){1.2}}
\put(1915,151){\makebox(0,0){1}}
\put(1539,151){\makebox(0,0){0.8}}
\put(1163,151){\makebox(0,0){0.6}}
\put(788,151){\makebox(0,0){0.4}}
\put(540,2109){\makebox(0,0)[r]{20}}
\put(540,1771){\makebox(0,0)[r]{18}}
\put(540,1433){\makebox(0,0)[r]{16}}
\put(540,1096){\makebox(0,0)[r]{14}}
\put(540,758){\makebox(0,0)[r]{12}}
\put(540,420){\makebox(0,0)[r]{10}}
\end{picture}
    \caption{Radius as function of central density $n_c$
            for various values of the bag parameter $B$ for
            the Maxwell constructed  
            EoS. For comparison we include also the results
            from the $pn$-matter EoS for $\beta$-stable with $\delta=0.2$.
            The region where $R$ is constant reflects the density region
            where the pressure is constant in the Maxwell construction.
            See also the figure caption to the previous figure.}
    \label{fig:sec5fig6}
\end{center}\end{figure}
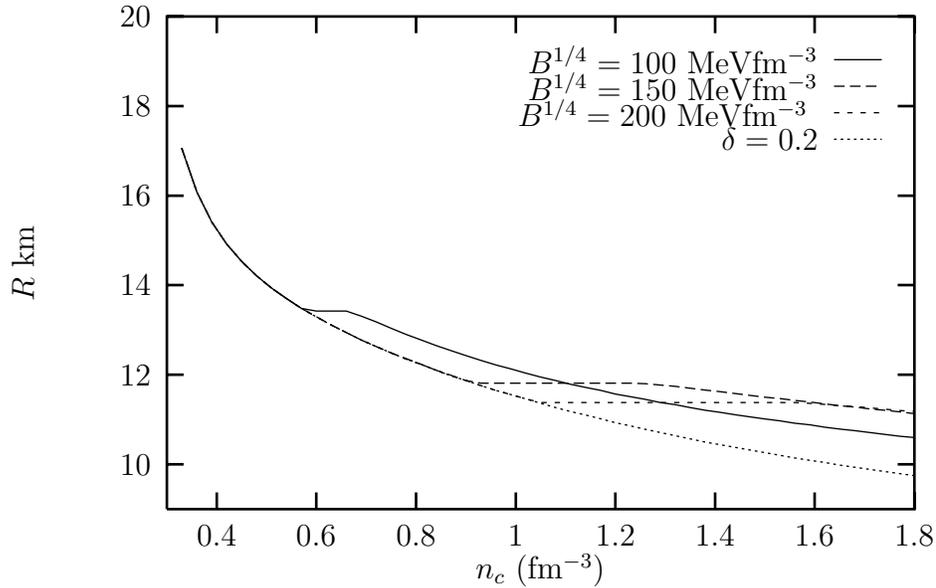
\begin{figure}\begin{center}
   % GNUPLOT: LaTeX picture with Postscript
\setlength{\unitlength}{0.1bp}
\special{!
%!PS-Adobe-2.0
%%Creator: gnuplot
%%DocumentFonts: Helvetica
%%BoundingBox: 50 50 770 554
%%Pages: (atend)
%%EndComments
/gnudict 40 dict def
gnudict begin
/Color false def
/Solid false def
/gnulinewidth 5.000 def
/vshift -33 def
/dl {10 mul} def
/hpt 31.5 def
/vpt 31.5 def
/M {moveto} bind def
/L {lineto} bind def
/R {rmoveto} bind def
/V {rlineto} bind def
/vpt2 vpt 2 mul def
/hpt2 hpt 2 mul def
/Lshow { currentpoint stroke M
  0 vshift R show } def
/Rshow { currentpoint stroke M
  dup stringwidth pop neg vshift R show } def
/Cshow { currentpoint stroke M
  dup stringwidth pop -2 div vshift R show } def
/DL { Color {setrgbcolor Solid {pop []} if 0 setdash }
 {pop pop pop Solid {pop []} if 0 setdash} ifelse } def
/BL { stroke gnulinewidth 2 mul setlinewidth } def
/AL { stroke gnulinewidth 2 div setlinewidth } def
/PL { stroke gnulinewidth setlinewidth } def
/LTb { BL [] 0 0 0 DL } def
/LTa { AL [1 dl 2 dl] 0 setdash 0 0 0 setrgbcolor } def
/LT0 { PL [] 0 1 0 DL } def
/LT1 { PL [4 dl 2 dl] 0 0 1 DL } def
/LT2 { PL [2 dl 3 dl] 1 0 0 DL } def
/LT3 { PL [1 dl 1.5 dl] 1 0 1 DL } def
/LT4 { PL [5 dl 2 dl 1 dl 2 dl] 0 1 1 DL } def
/LT5 { PL [4 dl 3 dl 1 dl 3 dl] 1 1 0 DL } def
/LT6 { PL [2 dl 2 dl 2 dl 4 dl] 0 0 0 DL } def
/LT7 { PL [2 dl 2 dl 2 dl 2 dl 2 dl 4 dl] 1 0.3 0 DL } def
/LT8 { PL [2 dl 2 dl 2 dl 2 dl 2 dl 2 dl 2 dl 4 dl] 0.5 0.5 0.5 DL } def
/P { stroke [] 0 setdash
  currentlinewidth 2 div sub M
  0 currentlinewidth V stroke } def
/D { stroke [] 0 setdash 2 copy vpt add M
  hpt neg vpt neg V hpt vpt neg V
  hpt vpt V hpt neg vpt V closepath stroke
  P } def
/A { stroke [] 0 setdash vpt sub M 0 vpt2 V
  currentpoint stroke M
  hpt neg vpt neg R hpt2 0 V stroke
  } def
/B { stroke [] 0 setdash 2 copy exch hpt sub exch vpt add M
  0 vpt2 neg V hpt2 0 V 0 vpt2 V
  hpt2 neg 0 V closepath stroke
  P } def
/C { stroke [] 0 setdash exch hpt sub exch vpt add M
  hpt2 vpt2 neg V currentpoint stroke M
  hpt2 neg 0 R hpt2 vpt2 V stroke } def
/T { stroke [] 0 setdash 2 copy vpt 1.12 mul add M
  hpt neg vpt -1.62 mul V
  hpt 2 mul 0 V
  hpt neg vpt 1.62 mul V closepath stroke
  P  } def
/S { 2 copy A C} def
end
}
\begin{picture}(3600,2160)(0,0)
\special{"
gnudict begin
gsave
50 50 translate
0.100 0.100 scale
0 setgray
/Helvetica findfont 100 scalefont setfont
newpath
-500.000000 -500.000000 translate
LTa
LTb
600 251 M
63 0 V
2754 0 R
-63 0 V
600 623 M
63 0 V
2754 0 R
-63 0 V
600 994 M
63 0 V
2754 0 R
-63 0 V
600 1366 M
63 0 V
2754 0 R
-63 0 V
600 1737 M
63 0 V
2754 0 R
-63 0 V
600 2109 M
63 0 V
2754 0 R
-63 0 V
600 251 M
0 63 V
0 1795 R
0 -63 V
1002 251 M
0 63 V
0 1795 R
0 -63 V
1405 251 M
0 63 V
0 1795 R
0 -63 V
1807 251 M
0 63 V
0 1795 R
0 -63 V
2210 251 M
0 63 V
0 1795 R
0 -63 V
2612 251 M
0 63 V
0 1795 R
0 -63 V
3015 251 M
0 63 V
0 1795 R
0 -63 V
3417 251 M
0 63 V
0 1795 R
0 -63 V
600 251 M
2817 0 V
0 1858 V
-2817 0 V
600 251 L
LT0
3114 1946 M
180 0 V
600 835 M
40 45 V
61 70 V
60 65 V
60 60 V
61 54 V
60 47 V
60 41 V
61 36 V
60 30 V
61 26 V
60 23 V
60 19 V
61 16 V
60 13 V
60 11 V
61 10 V
60 7 V
60 6 V
61 5 V
60 4 V
61 3 V
60 2 V
60 2 V
61 0 V
60 1 V
60 0 V
61 0 V
60 -1 V
60 0 V
61 -2 V
60 -1 V
61 -2 V
60 -1 V
60 -2 V
61 -2 V
60 -1 V
60 -3 V
61 -1 V
60 -2 V
60 -2 V
61 -1 V
60 -2 V
61 -1 V
60 -1 V
60 -1 V
61 -1 V
60 -1 V
LT1
3114 1846 M
180 0 V
600 676 M
40 54 V
61 75 V
60 66 V
60 59 V
61 50 V
60 45 V
60 38 V
61 32 V
60 29 V
61 23 V
60 20 V
60 17 V
61 14 V
60 12 V
60 9 V
61 8 V
60 6 V
60 5 V
61 4 V
60 3 V
61 2 V
60 1 V
60 1 V
61 1 V
60 0 V
60 -1 V
61 -1 V
60 -1 V
60 -2 V
61 -1 V
60 -2 V
61 -2 V
60 -2 V
60 -2 V
61 -2 V
60 -3 V
60 -2 V
61 -2 V
60 -2 V
60 -2 V
61 -2 V
60 -2 V
61 -1 V
60 -2 V
60 -1 V
61 -2 V
60 -1 V
LT2
3114 1746 M
180 0 V
600 678 M
40 49 V
61 68 V
60 63 V
60 57 V
61 52 V
60 47 V
60 41 V
61 37 V
60 34 V
61 29 V
60 26 V
60 22 V
61 20 V
60 17 V
60 16 V
61 12 V
60 12 V
60 9 V
61 8 V
60 6 V
61 6 V
60 4 V
60 4 V
61 2 V
60 2 V
60 1 V
61 1 V
60 0 V
60 -1 V
61 -2 V
60 0 V
61 -3 V
60 -2 V
60 -2 V
61 -2 V
60 -3 V
60 -3 V
61 -4 V
60 -3 V
60 -4 V
61 -3 V
60 -4 V
61 -4 V
60 -3 V
60 -5 V
61 -4 V
60 -4 V
LT3
3114 1646 M
180 0 V
600 847 M
40 38 V
61 61 V
60 60 V
60 57 V
61 53 V
60 49 V
60 45 V
61 40 V
60 36 V
61 32 V
60 28 V
60 26 V
61 23 V
60 20 V
60 17 V
61 16 V
60 13 V
60 12 V
61 9 V
60 9 V
61 8 V
60 6 V
60 6 V
61 4 V
60 4 V
60 3 V
61 2 V
60 2 V
60 2 V
61 1 V
60 1 V
61 0 V
60 1 V
60 -1 V
61 0 V
60 0 V
60 -1 V
61 0 V
60 -1 V
60 0 V
61 -2 V
60 0 V
61 -1 V
60 -1 V
60 -1 V
61 0 V
60 -1 V
stroke
grestore
end
showpage
}
\put(3054,1646){\makebox(0,0)[r]{Rotational mass for $\delta=0.2$}}
\put(3054,1746){\makebox(0,0)[r]{$\delta=0.2$}}
\put(3054,1846){\makebox(0,0)[r]{$B^{1/4}=200$ MeVfm$^{-3}$ }}
\put(3054,1946){\makebox(0,0)[r]{Rotational mass for $B^{1/4}=200$ MeVfm$^{-3}$ }}
\put(2008,21){\makebox(0,0){$n_c$ (fm$^{-3}$)}}
\put(100,1180){%
\special{ps: gsave currentpoint currentpoint translate
270 rotate neg exch neg exch translate}%
\makebox(0,0)[b]{\shortstack{$M/M_{\odot}$}}%
\special{ps: currentpoint grestore moveto}%
}
\put(3417,151){\makebox(0,0){1.8}}
\put(3015,151){\makebox(0,0){1.6}}
\put(2612,151){\makebox(0,0){1.4}}
\put(2210,151){\makebox(0,0){1.2}}
\put(1807,151){\makebox(0,0){1}}
\put(1405,151){\makebox(0,0){0.8}}
\put(1002,151){\makebox(0,0){0.6}}
\put(600,151){\makebox(0,0){0.4}}
\put(540,2109){\makebox(0,0)[r]{3}}
\put(540,1737){\makebox(0,0)[r]{2.5}}
\put(540,1366){\makebox(0,0)[r]{2}}
\put(540,994){\makebox(0,0)[r]{1.5}}
\put(540,623){\makebox(0,0)[r]{1}}
\put(540,251){\makebox(0,0)[r]{0.5}}
\end{picture}
   \caption{Rotational mass $M$ and gravitational mass for the pure $pn$ EoS
            with $\delta=0.2$ and equations of state based on the mixed
            phase construction for different values of $B$.}
    \label{fig:sec5fig7}
\end{center}\end{figure}
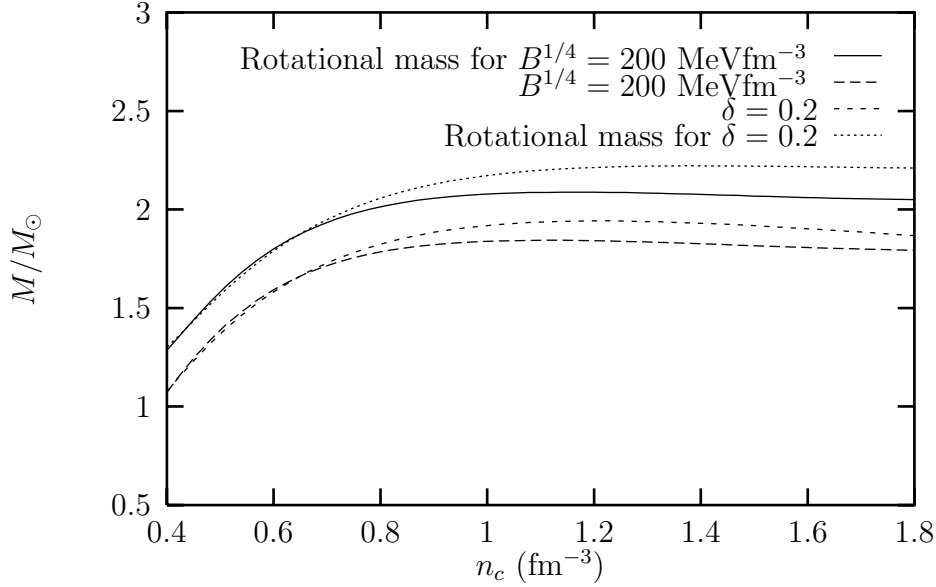
\begin{figure}\begin{center}
   % GNUPLOT: LaTeX picture with Postscript
\setlength{\unitlength}{0.1bp}
\special{!
%!PS-Adobe-2.0
%%Creator: gnuplot
%%DocumentFonts: Helvetica
%%BoundingBox: 50 50 770 554
%%Pages: (atend)
%%EndComments
/gnudict 40 dict def
gnudict begin
/Color false def
/Solid false def
/gnulinewidth 5.000 def
/vshift -33 def
/dl {10 mul} def
/hpt 31.5 def
/vpt 31.5 def
/M {moveto} bind def
/L {lineto} bind def
/R {rmoveto} bind def
/V {rlineto} bind def
/vpt2 vpt 2 mul def
/hpt2 hpt 2 mul def
/Lshow { currentpoint stroke M
  0 vshift R show } def
/Rshow { currentpoint stroke M
  dup stringwidth pop neg vshift R show } def
/Cshow { currentpoint stroke M
  dup stringwidth pop -2 div vshift R show } def
/DL { Color {setrgbcolor Solid {pop []} if 0 setdash }
 {pop pop pop Solid {pop []} if 0 setdash} ifelse } def
/BL { stroke gnulinewidth 2 mul setlinewidth } def
/AL { stroke gnulinewidth 2 div setlinewidth } def
/PL { stroke gnulinewidth setlinewidth } def
/LTb { BL [] 0 0 0 DL } def
/LTa { AL [1 dl 2 dl] 0 setdash 0 0 0 setrgbcolor } def
/LT0 { PL [] 0 1 0 DL } def
/LT1 { PL [4 dl 2 dl] 0 0 1 DL } def
/LT2 { PL [2 dl 3 dl] 1 0 0 DL } def
/LT3 { PL [1 dl 1.5 dl] 1 0 1 DL } def
/LT4 { PL [5 dl 2 dl 1 dl 2 dl] 0 1 1 DL } def
/LT5 { PL [4 dl 3 dl 1 dl 3 dl] 1 1 0 DL } def
/LT6 { PL [2 dl 2 dl 2 dl 4 dl] 0 0 0 DL } def
/LT7 { PL [2 dl 2 dl 2 dl 2 dl 2 dl 4 dl] 1 0.3 0 DL } def
/LT8 { PL [2 dl 2 dl 2 dl 2 dl 2 dl 2 dl 2 dl 4 dl] 0.5 0.5 0.5 DL } def
/P { stroke [] 0 setdash
  currentlinewidth 2 div sub M
  0 currentlinewidth V stroke } def
/D { stroke [] 0 setdash 2 copy vpt add M
  hpt neg vpt neg V hpt vpt neg V
  hpt vpt V hpt neg vpt V closepath stroke
  P } def
/A { stroke [] 0 setdash vpt sub M 0 vpt2 V
  currentpoint stroke M
  hpt neg vpt neg R hpt2 0 V stroke
  } def
/B { stroke [] 0 setdash 2 copy exch hpt sub exch vpt add M
  0 vpt2 neg V hpt2 0 V 0 vpt2 V
  hpt2 neg 0 V closepath stroke
  P } def
/C { stroke [] 0 setdash exch hpt sub exch vpt add M
  hpt2 vpt2 neg V currentpoint stroke M
  hpt2 neg 0 R hpt2 vpt2 V stroke } def
/T { stroke [] 0 setdash 2 copy vpt 1.12 mul add M
  hpt neg vpt -1.62 mul V
  hpt 2 mul 0 V
  hpt neg vpt 1.62 mul V closepath stroke
  P  } def
/S { 2 copy A C} def
end
}
\begin{picture}(3600,2160)(0,0)
\special{"
gnudict begin
gsave
50 50 translate
0.100 0.100 scale
0 setgray
/Helvetica findfont 100 scalefont setfont
newpath
-500.000000 -500.000000 translate
LTa
LTb
600 251 M
63 0 V
2754 0 R
-63 0 V
600 551 M
63 0 V
2754 0 R
-63 0 V
600 850 M
63 0 V
2754 0 R
-63 0 V
600 1150 M
63 0 V
2754 0 R
-63 0 V
600 1450 M
63 0 V
2754 0 R
-63 0 V
600 1749 M
63 0 V
2754 0 R
-63 0 V
600 2049 M
63 0 V
2754 0 R
-63 0 V
600 251 M
0 63 V
0 1795 R
0 -63 V
952 251 M
0 63 V
0 1795 R
0 -63 V
1304 251 M
0 63 V
0 1795 R
0 -63 V
1656 251 M
0 63 V
0 1795 R
0 -63 V
2009 251 M
0 63 V
0 1795 R
0 -63 V
2361 251 M
0 63 V
0 1795 R
0 -63 V
2713 251 M
0 63 V
0 1795 R
0 -63 V
3065 251 M
0 63 V
0 1795 R
0 -63 V
3417 251 M
0 63 V
0 1795 R
0 -63 V
600 251 M
2817 0 V
0 1858 V
-2817 0 V
600 251 L
LT0
2114 1946 M
180 0 V
927 251 M
85 71 V
271 224 V
243 200 V
222 179 V
197 159 V
176 142 V
159 124 V
137 111 V
123 96 V
106 83 V
95 72 V
81 60 V
74 53 V
60 42 V
53 36 V
45 28 V
36 23 V
31 15 V
28 12 V
18 8 V
18 3 V
10 0 V
11 -4 V
3 -6 V
4 -8 V
0 -11 V
-4 -12 V
-7 -14 V
-3 -15 V
-11 -16 V
-10 -18 V
-11 -18 V
-11 -19 V
-14 -20 V
-14 -20 V
-17 -20 V
-14 -21 V
-18 -21 V
-18 -21 V
-17 -21 V
-18 -22 V
-17 -22 V
-22 -21 V
-17 -22 V
-21 -22 V
LT1
2114 1846 M
180 0 V
856 251 M
216 172 V
278 217 V
239 190 V
212 164 V
179 143 V
155 123 V
134 102 V
113 89 V
95 73 V
77 61 V
67 51 V
57 40 V
42 33 V
39 27 V
28 20 V
24 15 V
18 10 V
14 8 V
11 3 V
7 2 V
3 -2 V
4 -2 V
0 -5 V
-4 -7 V
-3 -7 V
-7 -8 V
-7 -10 V
-7 -10 V
-8 -11 V
-10 -11 V
-11 -11 V
-10 -12 V
-11 -11 V
-10 -11 V
-11 -11 V
-10 -11 V
-11 -10 V
-7 -9 V
-11 -9 V
-7 -8 V
-7 -7 V
-7 -7 V
-7 -7 V
-7 -6 V
-7 -6 V
LT2
2114 1746 M
180 0 V
977 251 M
222 194 V
200 171 V
176 151 V
155 131 V
134 115 V
120 98 V
102 85 V
88 75 V
77 62 V
67 53 V
57 45 V
49 38 V
39 31 V
35 26 V
28 19 V
25 16 V
21 12 V
14 8 V
11 6 V
10 2 V
4 2 V
3 -2 V
4 -3 V
-4 -4 V
0 -6 V
-3 -6 V
-4 -7 V
-7 -8 V
-7 -8 V
-3 -7 V
-8 -8 V
-7 -9 V
-7 -7 V
-7 -8 V
-7 -7 V
-7 -7 V
-7 -6 V
-7 -6 V
-3 -6 V
-7 -5 V
-4 -5 V
-7 -6 V
-7 -5 V
-7 -7 V
LT3
2114 1646 M
180 0 V
600 269 M
783 404 L
994 554 L
191 134 V
169 118 V
151 105 V
137 92 V
120 80 V
106 69 V
91 59 V
78 49 V
70 42 V
56 35 V
50 29 V
42 22 V
35 19 V
28 14 V
25 10 V
18 7 V
14 5 V
14 2 V
7 1 V
7 -1 V
3 -2 V
4 -3 V
0 -4 V
0 -4 V
-4 -6 V
0 -6 V
-7 -6 V
-3 -7 V
-4 -8 V
-7 -8 V
-7 -10 V
-10 -10 V
-7 -11 V
-11 -12 V
-14 -12 V
-11 -13 V
-14 -14 V
-14 -14 V
-14 -14 V
-14 -14 V
stroke
grestore
end
showpage
}
\put(2054,1646){\makebox(0,0)[r]{$B^{1/4}=100$ MeVfm$^{-3}$ }}
\put(2054,1746){\makebox(0,0)[r]{$B^{1/4}=150$ MeVfm$^{-3}$ }}
\put(2054,1846){\makebox(0,0)[r]{$B^{1/4}=200$ MeVfm$^{-3}$ }}
\put(2054,1946){\makebox(0,0)[r]{$\delta=0.2$}}
\put(2008,21){\makebox(0,0){$M/M_{\odot}$}}
\put(100,1180){%
\special{ps: gsave currentpoint currentpoint translate
270 rotate neg exch neg exch translate}%
\makebox(0,0)[b]{\shortstack{$I$}}%
\special{ps: currentpoint grestore moveto}%
}
\put(3417,151){\makebox(0,0){2}}
\put(3065,151){\makebox(0,0){1.9}}
\put(2713,151){\makebox(0,0){1.8}}
\put(2361,151){\makebox(0,0){1.7}}
\put(2009,151){\makebox(0,0){1.6}}
\put(1656,151){\makebox(0,0){1.5}}
\put(1304,151){\makebox(0,0){1.4}}
\put(952,151){\makebox(0,0){1.3}}
\put(600,151){\makebox(0,0){1.2}}
\put(540,2049){\makebox(0,0)[r]{100}}
\put(540,1749){\makebox(0,0)[r]{90}}
\put(540,1450){\makebox(0,0)[r]{80}}
\put(540,1150){\makebox(0,0)[r]{70}}
\put(540,850){\makebox(0,0)[r]{60}}
\put(540,551){\makebox(0,0)[r]{50}}
\put(540,251){\makebox(0,0)[r]{40}}
\end{picture}
   \caption{Moment of Inertia $I$ in units of $M_{\odot}$km$^2$ as function
            of $M_{\odot}$ for the pure $pn$ EoS with $\delta=0.2$ and
            for the mixed phase construction with $B^{1/4}=200$ MeVfm$^{-3}$.}
    \label{fig:sec5fig8}
\end{center}\end{figure}
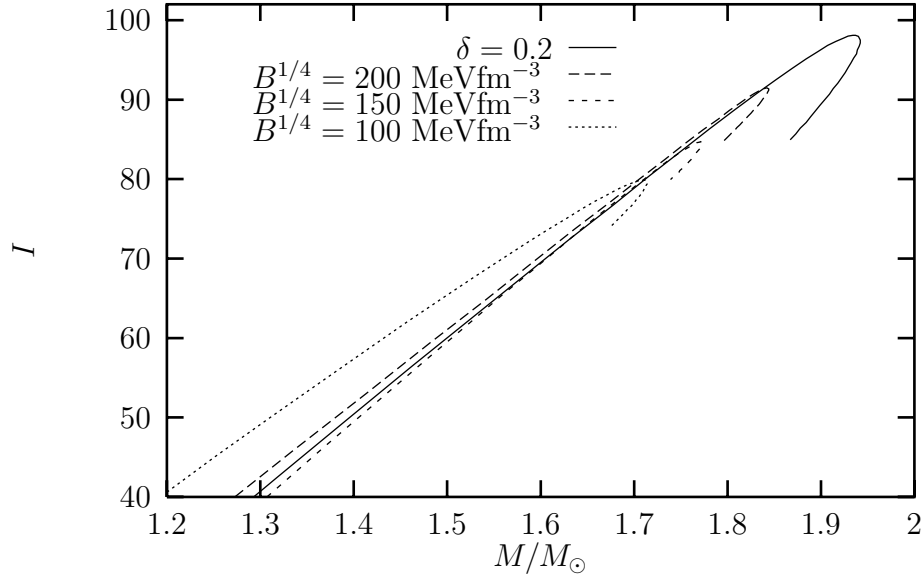
\begin{figure}\begin{center}
   % GNUPLOT: LaTeX picture with Postscript
\setlength{\unitlength}{0.1bp}
\special{!
%!PS-Adobe-2.0
%%Creator: gnuplot
%%DocumentFonts: Helvetica
%%BoundingBox: 50 50 770 554
%%Pages: (atend)
%%EndComments
/gnudict 40 dict def
gnudict begin
/Color false def
/Solid false def
/gnulinewidth 5.000 def
/vshift -33 def
/dl {10 mul} def
/hpt 31.5 def
/vpt 31.5 def
/M {moveto} bind def
/L {lineto} bind def
/R {rmoveto} bind def
/V {rlineto} bind def
/vpt2 vpt 2 mul def
/hpt2 hpt 2 mul def
/Lshow { currentpoint stroke M
  0 vshift R show } def
/Rshow { currentpoint stroke M
  dup stringwidth pop neg vshift R show } def
/Cshow { currentpoint stroke M
  dup stringwidth pop -2 div vshift R show } def
/DL { Color {setrgbcolor Solid {pop []} if 0 setdash }
 {pop pop pop Solid {pop []} if 0 setdash} ifelse } def
/BL { stroke gnulinewidth 2 mul setlinewidth } def
/AL { stroke gnulinewidth 2 div setlinewidth } def
/PL { stroke gnulinewidth setlinewidth } def
/LTb { BL [] 0 0 0 DL } def
/LTa { AL [1 dl 2 dl] 0 setdash 0 0 0 setrgbcolor } def
/LT0 { PL [] 0 1 0 DL } def
/LT1 { PL [4 dl 2 dl] 0 0 1 DL } def
/LT2 { PL [2 dl 3 dl] 1 0 0 DL } def
/LT3 { PL [1 dl 1.5 dl] 1 0 1 DL } def
/LT4 { PL [5 dl 2 dl 1 dl 2 dl] 0 1 1 DL } def
/LT5 { PL [4 dl 3 dl 1 dl 3 dl] 1 1 0 DL } def
/LT6 { PL [2 dl 2 dl 2 dl 4 dl] 0 0 0 DL } def
/LT7 { PL [2 dl 2 dl 2 dl 2 dl 2 dl 4 dl] 1 0.3 0 DL } def
/LT8 { PL [2 dl 2 dl 2 dl 2 dl 2 dl 2 dl 2 dl 4 dl] 0.5 0.5 0.5 DL } def
/P { stroke [] 0 setdash
  currentlinewidth 2 div sub M
  0 currentlinewidth V stroke } def
/D { stroke [] 0 setdash 2 copy vpt add M
  hpt neg vpt neg V hpt vpt neg V
  hpt vpt V hpt neg vpt V closepath stroke
  P } def
/A { stroke [] 0 setdash vpt sub M 0 vpt2 V
  currentpoint stroke M
  hpt neg vpt neg R hpt2 0 V stroke
  } def
/B { stroke [] 0 setdash 2 copy exch hpt sub exch vpt add M
  0 vpt2 neg V hpt2 0 V 0 vpt2 V
  hpt2 neg 0 V closepath stroke
  P } def
/C { stroke [] 0 setdash exch hpt sub exch vpt add M
  hpt2 vpt2 neg V currentpoint stroke M
  hpt2 neg 0 R hpt2 vpt2 V stroke } def
/T { stroke [] 0 setdash 2 copy vpt 1.12 mul add M
  hpt neg vpt -1.62 mul V
  hpt 2 mul 0 V
  hpt neg vpt 1.62 mul V closepath stroke
  P  } def
/S { 2 copy A C} def
end
}
\begin{picture}(3600,2160)(0,0)
\special{"
gnudict begin
gsave
50 50 translate
0.100 0.100 scale
0 setgray
/Helvetica findfont 100 scalefont setfont
newpath
-500.000000 -500.000000 translate
LTa
600 251 M
2817 0 V
LTb
600 251 M
63 0 V
2754 0 R
-63 0 V
600 499 M
63 0 V
2754 0 R
-63 0 V
600 746 M
63 0 V
2754 0 R
-63 0 V
600 994 M
63 0 V
2754 0 R
-63 0 V
600 1242 M
63 0 V
2754 0 R
-63 0 V
600 1490 M
63 0 V
2754 0 R
-63 0 V
600 1737 M
63 0 V
2754 0 R
-63 0 V
600 1985 M
63 0 V
2754 0 R
-63 0 V
788 251 M
0 63 V
0 1795 R
0 -63 V
1163 251 M
0 63 V
0 1795 R
0 -63 V
1539 251 M
0 63 V
0 1795 R
0 -63 V
1915 251 M
0 63 V
0 1795 R
0 -63 V
2290 251 M
0 63 V
0 1795 R
0 -63 V
2666 251 M
0 63 V
0 1795 R
0 -63 V
3041 251 M
0 63 V
0 1795 R
0 -63 V
3417 251 M
0 63 V
0 1795 R
0 -63 V
600 251 M
2817 0 V
0 1858 V
-2817 0 V
600 251 L
LT0
3114 1946 M
180 0 V
714 251 M
55 165 V
56 138 V
57 118 V
56 104 V
56 92 V
57 83 V
56 74 V
56 66 V
57 59 V
56 51 V
56 46 V
57 39 V
56 34 V
56 30 V
57 25 V
56 22 V
56 18 V
57 14 V
56 12 V
56 9 V
57 7 V
56 5 V
57 3 V
56 1 V
56 0 V
57 -1 V
56 -3 V
56 -3 V
57 -5 V
56 -5 V
56 -6 V
57 -6 V
56 -7 V
56 -7 V
57 -7 V
56 -8 V
56 -8 V
57 -9 V
56 -8 V
56 -9 V
57 -8 V
56 -9 V
56 -9 V
57 -9 V
56 -9 V
56 -9 V
57 -9 V
56 -9 V
LT1
3114 1846 M
180 0 V
706 251 M
7 29 V
56 175 V
56 140 V
57 119 V
56 104 V
56 89 V
57 79 V
56 68 V
56 59 V
57 50 V
56 43 V
56 36 V
57 31 V
56 25 V
56 21 V
57 17 V
56 13 V
56 11 V
57 8 V
56 7 V
56 4 V
57 3 V
56 2 V
57 0 V
56 0 V
56 -1 V
57 -3 V
56 -2 V
56 -3 V
57 -4 V
56 -4 V
56 -4 V
57 -4 V
56 -5 V
56 -5 V
57 -4 V
56 -5 V
56 -5 V
57 -4 V
56 -4 V
56 -5 V
57 -4 V
56 -3 V
56 -4 V
57 -2 V
56 -3 V
56 -3 V
57 -2 V
56 -3 V
LT2
3114 1746 M
180 0 V
722 251 M
47 165 V
882 672 L
994 868 L
113 157 V
113 125 V
112 97 V
113 73 V
113 55 V
112 40 V
113 26 V
113 16 V
113 7 V
112 0 V
113 0 V
113 0 V
112 0 V
113 0 V
113 0 V
112 0 V
113 0 V
113 -3 V
112 -12 V
113 -14 V
113 -16 V
stroke
grestore
end
showpage
}
\put(3054,1746){\makebox(0,0)[r]{Maxwell $B^{1/4}=200$ MeVfm$^{-3}$ }}
\put(3054,1846){\makebox(0,0)[r]{Mixed phase $B^{1/4}=200$ MeVfm$^{-3}$ }}
\put(3054,1946){\makebox(0,0)[r]{$\delta=0.2$}}
\put(2008,21){\makebox(0,0){$n_c$ fm$^{-3}$}}
\put(100,1180){%
\special{ps: gsave currentpoint currentpoint translate
270 rotate neg exch neg exch translate}%
\makebox(0,0)[b]{\shortstack{$I$}}%
\special{ps: currentpoint grestore moveto}%
}
\put(3417,151){\makebox(0,0){1.8}}
\put(3041,151){\makebox(0,0){1.6}}
\put(2666,151){\makebox(0,0){1.4}}
\put(2290,151){\makebox(0,0){1.2}}
\put(1915,151){\makebox(0,0){1}}
\put(1539,151){\makebox(0,0){0.8}}
\put(1163,151){\makebox(0,0){0.6}}
\put(788,151){\makebox(0,0){0.4}}
\put(540,1985){\makebox(0,0)[r]{140}}
\put(540,1737){\makebox(0,0)[r]{120}}
\put(540,1490){\makebox(0,0)[r]{100}}
\put(540,1242){\makebox(0,0)[r]{80}}
\put(540,994){\makebox(0,0)[r]{60}}
\put(540,746){\makebox(0,0)[r]{40}}
\put(540,499){\makebox(0,0)[r]{20}}
\put(540,251){\makebox(0,0)[r]{0}}
\end{picture}
   \caption{Moment of Inertia $I$ in units of $M_{\odot}$km$^2$ as function
            of central density $n_c$  for the $pn$ EoS with $\delta=0.2$
            and with quark degrees of freedom with $B^{1/4}=200$ MeVfm$^{-3}$ 
            for the mixed phase and Maxwell constructions. Note well that
            the Maxwell construction yields a constant $I$ since the pressure
            is constant in this case in the density region from 
            $1.04$ to $1.57$ fm$^{-3}$.}
    \label{fig:sec5fig9}
\end{center}\end{figure}
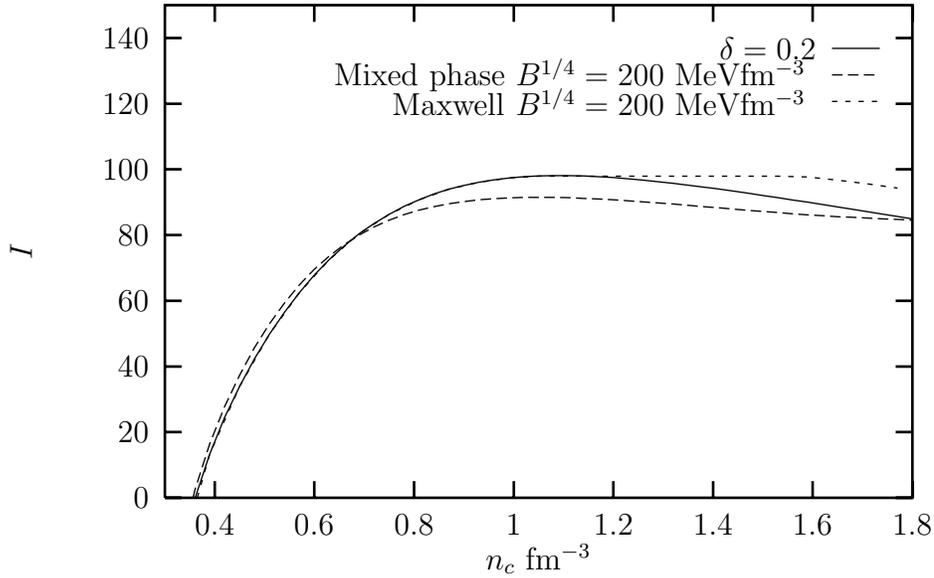

Finally, we end this subsection by listing in Table 
\ref{tab:maxstarproperties} the maximum values for masses, radii and
moments of inertia for several of the equations of state discussed in 
Figs.\ \ref{fig:sec5fig1}-\ref{fig:sec5fig9}.  
\begin{table}
\begin{center}
\caption{Maximum gravitational mass in $M_{\odot}$ and the 
         corresponding radius $R$, in units of km, 
         for the given central density
         $n_c$, in units of fm$^{-3}$, for various equations of state. 
         The maximum moment of Inertia $I$, in units of $M_{\odot}$km$^2$, 
         is also listed. Note that this occurs for another central  
         density than listed below. 
         All results are for $\beta$-stable matter and rotational
         corrections have not been included in the total mass.}         
\begin{tabular}{ccccc}
\hline\noalign{\smallskip}
 EoS & Max mass & Max $I$& $R$ & $n_c$\\
\noalign{\smallskip}\hline\noalign{\smallskip}
 $pn$ $\delta=0.13$    &2.07          & 110.1       & 11.0    &1.1    \\
 $pn$ $\delta=0.2$    &1.94          & 98.0       &10.8     &1.2  \\
 $pn$ $\delta=0.3$    &1.72          & 78.7       &10.3     & 1.4   \\
 $pn$ $\delta=0.4$    &1.58          & 66.9       &10.0     &1.6    \\
\noalign{\smallskip}\hline
 Mixed phase with $\delta=0.2$    &          &        &     &    \\
\noalign{\smallskip}\hline
 $B^{1/4}=200$ MeVfm$^{-3}$    &1.84          & 91.5       & 11.6    & 1.1   \\
 $B^{1/4}=150$ MeVfm$^{-3}$    &1.77          & 84.7       &  11.5   & 1.2   \\
 $B^{1/4}=100$ MeVfm$^{-3}$    &1.71          & 79.9       & 10.9    &1.3    \\
\noalign{\smallskip}\hline
\label{tab:maxstarproperties}
\end{tabular}
\end{center}
\end{table}
 From this table we see that the $pn$ EoS with the lowest value of
$\delta$ gives also the stiffest EoS, and thereby largest mass and smallest
central density. Similarly, the largest value for the bag constant
results also in the stiffest EoS.  
In connection with the discussion of QPO's, it is worth pointing 
out that in Kerr space the relation between the
Keplerian orbital frequency $\nu_K$ and the mass of the star is
$2.198M_{\odot}\left(\nu_K\mathrm{kHz}\right)^{-1}\left(1-0.748j\right)^{-1}$
with $j=I\omega/M^2$ a dimensionless measure of
the angular momentum of the star\footnote{Recall that in all
equations $G=c=\hbar=1$.}. Following Ref.\ \cite{kfc97} and
inserting the $1171$ Hz QPO from 4U 1636-536, a rotational frequency
$\omega/2\pi=272$ Hz and an assumed moment of inertia of $\sim
100M_{\odot}$km$^2$ results in a mass of $2.02M_{\odot}$ and a radius
of $9.6\pm 0.6$ km. From the above table, we see that
these results are fairly close to those which we
get for the pure $pn$ EoS with $\delta=0.2$, i.e.\  for the 
$\delta$ value which gave the best fit 
to the EoS of Akmal et al.\ \cite{apr98}. 
Thus, 
if QPO's occur near the innermost
stable orbits, then neutron star masses are $M\simeq 2.2M_\odot$.
This constrains the nuclear EoS including causality in a smooth way
to only the stiffest ones - specifically $\delta\la 0.2$ in the EoS
of Eq. (\ref{eq:EA}). Phase transitions in cores of neutron
stars softens the EoS and strong transitions can therefore be ruled out
except at very high densities $n\ga 5n_0$.
On the other hand, if it turns out that the QPO are not from the innermost
stable orbits and that even accreting neutron stars have small masses,
say like the binary pulsars $M\la 1.5M_\odot$, this indicates that
heavier neutron stars are not stable. Therefore the EoS must be
soft at high densities, i.e.\  $\delta\ga 0.4$ or that a phase transition
occurs at a few times nuclear matter densities.

The role of phase transitions and its possible link with observation 
will be discussed in the following subsections.

\subsection{Maximum masses} \label{Maxmass}

\begin{figure}
       \begin{center}
       \setlength{\unitlength}{1mm}
       \begin{picture}(140,100)
       \put(25,5){\epsfxsize=12cm \epsfbox{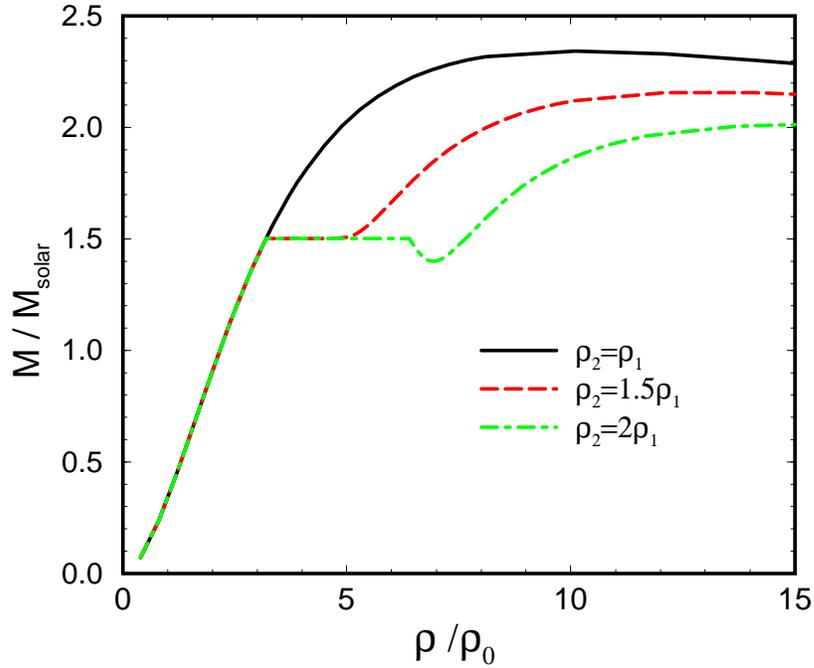}}
       \end{picture}
        \caption{Mass vs. central density for the Bethe-Johnson
polytrope ($\Gamma=2.54$) \protect\cite{BJ} with first order phase transitions
at $\varepsilon_1=3.2\varepsilon_0$ to $\varepsilon_2$. A region of
instability occurs when Eq. (\protect\ref{ecrit}) is fulfilled.}
       \end{center}
        \label{fig:doublepeak}
\end{figure}

A stellar object of mass $\sim 1.4M_\odot$ can either be an ordinary
star of type F, a white dwarf, a neutron star, a black hole, or
possibly a quark star, see Fig. \ref{fig:doublepeak}.
As shown by Fechner and
Joss \cite{qmref} a second branch of quark stars
are possible for certain equation of states (EoS) - specifically for
some parameter values of the Bag constant. The stars considered were
hybrid stars consisting of a quark matter core with a mantle
of nuclear matter around. A double maximum mass in the mass-density
plot for neutron and mixed phase quark stars have also been found
in \cite{Kettner} (see also \cite{BaymMM}).

The occurrence of a second maximum is a curious phenomenon that occurs
under specific conditions that can
be quantified. For that purpose we first consider a
simple model that can be solved analytically, namely an 
EoS consisting of two incompressible fluids with a first order 
phase transition between energy density
$\varepsilon_1$ and $\varepsilon_2$ ($\varepsilon_1<\varepsilon_2$) 
coexisting at a pressure $P_0$. We shall also first ignore effects of general
relativity, i.e.\  take the Newtonian limit.

The mass function, $M(r)=4\pi\int_0^r \varepsilon(r')r'^2dr'$, is very simple 
in the Newtonian limit and the boundary condition $M(R)=M$
relates the star mass $M$ and 
radius $R$ to the radius of the dense core, $R_0$, as
\begin{equation}
  M(R)=\frac{4\pi}{3}  \left(\varepsilon_1R^3+
  (\varepsilon_2-\varepsilon_1)R_0^3\right) \,,\label{MR}
\end{equation}
where $R_0$ is the radius of the dense core.
 From Newton's equation for hydrostatic equilibrium
\begin{eqnarray}
  \frac{dP}{dr} = -\varepsilon \frac{M(r)}{r^2} \,, \label{Pr}
\end{eqnarray}
the pressure is easily obtained. From the boundary condition $P(R)=0$ we
obtain
\begin{eqnarray}
  P_0= \frac{4\pi}{3}  \varepsilon_1^2
  \left( R^2+(2\frac{\varepsilon_2}{\varepsilon_1} -3)R_0^2 - 
 (\frac{\varepsilon_2}{\varepsilon_1}-1) \frac{R_0^3}{R} \right) \,.\label{P0}
\end{eqnarray}

When the dense core is small, $R_0\ll R$, the $R_0^3$ terms in Eqs.\
(\ref{MR},) and (\ref{P0}) can be ignored. From Eq.\ (\ref{P0}) we therefore
observe that when $2\varepsilon_2> 3\varepsilon_1$, the radius $R$ of the star
{\it decreases} with increasing size of the dense core $R_0$.
Correspondingly, its mass $M(R)$ of Eq. (\ref{MR}) decreases.
In other words, as the average density of the star increases, its
mass decreases and a stability analysis reveals that the star is
unstable. It will contract until $R_0$ is comparable to $R$ such that
the $R_0^3$ terms in Eqs.\ (\ref{MR}) and (\ref{P0}) stabilizes the star and
its mass again increases with increasing size.
We can therefore conclude that when
\begin{eqnarray}
  \varepsilon_2 \ge \frac{3}{2} \varepsilon_1 \quad \Rightarrow \quad
   {\rm instability\,\,  region} , \label{ecrit}
\end{eqnarray}
a second maximum mass appears.

Another case, that can be solved analytically, is the $\Gamma=2$
polytropic EoS, $P\propto \varepsilon^2$. In this case the Newtonian
version of the TOV equation is equivalent to the Schr\"odinger
equation in a square well (or the Klein-Gordon equation) with solution
$\varepsilon(r)\propto sin(\pi r/R)/r$. Including a first order phase
transition leads to a phase shift in the corresponding sine solution
for the outer mantle at
densities $\varepsilon\le\varepsilon_1$.  Curiously, one finds exactly the
same instability criteria as Eq. (\ref{ecrit}). However, varying the
polytropic index around $\Gamma=2$ does change this condition slightly.
Also, including general relativity affects the instability condition
of  Eq. (\ref{ecrit}) when $\varepsilon_2$ is close to the maximum central
density where the star becomes unstable with respect to a collapse to a
black hole, see Fig.\ \ref{fig:doublepeak}.
A second order phase transition as, e.g.\ the mixed nuclear and quark
matter phase, can also lead to instabilities and second maximum
masses when the EoS is sufficiently softened \cite{Kettner}.

\subsection{Phase transitions in rotating neutron stars} \label{subsec:PTR}

During the last years, and as discussed in the preceeding sections,
interesting phase transitions in nuclear matter 
to quark matter, 
mixed phases of quark and nuclear matter \cite{glendenning92,hps93}, kaon
\cite{kn87} or pion condensates \cite{apr98}, 
neutron and proton superfluidity \cite{eeho96},
hyperonic matter \cite{schulze97,sm96,kpe96}
crystalline nuclear matter, magnetized matter, etc., have
been considered.  Recently, Glendenning et al.\ \cite{gpw97} have
considered rapidly rotating neutron stars
and what happens as they slow down when the decreasing centrifugal
force leads to increasing core pressures.  They find that a drastic
softening of the equation of state, e.g.\  by a phase transition to
quark matter, can lead to a sudden contraction of the neutron star at
a critical angular velocity and shows up in a backbending moment of
inertia as function of frequency.  Here we consider another
interesting phenomenon namely how the star and in particular its moment
of inertia behaves near the critical angular velocity where the core
pressure just exceeds that needed to make a phase transition. We
calculate the moment of inertia, angular velocities, braking
index, etc.\ near the critical angular velocity and discuss
observational consequences for first and second order phase
transitions.

Here we will make
the standard approximation of slowly
rotating stars, i.e.\  the rotational angular velocity is
$\Omega^2 \ll M/R^3$. 
For neutron stars with mass $M=1.4M_\odot$ and
radius $R\sim 10$ km their period should thus be larger than a few
milliseconds, a fact which applies to all measured pulsars insofar.
The general relativistic equations for slowly rotating stars were 
presented by Hartle \cite{hartle67}
and reviewed in subsection \ref{subsec:generalprop}.
Hartle's equations are quite elaborate to solve
as they consist of six coupled differential equations as compared to
the single Tolman-Oppenheimer-Volkoff 
equation in Eq.\ (\ref{eq:tov}) in the non-rotating case.
In order to be able to analytically
extract the qualitative behavior near the critical angular velocity
$\Omega_0$, where a
phase transition occurs in the center, we will first solve the
Newtonian equations for a simple equation of state.
This will allow us to make general predictions on properties
of rotating neutrons stars when phase transitions occur in the interior
of a star.
The corrections from general relativity are typically of order
$M/R\simeq$10-20\% for neutron stars of mass $M\simeq 1.4M_\odot$.
The extracted analytical properties of a rotating star 
are then  checked below by actually solving 
Hartle's equations numerically for a realistic equation of state.
 
The simple Newtonian equation of
motion expresses the balance between the pressure gradient and 
the gravitational and centrifugal forces
\begin{equation}
   \nabla P = -\varepsilon\left( \nabla V +
        {\bf \Omega}\times{\bf \Omega}\times{\bf r} \right),
   \label{NOV}
\end{equation}
Here, $V({\bf r})$ is the gravitational potential for the deformed
star and $\varepsilon$ the energy ($\sim$mass) density.  We assume that friction 
in the 
(nonsuperfluid) matter insures
that the star is uniformly rotating.  Since cold neutron stars are
barotropes, i.e.\  the pressure is a function of density,
the pressure, density and effective gravitational potential, 
$\Phi=V-\frac{1}{2}({\bf \Omega}\times{\bf r})^2$, are all
constants on the {\it same} isobaric surfaces
for a uniformly rotating star \cite{hartle67}.
We denote these surfaces by the effective radius, $a$, and
for slowly rotating stars it is related to the distance $r$ from the
center and the polar angle $\theta$ from the rotation axis along
${\bf\Omega}$ by \cite{hartle67}
\begin{equation}
   r(a,\theta) = a\left[ 1-\varepsilon(a) P_2(\cos\theta) \right],
   \label{ra}
\end{equation}
where $P_2(\cos\theta)$ is the 2nd Legendre polynomial and
$\varepsilon(a)$ is the deformation of the star from spherical symmetry.

Inserting Eq.\ (\ref{ra}) in Eq.\ (\ref{NOV}) 
one obtains for small deformations 
\cite{hartle67} the $l=0$ Newtonian hydrostatic equation
\begin{equation}
  \frac{1}{\varepsilon}\frac{dP}{da} = -\frac{M(a)}{a^2}+\frac{2}{3}\Omega^2 a,
  \label{Pa} 
\end{equation}
where $M(a)=4\pi\int^a_0\varepsilon(a')a'^2da'$ is the mass
contained inside the mean radius $a$.
The factor 2/3 in the centrifugal force arises because it only acts in
two of the three directions. The equation $(l=2)$ for the deformation
$\epsilon(a)$ is \cite{Tassoul}.
\begin{eqnarray}
  \left(  \frac{1}{a^2}\frac{d\epsilon}{da}\right) M(a) &-&
   4\pi\int^R_a\rho(a')\frac{d\epsilon(a')}{da'}da' 
   = \frac{5}{3}\Omega^2 
\,.\label{ea}
\end{eqnarray}
The deformation generally increases
with decreasing density, i.e.\  the star is more deformed in its outer layers.

In order to discuss the qualitative behavior near critical angular
velocities we first consider a simple EoS with phase transitions for which
Eq.\ (\ref{Pa}) can be solved analytically namely that of
two incompressible fluids with a first order phase transition
between energy density
$\varepsilon_1$ and $\varepsilon_2$ ($\varepsilon_1<\varepsilon_2$) 
coexisting at a pressure $P_0$.
The mass function $M(a)$ is very simple 
in the Newtonian limit and the boundary condition $M(R)=M$
relates the star radius $R$ to the radius of the dense core, $R_0$, as
\begin{equation}
    R = 
\left(\bar{R}^3-(\frac{\varepsilon_2}{\varepsilon_1}-1)R_0^3\right)^{1/3}, 
\end{equation}
where $\bar{R}=(3M/4\pi\varepsilon_1)^{1/3}$ is the star radius
in the absence of a dense core.
Solving Eq.\ (\ref{Pa}) gives the pressure
\begin{equation}
   P(a) =P_0+\frac{1}{2}(R_0^2-a^2)\varepsilon_2(\frac{4\pi}{3}\varepsilon_2
        -\frac{2}{3}\Omega^2),
  \label{Pai1}
\end{equation}
for $0\le a\le R_0 $ and
\begin{eqnarray}
    P(a)&=&P_0+\frac{1}{2}(R_0^2-a^2)\varepsilon_1(\frac{4\pi}{3}\varepsilon_1
     -\frac{2}{3}\Omega^2)\nonumber \\ 
    && 
-\frac{4\pi}{3}R_0^2(\varepsilon_2-\varepsilon_1)\varepsilon_1(1-\frac{R_0}{a}),
     \label{eq:Pai2}
\end{eqnarray}
for $R_0\le a\le R$ \cite{rot}.
The boundary condition at the surface $P(R)=0$ in
Eq.\  (\ref{eq:Pai2}) gives 
\begin{eqnarray}
  \omega^2 & \equiv &\frac{\Omega^2}{2\pi \varepsilon_1}
  = 1-2\left(\frac{3}{4\pi}\frac{P_0}{\varepsilon_1^2R^2} \right. \nonumber \\
    & & \left. 
+(\frac{\varepsilon_2}{\varepsilon_1}-1)\frac{R_0^2}{R^2}(1-R_0/R) \right)
               (1-R_0^2/R^2)^{-1}. \label{om}
\end{eqnarray}
The phase transition occurs right at the center when $R_0=0$ corresponding to
the 
{\it critical angular velocity} $\Omega_0=\omega_0\sqrt{2\pi \varepsilon_1}$
where \cite{rot}
\begin{equation}
   \omega_0^2= 1-2\frac{P_0\bar{R}}{M} 
%= 1-(\frac{6}{\pi})^{1/3} \frac{P_0}{M^{2/3}\varepsilon_1^{1/3}}
   \, .\label{o0}
\end{equation}
Generally, for any EoS the critical angular velocity depends on $P_0$, $M$, 
and $\varepsilon_1$ but not on $\varepsilon_2$.

For angular velocities just below $\omega_0$ very little of 
the high density phase exists and $R_0\ll R$. Expanding (\ref{om}) we obtain
\begin{equation}
  \frac{R_0}{\bar{R}} \simeq \sqrt{\frac{\omega_0^2-\omega^2}
             {3-2\varepsilon_2/\varepsilon_1-\omega_0^2}}. 
  \label{R0}
\end{equation}
For $\omega\ge\omega_0$ the dense phase disappears and $R_0=0$.
Generaly, one can interprete $R_0$ as an order parameter 
in analogy to, e.g.\  magnetization, the BCS gap, or the Higgs
field in the standard model, however, as function of
angular velocity instead of temperature.
Note, that for large density differences, 
$\varepsilon_2/\varepsilon_1\ge (3-\omega_0^2)/2$, Eq. (\ref{R0})
is not valid. This is related to an instability (see Eq. (\ref{ecrit})
and will be discussed in the following subsection.

The corresponding moment of inertia is for $R_0\ll R$  
\begin{eqnarray}
   I &=& \frac{4\pi}{5} \left( 
\varepsilon_2R_0^5+\varepsilon_1(R^5-R_0^5)\right)
     (1+\frac{2}{5}\epsilon) \nonumber \\
     &\simeq& \frac{2}{5}M\bar{R}^2 
    \left( 1-\frac{5}{3}(\frac{\varepsilon_2}{\varepsilon_1}-1)
 \frac{R_0^3}{\bar{R}^3}\right)  (1+\frac{1}{2}\omega^2),
    \label{I}
\end{eqnarray}
where we used that the deformation from Eq.
(\ref{ea}) is $\epsilon=(5/4)\omega^2$ in the low
density phase \cite{Tassoul}. 
However, for the qualitative behavior near $\Omega_0$ only the contraction
of the star radius $R$ with the appearance of the dense core $R_0$
is important whereas the deformations can be ignored.
The contraction is responsible for the term in
the moment of inertia and is proportional to
$R_0^3\propto (\omega_0^2-\omega)^{3/2}$ near the critical angular
velocity. Consequently,  the derivative $dI/d\omega^2$ displays the same
non-analytic square root dependence as $R_0$ (see Eq.\ (\ref{R0})). 

Latent heat is generated in the phase transition can be ignored
because of rapid neutrino cooling which will be even
faster than in supernova explosions. Thus temperatures will
drop below $\sim 1$MeV in seconds. Such temperatures are negligible
compared to typical Fermi energies of nucleons or quarks and the
timescales are also much smaller than $t_0$.

Let us subsequently 
consider a more realistic EoS for dense nuclear matter at high
densities such as the Bethe-Johnson EoS \cite{BJ}. At
high densities it can be approximated by a polytropic relation between
the pressure and energy density:
$P=K_1\varepsilon^{2.54}$, where 
$K_1=0.021\varepsilon_0^{-1.54}$ and $\varepsilon_0=m_n 0.15$fm$^{-3}$
is normal nuclear matter mass density. 
As we are only interested in the dense core we will for simplicity employ 
this Bethe-Johnson polytrope (BJP) EoS.
The central density of a non-rotating
1.4$M_\odot$
mass neutron star with the BJP EoS is $\sim 3.4\varepsilon_0$.
Furthermore, we assume that a first order phase
transition occurs at density $\varepsilon_1=3.2\varepsilon_0$ to a high density 
phase
of density $\varepsilon_2=4\varepsilon_0$ with a similar polytropic 
EoS $P=K_2\varepsilon^{2.54}$. From the Maxwell
construction the pressure is the same at the interface, $P_0$,
which determines $K_2=K_1(\varepsilon_1/\varepsilon_2)^{2.54}$. 
We now generalize Eq.\   (\ref{Pa}) by including 
effects of general relativity. From Einstein's field equations for the metric
we obtain from the $l=0$ part 
\begin{equation}
  \frac{1}{\varepsilon+P}\frac{dP}{da} = - 
  \frac{M(a)+4\pi a^3P}{a^2(1-2M(a)/a)} + \frac{2}{3}\Omega^2 a , 
  \label{PaGR}
\end{equation}
where $m(a)=4\pi\int_0^a\varepsilon(a')a'^2da'$.
In the centrifugal force term
we have ignored frame dragging and other corrections of order
$\Omega^2M/R\sim 0.1\Omega^2$ for simplicity and 
since they have only minor effects in our case.
By expanding the pressure, mass function and gravitational potential
in the difference between the rotating and non-rotating case, 
Eq.\  (\ref{PaGR}) reduces to the $l=0$ part of
Hartle's equations (cf. Eq.\  (100) in \cite{hartle67}.)
Note also that Hartle's full 
equations cannot be used in our case because the first order
phase transition causes discontinuities in densities so that changes
are not small locally. This shows up, for example, in the divergent
thermodynamic derivate $d\varepsilon/dP$.

\begin{figure}
\begin{center}
{\centering
\mbox{\psfig{figure=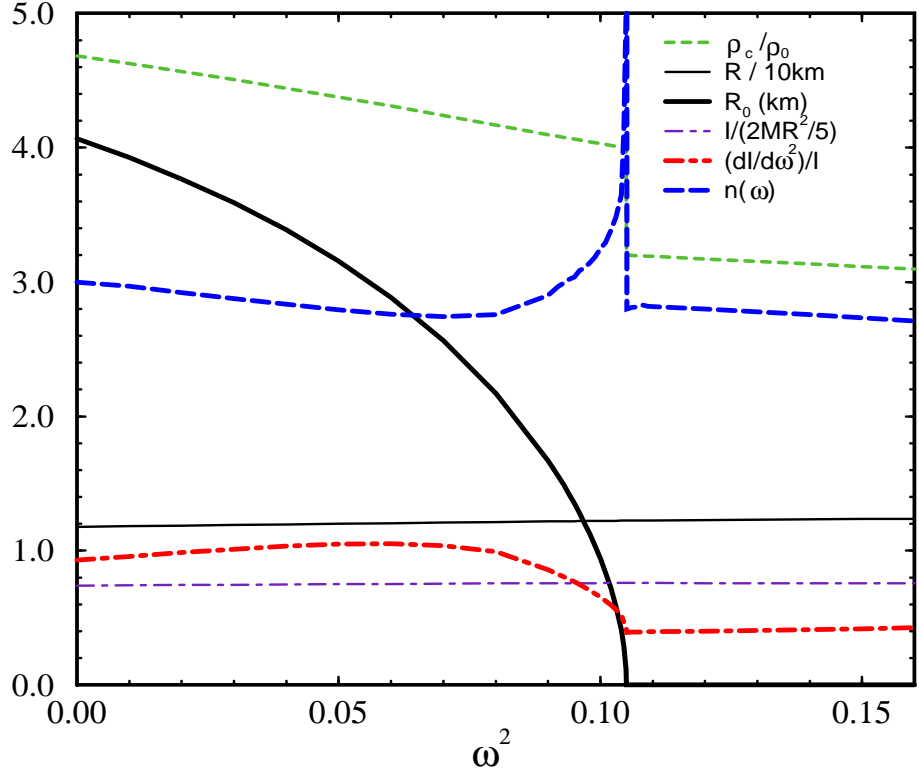,height=12cm,width=15cm,angle=-90}} }
\caption{Central density (in units of $\varepsilon_0$), radii of the neutron
star $R$ and its dense core $R_0$, moment of inertia, its derivative
$I'/I=dI/d\omega^2/I$ and the braking index are shown as function of
the scaled angular velocity $\omega^2=\Omega^2/(2\pi \varepsilon_1)$.
The rotating neutron star has mass $1.4M_\odot$ and a Bethe-Johnson
like polytropic equation of state with a first order phase transition
taking place at density $\varepsilon_1=3.2\varepsilon_0$ to
$\varepsilon_2=4\varepsilon_0$. }       \label{rot}
\end{center}
\end{figure}

The rotating version of the Tolman-Oppenheimer-Volkoff
equation (\ref{PaGR}) is now solved for a rotating neutron star of mass   
$M=1.4M_\odot$ with the BJP EoS including a first order phase transition. 
In Fig.\ \ref{rot} we show 
the central density, moment of inertia, braking index, star radius and
radius of the interface ($R_0$) as function of the scaled angular
velocity. It is important to note that $R_0\propto
\sqrt{\Omega_0^2-\Omega^2}$ for angular velocities just below the
critical value $\Omega_0$. The qualitative behavior of the
neutron star with the BJP EoS and a first order phase
transition is the same as for our simple analytic example of two
incompressible fluids examined above.  Generally, it is the finite
density difference between the phases that is important and leads to a
term in the moment of inertia proportional to
$(\Omega_0^2-\Omega^2)^{3/2}$ as in Eq.\  (\ref{I}).

The moment of inertia increases with angular velocity. Generally, for a
first order phase transition we find for 
$\Omega \raisebox{-.5ex}{$\stackrel{<}{\sim}$}\Omega_0$
(see also Eq.\  (\ref{I}) and Fig. (\ref{rot}))
\begin{equation}
  I = I_0\left( 1+\frac{1}{2}c_1\frac{\Omega^2}{\Omega_0^2} -\frac{2}{3}c_2 
                (1-\frac{\Omega^2}{\Omega_0^2})^{3/2} + ...
      \right) . 
  \label{Igen}
\end{equation}
For the two incompressible fluids with momentum of inertia given by
Eq.\  (\ref{I}), the small expansion
parameters are $c_1=\omega_0^2$ and 
$c_2=(5/2)\omega_0^3(\varepsilon_2/\varepsilon_1-1)/(3-2\varepsilon_2/
\varepsilon_1-\omega_0^2)^{3/2}$; for $\Omega>\Omega_0$ the $c_2$ term is 
absent. 
For the BJP we find from Fig.\ (1) that
$c_2\simeq 0.07\simeq 2.2\omega_0^3$. Generally, we find that the
coefficient $c_2$ is proportional to the density difference between the 
two coexisting phases and to the critical angular velocity to the third power,
$c_2\sim (\varepsilon_2/\varepsilon_1-1)\omega_0^3$. The scaled critical angular 
velocity
$\omega_0$ can at most reach unity for submillisecond pulsars.

To make contact with observation we consider the temporal behavior
of angular velocities of pulsars. The pulsars slow down at a rate
given by the loss of rotational energy which we shall assume is
proportional to the rotational angular velocity to some power
(for dipole radiation $n=3$)
\begin{equation}
  \frac{d}{dt} \left(\frac{1}{2}I\Omega^2\right) = 
                  -C \Omega^{n+1}. 
   \label{dE}
\end{equation}
With the moment of inertia given by Eq. (\ref{Igen})
the angular velocity will then decrease with time as
\begin{eqnarray}
  \frac{\dot{\Omega}}{\Omega} &=& -\frac{C\Omega^{n-1}}{I_0}
  \left( 1-c_1\frac{\Omega^2}{\Omega_0^2}
          -c_2\sqrt{1-\frac{\Omega^2}{\Omega_0^2}} \right) \nonumber\\
  &\simeq& -\frac{1}{(n-1)t}
  \left( 1-c_2 \sqrt{1-(\frac{t_0}{t})^{2/(n-1)}} +....\right), 
  \label{dOmega}
\end{eqnarray}
for $t\ge t_0$. Here,
the time after formation of the pulsar is, using Eq.\  (\ref{dE}),
related to the angular velocity as
$t\simeq t_0(\Omega_0/\Omega)^{n-1}$ and 
$t_0=I_0/((n-1)C\Omega_0^{n-1})$ for $n>1$, is the critical time where
a phase transition occurs in the center. For earlier times $t\le t_0$ there
is no dense core and Eq. (\ref{dOmega}) applies when setting $c_2=0$
The critical angular velocity is $\Omega_0=\omega_0\sqrt{2\pi 
\varepsilon_1}\simeq
6 kHz$ for the BJP EoS, i.e.\  comparable to a millisecond binary
pulsar. Applying these numbers to, for example, the Crab pulsar we find that
it would have been spinning with critical angular velocity approximately 
a decade after the Crab supernova explosion,
i.e.\  $t_0\sim 10$ years for the Crab. Generally,
$t_0\propto\Omega_0^{1-n}$ and the timescale for the transients in
$\dot{\Omega}$ as given by Eq. (13) may be months or centuries. In any
case it would not require continuous monitoring which would help
a dedicated observational program.

The braking index depends on the second derivative $I''=dI/d^2\Omega$
of the moment of inertia and thus diverges (see Fig. (1))
as $\Omega$ approaches $\Omega_0$ from below
\begin{eqnarray}
     n(\Omega) &\equiv& \frac{\ddot{\Omega}\Omega}{\dot{\Omega}^2} 
    \simeq n - 2c_1\frac{\Omega^2}{\Omega_0^2}
    +c_2\frac{\Omega^4/\Omega_0^4}{\sqrt{1-\Omega^2/\Omega_0^2}} \,.\label{n}
\end{eqnarray}
For $\Omega\ge\Omega_0$ the term with $c_2$ is absent.
The  {\it observational} braking index $n(\Omega)$ should be distinguished
from the {\it theoretical} exponent $n$ appearing in Eq. (12).
Although the results in Eqs. (13) and (14) were derived 
for the pulsar slow down assumed in Eq. (12) both
$\dot{\Omega}$ and $n(\Omega)$ will generally 
display the $\sqrt{t-t_0}$ behavior for $t\ge  t_0$ as long as the rotational
energy loss is a smooth function of $\Omega$.
The singular behavior will, however, be smeared on the pulsar
glitch ``healing'' time which in the case of the Crab pulsar is 
of order weeks only.

We now discuss possible phase transitions in interiors of neutron
stars.  The quark and nuclear matter mixed phase described in
\cite{glendenning92} has continuous pressures and densities. There are
no first order phase transitions but at most two second order phase
transitions. Namely, at a lower density, where quark matter first
appears in nuclear matter, and at a very high density (if gravitationally
stable), where all nucleons are finally dissolved into quark matter.
In second-order
phase transitions the pressure is a continuous function of density
and we find a continuous braking
index.  This mixed phase does, however, not include local surface and
Coulomb energies of the quark and nuclear matter structures. As shown
in \cite{hps93,HH} there can be an appreciable surface and Coulomb
energy associated with forming these structures and if the interface
tension between quark and nuclear matter is too large, the mixed phase
is not favored energetically. The neutron star will then have a core
of pure quark matter with a mantle of nuclear matter surrounding it and
the two phases are coexisting by a first order phase transition.  For
a small or moderate interface tension the quarks are confined in
droplet, rod- and plate-like structures \cite{hps93,HH} as found in the
inner crust of neutron stars \cite{lrp93}. 
Due to the finite Coulomb and
surface energies associated with forming these structures, the
transitions change from second order to first order at each
topological change in structure. 
If a Kaon condensate
appears it may also have such structures \cite{Schaffner}.
Pion condensates \cite{pion}, crystalline nuclear matter \cite{ap97,apr98},
hyperonic or magnetized matter, etc. may provide
other first order phase transitions.

If a neutron star cools
continuously, the temperature will decrease with time
and the phase transition boundary will move inwards.
The two phases could, e.g.\  be quark-gluon/nuclear matter
or a melted/solid phase. In the latter case the size of the hot
(melted) matter in the core is slowly reduced as the temperature drops
freezing the fluid into the solid mantle. 
Melting temperatures have been estimated in
\cite{lrp93,melt} for the crust and in \cite{hps93} for the quark matter
mixed phase.  
When the very core freezes we have a
similar situation as when the star slows down to the critical angular
velocity, i.e.\  a first order phase transition occurs right at the
center. Consequently, a similar behavior of moment of inertia, angular
velocities, braking index may occur as in
Eqs.\ (\ref{Igen}), (\ref{dOmega}) and (\ref{n}) replacing $\Omega(t)$ with
$T(t)$.

Thus, if a first order phase transitions is present at central
densities of neutron stars, it will show up in moments of inertia and
consequently also in angular velocities in a characteristic way.  For
example, the slow down of the angular velocity has a characteristic
behavior $\dot{\Omega}\sim c_2\sqrt{1-t/t_0}$ and the braking index
diverges as $n(\Omega)\sim c_2/\sqrt{1-\Omega^2/\Omega_0^2}$ (see
Eqs. (\ref{dOmega}) and (\ref{n})).  The magnitude of the signal
generally depends on the density difference between the two phases and
the critical angular velocity $\omega_0=\Omega_0/\sqrt{2\pi
\varepsilon_1}$ such that
$c_2\sim(\varepsilon_2/\varepsilon_1-1)\omega_0^3$.  The observational
consequences depend very much on the critical angular velocity
$\Omega_0$, which depends on the equation of state employed, at which
density the phase transition occurs and the mass of the neutron star.

We encourage a dedicated search for the characteristic
transients discussed above. As the pulsar slows down over a 
million years, its central densities spans a wide range of order
$1 n_0$ (see Fig. \ref{rot}). As we are interested in time scales
of years, we must instead study all $\sim 1000$ pulsars available.
By studying the corresponding 
range of angular velocities for the sample of different
star masses, the chance for encountering a critical angular velocity
increases.  Eventually, one may be able to cover the
full range of central densities and find all first order phase
transitions up to a certain size determined by the experimental
resolution.  Since the size of the signal scales with $\Omega_0^3$ the
transition may be best observed in rapidly rotating pulsars such as
binary pulsars or pulsars recently formed in supernova explosion and
which are rapidly slowing down. Carefully monitoring such pulsars may
reveal the characteristic behavior of the angular velocity or braking
index as described above which is a signal of a first order phase
transition in dense matter.

\subsection{Core quakes and glitches} \label{subsec:glitches}

The glitches observed in the Crab, Vela, and a few other pulsars are
probably due to quakes occurring in solid structures such as the
crust, superfluid vortices or possibly the quark matter lattice in the
core \cite{HH}.  As the rotating neutron star gradually slows down and
becomes less deformed, the rigid component is strained and eventually
cracks/quakes and changes its structure towards being more spherical.

The moment of inertia of the rigid component, $I_c$, decreases abruptly
and its rotation and pulsar frequency increases due to angular
momentum conservation resulting in a glitch. 
 The observed glitches are very small $\Delta\Omega/\Omega\sim
10^{-8}$.
The two components slowly
relaxate to a common rotational frequency on a timescale of days (healing time)
due to superfluidity of the other component (the neutron liquid).  
The
{\it healing parameter} $Q=I_c/I_{tot}$ measured in glitches reveals
that for the Vela and Crab pulsar about $\sim$3\% and $\sim$96\% of
the moment of inertia is in the rigid component respectively.

If the crust were the only rigid component the Vela neutron star
should be almost all crust.  This would require that the Vela is a
very light neutron star - much smaller than the observed ones which
all are compatible with $\sim 1.4M_\odot$.  If we by the lattice
component include not only the solid crust but also the protons in 
nuclear matter (NM)
(which is locked to the crust due to magnetic fields), superfluid
vortices pinned to the crust \cite{Pines} and the solid QM mixed phase
\begin{equation}
   I_c = I_{crust}+I_p+I_{sv}+I_{QM} \, ,
\end{equation}
we can better explain the large $I_c$ for the Crab.
The moment of inertia of the mixed phase is sensitive to the EoS's used.
For example, for a quadratic NM EoS \cite{hps93} 
decreasing the Bag constant from 110 to 95 MeVfm$^{-3}$ increases
$I_c/I_{total}$ from $\sim20\%$ to $\sim70\%$ 
for a 1.4$M_\odot$ neutron star - not including possible vortex pinning.
The structures in the mixed phase would exhibit anisotropic elastic properties,
being rigid to some shear strains but not others in much the same way as liquid
crystals. Therefore the whole mixed phase might not be rigid.

The energy released in glitches every few years
are too large to be stored in the crust only. 
The recurrence time for large quakes, $t_c$, is inversely
proportional to the strain energy \cite{Pines}, 
which again is proportional to the lattice density and the Coulomb energy
\begin{equation}
   t_c^{-1} \propto  \frac{1}{a^3} \frac{Z^2e^2}{a} \, .
\end{equation}
Since the lattice distance $a$ is smaller for the quark matter droplets 
and their charge larger 
than for atoms in the crust, the recurrence
time is shorter in better agreement with measurements of large glitches.

Detecting core and crust quakes
separately or other signs of three components in glitches, indicating
the existence of a crust, superfluid neutrons and a solid core, would
support the idea of the mixed quark and nuclear matter mixed
phase. However, magnetic field attenuation is expected to be small in
neutron stars and therefore magnetic fields penetrate through the
core. Thus the crust and core lattices as well as the proton liquid
should be strongly coupled and glitch simultaneously.

\subsection{Backbending and giant glitches}

In \cite{gpw97} the moment of inertia is found to ``backbend'' as
function of angular velocity.  The moment of inertia of some deformed
nuclei \cite{MV,Johnson} may also backbend when the coriolis force
exceeds the pairing force breaking the pairing whereby the nucleus
reverts from partial superfluidity to a rigid rotor. However, in the
limit of large nuclear mass number such backbending would
disappear. Instead pairing may lead to superfluidity in bulk
\cite{Pines}.  A backbending phenomenon  in neutron stars,
that appears to be similar to backbending in nuclei, can occur in
neutron stars although the physics behind is entirely different.
If we soften the EoS significantly at a density near the central density
of the neutron star, a non-rotating neutron star can have most of
its core at high densities where the soft EoS determines the
profile. A rapidly rotating star may instead have lower central
densities only probing the hard part of the EoS. Thus the star may at
a certain angular velocity revert from the dense phase to a more
dilute one and at the same time change its structure and moment of
inertia discontinuosly. Such a drastic change in moment of inertia
at some angular velocity will cause a giant glitch as found in \cite{gpw97}.

 The phenomenon of neutron star backbending is related to the
double maximum mass for a neutron and quark star respectively as shown
in Fig. (\ref{fig:doublepeak}).
The instabilities given in Eq. (\ref{ecrit}) are also 
evident when rotation is included as seen from 
Eq.\ (\ref{R0})
\begin{eqnarray}
  \varepsilon_2 \ge \frac{1}{2} \varepsilon_1 (3-\omega_0^2)
 \quad \Rightarrow \quad {\rm Giant \, glitch\, when\, \omega\simeq\omega_0}. 
 \label{ecrito}
\end{eqnarray}
The Bethe-Johnson EoS discussed in subsection \ref{Maxmass} also has
this discontinuity in the moment of inertia when $\rho_2/\rho_1\ga
3/2$. The neutron star may continue to slow down in its unstable
structure, i.e. ``super-rotate'', before reverting to its stable
configuration with a dense core.  As for the instabilities of Eq.\
(\ref{ecrit}) this condition changes slightly for a more general EoS and
when general relativity is included.  Neutron stars with a mixed phase
do not have a first order phase transition but may soften their EoS
sufficiently that a similar phenomenon occurs \cite{gpw97}.  We
emphasize, however, that the discontinuous jump in moment of inertia
is due to the drastic and sudden softening of the EoS near the central
density of neutron stars.  It is not important whether it is a phase
transition or another phenomenon that causes the softening.

\subsection{Cooling and temperature measurements}\label{subsec:cooling}

The thermal evolution of a neutron star may provide information
about the interiors of the star, and in recent years much effort 
has been devoted to measuring neutron star temperatures, especially
with the Einstein Observatory and ROSAT, see e.g.\ Ref.\ \cite{Oegelman95}.
Neutron stars are born with interior temperatures of the order
20-50 MeV, but cool rapidly via neutrino emission to temperatures 
of the order 1 MeV within minutes.
The only information on neutron star temperatures stems from 
surface temperatures, which typically are of the order
$10^6$ K for about $10^5$ yr, observed in X-ray or UV bands. 
However, the thermal radiation from a neutron star has yet 
to be identified unambiguously. Most observations
are for pulsars, and it is unclear how much of the observed radiation
is due to pulsar phenomena, to a synchotron-emitting nebula or to 
the neutron star itself.
Surface temperatures of neutron stars have been measured in a few cases
or upper limits have been set. 
Table \ref{tab:surfacetemps} collects some of these.
\begin{table}
\begin{center}
\caption{Luminosities, $L$, and spin-down ages, $\tau$, of pulsars}
\begin{tabular}{cccc}
\hline\noalign{\smallskip}
Pulsar & Name & $\log\tau$ [yr] & $\log L$ [erg/s]\\
\noalign{\smallskip}\hline\noalign{\smallskip}
  1706-44 & & 4.25 & $32.8\pm 0.7$ \cite{Becker92a} \\
  1823-13 & & 4.50 & $33.2\pm 0.6$ \cite{Finley93b}\\
  2334+61 & & 4.61 & $33.1\pm 0.4$ \cite{Becker93b} \\
  0531+21 & Crab & 3.09 & $33.9\pm 0.2$ \cite{Becker95a}\\
  1509-58 & SNR MSH 15-52 & 3.19 & $33.6\pm 0.4$  
  \cite{Seward83a} \\
  0540-69 & & 3.22 & $36.2\pm 0.2$ \cite{Finley93a} \\
  1951+32 & SNR CTB 80 & 5.02 & $33.8\pm 0.5$ \cite{SafiHarb95a} \\
  1929+10 & & 6.49 & $28.9\pm 0.5$ \cite{Yancopoulos93,Oegelman95} \\
  0950+08 & & 7.24 & $29.6\pm 1.0$ \cite{Seward88a} \\
  J0437-47 & & 8.88 & $30.6\pm 0.4$ \cite{Becker93c} \\
  0833-45 & Vela & 4.05 & $32.9\pm 0.2$ \cite{Oegelman93a} \\
  0656+14 & & 5.04 & $32.6\pm 0.3$ \cite{Finley92a} \\
  0630+18 & Geminga & 5.51 & $31.8\pm 0.4$ \cite{Halpern93a} \\
  1055-52 & & 5.73 & $33.0\pm 0.6$ \cite{Oegelman93b} \\
\noalign{\smallskip}\hline
\label{tab:surfacetemps}
\end{tabular}
\end{center}
\end{table}

The cooling history of the star, and energy loss mechanisms
from the interior are thus to be determined through various
theoretical models. The generally accepted picture is 
that the long-term cooling of a neutron star consists of two periods:
a neutrino cooling epoch which can last until $10^6$ yr and a photon
cooling period. 
If we now assume that
the main cooling mechanism in the early life of a neutron star is
believed to go through neutrino emissions in the core,
the most powerful energy losses are expected to be given by the so-called
direct Urca mechanism
\begin{equation}
    n\rightarrow p +l +\overline{\nu}_l, \hspace{1cm} p+l \rightarrow
    n+\nu_l ,
    \label{eq:directU}
\end{equation}
as rediscussed recently by several authors \cite{pethick92,pplp92,prakash94}.
The label $l$ refers to the leptons considered here, electrons and muons. 
However, in order to fullfil the 
momentum conservation $k_F^n < k_F^p + k_F^e$ and energy conservation
requirements, the process can only
start at densities $n$ several times
nuclear matter saturation density $n_0 =0.16$ fm$^{-3}$, see
e.g.\  Fig.\ \ref{fig:sec2fig19}, where the proton fraction exceeds
$x_p\ga 0.14$, see e.g.\  \cite{st83,pethick92,pplp92,prakash94}.

Thus, for long time the dominant processes for neutrino emission
have been the so-called modified Urca processes first discussed by
Chiu and Salpeter \cite{cs64}, in which the two reactions
\begin{equation}
    n+n\rightarrow p+n +l +\overline{\nu}_l,
    \hspace{0.5cm} p+n+l \rightarrow
    n+n+\nu_l ,
    \label{eq:ind_neutr}
\end{equation}
occur in equal numbers.
These reactions are just the usual processes of neutron
$\beta$-decay and electron and muon capture 
on protons of Eq.\ (\ref{eq:directU}),
with the addition of an extra bystander neutron. They produce
neutrino-antineutrino pairs, but leave the composition of matter constant
on average. Eq.\ (\ref{eq:ind_neutr}) is referred to as the
neutron branch of the modified Urca process. Another branch is the
proton branch
\begin{equation}
    n+p\rightarrow p+p +l +\overline{\nu}_l, \hspace{0.5cm} p+p+l
    \rightarrow
    n+p+\nu_l ,
    \label{eq:ind_prot}
\end{equation}
pointed out by Itoh and Tsuneto \cite{it72} and recently reanalyzed
by Yakovlev and Levenfish \cite{yl95}. The latter authors showed that
this process is as efficient as Eq.\ (\ref{eq:ind_neutr}).
In addition one also has the possibility of neutrino-pair
bremsstrahlung.
These processes form the basis for what is normally 
called the {\em standard cooling scenario}, i.e.\  no direct
Urca  processes are allowed. 

A {\em fast cooling scenario} would involve the direct
Urca process, similar direct
processes with baryons more massive than the nucleon
participating, such as isobars or hyperons \cite{pplp92,prakash94},
or neutrino emission from more exotic states like pion and kaon
condensates  \cite{BLRT,migdal90,toki} or quark matter
\cite{glendenning91,iwamoto82}. Actually, Prakash {\em et al.}
\cite{pplp92} showed that the hyperon direct Urca processes
gave a considerable contribution to the emissivity,
without invoking exotic states or the large proton fractions needed
in Eq.\ (\ref{eq:directU}).

If we now consider the neutron stars discussed in
subsection \ref{subsec:mass-radius}, we note that a
typical star with mass $\sim 1.4-2.0 M_{\odot}$ has central
densities ranging from $\sim 0.5-1.2$ fm$^{-3}$.
Depending on the value of the bag constant, the mixed phase could start
already at $\sim 0.2-0.3$ fm$^{-3}$. No pure quark phase
was found with the bag-model for stable neutron star
configurations. 
If one also recalls that our $\beta$-stable EoS allows for the direct
Urca process at densities starting from $0.8$  fm$^{-3}$, see 
Fig.\ \ref{fig:sec2fig19}, 
one clearly sees that, depending on the EoS and the adopted
model for dense matter, there is a considerable model dependence.
To give an example, the cooling  
could be strongly influenced by
the mixed phase. This could come about because nuclear 
matter in the droplet
phase has a higher proton concentration than bulk 
neutral nuclear matter and
this could make it easier to attain the threshold 
condition for the nucleon
direct Urca process.
Another possibility is that the presence of the spatial
structure of the droplet phase might allow 
processes to occur which would be
forbidden in a translationally invariant system. 
Also the mere presence
of quark matter can lead to fast cooling \cite{iwamoto82} 
when $\alpha_s\ne0$.
All these mechanisms would lead to faster cooling.

However, in order to compare with observation,
the structure of the star has to be computed in detail.
In particular the possible presence of superfluidity
in the interior has to be
considered. The superfluid would suppress the $\nu,\bar{\nu}$ emissivity and
would allow for 
reheating through friction with the crust.
In  the analysis
of Page as well \cite{page94}, it is hard to discriminate
between fast and slow cooling scenarios, though in both cases
agreement with the observed temperature of Geminga is
obtained if baryon pairing is present in most, if not all
of the core of the star. The recent analyses of Schaab et al.\ 
\cite{schaab97,schaab98} also seem to confirm the importance
of superfluidity in the interior of stars. 
There are also  
indications \cite{umeda,vanriper}
that temperatures of young ($\sim 10^4$ years old) neutron stars lie
below that obtained through the so-called modified  Urca processes.
One has also to note \cite{page94}
that the modified Urca processes are 
weakly dependent on the mass of the star,
i.e. on the central density, while
faster cooling mechanisms like the above direct
Urca processes are in general strongly
dependent on it. Thus,
the detection of two coeval stars, whose temperatures differ by a factor
of the order of 2 or larger
would allow to distinguish between traditional cooling
scenarios, like those discussed by Page \cite{page94}, and more exotic ones.
Such a huge variation in the
temperature of coeval stars could indicate the presence of a threshold
in the cooling mechanism, triggered by the density of the star.
\begin{figure}
       \begin{center}
       \setlength{\unitlength}{1mm}
       \begin{picture}(140,100)
       \put(25,5){\epsfxsize=8cm \epsfbox{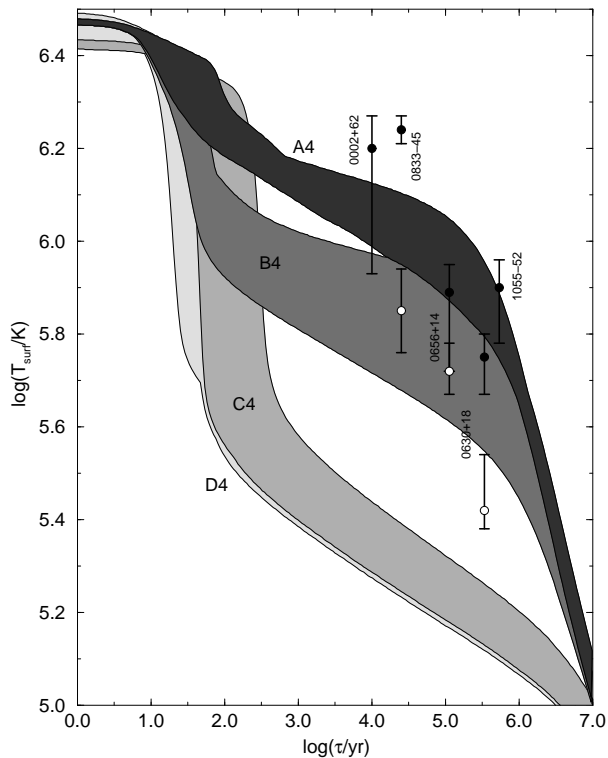}}
       \end{picture}
        \caption{ Thermal evolution of a $1.4M_{\odot}$ neutron star with
                  superfluidity. See text for further details. Taken from
                  Ref.\ \protect\cite{schaab98}.}
       \end{center}
        \label{fig:luminosity}
\end{figure}

In 
Fig.\ \ref{fig:luminosity} we display results from various cooling calculations
by Schaab et al.\ \cite{schaab98}. 
All calculations employ a 
superfluidity scenario which is described in section 
\ref{subsec:superf}. This corresponds to the label 4
in the above figure,  while the letters A,B,C,D represent
different equations of state employed to calculate the mass of the star. 
Label A corresponds to the  non-relativistic equations of state
from  Wiringa et al.\ \cite{WFF}, and is
similar to that of  Akmal et al.\ \cite{apr98} in section \ref{sec:eos}, and 
non-relativistic
equations of state based on two-body interactions only. The models encompassed 
by class A
allow for the standard cooling scenario only at higher densities.
These densities however,
are beyond the central density of a $1.4M_{\odot}$ neutron star. Models B and C 
allow
for faster cooling scenarios and the equations of state are based on 
relativistic
mean field models. Typically, see also the discussion of relativistic effects
and the proton fractions in Fig.\ \ref{fig:sec2fig16}, these models allow for 
the direct
Urca for nucleons at lower densities than the non-relativistic models. Pion 
and kaon condensantion could also lead to faster cooling scenarios. 
Model D allows for direct hyperon cooling as well but does not
include hyperon pairing, which reduces strongly the direct Urca for hyperons,
see Ref.\ \cite{nir98}.

The problem with most cooling calculations is that there is
no consistent calculation of properties entering the 
neutrino emissivities within the framework of say one given
EoS and  many-body approach. Typically, see e.g.\  Refs.\ 
\cite{schaab97,schaab98} and Fig.\ \ref{fig:luminosity}, 
the EoS is taken from one source, while
the pairing gap is taken from another calculation, with even entirely
different NN interactions or many-body approaches.
In addition, the expressions for the emissivities
of e.g.\  the modified Urca processes calculated by 
Friman and Maxwell \cite{fm79} treat in a rather cavalier way the role
of many-body correlations. Considering also the fact that
other severe approximations are made, these expressions, which enter typically
various cooling codes, could introduce errors at the level
of orders of magnitude. 

Thus, our message is that, before one attempts at a cooling calculation,
little can be learned unless the various
neutrino emissivities are reevaluated within the framework of
a given many-body scheme for dense matter.

We conclude this subsection with a demonstration
of the role of superfluidity
for the 
processes of Eqs.\ (\ref{eq:ind_neutr})-(\ref{eq:ind_prot})
at densities corresponding to the outer core of massive neutron stars
or the core of not too massive neutron stars when we have 
a superfluid phase. Here we limit ourselves to
study the role of
superconducting protons 
in the core of the star employing the 
gap for protons in the $^1S_0$ and
effective masses from lowest-order Brueckner-Hartree-Fock 
calculations discussed in section \ref{sec:eos} and subsections
\ref{subsec:nucdeg} and \ref{subsec:superf}.
The proton superconductivity  
reduces the energy losses considerably 
in the above reactions \cite{yl95},
and may have important consequences for the
cooling of young neutron stars.
The expressions for the
processes  in Eqs.\ (\ref{eq:ind_neutr})-(\ref{eq:ind_prot})
were derived by Friman and Maxwell \cite{fm79} and read \cite{yl95}
\begin{equation}
        Q_{n} \approx 8.5\times 10^{21}
        \left(\frac{m_n^*}{m_n}\right)^3
        \left(\frac{m_p^*}{m_p}\right)
        \left(\frac{n_e}{n_0}\right)^{1/3}
        T_9^8\alpha_n\beta_n,
        \label{eq:neutron_branch}
\end{equation}
in units of ergs cm$^{-3}$s$^{-1}$ where
$T_9$ is the temperature in units of $10^9$ K, and according
to Friman and Maxwell $\alpha_n$ describes the momentum transfer
dependence of the squared matrix element in the Born approximation
for the  production rate in the neutron branch. Similarly,
$\beta_n$ includes the non-Born corrections and corrections due
to the nucleon-nucleon interaction not described by one-pion
exchange. Friman and Maxwell \cite{fm79} used $\alpha_n \approx 1.13$
at nuclear matter saturation density
and $\beta_n=0.68$. In the results presented below, we will not
include $\alpha_n$ and $\beta_n$. For the reaction of
Eq.\ (\ref{eq:ind_neutr}) with muons, 
one has to replace $n_e$ with $n_{\mu}$ and add a factor
$(1+X)$ with $X=k_F^{\mu}/k_F^{e}$ \cite{tsuruta75,gs80}.
For the proton branch of Eq.\ (\ref{eq:ind_prot}) with electrons
we have the approximate equation \cite{yl95}
\begin{equation}
        Q_{p}\approx 8.5\times 10^{21}
        \left(\frac{m_p^*}{m_p}\right)^3
        \left(\frac{m_n^*}{m_n}\right)
        \left(\frac{n_e}{n_0}\right)^{1/3}
         T_9^8\alpha_p\beta_pF,
        \label{eq:proton_branch}
\end{equation}
with $F=\left(1-\frac{k_F^e}{4k_F^p}\right)\Theta$,
where $\Theta =1$ if $k_F^n < 3k_F^p+k_F^e$ and zero
elsewhere. Yakovlev and Levenfish \cite{yl95}
put $\alpha_n=\alpha_p$ and $\beta_p=\beta_n$. We will, due to the
uncertainty in the determination of these coefficients, omit them
in our calculations of the reaction rates. With muons, the same changes
as in Eq.\ (\ref{eq:neutron_branch}) are made.

The reaction rates for the
Urca processes are reduced due to the superconducting
protons. Here we adopt the results from Yakovlev and Levenfish
\cite{yl95}, their Eqs.\ (31) and (32) for the neutron
branch of Eq.\ (\ref{eq:neutron_branch}) and Eqs.\ (35) and (37)
for the proton branch of Eq.\ (\ref{eq:proton_branch}).
We single out proton singlet-superconductivity  only,
employing the approximation 
$k_BT_C=\Delta (0)/1.76$,
where $T_C$ is the critical temperature
and $\Delta (0)$ is the pairing gap at zero temperature
discussed in section 2. The critical temperature is then used to
obtain the temperature dependence of the corrections
to the neutrino reaction rates due to superconducting
protons. To achieve that we employ Eq.\ (23) of \cite{yl95}.
\begin{figure}
     \setlength{\unitlength}{1mm}
     \begin{picture}(80,70)
     \put(30,-30){\epsfxsize=10cm \epsfbox{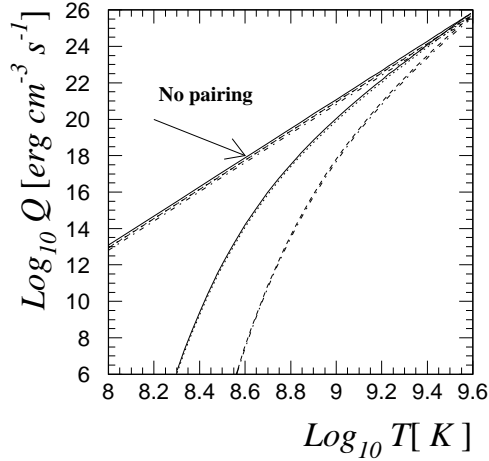}}
     \end{picture}
     \caption{Temperature dependence of neutrino energy loss
              rates in a neutron star core at a total baryonic
              density of $0.3$ fm$^{-3}$ with muons and electrons.  
              Solid line represents Eq.\ (2) with electrons, 
              dotted line is Eq.\ (2) with muons,
              dashed line is Eq.\ (3) with electrons 
              while the dash-dotted line
              is the processes of Eq.\ (3) with muons. 
              The corresponding results
              with no pairing are also shown.}
     \label{fig:modifiedUrca}
\end{figure}
We present results for the neutrino energy
rates at a density $n =0.3$ fm$^{-3}$.
The critical temperature is $T_C=3.993\times 10^{9}$ K, whereas without
muons we have $T_C=4.375\times 10^{9}$.
The implications for the final neutrino rates are shown in
Fig.\ \ref{fig:modifiedUrca}, where we show the results for the full case
with both muons and electrons for the 
processes of Eqs.\ (\ref{eq:ind_neutr}) and 
(\ref{eq:ind_prot}).
In addition, we also display the results when there is no reduction
due to superconducting protons.
At the density considered, $n=0.3$ fm$^{-3}$,
we see that the processes of Eqs.\
(\ref{eq:ind_neutr}) and (\ref{eq:ind_prot}) with muons are comparable
in size to those of Eqs.\
(\ref{eq:ind_neutr}) and (\ref{eq:ind_prot}) with electrons
However,
the proton pairing gap is still sizable at densities up to
$0.4$ fm$^{-3}$, and yields a significant suppression of the 
modified Urca processes discussed here, as seen in Fig.\ 
\ref{fig:modifiedUrca}.
We have omitted any discussion on neutron pairing in the $^3P_2$ state.
For this channel we find the pairing gap to be 
rather small, see again the discussion in subsection \ref{subsec:superf}, 
less
than $0.1$ MeV and close to that obtained in Ref.\ \cite{taka93}.
We expect therefore that the  major reductions of the neutrino rates 
in the core come from superconducting protons in the $^1S_0$ state.
The
contribution from neutrino-pair
bremsstrahlung in nucleon-nucleon collisions in the core
is also, for most temperature
ranges relevant for neutron stars,
smaller  than the contribution from modified Urca processes \cite{yl95}.
Possible candidates are then direct Urca processes due to hyperons
and isobars, as suggested in Ref.\ \cite{nir98,pplp92}, or neutrino production
through exotic states of matter, like kaon or pion condensation
\cite{migdal90,BLRT,toki} or quark matter
\cite{glendenning91,iwamoto82}. 

In conclusion, superfluidity reduces the neutrino emissivities
considerably for the standard cooling model. However, before a firm
conclusion from cooling calculations can be obtained, one needs much
more reliable estimates of various emissivity processes in dense
matter. In addition, such processes should be calculated consistently
within the same many-body approach and EoS model for the interior of a
star.

\subsection{Supernovae}

Neutron stars are born in type II or Ib 
supernovae at a rate of 1-3 per century in our galaxy.
Once the iron core in massive stars exceed the corresponding
Chandrasekhar mass, it collapses adiabatically until it is
stopped by the incompressibility of nuclear
matter, bounces, creates shock waves, stalls by infalling matter, and
presumably explodes by neutrinos blowing the infalling matter off after
a few tens of seconds. 

Measurements of isotope abundances of various elements (r- and
s-processes) give some insight in densities and temperatures during
certain stages of the explosions. Neutrinos were also detected from
SN1987A in the Large Magellan Cloud.

Neutron stars are normally treated as an input to
supernova calculations of the complicated transport processes of baryons,
photons, leptons and neutrinoes going on during the explosion.
Thus supernovae do not provide much information on details on neutron
star structure presently.
Recent supernova models include convection and spherical asymmetries
(see, e.g.\ \cite{Janka}).

The softer the equation of state the denser the matter is compressed
before it bounces and the deeper into the gravitational well the star has
fallen. Also a softer EoS creates a more coherent shock wave that excites the
matter less. The additional gravitational energy available can
be transferred to neutrino generation which is believed to power the
supernova explosion. Besides softening the EoS
a first order phase transition would store latent heat
which also could affect supernovae.

With luck we may observe a supernova nearby 
in the near future which produces a
rapidly rotating pulsar. Light curves and neutrino counts will test
supernova and neutron star models. The rapid spin down may be
exploited to test the structure and possible phase transitions in
the cores of neutron stars \cite{glendenning92,gpw97,rot}.

\subsection{Gamma ray bursters}

In the late 1960's the Vela satellite was launched carrying a gamma
ray detector in order to check the nuclear test ban treaty in space.
It was a huge success because numerous gamma ray bursts (GRB) were
observed \cite{Klebesadel} and (more importantly) they did not come
from Earth. Later (when declassified) the russians confirmed the
GRB. Numerous observations by the Pioneer Venus Orbiter, Compton Gamma
Ray Observatory, Burst And Transient Source Experiment, and later
gamma ray detectors now find bursts every day.  The GRB do not repeat
(except for a few soft gamma ray repeaters).  Their duration varies
from milliseconds to minutes.  The short bursts imply a small source
size $\la c\cdot 1$ ms$\sim 100$km, which points towards neutron stars or
black holes. The gamma rays have energy in the range 30 keV--2 MeV. The
high energy gamma rays are above the threshold for $\gamma\to e^+e^-$
photo production which implies that the radiation probably is
relativisticaly expanding.

The thousands of GRB observed show a high isotropy. Thus GRB cannot be
produced in the very anisotropic discs of our solar system or galaxy
or even an extended galactic halo.  The final kill to such (and many
other) models came from the Bepposax observations on may 8th, 1997,
where a burst could be pin-pointed on the sky within an arcminut,
which subsequently allowed ground based observations of an afterglow
in optical and radio wavelengths.  Fe--II and Mg--II absorption lines
were found at high redshifts of $z=0.835$. Assuming isotropic flux
gives an energy output of $E_\gamma\ga 10^{51}$ ergs within seconds -
much more powerful than supernovae.  In comparison supernova produce
$\sim10^{51}$ erg's mainly in neutrinos whereas a long distance
missile has (only) $\sim10^{31}$ ergs of explosive energy. At the time
of writing, two dozens of GRB with
afterglow in X-ray, one dozen in optical and half a dozen in radio
waves have been discovered.

Merging neutron stars may be responsible for Gamma Ray Bursts
\cite{Janka}.  Binary pulsars are rapidly spiralling inwards and will
eventually merge and create a gigantic explosion and perhaps collapse
to a black hole. From the number of binary pulsars and spiral rates,
one estimates about one merger per million year per galaxy. With $\sim
10^9$ galaxies at cosmological distances $z\la 1$ this gives about the
observed rate of GRB.  However, detailed calculations have problems
with baryon contamination, i.e.\  the baryons ejected absorb photons in
the relativistically expanding photosphere. Accreting black holes have
also been suggested to act as beaming GRB.

Sofar, the physics producing these GRB is not understood.  The time
scales and the enormous power output points towards neutron star or
black hole objects.  We hope to learn more about these objects and
maybe someday they can be used to gain information on neutron stars.

\section{Conclusions} \label{sec:conclusions}

The aim of this work has been to give a survey of recent progresses 
in the construction of the equation of state for dense
neutron star matter and possible implications for 
phase transitions inside neutron stars and the connections to
neutron star observables. 
We will here try to recapitulate several of the arguments presented.

\subsection{Many-body approaches to the equation of state}
In section \ref{sec:eos}
we attempted at a review of the present status of
microscopic many-body approaches to dense neutron star matter.
Several features emerged:
\begin{itemize}
\item Within non-relativistic lowest-order Brueckner theory (LOB),
      all the new phase-shift equivalent nucleon-nucleon
      potentials yield essentially similar equations of state up to
      densities of $3-4 n_0$ for both pure neutron matter
      and $\beta$-stable matter. Other properties like the 
      symmetry energy and proton fractions do also show
      a similar quantitative agreement.
\item The inclusion of more complicated many-body terms at the two-body
      level does not alter this picture and even the recent 
      summation of three-hole line diagrams of Baldo and co-workers
      \cite{baldo97,baldo98}
      results in an EoS which is too close to LOB when a continuous
      choice is used for the single-particle energies in matter.
\item At densities up to nuclear matter saturation density $n_0$, basically
      all many-body approaches discussed here give very similar
      equations of state.     
\item In symmetric nuclear matter the situation is however different, since 
there
      the nuclear tensor force is much more dominant due to presence of the 
$^3S_1$ and
      $^3D_1$ partial waves. This has also consequences for the three-body 
interactions,
      arising from both effective and real three-body force terms. 
      This is 
      expected since the strong isospin $T=0$ channel is present in the 
      three-body $nnp$ and $npp$ clusters. In neutron matter however, such 
clusters
      are in general  small at densities up to $n_0$. Similar behaviors are also
      seen in recent works on the energy of pure neutron drops \cite{ndrops97}
      and larges-scale shell-model calculations of Sn isotopes including 
      effective three-body interactions \cite{eh99}. However, for PNM and 
densities
      greater than $n_0$ 
      differences do however occur at densities 
      when one introduces real three-body forces and/or includes
      relativistic corrections. In a similar way, relativistic BHF calculations,
      yield also significant corrections above saturation density.
\item Based on the microscopic calculation of Akmal et al.\ \cite{apr98},
      a simple parametrization of the EoS was given in the previous
      subsection, where  the energy per particle could be written as
      \[
            {\cal E}={\cal E}_0 u\frac{u-2-\delta}{1+\delta u}
           +S_0 u^\gamma x^2,   
      \]
      with ${\cal E}_0=15.8$MeV,
      $\delta\simeq 0.2$, symmetry energy $S_0=32$MeV and $\gamma=0.6$.
      Causality is insured by $\delta\ge 0.13$.
\end{itemize}
In the case of non-nucleonic degrees of freedom we discussed also 
recent progresses in the construction of hyperon-hyperon interactions
\cite{stoks99a,stoks99b} and applications to neutron star matter
\cite{schulze97}. Results for $\beta$-stable matter with 
these recent hyperon-hyperon interactions are in progress \cite{isaac99}.
In this work however, we chose to focus on the Bag model in order to
deal with non-nucleonic degrees of freedom.

For studies of superfluidity in neutron star matter, one needs still
a careful analysis of polarization effects, especially for
the $^1S_0$ proton contaminant. For superfluid neutrons in the
$^3P_2$ partial wave, one needs nucleon-nucleon interactions which
are fitted up to at least 1 GeV in lab energy before firm conclusions
can be reached. These points leave clearly a large uncertainty for 
the role of superfluidity in cooling studies.  

\subsection{Phase transitions in neutron stars}

We have discussed a number of possible phase transitions in dense
nuclear matter such as pion, kaon and hyperon kondensation, superfluidity
and quark matter. We have specifically treated the nuclear to quark
matter phase transition and the possible mixed phase that can occur
in more than one component systems and replace the standard Maxwell
construction. The structure of the mixed phase is similar to the 
mixed phase in the inner crust of neutron stars of nuclei and the
neutron gas. However, as the mixed quark and nuclear matter phase
can occur already at a few normal nuclear matter densities, it
can soften the EoS in cores of neutron stars significantly and
lower the maximum mass. A number of numerical calculations of
rotating neutron stars with and without phase transitions were given.

The calculated maximum masses were discussed with the observed ones
and leave two natural options:

\begin{itemize}
 \item {\bf Case I}: {\it The large masses of the neutron stars in
QPO 4U 1820-30 ($M=2.3M_\odot$), PSR J1012+5307
($M=2.1\pm0.4 M_\odot$), Vela X-1 ($M=1.9\pm0.1 M_\odot$), and
Cygnus X-2 ($M=1.8\pm0.2 M_\odot$), turn out to be correct and are
complemented by other neutron stars with masses around $\sim 2M_\odot$.}

As a consequence, the EoS of dense nuclear matter is severely
restricted and only the stiffest EoS allowed by causality are allowed
(i.e.\ $\delta\sim0.2$).  Also any significant phase transition can be
excluded at densities below $\la 5n_0$.

 That the radio binary pulsars all have masses around $1.4M_\odot$ is
then probably due to the formation mechanism in supernovae and related
to the Chandrasekhar mass $M_{ch}\simeq 1.4M_\odot$ of white dwarfs.
Neutron stars can subsequently acquire larger masses by accretion.

 \item {\bf Case II}: 
{\it The heavy neutron stars proves erroneous by more detailed observations
and only the canonical $\sim 1.4M_\odot$ masses are found.}

If even in accreting neutron stars does not produce neutron
stars heavier than say $\ga 1.5M_\odot$, 
this indicates that heavier neutron stars simply
are not stable whic in turn implies a soft EoS, either $\delta\ga 0.4$ or
a significant phase transition.

\end{itemize}

Either way, the result is important information on neutrons stars,
the EoS of dense hadronic matter and possible phase transitions.
We impatiently await future observations and determinations of
neutron stars, that will answer these questions. 

\subsection*{Acknowledgement}

Needless to say, we have benefitted immensely from umpteenth 
interactions with many colleagues. Especially, we would like to thank
\O ystein Elgar\o y and Lars Engvik for providing us with several inputs
to various figures and many invaluable comments on the present
work. Moreover, 
we are much indebted to 
Marcello Baldo and Hans-Joseph Schulze for sending us the data for
Figs. 12 and 21 and to Vijay Pandharipande and Geoff Ravenhall for 
providing us with the data for the equation of state  from Ref.\
\cite{apr98} and for numerous discussions on nuclear many-body
theory. 
In addition, we wish to thank, Gordon Baym, Fabio de Blasio, Greg Carter,
Alessandro Drago,
Lex Dieperink, Jens Hjorth, Andy Jackson, Gianluca Lazzari,
Ruprecht Machleidt, Larry McLerran, Ben Mottelson,
Herbert M\"uther, Eivind Osnes, Erlend \O stgaard, 
Chris Pethick, Artur Polls, Angels Ramos, Ubaldo Tambini and 
Isaac Vida\~na
for the many discussions on the nuclear many-body problem
and neutron star physics.


\begin{thebibliography}{200}

%  section 1, introduction
\bibitem{Klis} R.\ Wijnands and M.\ van der Klis, Nature 394, 344 (1998).
              D.\ Chakrabarty and E.H.\ Morgan, Nature 395, 346 (1998).
\bibitem{Duncan} C. Thompson and R.C. Duncan, Astrophys.\ J.\ 408, 194 (1994);
 Astrophys.\ J. 473 (1996).
\bibitem{zss97} W.\ Zhang, T.E.\ Strohmayer and J.H.\ Swank,
	    Astrophys.\ J.\ 482 (1997) L167.
\bibitem{Walter} F.M.\ Walter and L.D.\ Metthews, Nature 389 (1997) 358;
F.M.\ Walter, S.J.\ Wolk, and R.\ Neuh\"auser, Nature 379 (1996) 233.
\bibitem{glendenning92}   N.K.\ Glendenning, Phys.\ Rev.\ D 46 (1992) 1274.
\bibitem{gpw97} N.K.\ Glendenning, S.\ Pei and F.\ Weber, 
              Phys.\ Rev.\ Lett.\ 79 (1997) 1603.
\bibitem{rot} H.\ Heiselberg and M.\ Hjorth-Jensen, Phys.\ Rev.\ Lett.\
              80 (1998) 5485.; A sign error has been corrected in Eq. 
(\ref{eq:Pai2}). 
\bibitem{eeho96} \O.\ Elgar\o y, L.\ Engvik, M\ Hjorth-Jensen and E.\ Osnes,
                 Phys.\ Rev.\ Lett.\ 77 (1996)  1421.
\bibitem{qmref} J.C.\ Collins and M.J.\ Perry, 
              Phys.\ Rev.\ Lett.\ 34 (1975) 1353;
              G.\ Baym and S.A.\ Chin, Phys.\ Lett.\ B 62 (1976) 241;
              G.F.\ Chapline and M.\ Nauenberg,
              Nature 264 (1976) and Phys.\ Rev.\ D 16 (1977) 450;
              M.B.\ Kislinger and P.D.\ Morley, 
              Astrophys.\ J.\ 219 (1978) 1017;
              W.B.\ Fechner and P.C.\ Joss, Nature 274 (1978) 347;
              B.\ Freedman and L.\ McLerran, Phys.\ Rev.\ D 17 (1978) 1109;
              V.\ Baluni, Phys. Rev.\ D 17 (1978) 2092;
              A.\ Rosenhauer, E.F.\ Staubo, L.P.\ Csernai, T.\ \O verg\aa rd 
              and E.\ \O stgaard, Nucl.\ Phys.\ A 540 (1992) 630.

\bibitem{hps93} H.\ Heiselberg, C.J.\ Pethick, and E.F.\ Staubo, 
              Phys.\ Rev.\ Lett.\ 70 (1993) 1355;
              Nucl.\ Phys.\ A 566 (1994) 577c.
\bibitem{kn87} D.B.\  Kaplan and A.E.\ Nelson, Phys.\ Lett.\ 
             B 291 (1986) 57.
\bibitem{ap97} A.\ Akmal and V.R.\ Pandharipande, Phys.\ Rev.\ C 56 (1997) 2261
\bibitem{apr98} A.\ Akmal, V.R.\ Pandharipande and D.G.\ Ravenhall,
Phys.\ Rev.\ C 58 (1998) 1804.
\bibitem{schulze97} H.-J.\ Schulze, M.\ Baldo, U.\ Lombardo,
J.\ Cugnon and A.\ Lejeune, Phys.\ Rev.\ C 57 (1998) 704;
M.\ Baldo, G.F.\ Burgio and H.-J.\ Schulze, Phys.\ Rev.\ C 58 (1998) 3688.
\bibitem{rueber93} A.\ Reuber, K.\ Holinde and J.\ Speth,
Nucl.\ Phys.\ A 585 (1994) 543. 
\bibitem{sm96} J.\ Schaffner and I.\ Mishustin, Phys.\ Rev.\
C 53 (1996) 1416. 
\bibitem{kpe96} R.\ Knorren, M.\ Prakash and P.J.\ Ellis,
Phys.\ Rev.\ C 52 (1995) 3470.
\bibitem{stoks99a} V.G.J.\ Stoks and T.-S.H.\ Lee, preprint nucl-th/9901030.
\bibitem{stoks99b} V.G.J.\ Stoks and Th.A.\ Rijken, preprint nucl-th/9901028.
\bibitem{kutschera} M.\ Kutschera, W.\ Broniowski and A.\ Kotlorz,
Phys.\ Lett.\ B 237 (1990) 159.
\bibitem{pal97} L.B.\ Leinson and A.\ P\'erez, JHEP 9 (1998) 20;
S.\ Chakrabarty, D.\ Bandyopadahay and S.\ Pal, Phys.\ Rev.\ Lett.\
78 (1997) 2898; ibid.\ 79 (1997) 2176.
\bibitem{Gies_Bolton_1986} D.R.\ Gies and C.T.\ Bolton, 
Astrophys.\ J.\ 304 (1986) 371.   
\bibitem{Cowley_etal_1983}
A.P.\ Cowley, D.\ Crampton, J.B.\ Hutchings, 
R.A.\ Remillard and J.P.\ Penfold, Astrophys.\ J.\ 272 (1983) 118.
\bibitem{McClintock_Remillard_1986}
J.E.\ Mc Clintock and R.A.\ Remillard, 
Astrophys.\ J.\ 308 (1986) 110.   
\bibitem{Casares_etal_1992}
J.\ Casares, P.A.\ Charles and T.\ Naylor, 
Nature 355 (1992) 614.   
\bibitem{pr95} C.J.\ Pethick and D.G.\ Ravenhall, 
Ann.\ Rev.\ Nucl.\ Part.\ Sci.\ 45 (1995) 429.
\bibitem{pbpelk97} M.\ Prakash, I.\ Bombaci, M.\ Prakash, P.J.\ Ellis, J.M.\ 
Lattimer and R.\
Knorren, Phys.\ Rep.\ 280 (1997) 1. 
\bibitem{tsuruta98} S.\ Tsuruta, Phys.\ Rep.\ 292 (1998) 1.

%  section 2, phases of dense matter

\bibitem{glendenning91} N.K.\ Glendenning, 
in: Structure of Hadrons and Hadronic
Matter, (World Scientific, Singapore, 1991), 275;
N.K.\ Glendenning, Compact Stars, (Springer, Berlin, 1997).
\bibitem{sw86} B.D.\ Serot and J.D.\ Walecka, 
                  Adv.\ Nucl.\ Phys.\ 16 (1986) 1.
\bibitem{bagmodel} See, e.g.\  J.\ Madsen, preprint astro-ph/9809032
                   and references therein.
\bibitem{cdm} A.\ Drago, U.\ Tambini and M.\ Hjorth-Jensen,
Phys.\ Lett.\ B 380 (1996) 13.
\bibitem{lrp93} C.P.\ Lorenz, D.G.\ Ravenhall and C.J.\ Pethick,
Phys.\ Rev.\ Lett.\  70 (1993) 379.
\bibitem{prl95} C.J.\ Pethick, D.G.\ Ravenhall and C.P.\ Lorenz,
Nucl.\ Phys.\  A 584 (1995) 675.
\bibitem{migdal90} A.B.\ Migdal, E.E.\ Saperstein, M.A. Troitsky and D.N.\ 
Voskresensky, Phys.\ Rep.\ 192 (1990) 179.
\bibitem{nim} V.G.J.\ Stoks, R.A.\ M.\ Klomp, C.P.F.\ Terheggen 
and J.J.\
de Swart, Phys.\ Rev.\ C 49  (1994) 2950.
\bibitem{v18} R.B.\ Wiringa, V.G.J.\ Stoks and R.\ Schiavilla, 
Phys.\ Rev.\ C 51 (1995) 38.
\bibitem{cdbonn} R.\ Machleidt, F.\ Sammarruca and Y.\ Song,
Phys.\ Rev.\ C 53 (1996) R1483.
\bibitem{mac89} R.\ Machleidt, Adv.\ Nucl.\ Phys.\ 19 (1989) 185. 
\bibitem{ehmmp97} L.\ Engvik, M.\ Hjorth-Jensen, R.\ Machleidt,
H.\ M\"uther and A.\ Polls, Nucl.\ Phys.\ A 627 (1997) 85.
\bibitem{pmmh98} A.\ Polls, H.\ M\"uther, R.\ Machleidt
 and M.\ Hjorth-Jensen, Phys.\ Lett.\ B 432 (1998) 1.
\bibitem{hko95} M.\ Hjorth-Jensen, T.T.S.\ Kuo and E.\ Osnes,
Phys.\ Rep.\ 261 (1995) 125.
\bibitem{mahaux} J.P.\ Jeukenne, A.\ Lejeune and C.\ Mahaux,
Phys.\ Rep.\  25 (1976) 83;
C.\ Mahaux, P.F.\ Bortignon, R.A.\ Broglia and C.H.\ Dasso,
 Phys.\ Rep.\ 120 (1985) 1.
\bibitem{nhkg97} A.\ Nogga, D.\ H\"uber, H.\ Kamada and W.\ Gl\"ockle, Phys.\ 
Lett.\ B 409 (1997) 19.
\bibitem{ehobo96} L.\ Engvik, E.\ Osnes, M.\ Hjorth-Jensen, G.\ Bao and
E.\ \O stgaard, Astrophys.\ J.\ 469  (1996) 794.
\bibitem{st83} see e.g.\  S.L.\ Shapiro and S.A.\ Teukolsky,
               Black Holes, White Dwarfs and Neutron Stars,
               (Wiley, New York, 1983).
\bibitem{ms92} C.\ Mahaux and R.\ Sartor, 
               Phys.\ Rep.\ 211 (1992) 53.
\bibitem{reid68} R.V.\ Reid, Ann.\ Phys.\ 50  (1968) 411.
\bibitem{angels} A.\ Ramos, PhD.\ Thesis, 
                 University of Barcelona, 1988, unpublished.
\bibitem{rpd89} A.\ Ramos, A.\ Polls and W.H.\ Dickhoff, 
                Nucl.\ Phys.\ A 503 (1989) 1.
\bibitem{shk86} S.D.\ Yang, J.\ Heyer and T.T.S.\ Kuo, 
                Nucl.\ Phys.\ A 448  (1986) 420.
\bibitem{shk87} W.H.\ Dickhoff and H.\ M\"{u}ther, 
                Rep.\ Prog.\ Phys.\ 55 (1992) 1947; 
                W.H.\ Dickhoff and H.\ M\"{u}ther, Nucl.\
                Phys.\ A 473 (1987) 394.
\bibitem{baldo97}H.Q.\ Song, M.\ Baldo, G.\ Giansiracusa, and
                  U.\ Lombardo, Phys.\ Lett.\ B 411 (1997) 237.
\bibitem{baldo98}H.Q.\ Song, M.\ Baldo, G.\ Giansiracusa, and
                  U.\ Lombardo, Phys.\ Rev.\ Lett.\ 81,
                   (1998) 1584; M.\ Baldo (private communication).
\bibitem{engvik97} L.\ Engvik, M.\ Hjorth-Jensen, E.\ Osnes
                   and T.T.S.\ Kuo, Nucl.\ Phys.\ A 622 (1997) 553. 
\bibitem{day81} B.D.\ Day, Phys.\ Rev.\ C 24 (1981) 1203.
\bibitem{day83} B.D.\ Day, Comments Nucl.\ Part.\ Phys.\ 11 (1983) 115.
\bibitem{jackson83} A.D.\ Jackson, Ann.\ Rev.\ Nucl.\ Part.\ Phys.\ 33
(1983) 105.
\bibitem{parquet} A.D.\ Jackson, A.\ Lande and R.A.\ Smith, 
Phys.\ Rep.\ 86 (1982) 55; A.\ Lande and R.A.\ Smith, Phys.\ Rev.\ 
A 45 (1992) 913 and references therein.
\bibitem{ccm} H.\ K\"{u}mmel, K.H.\ L\"{u}hrmann and J.G.\ Zabolitzky, Phys.\ 
Rep.\ 36 (1977) 1.
\bibitem{fhnc} A.\ Fabrocini, F.\ Arias de Savedra, G.\ C\'o and P.\ Folgarait,
Phys.\ Rev.\ C 57 (1998) 1668 and references therein.
\bibitem{pw} V.R.\ Pandharipande and R.B.\ Wiringa, Rev.\ Mod.\
             Phys.\ 51 (1979) 821.
\bibitem{dw} B.D.\ Day and R.B.\ Wiringa, 
               Phys.\ Rev.\ C 32 (1985) 1057.
\bibitem{forest} J.L.\ Forest, PhD.\ Thesis, University of Illinois
at Urbana-Champaign, 1997, unpublished.
\bibitem{pscppr} B.S.\ Pudliner, V.R.\ Pandharipande, 
                J.\ Carlson, S.C.\ Pieper and R.B.\
                Wiringa, Phys.\ Rev.\ C 56 (1997) 1720.
\bibitem{aryathesis98} A.\ Akmal, PhD.\ Thesis, 
University of Illinois at Urbana-Champaign, 1998, unpublished.
\bibitem{lp3} I.E.\ Lagaris and V.R.\ Pandharipande,
	Nucl.\ Phys.\ A 369 (1981) 470 
        and references therein.
\bibitem{serot92} B.D.\ Serot, Rep.\ Prog.\ Phys.\  55 (1992) 1855.
\bibitem{hs87} C.J.\ Horowitz and B.D.\ Serot, 
              Nucl.\ Phys.\ A 464 (1987) 613   .
\bibitem{brockmann78} R.\ Brockmann, Phys.\ Rev.\  C 18 (1978) 1510.
\bibitem{cs87} L.\ S.\ Celenza and C.\ Shakin,
Relativistic Nuclear Physics,  (World Scientific, Singapore, 1986).
\bibitem{bm90} R.\ Brockmann and R.\ Machleidt, 
               Phys.\ Rev.\ C 42 (1990) 1965.
\bibitem{tm87} B.\ ter Haar and R.\ Malfliet, Phys.\ Rep.\
 149 (1987) 207.
\bibitem{thompsson70} R.H.\ Thompson, 
                      Phys.\ Rev.\ D 1 (1970) 110. 
\bibitem{fredhorst98} F.\ de Jong and H.\ Lenske, Phys.\ Rev.\
                      C 58 (1998) 890; ibid.\ C 57 (1998) 3099. 
\bibitem{iz80} C.\ Itzykson and J.-B.\ Zuber, 
               Quantum Field Theory, 
               (McGraw-Hill, New York, 1980).
\bibitem{FPF95} J.L.\ Forest, V.R.\ Pandharipande and J.L.\ Friar,
	Phys.\ Rev.\ C 52 (1995) 568.
\bibitem{KF74} R.A.\ Krajcik and L.L.\ Foldy,
        Phys.\ Rev.\ D 10 (1974) 1777.
\bibitem{Fri75}   J.L.\ Friar, Phys. Rev. C 12 (1975) 695.
\bibitem{CPS93} J.\ Carlson, V.R.\ Pandharipande and R.\ Schiavilla,
        Phys. Rev. C 47 (1993) 484.
\bibitem{FPC95}J. L. Forest, V. R. Pandharipande, J. Carlson and 
	R. Schiavilla,
        Phys. Rev. C 52, 576 (1995).
\bibitem{brown87} G.E.\ Brown, W.\ Weise, G.\ Baym and 
                  J.\ Speth, 
                  Comments Nucl.\ Part.\ Phys.\  17 (1987) 39.
\bibitem{kalogera} V.\ Kalogera and G.\ Baym, 
        Astrophys.\ J.\ 470 (1996) L61.

\bibitem{WFF} R. B. Wiringa, V. Fiks and A. Fabrocini, Phys.\ Rev.\
              C 38 (1988) 1010.
\bibitem{vynnim} P.\ Maessen, Th.\ Rijken and J.\ de Swart, Phys.\ Rev.\
                 C 40 (1989) 2226.
\bibitem{isaac99} I.\ Vida\~na, A.\ Polls and A.\ Ramos, \O.\ Elgar\o y, 
                  L.\ Engvik and M.\ Hjorth-Jensen, in preparation.
\bibitem{BLRT} G. Brown, C. Lee, M. Rho and V. Thorsson, Nucl.\ Phys.\
               A 572 (1994) 693
\bibitem{Weise} T. Waas, M. Rho and W. Weise, Nucl.\ Phys.\
               A 617 (1997) 449-463. 
\bibitem{PPT} V.R. Pandharipande, C.J. Pethick and V. Thorsson,
              Phys.\ Rev.\ Lett.\ 75 (1995) 4567.
%\bibitem{Speth} R. B\"uttgen, K. Holinde, A. M\"uller-Groeling, J. Speth
%           and P. Wyborny, Nucl.\ Phys.\ A 506 (1990) 586.
\bibitem{kaon} J. Carlson, H. Heiselberg, V.R. Pandharipande and 
C.J. Pethick, to be published; 
H. Heiselberg, in:
Gross Properties of Nuclei and Nuclear Excitation: 
Nuclear Astrophysics, 
(preprint astro-ph/9802062). 
\bibitem{Schaffner} N.\  Glendenning and J.\ Schaffner, Phys.\ Rev.\ Lett.\
                    81 (1998) 4564.
\bibitem{toki} R.\ Tamagaki et al., 
Prog.\ Theor.\ Phys.\ Supplment 112 (1993)
and references therein.
\bibitem{glitch} 
J.A.\ Sauls, in: Timing Neutron Stars, 
eds.\ H. \"{O}gelman and E.P.J.\ van den Heuvel,   
(Dordrecht, Kluwer, 1989) p.\ 457. 
\bibitem{bcll90}
  M.\ Baldo, J.\ Cugnon, A.\ Lejeune and U.\ Lombardo,
  Nucl.\ Phys.\ A 515 (1990) 409.
\bibitem{kkc96} 
  V.A.\ Kodel, V.V.\ Kodel and J.W.\ Clark,
  Nucl. Phys. A 598 (1996) 390.
\bibitem{eh98}
  \O. Elgar\o y and M. Hjorth-Jensen,
  Phys. Rev.  C 57 (1998) 1174. 
\bibitem{sclbl96} 
  H.-J.\ Schulze, J.\ Cugnon, A.\ Lejeune, M.\ Baldo 
  and U.\ Lombardo, 
  Phys.\ Lett.\  B 375 (1996) 1. 
\bibitem{chen86}
  J.M.C.\ Chen, J.W.\ Clark, E.\ Krotschek and R.A.\ Smith,
  Nucl.\ Phys.\ A 451 (1986) 509;\\
  J.M.C.\ Chen, J.W.\ Clark, R.D.\ Dave and V.V.\ Khodel,
  Nucl.\ Phys.\ A 555 (1993) 59.

\bibitem{ains89}
  T.L.\ Ainsworth, J.\ Wambach and D.\ Pines,
  Phys.\ Lett.\ B 222 (1989) 173;
  J.\ Wambach, T.L.\ Ainsworth and D.\ Pines,
  Nucl.\ Phys.\ A 555 (1993) 128.
\bibitem{amu85}
  L.\ Amundsen and E.\ \O stgaard, 
  Nucl.\ Phys.\ A 437 (1985) 487.
\bibitem{bcll92}
  M.\ Baldo, J.\ Cugnon, A.\ Lejeune and U.\ Lombardo,
  Nucl.\ Phys.\ A 536 (1992) 349.
\bibitem{taka93}
  T.\ Takatsuka and R.\ Tamagaki,
  Prog. Theor. Phys. Suppl. 112 (1993) 27.
\bibitem{elga96}
  \O.\ Elgar\o y, L.\ Engvik, M.\ Hjorth-Jensen and E.\ Osnes,
  Nucl.\ Phys.\ A 607 (1996) 425.
\bibitem{khodel97} 
  V.V.\ Khodel, PhD.\ Thesis, Washington University, St. Louis, 
  unpublished (1997); 
  V.A.\ Khodel, V.V.\ Khodel and J.W.\ Clark, 
  Phys.\ Rev.\ Lett.\ 81 (1998) 3828. 
\bibitem{beehs98} M.\ Baldo, \O.\ Elgar\o y, L.\ Engvik, 
                  M.\ Hjorth-Jensen and H.-J.\ Schulze,
                  Phys.\ Rev.\ C 58 (1998) 1921.
\bibitem{am61} P.W.\ Anderson and P.\ Morel, 
               Phys.\ Rev.\ 123 (1961) 1911.
\bibitem{kr90} H.\ Kucharek and P.\ Ring, 
               Z.\ Phys.\ A 339 (1990) 23.
\bibitem{eeho96a} \O.\ Elgar\o y, L.\ Engvik, 
                  M.\ Hjorth-Jensen and E.\ Osnes, 
                  Nucl.\ Phys.\ A 604 (1996) 466.
\bibitem{eeho96c} \O.\ Elgar\o y, L.\ Engvik,
                   E.\ Osnes, F.\ V.\ De Blasio, 
                   M.\ Hjorth-Jensen and G.\ Lazzari,
                   Phys.\ Rev.\ Lett.\ 76 (1996) 1994.
\bibitem{nir98} Ch.\ Schaab, S.\ Balberg and J.\ Schaffner-Bielich,
                Astrophys.\ J.\ 504 (1998) L99.

\bibitem{Jackson} M.A. Halasz, A.D. Jackson, R.E. Shrock, M.A. Stephanov and 
J.J.M. Verbaarschot, preprint hep-ph/9804290.
\bibitem{RW}  M. Alford, K. Rajagopal and F. Wilczek,
Phys. Lett.\ B 422 (1998) 247.
\bibitem{Lattice} Y. Iwasaki, K. Kanaya, S. Kaya, 
S. Sakai and T. Yoshi, Phys.\ Rev.\ D 54 (1996) 7010.

\bibitem{SW98}  T. Schafer and F. Wilczek (hep-ph/9811473) 

\bibitem{kapusta} J.I.\ Kapusta, Finite Temperature Field Theory,
(Cambridge University Press, Cambridge, 1989).
\bibitem{BC76} G.A.\ Baym and S.A.\ Chin, Nucl.\ Phys.\ A 262
         (1976) 527.
\bibitem{CLS86} J.\ Cleymans, R.V.\ Gavai and E.\ Suhonen, 
        Phys.\ Rep.\  130 (1986) 217.
\bibitem{Sat82} H.\ Satz, Phys.\ Lett.\ B 113 (1982) 245.
\bibitem{pirner92} H.J.\ Pirner, Prog.\ Part.\ Nucl.\ Phys.\  29
(1992) 33.
\bibitem{birse90} M.C.\ Birse, Prog.\ Part.\ Nucl.\ Phys.\  25
(1990) 1.
\bibitem{dfb95} A.\ Drago, A.\ Fiolhais and U.\ Tambini,
Nucl.\ Phys.\  A 588 (1995) 801.
\bibitem{barone}V.\ Barone\ and A.\ Drago,\ Nucl.\ Phys.\  A 552
(1993) 479;  A 560 (1993) 1076;
V.\ Barone,\ A.\ Drago and M.\ Fiolhais,\ Phys.\ Lett.\ B 338 (1994) 433.
\bibitem{ff} M.\ Fiolhais, T.\ Neuber and K.\ Goeke,\ Nucl.\ Phys.\
 A 570 (1994) 782.
\bibitem{kurt} K.\ Br\"auer,\ A.\ Drago
and A.\ Faessler, Nucl.\ Phys.\  A 511 (1990) 558.
\bibitem{mitja}W.\ Broniowski,\ M.\ \^Cibej,\ 
M.\ Kutschera\ and M.\ Rosina,\
Phys.\ Rev.\ D  41 (1990) 285.
\bibitem{iwamoto82} N.\ Iwamoto,\ Ann.\ Phys.\  141 (1982) 1.

%\bibitem{Shuryak}  M.\ Stephanov, K.\ Rajagopal and E. Shuryak, 
%Phys.\ Rev.\ Lett.\ 81 (1998) 4816.


%  sections 3 and  4
\bibitem{wg91} F.\ Weber and N.K.\ Glendenning, in:
Hadronic matter and rotating relativistic neutron stars,
(World Scientific, Singapore, 1992).
\bibitem{PR88} C.J.\ Pethick and D.G.\ Ravenhall, Ann.\  Phys.\ 183 
(1988) 131.
\bibitem{Lasagna}  D.G.\ Ravenhall, C.J.\ Pethick and J.R.\ Wilson,
          Phys.\ Rev.\ Lett.\ 50  (1983) 2066;
    R. D. Williams and S. E. Koonin, Nucl.\ Phys.\ A 435 (1985) 844;
    M. Lassant, H. Flocard, P. Bonche, P. H. Heenen and E. Suraud,
    Astron.\ Astrophys.\ 183 (1987) L3.
\bibitem{MITSh83}E.V.\ Shuryak, Phys.\ Rep.\ 61 (1980) 71.

\bibitem{BJ87} M.S.\ Berger and R.L.\ Jaffe,
 Phys.\ Rev.\ C 35 (1987) 213; Phys.\ Rev.\ C 44 (1991) R566.
\bibitem{Latt}K. Kajantie, L. K{\"{a}}rk{\"{a}}inen and K. Rummukainen, Nucl.\
 Phys.\ B 357 (1991) 693;
 S.\ Huang, J.\ Potvion, C.\ Rebbi and S.\ Sanielevici, Phys.\ Rev.\
 D 43 (1991) 2056.
\bibitem{melt} W.L.\ Slattery, G.D.\  Doolen and H.E.\ Dewitt, 
              Phys. Rev.\ A 21 (1980) 2087.
\bibitem{Pe91} J.M.\ Lattimer, C.J.\ Pethick, M.\ Prakash 
and P.\ Haensel, Phys.\ Rev.\ Lett.\ 66 (1991) 2701.
\bibitem{HH} H.\ Heiselberg, in:
      Strangeness and Quark Matter, eds.\ G.\ Vassiliadis,
      A.\ D.\ Panagiotou, S.\ Kumar and J.\ Madsen, (World Scientific,
      Singapore,  1995), p.\ 298. 
% section 5

\bibitem{tc98} S.E.\ Thorsett and D.\ Chakrabarty, 
        preprint astro-ph/9803260 and Astrophys.\ J., in press.
\bibitem{Paradijs} J. van Paradijs, astro-ph/9802177 and 
in: The Many Faces of Neutron Stars, eds.\ R.\ Buccheri, J.\ van Paradijs and 
M.A.\
       Alpar, (Kluwer, Dordrecht, 1999) in press
\bibitem{mlp98} M.C.\ Miller, F.K.\ Lamb and D.\ Psaltis,
	    Astrophys.\ J.\ in press;
D.\ Psaltis, M.\ Mendez, R.\ Wijnands, J.\ Homan, P.G.\ Jonker, 
M.\ van der Klis, 
F.K.\ Lamb, E.\ Kuulkers, J.\ van Paradijs and W.H.G.\ Lewin,
astro-ph/9805084, Astrophys.\ J., in press.
\bibitem{kfc97} P.\ Kaaret, E.C.\ Ford and K.\ Chen,
	    Astrophys.\ J.\ 480 (1997) L27.
\bibitem{Orosz} J.A.\ Orosz and E.\ Kuulkers, 
Mon.\ Not.\ R.\ Astron.\ Soc., in press.
\bibitem{hartle67} J.B.\ Hartle, Astrophys.\ J.\ 150 (1967) 1005.
J.B.\ Hartle and M.W.\ Munn, Astrophys.\ J.\ 198 (1975) 467.
\bibitem{ov39} J.R.\ Oppenheimer and G.M.\ Volkoff, 
               Phys.\ Rev.\ 55 (1939) 374.
\bibitem{46} F.~Weber, N.K.~Glendenning and M.K.~Weigel, Astrophys.\ J.\
373 (1991) 579; in : Rotating Neutron Stars and the Equation of State of
Dense Matter,
eds.\ W.Y.~Pauchy Hwang, Shik-Chang~Lee,
Chin-Er~Lee and D.J.~Ernst, (Elsevier, Amsterdam, 1991) p.\ 309.
\bibitem{47} F.~Weber and N.K.~Glendenning, Z.~Phys.\ A 399 (1991) 211.
\bibitem{48} E.M.~Butterworth and J.R.~Ipser, Astrophys.\ J.\ 
204 (1976) 200.
\bibitem{49} J.L.~Friedmann, J.R.~Ipser and L.~Parker, 
Astrophys.\ J.\ 304 (1986) 115; Phys.\ Rev.\ Lett.\ 62 (1989) 3015.
%\bibitem{FJ} W.B.\ Fechner and P.C.\ Joss, Nature 274 (1978) 347.
\bibitem{Kettner}  N.K.\  Glendenning and C. Kettner, astro-ph/9807155.
\bibitem{BaymMM} G. Baym, Neutron Stars and the Properties of Matter
at High Density, Nordita lecture notes, 1977, p. 107.
\bibitem{BJ} H.A.\ Bethe and M.B.\ Johnson, 
             Nucl.\ Phys.\ A 230 (1974) 1.
\bibitem{Tassoul} J.\ Tassoul, 
Theory of Rotating Stars, (Princeton University Press, 1978).
\bibitem{MV} B.R.\ Mottelson and J.G.\ Valatin, Phys.\ Rev.\ Lett.\ 5
(1960) 511.
\bibitem{Johnson} A.\ Johnson, H.\ Ryde and S.A.\ Hjorth, 
Nucl.\ Phys.\ A 179 (1972) 753.
\bibitem{pion} G.\ Baym and C.J.\ Pethick, 
               Ann.\ Rev.\ Nucl.\ Sci.\ 25
               (1975) 27; Ann.\ Rev.\ Astron.\ 
               Astrophys.\ 17 (1979) 415.
\bibitem{Pines} D.\ Pines, in: Neutrons stars: 
        theory and observation of the Neutron Stars, 
        eds.\ J.~Alpar and D.\ Pines,
        (Kluwer, Dordrecht, 1991) p.\ 57, and references therein;
        P.W.\ Anderson and N.\ Itoh, Nature 256 (1975) 25.  
\bibitem{Oegelman95} H.\ {\"O}gelman, 
         in: The Lives of the Neutron Stars, 
        eds.\  M.~Alpar, {\"U}.~Kiziloglu and J.~van Paradijs,
        (Kluwer, Dordrecht, 1995) p.\ 101.
\bibitem{Becker92a}
W.\ Becker,  P.\ Predehl, J.\ Tr{\"u}mper and H.\ {\"O}gelman, 
            IAU Circular, No.\ 5554 (1992).
\bibitem{Finley93b}
J.P.\ Finley and H.\ {\"O}gelman, IAU Circ.\ No. 5787 (1993).
\bibitem{Becker93b}
W.\ Becker, IAU Circ.\ No. 5805 (1993).
\bibitem{Becker95a} W.\ Becker and B.\ Aschenbach, 
                    in: The Lives of the Neutron Stars, 
                    eds.\ M.~Alpar, {\"U}.~Kiziloglu and 
                    J.~van Paradijs,
                    (Kluwer, Dordrecht, 1995) p.\ 47.
\bibitem{Seward83a} F.\ Seward, F.\ Harnden, P.\ Murdin and
D.\ Clark, Astrophys.\ J.\ 267 (1983) 698.
\bibitem{Finley93a} J.P.\ Finley,  H.\ \"Ogelman, G.\ Hasinger and 
                    J.\ Tr\"umper, Astrophys.\ J,\. 410 (1993) 323.
\bibitem{SafiHarb95a} S.\ Safi-Harb and H.\ {\"O}gelman,  
         in: The Lives of the Neutron Stars, 
        eds.\  M.~Alpar, {\"U}.~Kiziloglu and J.~van Paradijs,
        (Kluwer, Dordrecht, 1995) p.\ 53. 
\bibitem{Yancopoulos93}
S.\ Yancopoulos, T.\ Hamilton and D.\ Helfland, Bull.\ American
  Astron.\  Soc.\ 25 (1993) 912.
\bibitem{Seward88a}
F.\ Seward and Z.-R.\ Wang, Astrophys.\ J.\ 332 (1988) 199.
\bibitem{Becker93c}
W.\ Becker and J.\ Tr{\"u}mper, Nature 365 (1993) 528.
\bibitem{Oegelman93a} H.\ {\"O}gelman, J.P.\ Finley 
                      and H.\ Zimmermann, Nature 361 (1993) 136.
\bibitem{Finley92a} J.P.\ Finley,  H.\ {\"O}gelman 
                     and {\"U}.\ Kiziloglu, 
                     Astrophys.\ J.\ 394 (1992) L21.
\bibitem{Halpern93a} J.\ Halpern and M.\ Ruderman, 
                     Astrophys.\ J.\  415 (1993) 286.
\bibitem{Oegelman93b} H.\ {\"O}gelman and J.P.\ Finley, 
                      Astrophys.\ J.\ 413 (1993) L31.
\bibitem{pethick92} C.J.\ Pethick, Rev.\ Mod.\ Phys.\ 64 (1992) 1133.
\bibitem{pplp92} M.\ Prakash, M.\ Prakash, J.M.\ Lattimer and C.J.\
Pethick, Astrophys.\ J.\  390 (1992) L77.
\bibitem{prakash94} M.\ Prakash, Phys.\ Rep.\ 242, 191 (1994).
\bibitem{cs64} H.-Y.\ Chiu and E.E.\ Salpeter, Phys.\ Rev.\ Lett.\
12 (1964) 413.
\bibitem{it72} I.\ Itoh and T.\ Tsuneto, Prog.\ Theor.\ Phys.\ 48 (1972) 149.
\bibitem{yl95} D.G.\ Yakovlev and K.P.\ Levenfish,
Astron.\ Astrophys.\ 297 (1995) 717.
\bibitem{page94} D.\ Page, Astrophys.\ J.\ 428 (1994) 250.
\bibitem{schaab97} Ch.\ Schaab, D.\ Voskresensky, A.D.\ Sedrakian,
                   F.\ Weber and M.K.\ Weigel, Astron.\ Astrophys.\
                   321 (1997) 591; Ch.\ Schaab, 
                   F.\ Weber, M.K.\ Weigel and N.K.\ Glendenning, 
                   Nucl.\ Phys.\ A 605 (1996) 531.
\bibitem{schaab98} Ch.\ Schaab, F.\ Weber and M.K.\ Weigel, 
                   Astron.\ Astrophys.\ 335 (1998) 596.
\bibitem{umeda}H.\ Umeda, N.\ Shibazaki, K.\ Nomoto and S.\ Tsuruta,
Astrophys.\ J.\  408 (1993) 186.
\bibitem{vanriper}K.\ VanRiper, R.\ Epstein and B.\ Link,
Astrophys.\ J.\ 448 (1995) 294.
\bibitem{fm79} B.L.\ Friman and O.V.\ Maxwell, 
               Astrophys.\ J.\ 232 (1979) 541.
\bibitem{tsuruta75} S.\ Tsuruta, Phys.\ Rep.\ 56 (1975) 237.
\bibitem{gs80} G.\ Glen and P.\ Sutherland, 
               Astrophys.\ J.\ 239 (1980) 671.
\bibitem{Janka}  E. M\"uller and H.-Th. Janka, Astron.\ Astrophys.\
317 (1997) 140;
M. Ruffert, H.-Th. Janka, astro-ph/9809280,
submitted to Astron.\ Astrophys.

% refs to section 6
\bibitem{Klebesadel} A.W. Klebesadel, I.B. Strong and R.A. Olsson, 
 Astrophys.\ J.\  182 1973 L85.

\bibitem{ndrops97} B.S.\ Pudliner, A.\ Smerzi, J.\ Carlson,
                   V.R.\ Pandharipande, S.C.\ Pieper  and
                   D.G.\ Ravenhall, Phys.\ Rev.\ Lett.\
                   76  (1996) 2416.
\bibitem{eh99} T.\ Engeland and M.\ Hjorth-Jensen, in preparation.

\end{thebibliography}
\end{document}